\documentclass{ws-ijmpa}
\usepackage[super,compress]{cite}
\usepackage{amsmath,amssymb,array,calc,epsfig}
\usepackage[
      colorlinks=true,
      linkcolor=red,  
      urlcolor=blue,    
      filecolor=red,     
      linkcolor=blue,
      citecolor = red,
      pdfstartview=FitV,
      pdftitle={},
        pdfauthor={Paolo Benincasa},
        pdfsubject={},
        pdfkeywords={},
        pdfpagemode=None,
        bookmarksopen=true    
      ]{hyperref}
\usepackage{mathrsfs}
\usepackage{graphicx}
\usepackage{ccaption}
\usepackage{mathpazo}
\usepackage{graphics}
\usepackage{mathtools}
\usepackage{stmaryrd}
\usepackage{pifont, bm}
\usepackage{colortbl}
\usepackage{multirow,bigdelim}
\usepackage{arydshln}
\usepackage{tikz, pict2e}
\usepackage{centernot}
\usepackage{lmodern}
\usepackage{tkz-euclide}
\usepackage{wrapfig}
\usepackage{float}
\usepackage{placeins}
\usepackage{enumitem}
\usepackage{xargs}
\usepackage{ifthen}
\usepackage{mwe}
\usepackage{framed}
\usepackage{amscd}
\usepackage{tocloft}

\usetikzlibrary{calc,backgrounds,intersections}
\usetikzlibrary{arrows,shapes,positioning,shadows,trees,shadows.blur,shadings}
\usetikzlibrary{mindmap}
\usetikzlibrary{decorations.pathreplacing, decorations.fractals, calligraphy}
\usetikzlibrary{decorations.pathmorphing}
\usetikzlibrary{decorations.markings}
\usetikzlibrary{decorations.text}
\usetikzlibrary{matrix}

\tikzstyle arrowstyle=[scale=1]
\tikzstyle directed=[postaction={decorate,decoration={markings,
    mark=at position .5 with {\arrow[arrowstyle]{stealth}}}}]
\tikzstyle reverse directed=[postaction={decorate,decoration={markings,
    mark=at position .5 with {\arrowreversed[arrowstyle]{stealth};}}}]
    
\definecolor{markcolor}{rgb}{.25,0,1}
\definecolor{darkpastelgreen}{rgb}{0.01, 0.75, 0.24}

\tikzstyle{vecArrow} = [thick, decoration={markings,mark=at position
   1 with {\arrow[semithick]{open triangle 60}}},
   double distance=1.4pt, shorten >= 5.5pt,
   preaction = {decorate},
   postaction = {draw,line width=1.4pt, white,shorten >= 4.5pt}]
\tikzstyle{innerWhite} = [semithick, white,line width=1.4pt, shorten >= 4.5pt]

\pgfdeclaredecoration{irregular fractal line}{init}
{
  \state{init}[width=\pgfdecoratedinputsegmentremainingdistance]
  {
    \pgfpathlineto{\pgfpoint{random*\pgfdecoratedinputsegmentremainingdistance}{(random*\pgfdecorationsegmentamplitude-0.02)*\pgfdecoratedinputsegmentremainingdistance}}
    \pgfpathlineto{\pgfpoint{\pgfdecoratedinputsegmentremainingdistance}{0pt}}
  }
}

\tikzset{
   paper/.style={draw=black!10, blur shadow, every shadow/.style={opacity=1, black}, shading=bilinear interpolation,
                 lower left=black!10, upper left=black!5, upper right=white, lower right=black!5, fill=none},
   irregular cloudy border/.style={decoration={irregular fractal line, amplitude=0.2},
           decorate,
     },
   irregular spiky border/.style={decoration={irregular fractal line, amplitude=-0.2},
           decorate,
     },
   ragged border/.style={ decoration={random steps, segment length=7mm, amplitude=2mm},
           decorate,
   }
}
    
\tikzset{
 path image/.style={
        	 path picture={
		  \node[opacity=.75] at (path picture bounding box.center) {
		    \includegraphics[width=\textwidth height=\textheight, keepaspectratio]{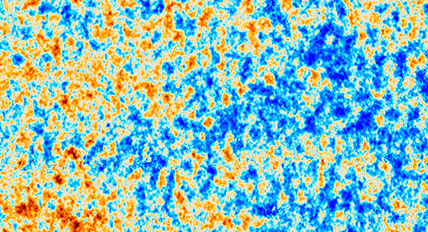}};}},
 path img/.style={
        	 path picture={
		  \node[opacity=.75] at (path picture bounding box.center) {
		    \includegraphics[width=\textwidth height=\textheight, keepaspectratio]{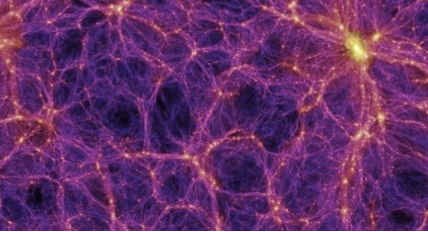}};}}
}

\DeclareMathOperator*{\Res}{Res}

\DeclareRobustCommand{\cross}[1]{%
\begin{tikzpicture}[cross/.style={cross out, draw, minimum size=.6em, inner sep=0}]
   \node[very thick, cross=4pt, rotate=0, color=#1, scale=.625] at (0,0) {};
\end{tikzpicture}%
}

\DeclareRobustCommand{\circle}[1]{%
\begin{tikzpicture}[ball/.style = {circle, draw, align=center, anchor=north, inner sep=0}]
   \node[ball, text width=.18cm, thick, color=#1] at (0,0) {};
\end{tikzpicture}%
}

\DeclareRobustCommand\bluecross{\cross{blue}}
\DeclareRobustCommand\redcross{\cross{red}}
\DeclareRobustCommand\purplecross{\cross{red!50!black}}
\DeclareRobustCommand\greencross{\cross{green!70!black}}
\DeclareRobustCommand\purplecircle{\circle{red!50!black}}
\DeclareRobustCommand\bluecircle{\circle{blue}}
\DeclareRobustCommand\redcircle{\circle{red}}

\def\dsqcup{\sqcup\mathchoice{\mkern-7mu}{\mkern-7mu}{\mkern-3.2mu}{\mkern-3.8mu}\sqcup}

\begin{document}

\renewcommand{\cftsecleader}{\cftdotfill{\cftdotsep}}

\markboth{Paolo Benincasa}{Amplitudes Meet Cosmology: A (Scalar) Primer}

%
\catchline{}{}{}{}{}
%

\title{Amplitudes meet Cosmology: A (Scalar) Primer}

\author{Paolo Benincasa}

\address{Max-Planck-Institut f{\"u}r Physisk, Werner-Heisenberg-Institut,\\
80805 M{\"u}nchen, Germany\\
pablowellinhouse@anche.no}

\maketitle


\begin{abstract}
We review the most recent progress in our understanding of quantum mechanical observables in cosmology in the perturbative regime. It relies on an approach that considers them directly as functions of the data at the space-like boundary at future infinity prescinding from the explicit time evolution. It takes inspiration from the on-shell formulation of perturbative scattering amplitudes developed in the past 20 years: starting with the requirement of consistency with some fundamental principles such as causality, unitarity and locality, it provides different ways of phrasing and extracting predictions. In this review, we aim to provide a pedagogical treatment of the most recent insights about the analytic structure of the perturbative quantum mechanical observables in cosmology, its relation to fundamental principles as well as physical processes, and how such observables and their features emerge from novel well-defined mathematical objects with their own first principle definition. The review is divided in three parts: Part 0 discusses the definition of quantum mechanical observables in cosmology and some general principles; Part I reviews the boundary approach to the analysis and computation of the perturbative wavefunction of the universe; Part II provides an introduction to the combinatorial-geometrical description of cosmological processes in terms of cosmological polytopes. 
\\
\vspace{.25cm}

\hspace{9.125cm} MPP-2022-9

\end{abstract}

\ccode{PACS numbers:}

\newpage

\tableofcontents

\newpage

\section{Introduction}	

Our understanding of the fundamental forces of nature relies on the languages of Quantum Field Theory (QFT) and General Relativity (GR). On one side QFT has provided a very successful description of physical phenomena at {\it accessible high} energies via the Standard Model of particle physics ~\cite{Weinberg:1995mt, Schwartz:2014sze}~ $\hspace{-.125cm}$, with the last (but definitely not least) success of the discovery in 2012 ~\cite{ATLAS:2012yve, CMS:2012qbp} of the Higgs boson predicted in the '60s ~\cite{Anderson:1963pc, Englert:1964et, Klein:1964ix, Gilbert:1964iy, Higgs:1964ia, Higgs:1964pj, Guralnik:1964eu, Weinberg:1967tq}~ $\hspace{-.175cm}$. On the other side, GR has allowed for a detailed account of gravitational phenomena at {\it sufficiently low} energies ~\cite{Weinberg:1972kfs, Wald:1984rg} and has passed numerous tests, the last being the {\it direct} detection of gravitational waves in 2015 ~\cite{LIGOScientific:2016lio, LIGOScientific:2016sjg} predicted one century earlier ~\cite{Einstein:1916cc}~.

One amazing fact about QFT and GR is that they are two different instances of the fundamental principles of quantum mechanics and special relativity simultaneously at play. In quantum mechanics the physical observables are transition amplitudes between certain quantum states, and encode the probabilities that a system undergoes a transition between them. One basic requirement is that the system has to evolve in time in such a way that such probabilities are individually positive and they are conserved, {\it i.e.} they have to sum up to one. This is the statement of {\it unitarity}. Special relativity instead comes with a notion of space-time with an inherent causal structure, with two events that can be causally related if and only if they are not space-like separated, and force the interactions to be local. All these features are encoded by the principles of {\it causality}, {\it locality} and {\it Lorentz invariance}. QFT turns out to reconcile all these principles together. Furthermore, their simultaneous validity {\it implies} the equivalence principle ~\cite{Weinberg:1964ew, Benincasa:2007xk}~, which is the heart of GR, with the Riemannian geometry that characterises the description of GR being a very convenient way to realise it. The equivalence principle is not the only condition that {\it emerges} from unitarity, causality, locality and Lorentz invariance: charge conservation ~\cite{Weinberg:1964ew}~, all the possible three-particle couplings ~\cite{Benincasa:2007xk, McGady:2013sga, Arkani-Hamed:2017jhn}~, the Yang and Weinberg-Witten theorems ~\cite{McGady:2013sga, Arkani-Hamed:2017jhn}~, the consistency of the interactions among a finite number of particles with spin less or equal to $2$ ~\cite{Benincasa:2007xk, Benincasa:2011pg, McGady:2013sga}~, the inconsistency of the interactions of a finite number of particles with spin higher than $2$ ~\cite{Weinberg:1964ew, Benincasa:2007xk, Benincasa:2011pg, McGady:2013sga}~, the graviton uniqueness theorem ~\cite{Benincasa:2007xk}~, are all instances of the simultaneous validity of these principles.

The strong constraints that these principles impose on the physics seem to make the latter fundamental. However there are several catches with this idea. First, it has been long known that GR is $2$-loop non-renormalisable ~\cite{Goroff:1985sz, Goroff:1985th}~. The renormalisability can be phrased as the following statement.  Let us consider a theory with divergences at high energies. In principle, they would seem to make it unsuitable for describing physical phenomena at such energies. However, it can still have high predictive power and describe the high energy phenomena if enlarging the set of input parameters were to eliminate such divergences: the number of predictions it can make is reduced by a small and finite amount, but it is still predictive at both low and high energies. In this case the theory is said to be {\it renormalisable}. If instead in order to make sense at all scales, the theory loses all its predictive power, than it is said to be non-renormalisable. Secondly, as we approach energies of the order of the Planck mass $M_{\mbox{\tiny Pl}}\,\sim\,10^{19}$ GeV, it is impossible to define local observables and to perform arbitrarily precise measurements (see Section \ref{sec:QMC}). Finally, our universe is expanding at an accelerated rate ~\cite{Perlmutter:1998np, Riess:1998cb} and also underwent a phase of accelerated expansion in its early stages -- the inflationary period (see ~\cite{Baumann:2009ds} and references therein). The accelerated expansion breaks Lorentz invariance and makes any type of quantum mechanical observable approximate as it gives an observer access to a small and finite amount of data.

All these issues point towards the idea that the way that our theories are current formulated does not allow to extrapolate our knowledge of physical processes at accessible high energies to arbitrarily high ones, and at least some of our basic principles need to be either modified or even replaced. Said differently, at least some of the principles we have been taking as fundamental should emerge from more fundamental ones. This also implies that our successful descriptions in terms of QFT and GR are effective theories, {\it i.e.} low energy approximations of a theory we do not know yet. Such a theory could need a very different language than the one we are accustomed to, which is based on the notions of fields and Riemannian geometry.

QFT and GR have the virtue of appearing as the most suitable language to implement their fundamental principles. Just as a matter of an example, if we consider a particle scattering in perturbation theory, its description in terms of Feynman diagrams makes Lorentz invariance, locality and unitarity manifest, as they appear as a function of Lorentz invariant combination of the momenta, they are analytic everywhere except at some points where they can have at most poles and branch cuts associated to propagators, and they make the factorisation theorems coming from unitarity manifest. However, if all these principle are really approximate, it would be reasonable to look for a description of the accessible high energy physics we already know in a language that does not make these principles manifest. Then, the resulting ideas can be tested in cosmology, whose quantum mechanical observables represent our window on the physics at energies way higher than the ones at play in accelerators on earth as the Hubble parameter can be as large as $10^{14}$ GeV, {\it i.e.} $10-11$ orders of magnitude higher than the energies reached at the LHC.

\begin{figure}[t]
	\centering
 	\includegraphics[width=.625\textwidth, height=.575\textheight, keepaspectratio]{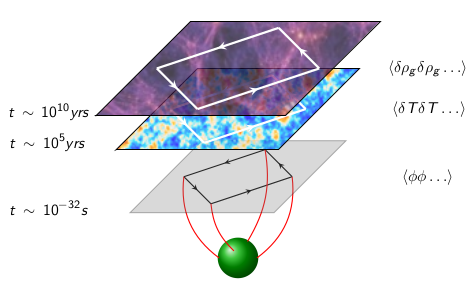}
    	\caption{Cartoon of the relation between correlations in the large scale structures or in the CMB and the quantum fluctuations generating them. Such correlations can be traced back to correlations of quantum fields at the end of inflation.}
	\label{fig:CC}
\end{figure}

A general assumption in cosmology is that all the patterns we can observe in the cosmic microwave background (CMB) and in the distribution of large scale structures (LSS) originated by quantum fluctuations in the infancy of the universe ~\cite{Starobinsky:1985ibc, Salopek:1990jq, Sasaki:1995aw, Maldacena:2002vr, Weinberg:2005vy}~, during the period of inflationary expansion that came to a sudden end. Thus, any correlation that can be measured at later times, such as for temperature fluctuations in the CMB or galaxies distributions, can be traced back to correlation functions of quantum fields computed at the end of inflation. Hence, understanding the physics at higher energies than the ones that can be reached with experiments on earth, goes through the achievement of a deeper understanding of the mathematical structure of these cosmological quantum correlations and the principles which determine (or at least constrain) it, consequently shedding light on the physics of inflation. Despite the fact that consistency conditions for inflationary correlation functions have been studied ~\cite{Maldacena:2002vr, Seery:2008ax, Leblond:2010yq, Creminelli:2011rh, Creminelli:2012ed, Senatore:2012wy, Assassi:2012zq, Goldberger:2013rsa, Hinterbichler:2013dpa, Pimentel:2013gza, Creminelli:2013mca, Bordin:2016ruc, Bordin:2017ozj, Finelli:2017fml, Pajer:2019jhb}~, very little is known about their general structure and the general rules that they follow started just recently, and it will be the main subject of this review.

Specifically, this review reports on a very recent program which aims to deepen our understanding of quantum mechanical observables in cosmology and the physics they encode by importing methods and philosophy developed with these formulations in the context of perturbative scattering amplitudes. Scattering amplitudes have been the central objects for a reformulation of perturbative quantum field theory which can prescind from all the redundancies associated to the notion of fields. While actions and the Feynman graphs we can read off from them, are not defined uniquely, rather up to field redefinitions and gauge transformations, scattering amplitudes do not suffer of such ambiguities. Such redundancies are needed in the QFT language as fields introduce unphysical degrees of freedom. This fact translates in an unnecessary complexity in the computation of scattering processes, hiding a lot of structures and simplicity. Prototypical examples of this phenomenon are given by: the maximally-helicity-violating amplitudes in Yang-Mills theory, whose expression from Feynman graphs seems extremely complicated but it reduces to the one-line Parke-Taylor formula ~\cite{Parke:1986gb}~; the dual conformal invariance and the, thus, overall Yangian symmetry of $\mathcal{N}=4$ Supersymmetric Yang-Mills theory (SYM) ~\cite{Drummond:2006rz}~. Such features becomes nevertheless manifest if we abandon the usual field language and we describe our scattering processes in terms of {\it on-shell data} only (see ~\cite{Elvang:2013cua, Benincasa:2013faa}~ and references therein). In fact, the simplicity of the Parke-Taylor formula becomes manifest via the BCFW recursion relation for Yang-Mills tree level amplitudes ~\cite{Britto:2005fq}~, which relates higher-point amplitudes to lower points.  Importantly, it has been shown that {\it any} tree-level scattering amplitude satisfy either a BCFW recursion relation or one of its generalisations ~\cite{Benincasa:2007qj, Cheung:2008dn, Cohen:2010mi, Benincasa:2011kn, Kampf:2013vha, Benincasa:2013faa}~, even in effective field theories ~\cite{Cheung:2015ota}~. For supersymmetric Yang-Mills theory such recursions are valid also for the all-loop {\it integrands} ~\cite{Arkani-Hamed:2010kv, Benincasa:2015zna}~. The validity of a recursion relation implies that a scattering amplitude can be fully reconstructed knowing the amplitude for lower-particle amplitudes: iterating the recursion, then an $n$-point amplitude can be expressed as a sum of products of the smallest possible process, the three-particle amplitudes. Consequently, if some principle provides us with the basic building blocks, the recursion instructs us how to glue them to build amplitudes with a higher number of particles. Such a principle is the isometry group of Minkowski space, the Poincar{\'e} group, and allows to classify all the allowed three-particle couplings ~\cite{Benincasa:2007xk, Arkani-Hamed:2017jhn}~. This provides a full-fledge on-shell treatment. Interestingly, the BCFW recursion relation makes both unitarity and locality hidden: just a subset of the factorisation channels are manifest (the others appear as certain soft limits ~\cite{Schuster:2008nh, Benincasa:2011pg}~) while spurious poles appear. Demanding the consistency of all factorisations led to rediscover a great deal of theorems and properties, including the already mentioned equivalence principle ~\cite{Benincasa:2007xk, Benincasa:2011pg, McGady:2013sga}~. 

In the case of $\mathcal{N}=4$ SYM theory such an on-shell formulation was found to be encoded into the geometry and combinatorics of the positive Grassmannian ~\cite{ArkaniHamed:2012nw}~, in which the BCFW recursion appears as triangulations of certain polytopes ~\cite{Hodges:2009hk, Arkani-Hamed:2010wgm}~, while the Yangian as diffeomorphisms of the Grassmannian preserving the positive structure\footnote{A similar formulation is also available for less supersymmetric Yang-Mills theories ~\cite{Benincasa:2015zna, Benincasa:2016awv}~.}. The description of scattering amplitudes as volumes of certain polytopes in momentum twistor space ~\cite{Hodges:2009hk}~ as well as the introduction of the positive Grassmannian ~\cite{Postnikov:2006kva}~ to describe the planar all-loop integrand for $\mathcal{N}=4$ SYM, suggested that geometrical and combinatorial ideas could play a crucial role for a novel understanding of scattering amplitudes. This perspective led to the formulation of the {\it amplituhedron} ~\cite{Arkani-Hamed:2013jha}~ for the very same planar $\mathcal{N}=4$ SYM amplitudes, ABHY associahedron ~\cite{Arkani-Hamed:2017mur, Frost:2018djd}~ for the bi-adjoint $\phi^3$ theory, the Stokes polytopes for planar $\phi^4$ theory ~\cite{Banerjee:2018tun, Salvatori:2019phs}~, the accordiohedra for planar $\phi^k$ theories ~\cite{Raman:2019utu, Aneesh:2019ddi}~. The novelty in all these {\it positive geometries} is that they are mathematical structures with their own first principle definition which makes no reference to any physical property -- see ~\cite{Ferro:2020ygk, Herrmann:2022nkh} and references therein. Nevertheless they turn out to encode the physics of flat-space scattering and hence all the fundamental physical properties, such as unitarity, causality and locality, can be thought of as emergent from these mathematical principles. In our search for a reformulation of quantum field theory which make no reference to all those properties which are likely to get either modified or replaced at very high energies, the mathematics of positive geometries seem to be a promising direction to pursue.

The question about the emergence of physical properties is particular pressing in the context of cosmology. As emphasised earlier, quantum mechanical observables in cosmology are naturally defined at the space-like future boundary of our expanding space-time. Hence, the time evolution is integrated out, but nevertheless they must carry its imprint. So we can ask the following question: 
\begin{quote} 
\centering 
	{\it what are the conditions that a quantum mechanical observable ought to satisfy in order to come from a causal and unitary evolution in a cosmological space-time?}
\end{quote}
In first instance we can look for a reformulation of perturbation theory in expanding backgrounds in terms of boundary data only and that prescinds of an explicit time evolution. Said differently, we can look for a cosmological analogous of the on-shell description for flat-space scattering amplitudes. Restricting to de Sitter space, its isometry group $SO(1,d+1)$ fixes the simplest correlators up to constants\footnote{The isometry group of de Sitter space $dS_{1+d}$ is nothing but the conformal group in $d$ dimensions. The associated conformal Ward identities were used to compute three-point correlation functions in momentum space with currents and stress-tensor ~\cite{Coriano:2013jba, Bzowski:2013sza, Bzowski:2015pba, Bzowski:2017poo, Coriano:2018zdo, Coriano:2018bbe, Isono:2018rrb, Bzowski:2018fql, Coriano:2018bsy}~.} ~\cite{Polyakov:1970xd, Osborn:1993cr, Maldacena:2011nz, Creminelli:2011mw, Bzowski:2013sza, Pajer:2016ieg}~. Interestingly, the three-point correlators also can be constrained also for more general FRW, provided that, together with the symmetries, extra conditions are imposed \cite{Pajer:2020wxk}. Once it is possible to classify the simplest correlators, we need principle to dictate how to systematically glue them to form higher point objects. As for the scattering amplitudes, such a principle is tied to their analytic properties as well as the imprint of the expected principle which should govern the time evolution, such that causality and unitarity. 

In recent years, a big deal of progress has been made in this respect. First, it has been understood a very interesting relation between cosmological correlations and flat-space physics: the lack of time-translation invariance reflects in the analytic structure of the cosmological correlations as a singularity in the sheet in kinematic space where the total energy vanishes, with the coefficient of such a singularity being the high-energy limit of the flat-space amplitudes ~\cite{Raju:2012zr, Maldacena:2011nz}~. For cases in which either the flat-space scattering is trivial or the states of interest have no flat-space counterpart, then this singularity is smoother and the coefficient turns out to be a purely cosmological effect ~\cite{Grall:2020ibl}~. Further singularities are associated to the vanishing of the total energies of certain subprocesses: in this case there is a factorisation into a product of a lower-point/lower-level scattering amplitudes and a lower-point wavefunction ~\cite{Arkani-Hamed:2017fdk}~. This information has been used to bootstrap tree-level four point correlation functions with external conformally-coupled scalars and propagating massive scalars ~\cite{Arkani-Hamed:2018kmz}~ as well as with propagating spinning particles ~\cite{Baumann:2020dch}~. A further relation between higher- and lower-point objects is provided by the cosmological optical theorem and the related cosmological cutting rules ~\cite{Goodhew:2020hob, Melville:2021lst, Goodhew:2021oqg}~, which codify perturbative unitarity -- interestingly, progress has also been made on the consequence of non-perturbative unitarity, which is codified in positivity of the spectral density ~\cite{Hogervorst:2021uvp, DiPietro:2021sjt}~. The cosmological cutting rules, together with constraints on the singularities and, in the case of massless states, with the requirement of {\it manifest locality}\footnote{Manifest locality is the statement that no singularity in the observable can come from the interactions ~\cite{Jazayeri:2021fvk}~.}, can allow to bootstrap the correlations of interest ~\cite{Jazayeri:2021fvk, Baumann:2021fxj, Cabass:2021fnw}~.

There has been an other approach to the computation of the correlation functions in the specific case of de Sitter space, which exploits its relation via analytic continuation between Euclidean Anti de Sitter space and computes the correlators as Mellin amplitudes ~\cite{Sleight:2019mgd, Sleight:2019hfp, Sleight:2020obc, Sleight:2021iix, Sleight:2021plv}~ -- an account of such developments can be found in ~\cite{Baumann:2022jpr}~.

Simultaneously, focusing on the wavefunction of the universe which generates the correlations rather than on the correlations themselves, it has been understood that, at least for conformally coupled scalars, there exist a {\it wavefunction universal integrand} from which, upon integration in the space of external energies with suitable measure, the wavefunction for any FRW cosmologies can be extracted ~\cite{Arkani-Hamed:2017fdk}~. Furthermore, acting with certain differential operators one can map these wavefunction integrands to wavefunction with massless ~\cite{Arkani-Hamed:2018kmz, Benincasa:2019vqr, Baumann:2019oyu}~ or, more generally, light states ~\cite{Benincasa:2019vqr}~. Such  wavefunction universal integrands turn out to enjoy a description in terms of combinatorial-geometrical objects named {\it cosmological polytopes} ~\cite{Arkani-Hamed:2017fdk, Benincasa:2019vqr}~. They allow to sharpen questions about the relation between cosmology and flat-space physics, providing a concrete framework that makes precise the sense in which Lorentz invariance as well as flat-space unitarity and causality emerge ~\cite{Arkani-Hamed:2018ahb, Benincasa:2020aoj}~. This led to novel, combinatorial, proofs of the flat-space cutting rules ~\cite{Arkani-Hamed:2018ahb}~ as well as Steinmann relations ~\cite{Benincasa:2020aoj}~ and the causal representation for scattering amplitudes ~\cite{Benincasa:2021qcb}~. The cosmological polytopes are objects living in projective space and are endowed with a differential form, called {\it canonical form}, with just logarithmic singularities on the boundary of the polytopes. As the wavefunction universal integrand is associated to the canonical form, all the singularities of the former are encoded in the boundaries of the latter. This allowed to further understand the physics encoded in such singularities and formulate novel constraints on the wavefunction in terms of compatibility conditions on multiple singularities ~\cite{Benincasa:2020aoj, Benincasa:2021qcb}~. Furthermore, the cosmological polytope description allowed to realise a deeper relation of the universal integrand to flat-space scattering amplitudes. It allowed not only to express all the residues in terms of (sum of) products of scattering amplitude, but also, starting from the flat-space amplitude and the notion of Bunch-Davies vacuum, to fully reconstruct the universal integrands at tree-level and extract the loop contributions out of them via a Feynman tree-like theorem ~\cite{Benincasa:2018ssx}~.

All the different approaches described provides an understanding of the wavefunction as well as methods for computing it at individual graph level. This has both positive aspects and drawback. The main positive aspect is that a number of statements if true at individual graph level straightforwardly extend to the sum of graphs which return the full desired wavefunction. The main drawback is that for general states an individual graph is not physical as it is not gauge-invariant. Hence we necessarily need to sum over all the graphs. However, as it happened for flat-space scattering amplitudes, it is not obvious from such a sum to read off eventual simplification and the underlying properties that leads to it. So, it would be ideal to have a completely gauge invariant approach to such wavefunctions, {\it i.e.} the cosmological equivalent of the on-shell formulation for scattering amplitude. This is the main reason for which the review will focus just on scalar theories, where there is no gauge-invariance issue and an individual graph can be taken as physical. In this case, the only ambiguity left is associated to field redefinitions.


The review aims to provide a pedagogical introduction to the bootstrap philosophy and methods as well as to the combinatorial-geometrical formulation of the perturbative wavefunction. As it is not meant to be a course neither in scattering amplitudes nor in cosmology, we refer to ~\cite{Elvang:2013cua, Benincasa:2013faa, Henn:2014yza}~ and to ~\cite{Weinberg:2008zzc, Baumann:2018muz}~ for more standard references and background material on the two subjects. Also we refer to ~\cite{Anninos:2012qw}~ for a detailed account of the geometry of asymptotically de Sitter space-times and additional issues associated to their physics. The review is organised in three parts.\\
\noindent
\underline{\bf Part 0} provides a general discussion of the quantum mechanical observables in cosmology and the general properties they should satisfy. Among the observables, three of them have been analysed in some detail: the wavefunction of the universe, the spatial correlators of fields, and the mean square displacement distribution in field space. In particular, the Feynman rules for the wavefunction are discussed as well as their relation to the diagrammatics for the spatial correlators. The mean square displacement distribution instead sharpens the question about the branching diffusion process that the wavefunction undergoes, its relation to cluster decomposition and the property of hypermetricity. In this part, we also provide a contained discussion of the expected factorisation properties for the wavefunction as well as unitarity, whose avatar in the wavefunction is codified by the cosmological optical theorem and the associated cutting rules. We end this part with a brief discussion of causality, whose imprint in the wavefunction is probably the least understood.\\
\noindent
\underline{\bf Part I} discusses the boundary approaches to the wavefunction, based on symmetries, the knowledge of its analytic properties as well as the cosmological optical theorem. It concerns both wavefunction in de Sitter space as well as in more general FRW cosmologies. Symmetries constrain the three-point wavefunction, which can be fixed up to a constant in the case of de Sitter, while for FRW cosmologies needs more constraints coming from the Bunch-Davis condition, the flat-space limit and the manifest locality. Symmetries as well as the cosmological cutting rules can allow to reconstruct four-point wavefunctions. A section deals with a recursive approach to the wavefunction, which provides a more algorithmic way of computing the wavefunction from lower point wavefunctions as well as the flat-space amplitude. We also focus on the so-called wavefunction universal integrand, its properties and how it is possible to extract the actual wavefunction, either via differential operators and/or in terms of the so-called symbols.\\
\noindent
\underline{\bf Part II} is devoted to the combinatorial description of the wavefunction of the universe. In order to be self-contained, it contains the basics of projective geometry and provides a pedagogical introduction to the cosmological polytopes. Through the analysis of their structure, we will get insights on the physics encoded into the singularities of the wavefunction, their relation to flat-space processes, as well as we derive novel constraints on the wavefunction will be discussed as well as novel ways of organising perturbation theory. Interestingly, it allows to sharpen the question of how flat-space properties emerge from the more fundamental cosmological context, in particular Lorentz invariance, unitarity (in terms of cutting rules) and causality (as Steinmann relations). As the cosmological polytope provides a description of the wavefunction universal integrand, we end this part with a discussion of how it also encodes the actual wavefunction in terms of its symbols.


\newpage

\addcontentsline{toc}{section}{\underline{Part 0: General Principles}}
\section*{\hfill \underline{Part 0: General Principles}\hfill}

We begin with a general discussion of the quantum mechanical observables in cosmology and some general principles that they should encode. It will focus on the wavefunction of the universe and its relation to other observables such as equal-time correlation functions and mean square displacement distributions, as well as to the current understanding of the general analytic structure, clustering, unitarity, locality and causality.


\section{Quantum mechanical observables in cosmology}\label{sec:QMC}

In a universe which undergoing a phase of accelerated expansion, it is not possible, not even in principle, to define a precise quantum mechanical observable. In order to have a well-defined quantum mechanical observable leading to arbitrarily precise predictions, there are two fundamental requirements that needs to hold:
{
\renewcommand{\theenumi}{\roman{enumi}}
\begin{enumerate}
    \item there should exist a sharp distinction between the {\it infinitely large} measuring apparatus and the {\it finite} system which need to be measured. This requirement prevents the measurement apparatus from fluctuating and, consequently, introducing finite errors to the measure;
    \item it should be possible to access an arbitrarily large amount of data, so that the correct probabilities can emerge.
\end{enumerate}
}
In presence of gravity, we could think to place a measuring apparatus at some finite location, {\it i.e.} we could try to define local observables. However, as we make it infinitely larger than the system we want to measure while keeping it at a finite location, it would also become heavier and heavier until it would collapse to form a black hole as its mass approaches the order of magnitude of the Planck mass, making it impossible to probe the system with arbitrary precision. Hence, there is always a certain {\it finite} degree of errors associated to any {\it local} measurement in presence of gravity. The only way that we can talk about precise quantum mechanical observables is to place the apparatus at the boundary, where it can be made infinitely large without forming a black hole. 

Now, in cosmology such a boundary is a space-like surface at future infinity. However, because of the accelerated expansion, just a finite size region of such a surface is causally accessible to a given observer, implying that it can only access the related {\it finite} amount of data. Thus, there is always a {\it finite}, {\it classical} uncertainty associated to measurements in a universe which is expanding with an accelerated rate and, consequently, no precise quantum mechanical observable can be defined.

\begin{figure}
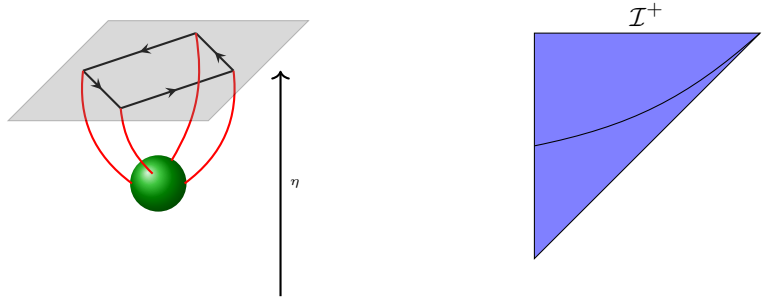

    \centering

    \caption{Correlations in the CMB and in the LSS can be traced back to correlations of quantum fields at the end of inflation. On the left, the cartoon shows a four-point correlation in momentum space, with the polygon implementing spatial momentum conservation, which is generated by processes at early times, represented by the green blob. On the right, the Penrose diagram for the future directed Poincar{\'e} patch of the special case of $dS_{1+3}$. The line in the diagram depicts a constant conformal-time slice.}
\end{figure}

However, if we assume that the universe becomes infinitely large and flat  at late enough time, so that we can access an infinitely amount of data, then we can define spatial correlations -- in other words, it is possible to perform a measurement of a given quantity at different locations on the space-like flat boundary and, then, average over all the possible different locations keeping the relative distance fixed because of the assumption of space-translation invariance. 

While a universe which becomes infinitely large and flat at sufficiently late times is indeed not the universe we live in precisely because of the accelerated expansion, this setting becomes a good approximation for modes which are sufficiently smaller than the Hubble radius.

Any type of correlation is generated by a probability distribution, which is given by the squared modulus of a wavefunction for the modes of interest. So, when we talk about precise quantum mechanical observables in cosmology, in the sense we just specified, we can equivalently talk about either spatial correlations or the {\it wavefunction of the universe} which generates them.

In this review we will mostly discuss the wavefunction of the universe because its squared modulus provides the density distribution which is needed to compute any type of spatial average:  together with the usual correlations of fields $\langle\Phi(\vec{p}_1)\ldots\Phi(\vec{p}_n)\rangle$, it is worth to mention the mean square displacement distributions in field space used to sharpen the question of how to detect and describe the dynamical branching of the wavefunction for massless ~\cite{Anninos:2011kh, Benna:2011as, Shaghoulian:2013qia} and light states ~\cite{Roberts:2012jw} and its relation to the cluster decomposition principle.


\subsection{The wavefunction of the universe}\label{subsec:WF}

Let us begin with considering some generic quantum mechanical system described by a given (Hermitian) Hamiltonian $H(t)$, and let $\Phi$ and $\Phi'$ be two configurations at time $t\,=\,0$ and $t\,=\,-T$ respectively. The transition amplitude between such configurations is then given by
\begin{equation}\label{eq:TA}
    \langle\Phi|U|\Phi'\rangle\: :=\:\langle\Phi|\hat{\mathcal{T}}\{e^{-i\int_{-T}^0dt\,H(t)}\}|\Phi'\rangle\:=\:
                                        \mathcal{N}\int\limits_{\phi(-T)=\Phi'}^{\phi(0)=\Phi}\mathcal{D}\phi\,e^{iS[\phi]},
\end{equation}
where $U\,:=\,\hat{\mathcal{T}}\{\mbox{exp}\{-i\int dt\,H(t)\}\}$ is the evolution operator such that $UU^{\dagger}\,=\,\mathbb{1}\,=\,U^{\dagger}U$, $S[\phi]$ is the action describing the system in terms of fields configurations, and $\mathcal{N}$ is a normalisation constant. Given the transition amplitude \eqref{eq:TA}, we are now interested in extracting the vacuum wavefunction out of it. The vacuum is the dominant contribution when the limit $T\,\longrightarrow\,\infty(1-i\varepsilon)$ is taken, with the other states being exponentially suppressed -- the $i\varepsilon$ prescription is needed for the convergence of the Lorentzian path integral as it contains oscillatory phases. Thus:
\begin{equation}\label{TAlim}
    \mathcal{N}\int\limits_{\phi(-\infty(1-i\varepsilon))=\Phi'}^{\phi(0)=\Phi}\mathcal{D}\phi\,e^{iS[\phi]}\:\sim\:
    \Psi_{\circ}^{\star}[\Phi']\Psi_{\circ}[\Phi]
\end{equation}
where $\Psi_{\circ}^{\star}[\Phi']$ and $\Psi_{\circ}[\Phi]$ are respectively the (complex conjugate of the) ground state wavefunction for the configuration $\Phi'$, and the ground state wavefunction for $\Phi$. If we now set $\Phi'$ to be the vacuum, {\it i.e.} $\Phi'\,=\,0$, then the transition amplitude reduces to the vacuum wavefunction:
\begin{equation}\label{eq:WFdef}
    \mathcal{N}\int\limits_{\phi(-\infty(1-i\varepsilon))=0}^{\phi(0)\,=\,\Phi}\mathcal{D}\phi\,e^{iS[\phi]}\:=\:\Psi_{\circ}[\Phi],
\end{equation}
where $\mathcal{N}$ is a suitable normalisation. 

In a cosmological context, $\phi$ represents the collection of all the modes contributing, including the gravitational ones. 

Some comments are now in order. First, notice that the wavefunction \eqref{eq:WFdef} depends on the configuration $\Phi$ at the fixed time slice $t\,=\,0$, with the time evolution which has been completely integrated out. Living on a fixed time (space-like) slice, time translation invariance is broken as well as Lorentz invariance is broken down to spatial rotations. The lack of time-translation invariance also implies not only the lack of energy conservation, but also that the concept of energy itself is not well-defined and, hence, no positive energy theorem is available. Secondly, the choice of boundary conditions in \eqref{eq:WFdef} reflects the choice of boundary conditions in cosmology: if $S[\phi]$ in \eqref{eq:WFdef} is an action describing some mode $\phi$ in a cosmological background, while $\phi(0)\,=\,\Phi$ represents the mode configuration at late time $t\,=\,0$, $\phi(-\infty(1-i\varepsilon))$ selects a state at early times, the Bunch-Davies state \cite{Bunch:1978yq} with positive frequency only.


\subsubsection{A simple example}\label{subsubsec:ExWF}

In order to fix ideas let us discuss a simple but yet illustrative example. Let us consider the quantum system given by a harmonic oscillator with time-dependent coefficients:
\begin{equation}\label{eq:thS}
    S[\phi]\:=\:-\int_{-\infty}^0d\eta\,
        \left[
            \frac{1}{2}\kappa(\eta)\dot{\phi}^2-\frac{1}{2}\omega^2(\eta)\phi^2,  
        \right]
\end{equation}
where $\kappa(\eta)$ and $\omega(\eta)$ are arbitrary functions of $\eta$. Let us compute the vacuum wavefunction. The path integral defining the wavefunction \eqref{eq:WFdef} is Gaussian and, consequently, its solution is simply the saddle point:
\begin{equation}\label{eq:thWF}
    \Psi_{\circ}[\Phi]\:=\:\mathcal{N}\,e^{iS_{cl}[\Phi]},
\end{equation}
with $S_{cl}$ being the action computed at the solution $\phi_{cl}$ of the equation of motion and $\Phi$ being the value of the field $\phi$ at $\eta\,=\,0$: 
\begin{equation}\label{eq:thScl}
    \begin{split}
        S_{cl}[\Phi]\:&=\:\int_{-\infty}^0 d\eta\,
            \left\{
                \partial_{\eta}\left[\frac{1}{2}\kappa(\eta)\phi_{cl}\dot{\phi}_{cl}\right]-
                \phi_{cl}\overbrace{\left[\partial_{\eta}\left[\frac{1}{2}\kappa(\eta)\dot{\phi}_{cl}\right]+\frac{1}{2}\omega^2(\eta)\phi_{cl}\right]}^{=0}
            \right\}\:=\\
            &=\:\left.\frac{1}{2}\kappa(\eta)\phi_{cl}\dot{\phi}_{cl}\right|_{\phi_{cl}(-\infty)=0}^{\phi_{cl}(0)=\Phi},
    \end{split}
\end{equation}
where the second term in the first line vanishes because it is the equation of motion, at which solution the action is computed, while the term in the second line comes purely from the boundary information. Being the equation of motion of second order, a general solution is given by the linear combination of its two solutions. The one of interest in this case has to vanish as $\eta\,\longrightarrow\,-\infty(1-i\varepsilon)$ ($\varepsilon\,>\,0$), and acquire the value $\Phi$ at the boundary:
\begin{equation}\label{eq:thFcl}
    \phi_{cl}(\eta)\:=\:\Phi\frac{\phi_{+}(\eta)}{\phi_{+}(0)},\qquad
    \mbox{with }\lim_{\eta\longrightarrow-\infty(1-i\varepsilon)}\phi_{+}(\eta)\:\sim\:e^{i\tilde{\omega}\eta}\:=\:0,
\end{equation}
$\tilde{\omega}$ being a positive constant which depends on the specific expression for the functions $\kappa(\eta)$ and $\omega(\eta)$. Hence, the final expression for the vacuum wavefunction can be written as
\begin{equation}\label{eq:thWF2}
    \Psi_{\circ}[\Phi]\:=\:\mathcal{N}\,e^{i\kappa(0)\left.\partial_{\eta}\log(\phi_{+}(\eta))\right|_{\eta=0}\Phi^2}.
\end{equation}
As we will show later, upon suitable choices for the functions $\kappa(\eta)$ and $\omega(\eta)$ and up to an overall spatial momentum conserving $\delta$-function, the vacuum wavefunction describes a free conformally-coupled scalar ~\cite{Arkani-Hamed:2017fdk} as well as a scalar with a general mass ~\cite{Benincasa:2019vqr} in FRW cosmologies.

We can further generalise this example by introducing terms containing higher powers of the field $\phi$ and its derivatives. We will discuss these cases in the next subsections in the perturbative approximation, under the assumption that there exist a regime where such an approximation holds. This seems to be a reasonable assumption as density perturbations $d\rho/\rho$ arising from inflationary physics are known to be of order $10^{-5}$.


\subsection{Feynman rules for the perturbative wavefunction}\label{subsec:FRWF}

Let us focus on the perturbative regime and let us consider a generic theory described by an action involving certain modes $\phi$ in a space-time with a space-like boundary located at time $\eta\,=\,0$ ($\eta\,\in\,]-\infty,\,0]$), with the following structure
\begin{equation}\label{eq:Sgen}
    S[\phi]\:=\:S_2[\phi]+S_{\mbox{\tiny int}}[\phi],
\end{equation}
where $S_2[\phi]$ is its free part while the interactions are encoded into $S_{\mbox{\tiny int}}[\phi]$. Depending on the specific form of the action \eqref{eq:Sgen}, it might be convenient to consider the boundary at some cut-off time $\eta\,=\,\eta_{\circ}$ and then take the late-time limit $\eta_{\circ}\longrightarrow0^{-}$ at the very end of the computation.

As we are interested in perturbatively computing  the wavefunction of the universe for a theory described by such an action, we can split the fields $\phi(\eta,\vec{x})$ into its classical free solution $\phi_{\circ}$ and its fluctuations $\varphi$:
\begin{equation}\label{eq:phi0var}
    \phi(\eta,\vec{x})\:=\:\phi_{\circ}(\eta,\vec{x})+\varphi(\eta,\vec{x}).
\end{equation}
In order to compute the path integral \eqref{eq:WFdef}, we also need to understand how the boundary conditions on the original field $\phi$ translates into boundary conditions on the pair $(\phi_{\circ},\,\varphi)$. This is pretty straightforward and the boundary conditions in the terms of the classical free solution and its fluctuation are given by
\begin{equation}\label{eq:BCpvp}
    \begin{split}
        &\phi_{\circ}(-\infty(1-i\varepsilon),\vec{x})\:=\:0,\qquad
         \phi_{\circ}(0,\vec{x})\:=\:\Phi(\vec{x});\\
        &\varphi(-\infty(1-i\varepsilon),\vec{x})\:=\:0,\hspace{.875cm}
         \varphi(0,\vec{x})\:=\:0,
    \end{split}
\end{equation}
{\it i.e.} the original boundary conditions are reflected into the free classical solution $\phi_{\circ}$ and the quantum fluctuations has to vanish both at the infinite past and at the boundary. Importantly, as a consequence, the action gets organised into three terms
\begin{equation}\label{eq:SgenExp}
    S[\phi]\:=\:S_2[\phi_{\circ}]+S_2[\varphi]+S_{\mbox{\tiny int}}[\phi_{\circ},\varphi],
\end{equation}
with the first two terms coming from the free part of the original action and describing the free classical propagation and the free contribution for the fluctuations respectively (notice that the mixed term $S_2[\phi_{\circ},\varphi]$ vanishes because of the equation of motion as well as the boundary conditions \eqref{eq:BCpvp} on the fluctuations $\varphi$); the third term describes the interactions and inevitably mixes $\phi_{\circ}$ and $\varphi$. The classical free mode $\phi_{\circ}(\eta,\vec{x})$ is fixed by the equation of motion of $S_2[\phi_{\circ}]$ endowed with the boundary conditions \eqref{eq:BCpvp}, which is assumed to be given in terms of a second order differential operator. Because of the assumption of invariance under spatial translation, we can formulate our problem in momentum space
\begin{equation*}
    \phi_{\circ}(\eta,\vec{x})\:=\:\int\frac{d^d p}{(2\pi)^d}\,e^{i\vec{p}\cdot\vec{x}}\tilde{\phi}_{\circ}(\eta,\vec{p})
\end{equation*}
and the second order differential operator gets mapped into another second order differential operator $\hat{\mathcal{O}}$ depending just on time derivatives as well as on $E\,:=\,|\vec{p}|\,>\,0$, which with a bit abuse of language we will refer to as {\it energy}:
\begin{equation}\label{eq:eom}
    \hat{\mathcal{O}}(\partial_{\eta}^2,\partial_{\eta},E)\tilde{\phi}_{\circ}(\eta,\vec{p})\:=\:0
\end{equation}
The boundary conditions \eqref{eq:BCpvp} selects the solution of \eqref{eq:eom} such that
\begin{equation}\label{eq:eoms}
    \tilde{\phi}_{\circ}(0,\vec{p})\:=\:\tilde{\Phi}(\vec{p}),\qquad
    \lim_{\eta\longrightarrow -\infty(1-i\varepsilon)}\tilde{\phi}_{\circ}(\eta,\vec{p})\:\sim\:e^{iE\eta},
\end{equation}
with $\tilde{\Phi}$ being the Fourier transform of $\Phi$. The boundary condition in the infinite past, together with the chosen $i\varepsilon$ prescription, selects the solution with positive frequency. We can indicate such a solution of \eqref{eq:eom} as 
\begin{equation}
    \tilde{\phi}_{\circ}(\eta,\vec{p})\:=\:\tilde{\Phi}(\vec{p})\frac{\phi_{+}(\eta,E)}{\phi_{+}(0,E)}
\end{equation}
such that $\phi_{+}$ vanishes in the infinite past. The mode function $\phi_{+}$ encodes the bulk-to-boundary propagators for our scalars. For simplicity we can normalise $\phi_{+}(\eta,E)$ such that $\phi_{+}(0,E)=1$. 

Thus, using the split \eqref{eq:phi0var} of the scalar field into the classical free solution $\phi_{\circ}$ and the fluctuation $\varphi$ the path integral \eqref{eq:WFdef} defining the wavefunction can be rewritten as:
\begin{equation}\label{eq:WFpert1}
    \Psi_{\circ}[\Phi]\:=\:\mathcal{N}e^{iS_2[\Phi]}\int_{\varphi(-\infty(1-i\varepsilon))=0}^{\varphi(0)=0}\mathcal{D}\varphi\,
        e^{iS_2[\varphi]+iS_{\mbox{\tiny int}}[\Phi,\varphi]},
\end{equation}
where
\begin{equation*}
    iS_2[\Phi]\:=\:-\int\left[\prod_{r=1}^2\frac{d^d p_r}{(2\pi)^d}\tilde{\Phi}(\vec{p}_r)\right]
        \delta^{\mbox{\tiny $(d)$}}(\vec{p}_1+\vec{p}_2)\psi_2(E).
\end{equation*}
with $\psi_2(E)$ being the two-point wavefunction which depends only on the energy $E\,:=\,|\vec{p}_1|=|\vec{p}_2|$ and whose precise form depends only on the particular states one is considering. Perturbatively expanding in the interactions, we can further obtain:
\begin{equation}\label{eq:WFper2}
    \begin{split}
        \Psi_{\circ}[\Phi]\:&=\:\mathcal{N}e^{iS_2[\Phi]}\int_{\varphi(-\infty(1-i\varepsilon))=0}^{\varphi(0)=0}\mathcal{D}\varphi\,e^{iS_2[\varphi]}
                         \sum_{j=0}^{\infty}\frac{i^j}{j!}S_{\mbox{\tiny int}}^j[\Phi,\varphi]\:=\\
                      &=\:e^{iS_2[\Phi]}\sum_{j=0}^{\infty}\frac{i^j}{j!}\langle S^j_{\mbox{\tiny int}}\rangle[\Phi],
    \end{split}
\end{equation}
where $\langle S_{\mbox{\tiny int}}^j\rangle$ just indicates the path integrated $j$-th term of the sum in the first line, including the normalisation $\mathcal{N}$, which is taken to be
\begin{equation}\label{Ndef}
    \mathcal{N}^{-1}\:=\:\int_{\varphi(-\infty(1-i\varepsilon))=0}^{\varphi(0)=0}\mathcal{D}\varphi\,e^{iS_2[\varphi]}
\end{equation}
A pretty generic form for $S_{\mbox{\tiny int}}$ can be taken to be
\begin{equation}\label{eq:Sint}
    S_{\mbox{\tiny int}}[\phi_{\circ},\varphi]\:=\:\int d^d x\int_{-\infty}^0 d\eta\,\lambda_k(\eta)V_k(\phi_{\circ},\varphi,\partial_{\eta},\partial_i),
\end{equation}
where $\lambda_k(\eta)$ is a generic function of the time $\eta$, and $V_k$ is a polynomial function of an overall number $k$ of fields $\phi_{\circ}$ and $\varphi$ and their time and spatial derivatives. Hence:
\begin{equation}\label{eq:SintAv}
    \langle S_{\mbox{\tiny int}}^j\rangle\:=\:\mathcal{N}\int\limits_{\varphi(-\infty(1-i\varepsilon))=0}^{\varphi(0)=0}\mathcal{D}\varphi\,e^{iS_2[\varphi]}\,
        \prod_{r=1}^j\int d^d x_j\int_{-\infty}^0 d\eta_j\,\lambda_k(\eta_j)\,V_k(\phi_{\circ},\varphi,\partial_{\eta_j},\vec{\partial}_j)
\end{equation}
Having assumed $V_k$ to depend polynomially on the fluctuations $\varphi$ and their derivatives and being $S_2[\varphi]$ typically Gaussian in $\varphi$, just the terms with an even number of $\varphi$ are non-zero, while the path integration over the fluctuations produces bulk-to-bulk propagators $G(y_{e};\,\eta_{e_v},\eta_{e_{v'}})$, with has to vanish as any of $\eta_{r_i}\,\longrightarrow\,0^{-}$ and which depends on the energy $y_e$ running through it:
\begin{equation}\label{eq:G}
    \begin{split}
    G(y_e;\,\eta_{v_e},\eta_{v'_e})\:=\:\frac{1}{\mbox{Re}\{2\psi_2(y_e)\}}
        \Big[
            &\bar{\phi}_{+}(-y_e\eta_{v_e})\phi_{+}(-y_e\eta_{v'_e})\vartheta(\eta_{v_e}-\eta_{v'_e})+\\
            +\:&\phi_{+}(-y_e\eta_{v})\bar{\phi}_{+}(-y_e\eta_{v'_e})\vartheta(\eta_{v'_e}-\eta_{v_e})-\\
            -\:&\phi_{+}(-y_e\eta_{v_e})\phi_{+}(-y_e\eta_{v'_e})
        \Big].
    \end{split}
\end{equation}
\begin{figure}[t]
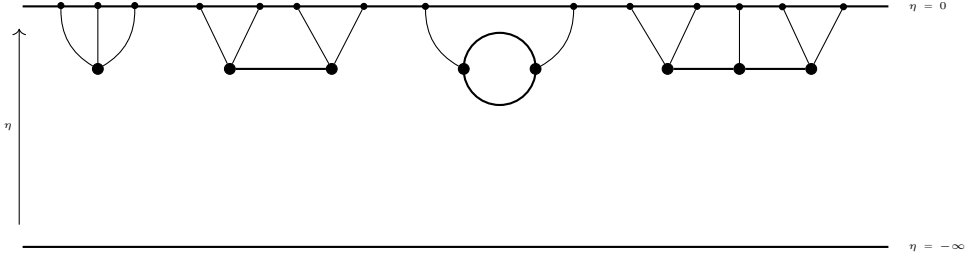

  \centering
 
  \caption{Examples of Feynman graphs associated to the wavefunction. The external lines, which connect point at the boundary $\eta\,=\,0$ to the sites at $\eta\,=\,\eta_{v}$ represent the bulk-to-boundary propagators, given by $\phi_{+}$. The sites are associated to the vertex function $V_v(\vec{p},\eta_{v})$, which encodes the interaction and it is a function of product of spatial momenta. Finally, the edges connecting the sites are associated to the three-term bulk-to-bulk propagators.}
  \label{fig:Graphs}
\end{figure}
Thus, the wavefunction of the universe \eqref{eq:WFper2} can be written as
\begin{equation}\label{eq:WFpert3}
    \Psi_{\circ}[\Phi]\:=\:e^{iS_2[\Phi]}
        \left\{
            1+\sum_{k\ge2}\int\prod_{j=1}^k
            \left[
                \frac{d^d p^{\mbox{\tiny $(j)$}}}{(2\pi)^d}\Phi(\vec{p}^{\mbox{\tiny $(j)$}})
            \right]
            \sum_{L=0}^{\infty}\psi_k^{\mbox{\tiny $(L)$}}
        \right\}
\end{equation}
where $\psi_k^{\mbox{\tiny $(L)$}}$ is a function of the spatial momenta $\{\vec{p}^{\mbox{\tiny $(j)$}}\,|\,j=1,\ldots,k\}$, and can be represented by a sum of connected graphs with the same numbers $k$ and $L$ of bulk-to-boundary propagators and loops respectively. If $\mathcal{G}_{k}^{\mbox{\tiny $(L)$}}:=\{\mathcal{G}\}$ is the collection of all the connected graphs contributing to $\psi_k^{\mbox{\tiny $(L)$}}$, then 
\begin{equation}\label{eq:PkLPG}
    \psi_k^{\mbox{\tiny $(L)$}}\:=\:\sum_{\mathcal{G}\subset\mathcal{G}_k^{\mbox{\tiny $(L)$}}}\psi_{\mathcal{G}},
\end{equation}
where $\psi_{\mathcal{G}}$ is the contribution of a specific graph $\mathcal{G}$. Each graph $\mathcal{G}$ is defined by the sets $\mathcal{V}$ and $\mathcal{E}$ of sites and edges connecting the sites respectively, and external lines connecting the sites to the late-time space-like boundary. Then the functional form of $\psi_{\mathcal{G}}$ can be written as
\begin{equation}\label{eq:Feyn1}
    \psi_{\mathcal{G}}\:=\:\delta^{\mbox{\tiny $(d)$}}\left(\sum_{j=1}^k\vec{p}_j\right)
        \int_{-\infty}^0\prod_{v\in\mathcal{V}}\left[d\eta_v\phi_{+}^{\mbox{\tiny $(v)$}}V_v\right]\prod_{e\in\mathcal{E}}G(y_e,\,\eta_{v_e},\eta_{v'_e})
\end{equation}
with bulk-to-boundary propagators $\phi_+$ associated to the external lines, bulk-to-bulk propagators $G(y_e;\,\eta_{v_e},\eta_{v'_e})$ associated to the edges connecting the interaction sites, the latter given by the function $V_v(\vec{p},\eta_v)$ -- see Figure \ref{fig:Graphs} for some examples. 

The spatial momentum conserving $\delta$-function appearing in \eqref{eq:Feyn1} is the result of the integration over the spatial coordinate and it is a consequence of the assumption of space-translation invariance. Notice that, given a graph $\mathcal{G}$ with $n_s$ sites and $n_e$ edges, because of the three-term expression \eqref{eq:G}, the related wavefunction contribution has $3^{n_e}$ terms.


\subsection{Probability distribution of field configurations}\label{subsec:PDfc}

The wavefunction of the universe can be thought to be the most primitive among the observables in cosmology. As already remarked, if it is indeed true that it is not an observable in the strict sense -- no observation would directly measure the wavefunction in and of itself --, on one side it is characterised by {\it sufficiently physical} features, {\it e.g.} gauge invariance, and on the other side its squared modulus $|\Psi[\Phi]|^2$ provides the probability distribution which allows to define any type of correlation. It is therefore both instructive and useful to have an explicit discussion about $|\Psi[\Phi]|^2$ in perturbation theory.

Organising the perturbative expansion in terms of the number of external boundary fields, the probability distribution for a certain field configuration can be written as
\begin{equation}\label{ProbDistrPert}
    \begin{split}
        |\Psi_{\circ}[\Phi]|^2\:&=\:
        \resizebox{0.675\hsize}{!}{$\displaystyle
        e^{-2\mbox{\footnotesize Im}\{S_2[\Phi]\}}
        \left\{
            1+
            \int
            \left[
                \prod_{j=1}^2\frac{d^d p^{\mbox{\tiny $(j)$}}}{(2\pi)^d}\,\Phi(\vec{p}^{\mbox{\tiny $(j)$}})
            \right]
            \sum_{L=1}^{\infty}[\psi_2^{\mbox{\tiny $(L)$}}+\psi_2^{\dagger\mbox{\tiny $(L)$}}]+
        \right.$}\\
        &\resizebox{0.875\hsize}{!}{$\displaystyle
         \left.\hspace{-1.5cm}
            +\:\sum_{k\ge3}
            \int
            \left[
                \prod_{j=1}^k\frac{d^d p^{\mbox{\tiny $(j)$}}}{(2\pi)^d}\,\Phi(\vec{p}^{\mbox{\tiny $(j)$}})
            \right]
            \sum_{L\ge0}
            \left[
                \psi_k^{\mbox{\tiny $(L)$}}+\psi_k^{\dagger\mbox{\tiny $(L)$}}+
                \sum_{M=0}^L\sum_{r=2}^{k-2}
                \left(
                    \psi_r^{\mbox{\tiny $(M)$}}\psi_{k-r}^{\dagger\mbox{\tiny $(L-M)$}}+
                    \psi_{k-r}^{\mbox{\tiny $(L-M)$}}\psi_r^{\dagger\mbox{\tiny $(M)$}}
                \right)
            \right]
        \right\}.$}
    \end{split}
\end{equation}

Let $\mathcal{O}[\Phi]$ be some function of the boundary field configuration $\Phi$. Then, the graphical rules for its perturbative computation can be readily read off from \eqref{eq:ProbDistr} and the Feynman rules for the wavefunction coefficients, by taking
\begin{equation}\label{eq:Ovev}
    \langle\mathcal{O}[\Phi]\rangle\:=\:\mathcal{N}\,\int\mathcal{D}\Phi\,|\Psi_{\circ}[\Phi]|^2\mathcal{O}[\Phi],
\end{equation}
where $\mathcal{N}$ is the normalisation factor
\begin{equation}\label{eq:NO}
    \mathcal{N}^{-1}\:=\:\int\mathcal{D}\Phi\,|\Psi_{\circ}[\Phi]|^2.
\end{equation}
%


\subsection{Correlation functions and the wavefunction}\label{subsec:CF}

A typical example of spatial averages we can compute at the future boundary of an expanding universe, is given by correlation functions for $n$ fields $\Phi$. While they are usually computed via the Keldysh-Schwinger formalism \cite{Weinberg:2005vy}, they can also be obtained by exploiting their direct relation \ref{eq:Ovev} with the probability distribution $|\Psi(\Phi)|^2$ taking $\mathcal{O}[\Phi]\,=\,\prod_{j=1}^n\Phi(\vec{p}^{\mbox{\tiny $(j)$}})$. Explicitly, 
\begin{equation}\label{eq:CorrFct}
    \langle\prod_{j=1}^n\Phi(\vec{q}^{\mbox{\tiny $(j)$}})\rangle \:=\: \mathcal{N}\int\mathcal{D}\Phi\,|\Psi_{\circ}(\Phi)|^2\,\prod_{j=1}^n\Phi(\vec{q}^{\mbox{\tiny $(j)$}}).
\end{equation}

Such formula, together with the Feynman rules for the wavefunction, allows to obtain directly graphical rules for such correlation functions as well.

First of all, notice that the path integration \eqref{eq:CorrFct} provides non-vanishing contributions to the correlation functions for $n+k$ even. However, connected graphs are obtained for $k\ge n+2m$ and the non-connected graphs are moded out by the normalisation. Hence, the non-vanishing connected contributions to the $n$-point correlation function \eqref{eq:CorrFct} acquires the following schematic form
\begin{equation}\label{eq:CorrFct2}
    \begin{split}
        \langle\prod_{j=1}^n\Phi(\vec{q}^{\mbox{\tiny $(j)$}})\rangle \:\sim\:
        \mathcal{N}&\int\mathcal{D}\Phi\,\prod_{j=1}^n\Phi(\vec{q}^{\mbox{\tiny $(j)$}})\,e^{-2\mbox{\footnotesize Im}\{S_2[\Phi]\}}
        \sum_{L\ge0}\sum_{m\ge0}\prod_{j=1}^{n+2m}\Phi(\vec{p}^{\mbox{\tiny $(j)$}})\times\\
        &\hspace{-1cm}\resizebox{0.75\hsize}{!}{$\displaystyle
        \times
        \left\{
            2\mbox{Re}\{\psi_{n+2m}^{\mbox{\tiny $(L-m)$}}\}+
            \sum_{r\ge2}\sum_{M\ge0}
            \left[
                \psi_{r+m}^{\mbox{\tiny $(M)$}}\psi_{n+m-r}^{\dagger\mbox{\tiny $(L-M)$}}+
                \psi_{n+m-r}^{\mbox{\tiny $(L-M)$}}\psi_{r+m}^{\dagger\mbox{\tiny $(M)$}}
            \right]
        \right\}
        $}
    \end{split}
\end{equation}
The path integration returns Wick contractions among the boundary fields producing inverses of the real part of the $2$-point wavefunction:
\begin{equation}\label{eq:CorrFct3}
    \begin{split}
        \langle\prod_{j=1}^n\Phi(\vec{q}^{\mbox{\tiny $(j)$}})\rangle \:=\:
        &\prod_{j=1}^n\frac{1}{2\mbox{Re}\{\psi_2(E_j)\}}\sum_{L\ge0}\sum_{m\ge0}
         \prod_{\substack{a,b=1\\a<b}}^m\left(\frac{1}{2\mbox{Re}\{\psi_2(y_{ab})\}}\right)\times\\
        &\times\left\{
            2\mbox{Re}\{\psi_{n+2m}^{\mbox{\tiny $(L-m)$}}\}+
            \sum_{r=2}^{n+2m}\sum_{M=0}^L 2\mbox{Re}\{\psi_{r+m}^{\mbox{\tiny $(M)$}}\psi_{n+m-r}^{\dagger\mbox{\tiny $(L-M)$}}\}
        \right\}
    \end{split}
\end{equation}
where $E_j:=|\vec{q}^{\mbox{\tiny $(j)$}}|$ and $y_{ab}:=|\vec{p}^{\mbox{\tiny $(a)$}}|=\vec{p}^{\mbox{\tiny $(b)$}}|$. Each contribution at a given loop order $L$ and for a given number $n$ of boundary states, can be obtained via rules on graphs for the wavefunction contributions.

Let $\psi_{\mathcal{G}}^{\mbox{\tiny $(L)$}}$ be the contribution of the graph $\mathcal{G}$ to the wavefunction. Then the contribution to the correlation function is obtained by summing the wavefunction to all the possible ways of deleting the internal edges of the graph, replacing the bulk-to-bulk propagators associated to the deleted edges with the real part of the bulk-to-boundary one, further taking the real part of this sum and multiplying it by the product of $(2\mbox{Re}\{\psi_2(E_j)\})^{-1}$ for all the external states. As an example, let us consider the bubble graph for a quartic interaction. Then
\begin{equation}\label{eq:CorrFctEx}

\end{equation}
where each graph represents a wavefunction coefficient and each crossed-out edge represents a bulk-to-boundary propagator and the right-hand-side is understood to be (twice) the real part of each term.

Thus, these graphical rules allow to straightforwardly extract the perturbative correlation function for $n$ fields from the knowledge of the wavefunction coefficients.


\subsection{Mean square displacement distributions in field space}\label{subsec:DisDis}

There is an important point to be made. Let us begin with considering a massless scalar in a fixed expanding background and its evolution. The two-point correlation function of two scalar states grows logarithmically as they are taken spatially far away from each other -- see Section \ref{subsec:CD}. Furthermore, some of the quantum fluctuations of a scalar state can become large with respect to the horizon scale, freeze and classicalise. Then the fluctuations of such modes can in turn grow large, freeze and classicalise, in a branch diffusion process until the late-time future infinity boundary is reached \cite{Starobinsky:1982ee}. This implies that a given mode does not have a definite value as, starting from the Bunch-Davies vacuum, it can freeze out in a large number of possible late-time configurations. 

In the case of massive states, the two-point correlator vanishes, following a power-law decay, as the separation between the two states is taken arbitrarily large -- see Section \ref{subsec:CD}. Furthermore, it depends on the cut-off surface location $\eta\,=\,\eta_{\circ}$ through a positive power and, consequently, as the late-time limit is taken, the two-point correlator vanishes: this is just the statement that massive states do not get to the boundary at future infinity. However, we can still consider all the configurations that are generated and reach the cut-off surface.

How can we probe such diffusion process? Is there an avatar of it in the wavefunction of the universe? Given the existence of a plethora of late-time configurations, it makes sense to consider the distances between two (or more) of them, and compute their probability distributions \cite{Anninos:2011kh, Roberts:2012jw, Shaghoulian:2013qia}. Let us begin with considering a generic, conformally flat, FRW metric in $(d+1)$ dimensions with periodic boundary conditions along the spatial dimensions which serve to introduce an explicit IR cutoff $L$:
\begin{equation}\label{eq:FRWper}
    ds^2\:=\:a^2(\eta)\left[-d\eta^2+\delta_{ij}dx^i dx^j\right],\qquad
    x_i\:\sim\:x_i+L,
\end{equation}
where $a(\eta)$ is a warp factor depending on the conformal-time $\eta$ only and such that it increases as it approaches the late-time boundary at $\eta\,=\,0$, while it vanishes as we go in the far past as $\eta\,\longrightarrow\,-\infty$. The latin letters $i,\,j,\ldots$ are used as indices for the spatial directions and run from $1$ to $d$, while $L$ is the compactification radius for each of the spatial directions.

Given two field configurations $\phi_1(\eta_{\circ},\vec{x})$ and $\phi_2(\eta_{\circ},\vec{x})$, their mean square displacement at at a fixed-time slice $\eta\,=\,\eta_{\circ}$ -- or, more properly, their mean square displacement -- can be defined as ~\cite{Anninos:2011kh, Roberts:2012jw, Shaghoulian:2013qia}
\begin{equation}\label{eq:disf12}
    \delta^{\mbox{\tiny $(12)$}}\:=d^{\mbox{\tiny $(12)$}}-\langle d^{\mbox{\tiny $(12)$}}\rangle,
\end{equation}
where 
\begin{equation}\label{eq:disf12a}
    d^{\mbox{\tiny $(12)$}}\:=\:f
        \int\frac{d^dx}{L^d}
        \left[
            \hat{\phi}_1(\eta_{\circ},\vec{x})-\hat{\phi}_2(\eta_{\circ},\vec{x})
        \right]^2,
\end{equation}
with $\hat{\phi}_j$ being a field configuration with the zero modes removed and a physical size UV cut-off, and $f$ being a factor that depends on the characteristic scales of the problem, {\it i.e.} the space-time radius $\ell_{\gamma}$, the IR cut-off $L$ and time $\eta_{\circ}$ at which the correlations are computed: 
\begin{equation}\label{eq:fdef}
    f\,:=\,
    \left\{
        \begin{array}{l}
            \displaystyle c_{\circ}\left(\frac{\ell_{\gamma}}{L}\right)^{(\gamma-1)(d-1)} 
            \hspace{3.675cm} \mbox{(massless states, $\gamma\neq0$)}\\
            \phantom{\ldots}\\
            \displaystyle c_{\Delta_{-}}\left[\frac{L}{\ell_1} a(\eta_{\circ})\right]^{2\alpha_{\Delta_{-}}}
                            \left[\log{\left(-\frac{L}{\eta_{\circ}}\right)}\right]^{-\delta_{\Delta_{-},\frac{d}{4}}} 
            \quad \mbox{(massive states)}
        \end{array}
    \right.
    ,
\end{equation}
$\delta_{\Delta_{-},d/4}$ being a Kronecker delta, $2\Delta_{-}:=d-\sqrt{d^2-(2m\ell_{1})^2}$ and
\begin{equation}\label{eq:aDm}
    \alpha_{\Delta_{-}}\,:=\,\Delta_{-}\vartheta\left(\Delta_{-}-\frac{d}{4}\right)+
                              \frac{d}{4}\vartheta\left(\frac{d}{4}-\Delta_{-}\right).
\end{equation}
Notice that the definition of the prefactor $f$ in \eqref{eq:fdef} holds for massless states in arbitrary FRW cosmologies (parametrised by $\gamma$)\footnote{In flat space (\it{i.e.} $\gamma\,=\,0$), $f\,=\,1$.} and for massive states in $dS_{1+d}$ as the explicit expression for the mode functions for arbitrary masses and arbitrary FRW cosmologies are not known.

The subtraction form \eqref{eq:disf12a} provides a definition for the mean squared displacement free of divergences as the late time boundary is approach. Importantly, the mean square displacement measures the similarity between the two field configurations.

Then,  the probability distribution that two field configurations are at some distance $\Delta$, {\it i.e.} that $\delta^{\mbox{\tiny $(12)$}}\,=\,\Delta$ can be written as
\begin{equation}\label{eq:ProbDistr}
    \mathcal{P}(\Delta)\: :=\:\langle\delta(\Delta-d^{\mbox{\tiny $(12)$}})\rangle\:=\:
    \mathcal{N}_{12}\int\prod_{j=1}^2\left[\mathcal{D}\hat{\phi}_j|\Psi_{\circ}[\hat{\phi}_j]|^2\right]
        \delta(\Delta-\delta^{\mbox{\tiny $(12)$}})
\end{equation}
where $|\Psi_{\circ}[\hat{\phi}_j]|^2$ represents the probability distribution of the field configuration $\hat{\phi}_j$ on the space-like surface at $\eta\,=\,\eta_{\circ}$. It is important to pause one moment and make some remarks on \eqref{eq:ProbDistr}. One considers two copies of the system, represented by the two independent probability distributions, one each field configuration $\hat{\phi}_j$, while the delta-function $\delta(\Delta-d^{\mbox{\tiny $(12)$}})$ enforces the mean square displacement in field space between two field configurations to be equal to $\Delta$. 

Similarly, the mean square displacement probability distribution among configurations of three replicas has the form
\begin{equation}\label{eq:ProbDistr3}
    \begin{split}
        \mathcal{P}(\Delta_1,\Delta_2,\Delta_3)\: :&=\:
            \langle\prod_{j=1}^3\delta(\Delta_{j-1}-\delta^{\mbox{\tiny $(j,j+1)$}})\rangle\:=\\
        &=\:\mathcal{N}_{123}\int\prod_{j=1}^3\left[\mathcal{D}\hat{\phi}_j|\Psi_{\circ}[\hat{\phi}_j]|^2\right]
            \prod_{r=1}^3\delta(\Delta_{r-1}-\delta^{\mbox{\tiny $(r,r+1)$}})
    \end{split}
\end{equation}

The probability distributions for the mean square displacement is an observable which allows to probe a certain type of non-localities intrinsic in processes occurring in an expanding universe: as the wavefunction evolves, it branches out in such a way that while each branch is still characterised by cluster decomposition, globally cluster decomposition is lost. This can be observed even at the level of free states, for which the wavefunction is a simple Gaussian. So a general question is what is the avatar of this branch diffusion process at the level of the wavefunction itself? While this question is still unanswered, observables such as the probability distributions for the mean square displacement \eqref{eq:ProbDistr} and \eqref{eq:ProbDistr3} allow to sharpen this question, as we will see in Section \ref{subsec:CD}.


\section{Fundamental properties}\label{sec:FunProp}

Physical processes are governed by basic principles. Knowing them and how they constrain the quantum mechanical observables, allow to both compute them and understand how interactions are constrained. In flat-space, we know how precisely this happen and that a number of theorems are just a consequence of the validity of such principles. Remarkable examples are the the conservation of charge and the equivalence principle, which can be obtained by imposing Lorentz invariance and unitarity on the S-matrix \cite{Weinberg:1964ew}. 

In this section we review our current understanding of some properties and basic principles in an expanding universe, such as the general singularity structure, factorisations, cluster decomposition, unitarity and causality.


\subsection{Cosmic clustering}\label{subsec:CD}

One of the crucial principles in our understanding of particle physics is the idea that experiments which are sufficiently separated in space, yield uncorrelated results. It goes under the name of cluster decomposition principle. It implies that scattering amplitudes ought to be sufficiently analytic functions, with at most poles and branch cuts, modulo the overall momentum conserving delta-function \cite{Weinberg:1995mt}, while connected correlation functions in position space have to decay polynomially sufficiently fast\footnote{Equivalently, cluster decomposition reflects into the full correlation via its factorisation as any of the relative distances as taken to be arbitrarily large.}.

In expanding universes, this is not generally true, as the two-point functions for massless states grow logarithmically as the spatial separation increases. Let us consider the wavefunction of the universe for a generic free scalar. It has the general form
\begin{equation}\label{eq:Psi2pt}
    \Psi_{\circ}[\Phi]\: =\:\mbox{exp}
        \left\{ 
            -\int\prod_{j=1}^2\left[d^d p_j\,\Phi(\vec{p}_j)\right]
            \psi_2(E,\eta_{\circ})
        \right\},
\end{equation}
with $\psi_2$  being the generic form  for the two-point function, which has support on $\delta^{\mbox{\tiny $(d)$}}(\vec{p}_1+\vec{p}_2)$ and depends just on $E\, :=\,|\vec{p}_1|\,=\,|\vec{p}_2|$ and on the location $\eta\,=\,\eta_{\circ}$ of the space-like surface where it is computed. Therefore, the probability distribution for a certain field configuration $\Phi$ is given by
\begin{equation}\label{eq:ProbDistr2pt}
    |\Psi_{\circ}[\Phi]|^2\: =\:\mbox{exp}
        \left\{
            -\int\prod_{j=1}^2\left[d^d p_j\,\Phi(\vec{p}_j)\right]
            \mbox{Re}\{2\psi_2(E,\eta_{\circ})\}
        \right\}.
\end{equation}
In the late-time limit $\eta_{\circ}\,\longrightarrow\,0^{-}$, the real part of the two-point wavefunction behaves as
\begin{equation}\label{eq:2ptltwf}
    \mbox{Re}\{2\psi_2(E,\,\eta_{\circ})\}\:\overset{\eta_{\circ}\longrightarrow0^{-}}{\sim} E^{\alpha}\eta_{\circ}^{\beta},\qquad \alpha\in\mathbb{R}_{+},
\end{equation}
with two-point correlator being
\begin{equation}\label{eq:2ptcf}
    \langle\Phi(\vec{p}_1,\vec{p}_2)\rangle\:=\:
        \frac{\delta^{\mbox{\tiny $(d)$}}(\vec{p}_1+\vec{p}_2)}{\mbox{Re}\{2\psi_2(E,\eta_{\circ})\}}
\end{equation}
which therefore behaves as in $E^{-\alpha}$ in the late-time limit. Consequently, from the Fourier transform mapping the two-point function in momentum space to position space, the power-law suppression as $|\vec{x}_1-\vec{x}_2|\,\longrightarrow\,\infty$ required by cluster decomposition is obtained if and only if $\alpha\,\in\,]0,\,d[$, while for $\alpha\,\ge\,d$ the two-point correlator would blow up as the relative distance is taken to be large, with the logarithmic growth for the case $\alpha\,=\,d$. Thus, the avatar in the two-point wavefunction for the presence or not of cluster decomposition is its behaviour with the energy in the late-time limit: in order for cluster decomposition to hold, the real part of the two-point wavefunction should be polynomially bounded in momentum space by $E^d$, {\it without} saturating it. Notice that the real part of the two-point wavefunction for a free scalar in flat-space -- and therefore a conformally coupled-scalar in any conformally flat universe -- behaves as $E$ in the late-time limit, thus satisfying cluster decomposition as long as $d\,>\,1$, while for a minimally-coupled massless scalar in de Sitter space, it behaves as $E^d$.

As a further comment, if we consider FRW cosmologies with warp factor $a(\eta)\,\sim\,(-\eta)^{-\gamma}$ ($\gamma\,\in\,\mathbb{R}_{+}$), for which the free solution is given in terms of Hankel function of second type whose order $\nu$ is related to the propagating state, then, for $\nu\in\mathbb{R}_+$, the late-time behaviour \eqref{eq:2ptltwf} is characterised by $\alpha\,=\,2\nu$, so that the two-point correlator \eqref{eq:2ptcf} shows a power-law suppression as the relative distance is taken to be arbitrarily large if and only if $\nu<d/2$. If instead $\nu$ is purely imaginary -- as it is the case for the principal series of $dS_4$, then also $\alpha$ is imaginary and the two-point correlator for these states still exhibit the cluster decomposition property.

As we emphasised earlier, there is a plethora of possible late-time configurations, it can be instructive to look at the probability distributions for the mean square displacement for two \eqref{eq:ProbDistr} and three \eqref{eq:ProbDistr3} of them.

An explicit and exact solution for $\psi_2$ can be found for minimally-coupled massive scalars in $dS$, FRW cosmologies with $\gamma=1/2$ and flat-space, as well as a massless scalar in FRW cosmology with arbitrary $\gamma$. The following discussion will be restricted to these cases.

Let us begin with the probability distribution for the mean square displacement of two field configurations $\Phi_1$ and $\Phi_2$ as defined in \eqref{eq:ProbDistr}, for which we consider an explicit IR cut-off by compactifying the spatial directions on a torus of typical size $L$, {\it i.e.} $x_i\,\sim\,x_i+L$, for a generic conformally flat metric with time-dependent warp factor $a(\eta)$ -- see \eqref{eq:FRWper}
\begin{equation}
    \begin{split}
        \mathcal{P}(\Delta)\: 
            :&=\:\langle\delta(\Delta-d^{\mbox{\tiny $(12)$}})\rangle\:=\:
            \mathcal{N}_{\mbox{\tiny $12$}}\int\prod_{j=1}^2\left[\mathcal{D}\Phi_j|\Psi_{\circ}[\Phi_j]|^2\right]
            \delta(\Delta-\delta^{\mbox{\tiny $(12)$}})\: =\\
            &\hspace{-1cm}=\:\int_{-i\infty}^{+i\infty}\frac{d\alpha}{2\pi i}\,e^{\alpha\Delta}\,
            e^{\alpha\langle d^{\mbox{\tiny $(12)$}}\rangle}
            \mathcal{N}_{\mbox{\tiny $12$}}\int\prod_{j=1}^2
            \Big[
                \mathcal{D}\Phi_{j}\,
                e^{\mbox{\tiny $\displaystyle -\frac{1}{L^d}\sum_{\vec{p}}\mbox{Re}\{2\psi_2\}|\Phi_j|^2$}}
            \Big]
                e^{\mbox{\tiny $\displaystyle -\frac{2\alpha f}{L^{2d}}\sum_{\vec{p}}|\Phi_1-\Phi_2|^2$}}
    \end{split}
\end{equation}
where the compactification on the torus allows to map the integrals to sums:
\begin{equation}\label{eq:compactp}
    \int d^d p\:\longrightarrow\:\frac{1}{L^d}\sum_{\vec{p}},\qquad
    \mbox{with } LE\:=\:2\pi n \quad (n\,\in\,\mathbb{Z}_+),
\end{equation}
while $f$ is the prefactor in \eqref{eq:fdef}. The probability distribution for the mean square displacement for two field configuration then acquires the form
\begin{equation}\label{eq:PDf}
    \mathcal{P}(\Delta)\:=\:\lim_{L\longrightarrow\infty}
        \int_{-i\infty}^{+i\infty}\frac{d\alpha}{2\pi i}\,e^{\alpha\Delta}
        \prod_{\vec{p}\neq0}\frac{e^{\frac{2\alpha f}{L^d\mbox{\footnotesize Re}\{2\psi_2\}}}}{1+e^{\frac{2\alpha f}{L^d\mbox{\footnotesize Re}\{2\psi_2\}}}}
\end{equation}
where the zero modes have been mode out and the limit $L\,\longrightarrow\,\infty$ has to be understood as the compactification radius $L$ is larger than any other scales in the game, such as $\eta_{\circ}$ and the characteristic length $\ell_{\gamma}$ of the space-time.

\begin{figure}
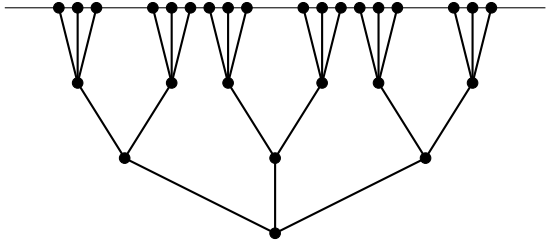

    \centering

    \caption{Tree-like structure in field space. There is a plethora of late-time configurations which can be thought to organise in a tree-like structure where the leaves correspond to the different configurations and the distance between two configurations measure the distance between the corresponding leaves of the tree and a common branch.}
    \label{fig:TLS}
\end{figure}
Now, the real part of the two-point function in the late-time limit can be straightforwardly computed to be \cite{Anninos:2011kh, Roberts:2012jw, Shaghoulian:2013qia}
\begin{equation}\label{eq:ReWF2}
    \mbox{Re}\{2\psi_2\}\:=\:
    \left\{
        \begin{array}{l}
            \displaystyle E^{1+(d-1)\gamma}\ell^{\gamma(d-1)}_{\gamma},\hspace{1cm}\mbox{(massless states in FRW cosmologies)}  \\
            \phantom{\ldots}\\
            \displaystyle E^{d-2\Delta_{-}}\ell_1^{d-1}\eta_{\circ}^{-2\Delta_{-}},\qquad\mbox{(massive states in $dS_{1+d}$)}
        \end{array}
    \right.
\end{equation}
where $2\Delta_{-}\,=\,d-\sqrt{d^2-(2m\ell_1)^2}$ and $\gamma\,=\,0,\,1$ respectively correspond to the flat- and $dS$-space cases in $(d+1)$-dimensions. 

Let us first focus on the case of massless states. Using the quantisation condition $LE\,=\,2\pi n$, the argument of the exponential in \eqref{eq:PDf} can be written as
\begin{equation}
    \frac{2\alpha\ell^{d-1}_{\gamma}f}{L^d\mbox{Re}\{2\psi_2\}}\: :=\:
    \frac{2\alpha f}{(2\pi n)^{1+(d-1)\gamma}}\left(\frac{\ell_{\gamma}}{L}\right)^{(d-1)(1-\gamma)}
\end{equation}
where $\alpha$ has been rescaled by $\ell_{\gamma}^{d-1}$ to make it dimensionless\footnote{In flat space, there is not a scale such as $\ell_{\gamma}$. In this case, we can keep $\alpha$ as dimensionful.}. In flat space, for which $\gamma\,=\,0$, $f\,=\,1$, the product in \eqref{eq:PDf} can be easily performed to give a gamma function which approaches $1$ when the IR cut-off is removed, and hence the mean square displacement distribution returns a $\delta$-function
\begin{equation}\label{eq:MSDflat}
    \left.\mathcal{P}(\Delta)\right|_{\mbox{\tiny flat}}\:=\:
        \lim_{L\longrightarrow\infty}\int_{-i\infty}^{+i\infty}\frac{d\alpha}{2\pi i}\,e^{\alpha\Delta}
            e^{\frac{\alpha\gamma_{\mbox{\tiny EM}}}{\pi L^{d-1}}}\Gamma\left(1+\frac{\alpha}{\pi L^{d-1}}\right)\:=\:\delta(\Delta).
\end{equation}
This implies that in the ensemble of the fluctuations determined by a flat-space wavefunction, the distance between two independently chosen field configuration is picked in zero: the production of a plethora of boundary configurations does not occur in flat-space.

For massless states in an FRW cosmology with warp factor $a(\eta)\,=\,(-\ell_{\gamma}/\eta)^{\gamma}$, the argument of the exponentials in \eqref{eq:PDf} no longer depends on the IR cut-off scale $L$. The integral \eqref{eq:PDf} can be solved exactly for $d=1$ and $\gamma=1$, {\it i.e} in $dS_{1+1}$, while for $dS_{1+d}$ it can be computed via the saddle point approximation ~\cite{Anninos:2011kh}~.  The most general case can be just treated numerically ~\cite{Shaghoulian:2013qia}~. However, in all cases the result is an asymmetric distribution: for de Sitter ($\gamma\,=\,1$), it is a Gumbel distribution ~\cite{Gumbel:1954evd}
\begin{equation}\label{eq:Gumbel}
    \left.\mathcal{P}(\Delta)\right|_{\mbox{\tiny dS}} \:\sim\: e^{-\kappa[a(\Delta+b)+e^{-a(\Delta+b)}]},
\end{equation}
where $\kappa$, $a$ and $b$ are suitable constants that depend on $d$. For FRW cosmologies ($\gamma\neq0,1$), the mean square displacement follows a Weibull distribution ~\cite{Shaghoulian:2013qia}, which has the form ~\cite{Weibull:1951evd}
\begin{equation}\label{eq:Weibull}
    \left.\mathcal{P}(\Delta)\right|_{\mbox{\tiny FRW}} \:\sim\: a(\Delta+\Delta_{\gamma})^{b-1} \, e^{-a(\Delta+\Delta_{\gamma})^b}\vartheta(\Delta+\Delta_{\gamma}),
\end{equation}
with $a$, $b$ and $\Delta_{\gamma}>0$ depending on $d$ and $\gamma$ and $\vartheta(\Delta)$ being the Heaviside step function. In both cases, it falls exponentially for large and positive $\Delta$. For negative $\Delta$, the Gumbel distribution still falls off  but faster than for large and positive $\Delta$, while the Weibull distribution has an exponential tail until some finite point $-\Delta_{\gamma}$, after which it no longer has support. Such a cut off becomes less negative as the space-time expansion  acceleration, parametrised by $\gamma$, increases ~\cite{Shaghoulian:2013qia}~. The asymmetry of the fall-off for positive and negative values of $\Delta$ for both the Gumbel and Weibull distributions, is the manifestation of the fact that there is a plethora of dissimilar configurations.

\begin{figure}[t]
    \centering
    \includegraphics[width=.625\textwidth, height=.625\textheight, keepaspectratio]{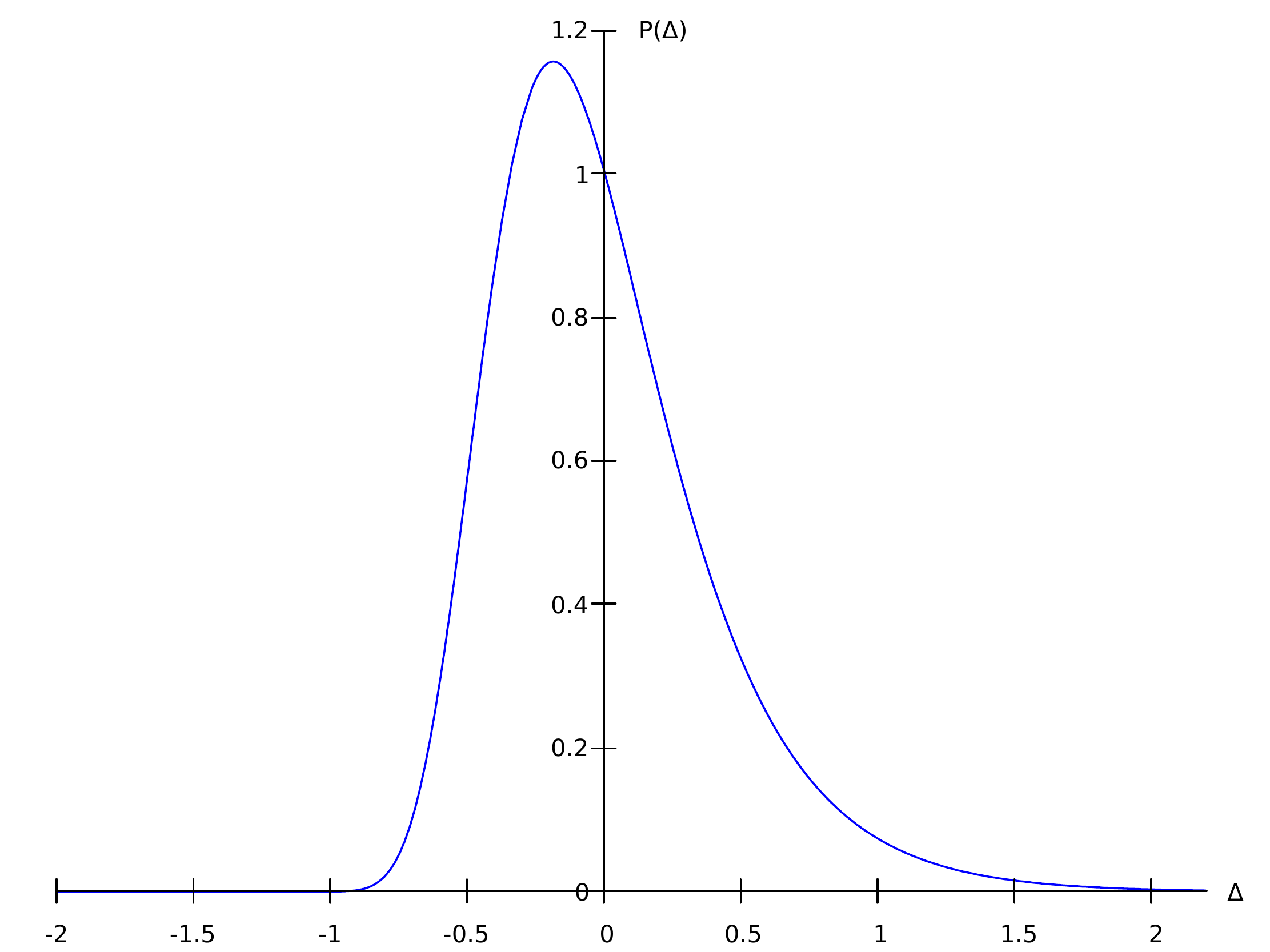}
    \caption{Mean square displacement distribution for two field configurations. It represents the Gumbel distribution emerging for a massless scalar in $dS_2$.}
    \label{fig:Gumbel}
\end{figure}

Similar results can be found for ultra-light states in $dS_{1+d}$, {\it i.e.} massive states such that $\Delta_{-}<d/4$: the mean square displacement distribution $\mathcal{P}(\Delta)|_{\mbox{\tiny u.l.}}$ turns out to be an asymmetric distribution which smoothly connects to the Gumbel distribution as the massless limit is taken ~\cite{Roberts:2012jw}~. For heavier states, the product in \eqref{eq:PDf} is dominated by UV modes and the mean square displacement distribution is a Gaussian.

Let us further consider the mean square displacement distribution among three fields configurations \eqref{eq:ProbDistr3}:
\begin{equation}\label{eq:ProbDistr3b}
    \begin{split}
        \mathcal{P}(\Delta_1,\Delta_2,\Delta_3)&\:=\:
        \lim_{L\longrightarrow\infty}\int_{-i\infty}^{+i\infty}
            \prod_{j=1}^3\left[\frac{d\alpha_j}{2\pi i}\,e^{\alpha_j\Delta_j}\right]\times\\
        &\times
        \prod_{\vec{p}\neq0}\,\frac{\displaystyle e^{2\sum_{j=1}^3\frac{\alpha_j f_j}{\mbox{\footnotesize Re}\{2\psi_{2}\}}}}{
            \displaystyle 1+
            2\sum_{j=1}^3\frac{\alpha_j f_j}{\mbox{\footnotesize Re}\{2\psi_2\}}+3\sum_{j=1}^3\frac{s_js_{j+1}f_jf_{j+1}}{\mbox{\footnotesize Re}\{2\psi_2\}}
            }
    \end{split}
\end{equation}
Using the quantisation condition \eqref{eq:compactp}, one can try to infer the solution: a closed form for the product in the integrand as well as the exact integrations can be carried out just in a very limited amount of cases, such as  heavy states in $dS_{1+d}$ and both massless and massive states in flat-space in $d$-dimensions. In the other cases, it is either necessary to resort to the saddle point approximation, as for massless states in $dS_{1+d}$ or to numerics as for ultra-light states in $dS_{1+d}$ and massless states in FRW cosmologies.

It is useful to consider the conditional probability that two fields configurations, namely $\Phi_1$ and $\Phi_2$, are at a distance $\Delta_{3}$ when the distances between each pair $(\Phi_2,\,\Phi_3)$ and $(\Phi_{3},\,\Phi_{1})$ are respectively $\Delta_1$ and $\Delta_2$:
\begin{equation}\label{eq:CondProbMSD}
    \mathcal{P}(\Delta_3\,|\,\Delta_1,\,\Delta_2)\:=\:\frac{\mathcal{P}(\Delta_1,\Delta_2,\Delta_3)}{\displaystyle
        \int d\Delta_3\,\mathcal{P}(\Delta_1,\Delta_2,\Delta_3)}
\end{equation}
It turns out that the distribution \eqref{eq:CondProbMSD} when considered for massless states in $dS_{1+d}$, FRW cosmologies as well as ultra light states, is picked for configurations with $\Delta_3\,=\,\mbox{max}\{\Delta_1,\Delta_2\}$, {\it i.e.} for isosceles triangles, with the unequal side -- $\mbox{min}\{\Delta_1,\Delta_2\}$ -- being the shortest among the three. This property goes under the name of {\it ultrametricity}\footnote{It was originally found and discussed in the context of spin glasses ~\cite{Parisi:1983dx, Rammal:1986zz}~.}. However, what does this mean? One can imagine that the space of configurations forms a tree-like structure, with its leaves corresponding to the different configurations and the distance between two configurations measure the distance between the corresponding leaves of the tree and a common branch.

As the mean square displacement distributions were originally introduced to sharpen the question about the relation between cluster decomposition and the branch-diffusion process that the wavefunction undergoes as it evolves towards future infinity, it is useful to compare the existence of the ultrametric structure in field space with the cluster decomposition properties in $dS_{1+d}$.

Recall that the condition for cluster decomposition on the two-point wavefunction $\psi_2$ of free states is that in the late-time limit its real part scales with energy as $E^{\alpha}$ with $\alpha\in]0,\,d[$. Notice that in $dS_{1+d}$ one can relate $\alpha$ to the scaling dimension $\Delta_{-}$ and, hence, to the mass of the states. Thus, in terms of $\Delta_{-}$, cluster decomposition occurs if $\Delta_{-}\in]0,d/2[$. However, the analysis of the mean square displacement distributions for massive states in $dS_{1+d}$ shows the emergence of extreme value distributions and ultrametricity for $\Delta_{-}<d/4$ (ultra light states), while for $\Delta_{-}>d/4$ (heavier states) they do not occur. So, interestingly enough the existence of a plethora of late-time configurations as well as a tree-like structure in such a space does not seem to be in a one-to-one correspondence with the lack of global cluster decomposition.

\subsection{Singularities and Factorisations}\label{subsec:SingFact}

The choice of the Bunch-Davies vacuum as the initial state reflects in the way that the wavefunction depends on the momenta and, consequently, on the way that the kinematic space for a given process can be parametrised. Concretely, the Bunch-Davies condition selects the positive energy solutions for the mode functions such that they vanish at early times. This means that the physical region of kinematic space for the Bunch-Davies wavefunction is defined by all the energies being positive:
\begin{equation}\label{eq:physreg}
    \{E_j\,\ge\,0\:|\:j=1,\ldots,n\} \mbox{ and } 
    \left\{
      y_\mathcal{I}:=\left|\sum_{k\in\mathcal{I}}\vec{p}_k\right|\,\ge\,0,\:\Big|\:\forall\,\mathcal{I}\subset\{1,\ldots,n\}
    \right\} 
\end{equation}
Being a vacuum wavefunction, no particle production should be allowed in the physical region, which implies that no singularity of the form $\sum_{j}\sigma_jE_j+\sum_k\sigma_{\mathcal{I}} y_{\mathcal{I}}=0$ are allowed, $E_j$'s, $y_{\mathcal{I}}$'s being the (non-zero) energies of the external and internal states and the $\sigma$'s being suitable signs such that they cannot be all equal. The singularities of this type are called {\it folded singularities}, and the above property can be stated by saying that the Bunch-Davies wavefunction does not have folded singularities. This is precisely the signature that cosmic structure of the universe has originated by quantum fluctuations ~\cite{Green:2020whw}~.

As a consequence, the singularities can be only of the form $\sum_jE_j+\sum_{\mathcal{I}} y_{\mathcal{I}}\,=\,0$, {\it i.e.} taking all the $\sigma$'s above to be equal. This sheet in kinematic space can be reached in the physical region just if all the energies $E_j$'s and $y_k$'s all vanish. This type of combinations of energies are referred to as {\it partial energies}.

Also, notice that the wavefunction has an overall spatial momentum conservation associated to the invariance under spatial translations, but the expansion breaks time-translation invariance. Consequently, rather than having support on an energy conserving delta-function, the wavefunction will also depend on the {\it total energy}, {\it i.e.} the sum of all the energies of the external states. Such a sum can vanish in the physical region of kinematic space if and only if all the external energies vanish.

If on one side in the physical region the sheet where the total and partial energies vanish is reached in a trivial way, on the other it is possible to perform an analytic continuation such that some energies become negative and other stays positive and hence we can reach these sheets for non-zero external and internal energies. In these sheets outside the physical region, the wavefunction can develop singularities. One lesson learned in the context of scattering amplitudes is that their physical content is encoded in the coefficients of their singularities, which are typically constrained by unitarity (see $\:$ \cite{Elvang:2013cua, Benincasa:2013faa, Henn:2014yza} for an extensive account on this subject). So, what happens to the wavefunction when the total and partial energies singularities are reached?


\subsubsection{Total energy singularity and the flat-space limit}\label{subsubsec:EtotFS}

Let us begin with considering the total energy singularity first and let us try to get some intuition. Let us label to sum of the energies of all the external states as $E_{\mbox{\tiny tot}}$. In order to take the limit $E_{\mbox{\tiny tot}}$ in a non-trivial way,
we need to analytic continue the energies outside the physical region in a region where some of them can take negative value. This implies that some states can be considered as {\it in-states} and others as {\it out-states}. As $E_{\mbox{\tiny tot}}\,\longrightarrow\,0$, energy conservation get imposed: we are in the situation where, together with in- and out-states we have also conservation of the total energy, {\it i.e.} in the very same situation of a more familiar flat-space scattering process. Hence, we would expect that the coefficient of the leading term in the expansion of the wavefunction around $E_{\mbox{\tiny tot}}\,=\,0$ returns a flat-space scattering amplitude ~\cite{Maldacena:2011nz, Raju:2012zr}~.

Let us consider the time-integral representation of the wavefunction as given by \eqref{eq:Feyn1}. Recall that as $\eta\longrightarrow-\infty(1-i\varepsilon)$ the mode functions are exponentially suppressed, see \eqref{eq:eom}. If we consider
a ``center-of-mass time'' $\bar{\eta}$ and take $\bar{\eta}\longrightarrow-\infty(1-i\varepsilon)$, then the wavefunction $\psi_{\mathcal{G}}$ can be written as
\begin{equation}\label{eq:WFet}
    \psi_{\mathcal{G}}\:\sim\:\int\limits_{-\infty(1-i\varepsilon)}d\bar{\eta}\,f(\bar{\eta})\,e^{iE_{\mbox{\tiny tot}}\bar{\eta}}\,
        \tilde{\psi}_{\mathcal{G}}
\end{equation}
where the explicit form of the function $f(\bar{\eta})$ depends on both the specific cosmology and the modes considered and determines whether the total energy singularity is a (multiple) pole or a branch point, while $\tilde{\psi}_{\mathcal{G}}$ is the part of wavefunction obtained via the change of variables and that can be expressed in terms of time-integrals other than the one over the ``center-of-mass time'' $\bar{\eta}$. In the physical region, $E_{\mbox{\tiny tot}}\,>\,0$
and hence there is no contribution from early time. However, if perform an analytic continuation such that we can reach the sheet $E_{\mbox{\tiny tot}}\,=\,0$ which precisely where the integral gets a non-trivial contribution: hence it has support on $\delta(E_{\mbox{\tiny tot}})$.  Taking the ``center-of-mass'' time to early times also moves the process infinitely away from the space-like boundary at future infinity, which together with the restoration of time-translation invariance and the presence of in- and out-states return the conditions of a flat-space scattering process. 

\begin{figure}[t]
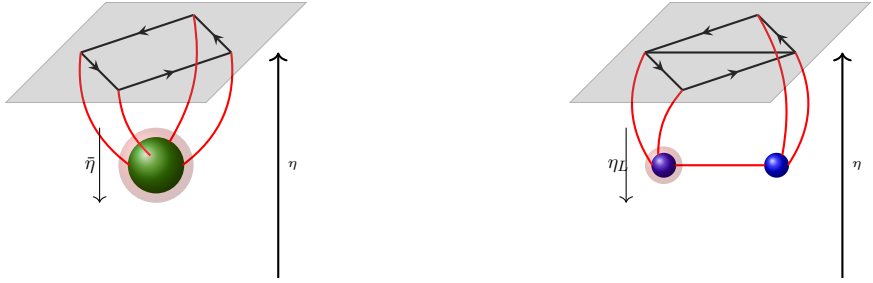

    \centering

    \caption{Behaviour of the Bunch-Davies wavefunction as the total and partial energy singularities are approached. In the first case (on the left), the coefficient of the leading term is a flat-space scattering amplitudes. For states which do not have a flat-space counterpart the singularity is milder and its leading coefficient is a purely cosmological effect. When the partial energies are approached (on the right), the Bunch-Davies wavefunction factorises into a product of a lower-point/lower-loop 
    scattering amplitude and a sum of the same lower-point/lower-level wavefunction over the signs of the energies associated to the
    propagators along which the original Bunch-Davies wavefunction factorises.}
    \label{fig:WFsings}
\end{figure}

As a matter of an example, let us consider as a toy model of a massless scalar with a polynomial interaction $\phi^k$ in flat space-time with a space-like boundary at $\eta\,=\,0$ and let us compute the contribution to the wavefunction of the universe of a contact graph:
\begin{equation}\label{eq:PsiPhik}
    \psi_{\mathcal{G}}\:=\:i\lambda_k
        \delta^{\mbox{\tiny $(d)$}}\left(\sum_{j=1}^k\vec{p}^{\mbox{\tiny $(j)$}}\right)
        \int\limits_{-\infty(1-i\varepsilon)}^0d\eta\,e^{iE_{\mbox{\tiny tot}}\eta}\:=\:
    \delta^{\mbox{\tiny $(d)$}}\left(\sum_{j=1}^k\vec{p}^{\mbox{\tiny $(j)$}}\right)
    \frac{\lambda_k}{E_{\mbox{\tiny tot}}}.
\end{equation}
Even in this very toy example, it is manifest how the lack of time-translation invariance -- time translation invariance is broken by the presence of a space-like boundary at $\eta\,=\,0$ -- produces a simple pole in $E_{\mbox{\tiny tot}}=0$. Its residue is nothing but the scattering amplitude $\mathcal{A}_k\,=\,\lambda_k$ for a $\phi^k$ interaction. Notice that \eqref{eq:PsiPhik} holds for a conformally coupled scalar in any FRW cosmology provided that the space dimension $d$ and the degree $k$ of the interaction are related to each other via $d=(k+2)/(k-2)$ -- there are all cases of a conformal theory in a conformally flat background. If we still consider a conformally-coupled scalar away from the conformal point in FRW cosmologies of the type $a(\eta)\,=\,(-\ell_{\gamma}/\eta)^{\gamma}$ ($\gamma\in\mathbb{R}_{+}$), then
\begin{equation}\label{eq:PsiPhikcc}
    \psi_{\mathcal{G}}\:=\:i\ell_{\gamma}^{\gamma}\lambda_k
        \delta^{\mbox{\tiny $(d)$}}\left(\sum_{j=1}^k\vec{p}^{\mbox{\tiny $(j)$}}\right)
        \int\limits_{-\infty(1-i\varepsilon)}^0\frac{d\eta}{\eta^{\gamma[2+\frac{(2-k)(d-1)}{2}]}}\,e^{iE_{\mbox{\tiny tot}}\eta}.
\end{equation}
If the power of $\eta$ in the integrand is negative, {\it i.e.} $d<(k+2)/(k-2)$, then the wavefunction coefficient in question has a multiple pole in the total energy of multiplicity $\mu=\gamma[(k-2)(d-1)/2-2]+1$. Notice that the multiplicity are tied to the detailed of the interactions and specifically to its mass dimension (here parametrised by the number of points $k$ of the interaction and the spatial dimension $d$ of the space-time) as well as of the expansion acceleration (here parametrised by $\gamma$). For more general graph contributions than the contact one, the multiplicity $\mu$ of the total energy pole is then given by
\begin{equation}\label{eq:mulpol}
    \mu\:=\:1+\sum_{v\in\mathcal{V}}\gamma\left[\frac{(k_v-2)(d-1)}{2}-2\right]
\end{equation}
where $k_v$ is the number of points of the interaction at the vertex $v$, and the right-hand-side is tied to the mass dimension of the interaction at each vertex. The relation between the multiplicity of the total energy pole and the mass dimension of the interactions were first noticed in the context of the actual correlators and was deduced both via the analysis of the time-integral representation ~\cite{Goodhew:2020hob} and using scale invariance and dimensional analysis ~\cite{Pajer:2020wxk}~. 

If instead the power of $eta$ in the integrand \eqref{eq:PsiPhikcc} is positive, {\it i.e.} $d>(k+2)/(k-2)$, the integral returns a transcendental function. In both cases, the coefficient of the leading total energy singularity is (proportional to) the flat-space scattering amplitude associated to the same graph.

This very simple example illustrates quite neatly how the total energy singularity is a consequence of time-translation invariance been broken, and the details of the type of singularity (simple pole, multiple pole, branch-point) are a consequence of the time-dependence of the interaction, with simple poles arising either in flat-space or in conformal cases. We will come back to this later on.

As an important remark, the expansion of the universe not only breaks time-translation invariance but also Lorentz boosts. When we consider the flat-space limit by going to the total energy singularity, while it implies time-translation invariance, Lorentz symmetry may or may not be restored and the coefficient of the leading term may be a usual Lorentz-invariant scattering amplitude ~\cite{Arkani-Hamed:2018ahb} or a boostless amplitude ~\cite{Pajer:2020wnj, Pajer:2020wxk}~. Finally, if the process does not have a non-trivial scattering amplitude as the flat-space limit is taken, the total energy singularity turns out to be milder and its coefficient is a purely cosmological effect ~\cite{Grall:2020ibl}~.


\subsubsection{Partial energy singularities and factorisations}\label{subsubsect:EparFct}

Let us move on to consider the partial energy singularities. A large part of the previous discussion applies. Any partial energy singularity can only be reached upon an analytic continuation that makes some of the energies of the external states negative. In such a sheet, energy conservation is imposed on a subprocess, for which we can mimic the entire discussion of the total energy pole. 

Let $\mathcal{L}$ and $\mathcal{R}$ be two sets of external states such that $\mathcal{L}\cup\mathcal{R}\,=\,\{1,\ldots,n\}$. We are interested in studying the behaviour of the Bunch-Davies wavefunction when $E_{\mbox{\tiny tot}}^{\mbox{\tiny $(\mathcal{L})$}} := E_{\mathcal{L}} + \sum_{e\in\centernot{\mathcal{E}}}y_e$ it taken to zero, $E_{\mathcal{L}}$ and $\centernot{\mathcal{E}}\subset\mathcal{E}$ being respectively the sum of all the energies of the external states in the subset $\mathcal{L}$ and a subset of the internal states. First, notice that taking $E_{\mbox{\tiny tot}}^{\mbox{\tiny $(\mathcal{L})$}} \longrightarrow 0$ means imposing energy conservation on the subprocess involving the elements of $\mathcal{L}\cup\centernot{\mathcal{E}}$ as external states. Let us explicitly consider the time-integral representation for the wavefunction \eqref{eq:Feyn1} and organising it according to the division of all the states into the two mutually complementary subsets $\mathcal{L}$ and $\mathcal{R}$:
\begin{equation}\label{eq:WFtintLR}
    \begin{split}
        \psi_{\mathcal{G}}\:=\:\int\prod_{e\in\mathcal{E}}\frac{d^d q_e}{(2\pi)^d}
            &\left[
                \int_{-\infty}^0\prod_{v_{\mbox{\tiny $\mathcal{L}$}}\in\mathcal{V}_{\mbox{\tiny $\mathcal{L}$}}} 
                    \left[d\eta_{v_{\mbox{\tiny $\mathcal{L}$}}}\phi_+^{\mbox{\tiny $(v)$}}V_{v_{\mbox{\tiny $\mathcal{L}$}}}\right]
                    \prod_{e_{\mbox{\tiny $\mathcal{L}$}}\in\mathcal{E}_{\mbox{\tiny $\mathcal{L}$}}}
                    G(y_{e_{\mbox{\tiny $\mathcal{L}$}}};\eta_{v_{e_{\mbox{\tiny $\mathcal{L}$}}}},\eta_{v'_{e_{\mbox{\tiny $\mathcal{L}$}}}})
        \right]\times\\
        &\hspace{-.75cm}\times
            \left[
                \int_{-\infty}^0\prod_{v_{\mbox{\tiny $\mathcal{R}$}}\in\mathcal{V}_{\mbox{\tiny $\mathcal{R}$}}} 
                    \left[d\eta_{v_{\mbox{\tiny $\mathcal{R}$}}}\phi_+^{\mbox{\tiny $(v)$}}V_{v_{\mbox{\tiny $\mathcal{R}$}}}\right]
                    \prod_{e_{\mbox{\tiny $\mathcal{R}$}}\in\mathcal{E}_{\mbox{\tiny $\mathcal{R}$}}}
                    G(y_{e_{\mbox{\tiny $\mathcal{R}$}}};\eta_{v_{e_{{\mbox{\tiny $\mathcal{R}$}}}}},\eta_{v'_{e_{\mbox{\tiny $\mathcal{R}$}}}})
        \right]\times\\
        &\hspace{-.75cm}\times
        \prod_{\centernot{e}\in\centernot{\mathcal{E}}}G(y_{\centernot{e}};
            \eta_{v_{\mathcal{L}_{\centernot{e}}}},\eta_{v'_{\mathcal{R}_{\centernot{e}}}})
    \end{split}
\end{equation}
where $\vec{q}_e$ is the momentum flowing the edge $e\in\mathcal{E} = \mathcal{E}_{\mathcal{L}} \cup \mathcal{E}_{\mathcal{R}} \cup \centernot{\mathcal{E}}$, with the relevant momentum conserving delta-functions left implicit, and the propagators in the last line connect the two subsets $\mathcal{L}$ and $\mathcal{R}$. Let us now perform a change of variables in order to introduce a ``center-of-mass time'' $\bar{\eta}_{\mathcal{L}}$ for the subprocess involving the states in the set $\mathcal{L}\cup\centernot{\mathcal{E}}$ whose total energy is precisely $E_{\mbox{\tiny tot}}^{\mbox{\tiny $(\mathcal{L})$}}$. The propagators associated to the states $\centernot{\mathcal{E}}$ connecting $\mathcal{L}$ and $\mathcal{R}$ which appear in the last line of \eqref{eq:WFtintLR} depends on $\bar{\eta}_{\mbox{\tiny $\mathcal{L}$}}$. For the sake of clarity, let us consider a single bulk-to-bulk propagator $G(y_{\centernot{e}},\eta_{v_{\centernot{e}}},\eta_{v'_{\centernot{e}}})$ as given by its general expression \eqref{eq:G}, and let us take $\eta_{v_{\centernot{e}}}\longrightarrow-\infty(1-i\varepsilon)$. Then
\begin{equation}\label{eq:GmInf}
    \lim_{\eta_{v_{\centernot{e}}}\longrightarrow-\infty(1-i\varepsilon)}
        G(y_{\centernot{e}},\eta_{v_{\centernot{e}}},\eta_{v'_{\centernot{e}}})\:\sim\:
        \frac{f_{\centernot{e}}(\eta_{v_{\centernot{e}}})e^{iy_{\centernot{e}}\eta_{v_{\centernot{e}}}}}{\mbox{Re}\{2\psi_2(y_{\centernot{e}})\}}
        \left[
            \bar{\phi}_{+}(-y_{\centernot{e}}\eta_{v'_{\centernot{e}}})-\phi_{+}(-y_{\centernot{e}}\eta_{v'_{\centernot{e}}})
        \right]
\end{equation}
and the bulk-to-bulk propagator $G(y_{\centernot{e}},\eta_{v_{\centernot{e}}},\eta_{v'_{\centernot{e}}})$ returns a linear combination of bulk-to-boundary propagators for the same state, which differ for the sign of the energy $y_{\centernot{e}}$.

Let us now turn to the bulk-to-bulk propagators $G(y_{e_{\mbox{\tiny $\mathcal{L}$}}};\eta_{v_{e_{\mbox{\tiny $\mathcal{L}$}}}},\eta_{v'_{e_{\mbox{\tiny $\mathcal{L}$}}}})$ contributing to the subprocess $\mathcal{L}\cup\centernot{\mathcal{E}}$. If we take $\eta_{v_{e_{\mbox{\tiny $\mathcal{L}$}}}},\,\eta_{v'_{e_{\mbox{\tiny $\mathcal{L}$}}}}\,\longrightarrow\,-\infty(1-i\varepsilon)$, then the term due to the boundary condition that the fluctuation vanishes at the boundary, is exponentially suppressed, and the bulk-to-bulk propagator reduces to the flat-space Feynman propagator, up to an overall function which depends on the two times and it is due to the behaviour of the mode functions at early times.

Thus, when we take the center-of-mass time $\bar{\eta}_{\mathcal{L}}$ to the infinite past, all the bulk-to-boundary propagators associated to the external states in $\mathcal{L}$ behave as exponential, while the bulk-to-bulk propagators behave as \eqref{eq:GmInf}. So, we can write
\begin{equation}\label{eq:PsiEL}
    \psi_{\mathcal{G}}\:\sim\:
        \left(
            \int\limits_{-\infty(1-i\varepsilon)}d\bar{\eta}_{\mathcal{L}}f_{\mathcal{L}}(\bar{\eta}_{\mathcal{L}})
            e^{iE_{\mbox{\tiny tot}}^{\mbox{\tiny $(\mathcal{L})$}}\bar{\eta}_{\mathcal{L}}}
        \right)
        \mathcal{A}(\mathcal{L},\centernot{\mathcal{E}})\times
        \sum_{\{\sigma_{\centernot{e}}\}=\{\mp1\}}
        \frac{\psi(\centernot{\mathcal{E}}(\sigma_{\centernot{e}}),\mathcal{R})}{\mbox{Re}\{2\Psi_2(y_{\centernot{e}})\}}
\end{equation}
with the term in brackets not vanishing if and only if $E_{\mbox{\tiny tot}}^{\mbox{\tiny $(\mathcal{L})$}}\,=\,0$, the function $f_{\mathcal{L}}(\bar{\eta}_{\mathcal{L}})$ determining the type of singularity, $\psi(\centernot{\mathcal{E}}(\sigma_{\centernot{e}}),\mathcal{R})$ being the wavefunction with external states given by the set $\centernot{\mathcal{E}}\cup\mathcal{R}$, and $\sigma_{\centernot{e}}$ being the sign of the energy $y_{\centernot{e}}$ of the state labelled by $\centernot{e}\in\centernot{\mathcal{E}}$. Hence, when a partial energy singularity is approached, the wavefunction factorises in a flat-space scattering amplitude and a linear combination of the wavefunctions for the same process but with different signs of the energies for the states $\centernot{e}$ which goes to the boundary.

This factorisation property is very similar to the factorisation theorems for scattering amplitudes, where as a singularity is approached, some internal states go on-shell and the scattering amplitude factorise in lower-point/lower-loop scattering amplitudes. However, while in the latter case they are a direct consequence of unitarity (together with the positivity of the coefficient of the singularity), the cosmological optical theorem ~\cite{Goodhew:2020hob} return different cutting rules ~\cite{Melville:2021lst, Goodhew:2021oqg, Meltzer:2021zin} than \eqref{eq:PsiEL} and no actual singularity of the wavefunction is crossed.



\subsection{Unitarity}\label{subsec:UN}

Loosely speaking, unitarity is the statement that the probabilities have to sum to $1$. This translates into the requirement that the evolution operator $\hat{U}$, defined in Section \ref{subsec:WF}, has to be a unitary operator, {\it i.e.}:
\begin{equation}\label{eq:UnU}
    \hat{U}\hat{U}^{\dagger}\:=\:\hat{\mathbb{1}}\:=\:\hat{U}^{\dagger}\hat{U}.
\end{equation}
Before discussing the constraints that the condition \eqref{eq:UnU} imposes on the analytic structure of the quantum mechanical observables in cosmology, it is worth to make an important remark.

For the sake of clarity, let us recall the definition of the evolution operator $\hat{U}$
\begin{equation}\label{eq:UnDef}
    \hat{U}(\eta_{\circ})\: :=\:e^{-i\int_{-\infty}^{\eta_{\circ}} d\eta\,\hat{H}(\eta)}.
\end{equation}
The definition \eqref{eq:UnDef} needs to be regularised in such a way to pick the positive energy solution in the infinite past. A general way to perform such a regularisation is via a standard $i\varepsilon$-prescription in the infinite past, $\eta\,\longrightarrow\,-\infty(1-i\varepsilon)$,
\begin{equation}\label{eq:UnEps}
    \hat{U}_{\varepsilon}(\eta_{\circ})\: :=\: e^{-i\int_{-\infty(1-i\varepsilon)}^{\eta_{\circ}} d\eta\,\hat{H}(\eta)}.
\end{equation}
However, the $i\varepsilon$-regularised evolution operator $\hat{U}_{\varepsilon}$ is no-longer unitary, as $\hat{U}_{\varepsilon}\hat{U}^{\dagger}_{\varepsilon}\,\neq\,\hat{\mathbb{1}}$. In order to deduce unitarity constraints on the observables from the unitarity of the evolution operator, it is crucial to regularise the latter in the infinite past in such a way that the regularised evolution operator is still unitary.

Rather than modifying the contour of integration, we can consider an $\varepsilon$-deformed Hamiltonian $\hat{H}_{\varepsilon}(\eta)$ ~\cite{Baumgart:2020oby}~,
\begin{equation}\label{eq:Hdef}
    \hat{H}_{\varepsilon}(\eta)\: :=\:\hat{H}(\eta)e^{\varepsilon\eta}.
\end{equation}
Importantly, $\hat{H}_{\varepsilon}(\eta)$ is still Hermitian, and the $\varepsilon$-deformed evolution operator
\begin{equation}\label{eq:UnEpsU}
    \hat{U}_{\varepsilon}(\eta_{\circ})\: :=\: 
        e^{-i\int_{-\infty}^{\eta_{\circ}} d\eta\,\hat{H}_{\varepsilon}(\eta)}\:=\:
                                 e^{-i\int_{-\infty}^{\eta_{\circ}} d\eta\,e^{\varepsilon\eta}\hat{H}(\eta)}
\end{equation}
is manifestly unitary. With such a deformed evolution operator at hand, it is possible to compute both the bulk-to-boundary $\tilde{\phi}_{\circ}$ and bulk-to-bulk propagators $G(y;t,t')$. Such ``unitary'' propagators turn out to pick an overall $\varepsilon$-dependent exponential factor
\begin{equation}\label{eq:UnProps}
    \tilde{\phi}_{\circ}^{\mbox{\tiny $(\varepsilon)$}}(-E\eta)=e^{\varepsilon\eta}\tilde{\phi}_{\circ}(-E\eta),\qquad
    G_{\varepsilon}(y;\,\eta,\,\eta')\:=\:e^{\varepsilon(\eta+\eta')}G(y;\,\eta,\,\eta').
\end{equation}
It is possible to show that the correlators computed using the deformed propagators \eqref{eq:UnProps} are equivalent to the ones obtained via the in-in formalism. We will not go through the proof of this statement, rather we refer to the original paper ~\cite{Baumgart:2020oby} as the focus of the present discussion is the avatar of unitarity in the analytic structure of the quantum mechanical observables in cosmology.


\subsubsection{The cosmological optical theorem}

Let us consider the $\varepsilon$-deformed evolution operator \eqref{eq:UnEpsU}  with some fixed early time $\eta\,=\,-T$. Let us split it as $\hat{U}_{\varepsilon}^{\mbox{\tiny $(T)$}}\:=\:\hat{\mathbb{1}}+\hat{V}^{\mbox{\tiny $(T)$}}_{\varepsilon}$, such that the unitary condition $\hat{U}_{\varepsilon}^{\mbox{\tiny $(T)$}}\hat{U}_{\varepsilon}^{\mbox{\tiny $(T)$}\dagger}\,=\,\hat{\mathbb{1}}$ acquires the form
\begin{equation}\label{eq:Un2}
    \hat{V}^{\mbox{\tiny $(T)$}}_{\varepsilon}+\hat{V}_{\varepsilon}^{\mbox{\tiny $(T)$}\dagger}\:=\:
    -\hat{V}_{\varepsilon}^{\mbox{\tiny $(T)$}}\hat{V}_{\varepsilon}^{\mbox{\tiny $(T)$}\dagger}.
\end{equation}
Such a split separates the free part and the pure interactions, the latter being encoded in the operator $\hat{V}$. Let $\Phi$ and $\Phi'$ be two field configuration, then
\begin{equation}\label{eq:Un2a}
    \langle\Phi|\hat{V}_{\varepsilon}^{\mbox{\tiny $(T)$}}|\Phi'\rangle+\langle\Phi|\hat{V}_{\varepsilon}^{\mbox{\tiny $(T)$}\dagger}|\Phi'\rangle\:=\:
    -\langle\Phi|\hat{V}_{\varepsilon}^{\mbox{\tiny $(T)$}}\hat{V}_{\varepsilon}^{\mbox{\tiny $(T)$}\dagger}|\Phi'\rangle.
\end{equation}
Inserting a complete set of field configurations, the unitary equation above further becomes
\begin{equation}\label{eq:Un3}
    \langle\Phi|\hat{V}_{\varepsilon}^{\mbox{\tiny $(T)$}}|\Phi'\rangle+\langle\Phi|\hat{V}_{\varepsilon}^{\mbox{\tiny $(T)$}\dagger}|\Phi'\rangle\:=\:
    -\tilde{\mathcal{N}}\int\mathcal{D}\tilde{\Phi}\,\langle\Phi|\hat{V}_{\varepsilon}^{\mbox{\tiny $(T)$}}|\tilde{\Phi}\rangle\langle\tilde{\Phi}|\hat{V}_{\varepsilon}^{\mbox{\tiny $(T)$}\dagger}|\Phi'\rangle
\end{equation}
As $T\,\longrightarrow\,\infty$, the vacuum is the dominant contribution while the other states are suppressed and setting $\Phi'\,=\,0$
\begin{equation}\label{eq:Un4}
    \langle\Phi|\hat{V}_{\varepsilon}|0\rangle+\langle\Phi|\hat{V}_{\varepsilon}^{\dagger}|0\rangle\:=\:
    -\tilde{\mathcal{N}}\int\mathcal{D}\tilde{\Phi}\,\langle\Phi|\hat{V}_{\varepsilon}|\tilde{\Phi}\rangle\langle\tilde{\Phi}|\hat{V}_{\varepsilon}^{\dagger}|0\rangle
\end{equation}
Notice that
\begin{equation}\label{eq:PsiInt}
    \Psi_{\circ}^{\mbox{\tiny (int)}}[\Phi]\: :=\: 
        \lim_{\varepsilon\longrightarrow0}\langle\Phi|\hat{V}_{\varepsilon}|0\rangle\:=\:
        \Psi_{\circ}[\Phi]-\Psi_{\circ}^{\mbox{\tiny (free)}}[\Phi]
\end{equation}
and the unitarity relation \eqref{eq:Un4} provides constraints on the analytic structure of the wavefunction of the universe
\begin{equation}\label{eq:PsiInt2}
    \Psi_{\circ}^{\mbox{\tiny (int)}}[\Phi;\varepsilon]+\Psi_{\circ}^{\dagger\mbox{\tiny (int)}}[\Phi;\varepsilon]\:=\:
        -\tilde{\mathcal{N}}\int\mathcal{D}\tilde{\Phi}\,\langle\Phi|\hat{V}_{\varepsilon}^{\mbox{\tiny $(T)$}}|\tilde{\Phi}\rangle\langle\tilde{\Phi}|\hat{V}_{\varepsilon}^{\mbox{\tiny $(T)$}\dagger}|0\rangle.
\end{equation}
The equation \eqref{eq:PsiInt2} constitutes the cosmological optical theorem ~\cite{Goodhew:2020hob, Benincasa:2022cot}~\footnote{In the original derivation of the cosmological optical theorem ~\cite{Goodhew:2020hob}~, the issue of how correctly perform the regularisation and the selection of the positive frequency solutions has been overlooked: the relation of the same type as \eqref{eq:PsiInt2} has been found using the unitarity of the undeformed evolution operator, while the usual $i\varepsilon$ prescription is introduced just when computing the time integrals in the perturbative expansion of \eqref{eq:PsiInt2}. The correctness of the results of ~\cite{Goodhew:2020hob} is a consequence of the fact that the correct regularisation \eqref{eq:UnEpsU} is equivalent to shifting the energies by $-i\varepsilon$ both for the wavefunction and its hermitian conjugate, and also the regularisation as $\eta\longrightarrow -\infty(1-i\varepsilon)$ at the level of the perturbative time-integrals can be thought the same way ~\cite{Benincasa:2022cot}~.} and its formulation is complete analogy with the optical theorem in flat-space. Notice that because of \eqref{eq:UnEpsU}, the regularisation prescription is the same for $\Psi_{\circ}$ and $\Psi_{\circ}^{\dagger}$. Furthermore, we can write \eqref{eq:PsiInt2} for the actual wavefunction coefficients in momentum space by using the reality conditions $\Phi^{\dagger}(\vec{p})\,=\,(-1)^d\Phi(-\vec{p})$ and $\phi^{\dagger}(-E\eta)\,=\,\phi(E\eta)$:
\begin{equation}\label{eq:PsiInt3}
    \resizebox{0.9\hsize}{!}{$\displaystyle
        \psi_{k}^{\mbox{\tiny $(L)$}}(E-i\varepsilon, y)+\psi_{k}^{\dagger\mbox{\tiny $(L)$}}(-(E-i\varepsilon),y)\:=\:
        -\int\prod_{l\in I}\left[\frac{d^dq^{\mbox{\tiny $(l)$}}}{(2\pi)^d}\right]
        \langle\{p^{\mbox{\tiny $(j)$}}\}|\hat{V}_{\varepsilon}|\{q^{\mbox{\tiny $(l)$}}\}\rangle
        \langle\{q^{\mbox{\tiny $(l)$}}\}|\hat{V}_{\varepsilon}^{\dagger}|0\rangle
    $}
\end{equation}
where $y$ is the collection of the energies of the internal states and parametrises the angles among the momenta. However, while \eqref{eq:PsiInt2} and \eqref{eq:PsiInt3} are a general statement, its right-hand-side does not have a clear general interpretation, except in perturbation theory in which context they allow to extract cutting rules for both the wavefunction ~\cite{Melville:2021lst, Baumann:2021fxj, Meltzer:2021zin} and the actual cosmological correlators ~\cite{Goodhew:2021oqg}~.


\subsubsection{Cosmological cutting rules}\label{subsubsec:Ccr}

In perturbation theory, the right-hand-side of \eqref{eq:PsiInt3} can be computed explicitly ~\cite{Goodhew:2020hob}~. A more elegant way of proceeding is to rely on an argument, as in ~\cite{Melville:2021lst}~, which follows the same lines of largest time equation\footnote{The largest time equation was also extended to correlation function in $AdS$, leading to cutting rules for $AdS$ correlators ~\cite{Meltzer:2020qbr}~.} used to prove the cutting rules for scattering amplitudes as a consequence of the flat-space unitarity ~\cite{Veltman:1994wz}~, or dispersion representations for the bulk-to-bulk propagators ~\cite{Meltzer:2021zin}~.

Let $\mathcal{G}_k^{\mbox{\tiny $(L)$}}:=\{\mathcal{G}\}$ be the collection of all the graphs contributing to $\psi_k^{\mbox{\tiny $(L)$}}$. Then the cosmological optical theorem \eqref{eq:PsiInt3} can be applied on an individual graph, and the all the methods just mentioned translate its right-hand-side into cutting rules on the individual graphs, {\it i.e.} the left-hand-side $\Delta\psi_{\mathcal{G}}:=\psi_{\mathcal{G}}+\psi_{\mathcal{G}}^{\dagger}$ of \eqref{eq:PsiInt3} is expressed in terms of all the possible ways of erasing an edge of the graph associating $\Delta\psi_{\mathfrak{g}}:=\psi_{\mathfrak{g}}+\psi_{\mathfrak{g}}^{\dagger}$ to each subgraph $\mathfrak{g}$ which $\mathcal{G}$ gets divided into :
\begin{equation}\label{eq:CCR}
    \Delta\psi_{\mathcal{G}}\:=\:\sum_{\{\mathcal{E}_c\}}
        \left[
            \prod_{e\in\mathcal{E}_c}\int\frac{d^dq_{v_e}}{(2\pi)^2}\int\frac{d^dq_{v'_e}}{(2\pi)^2}
            \frac{1}{\mbox{Re}\{2\psi_2(y_e)\}}
        \right]
        \prod_{\mathfrak{g}\subset\mathcal{G}_c}\Delta\psi_{\mathfrak{g}},
\end{equation}
where $\{\mathcal{E}_c\}$ is the collection of all the inequivalent subsets of $\mathcal{E}$ (except the empty set), $\mathcal{G}_c$ is the collection of subgraphs in which $\mathcal{G}$ gets divided when erasing the edges in a given $\mathcal{E}_c$ ($\mathcal{G}=\bigcup_{\mathfrak{g}\in\mathcal{G}_c}\mathfrak{g}$), $v_e$ and $v'_e$ are the vertices at the endpoints of the edge $e$, and all the $\psi^{\dagger}$'s have to be understood has having all the energies reversed in sign (except for the one associated to the edge $e$).

Two comments are in order. First, it is straightforward to notice that the leading coefficient for $\psi_{\mathcal{G}}$ and $\psi_{\mathcal{G}}^{\dagger}$ is the same up to a sign: $\Delta\psi_{\mathcal{G}}$ does not show the typical flat-space behaviour as the total energies are taken to zero. Also, the cutting rules \eqref{eq:CCR} provide a representation for $\Delta\psi_{\mathcal{G}}$ which, lacking of the expected flat-space behaviour, does not reproduce the flat-space cutting rules. Secondly, it is possible to consider $\Delta_{\mathfrak{g}}\psi_{\mathcal{G}}$ defined still as the sum of $\psi_{\mathcal{G}}$ and $\psi_{\mathcal{G}}^{\dagger}$, but now $\psi_{\mathcal{G}}^{\dagger}$ is defined by the reversing the energies for all the edges contained in the complementary graph $\bar{\mathfrak{g}}$ of $\mathfrak{g}$. Then \eqref{eq:CCR} still holds, but now $\{\mathcal{E}_c\}$ is the collection of all the inequivalent ways of grouping the edges of $\bar{\mathfrak{g}}$.


\subsection{Manifest locality}\label{subsec:ML}

The concept of locality  is tied to the cluster decomposition principle discussed in Section \ref{subsec:CD}, at least in flat-space: scattering amplitudes are sufficiently analytic functions with at most poles and branch-cuts all of them corresponding to the propagating particles which can transport information between two events provided that they are not space-like separated. Consequently, in the usual Lagrangian language, this means that no singularity can come from the actual interactions, which ought to be polynomial in the momenta.

However, in an expanding universe the cluster decomposition principle does not hold in general as even free massless fields are correlated at future infinity and, consequently, the physics is inherently non-local. However, one can still restrict to interactions which do not contribute to the singularity structure of the wavefunction. Such interactions are referred to in the literature  as {\it manifestly local} ~\cite{Jazayeri:2021fvk}~, with reference to how locality manifest in the analytic structure of scattering amplitudes.

The requirement of manifest locality selects a very special class of all the admissible interactions, as there exist some whose Lagrangian description involve non-local operators ~\cite{Maldacena:2002vr}~. In terms of the analytic structure of the wavefunction coefficients, it translates into the statement that they are regular as the energy of the internal states are taken to be soft ~\cite{Jazayeri:2021fvk}~. More precisely, for massless\footnote{In this review we have been focusing on scalars only. However, the manifestly local condition \eqref{eq:ML} as for the massless scalar applies also to gravitons ~\cite{Jazayeri:2021fvk}~.} and conformally-coupled scalars in $d=3$ it reads
\begin{equation}\label{eq:ML}
    \left.\frac{\partial}{\partial E_j}\psi_{\mathcal{G}}(\{E_j\}\})\right|_{E_j=0}\:=\:
    \left\{
        \begin{array}{l}
             0  \hspace{2cm}              \mbox{massless}\\
             \phantom{\ldots}\\
             \mbox{finite} \hspace{1.5cm} \mbox{conformally coupled}
        \end{array}
    \right.
    ,
\end{equation}
while for massive states
\begin{equation}\label{eq:ML2}
    \lim_{y_e\longrightarrow0^{+}}
    \left[
        \psi_{\mathcal{G}}(\{E_j\},y_e)+\psi_{\mathcal{G}}^{\dagger}(\{-E_j\},y_e)
    \right]\:=\:\mathcal{O}\left(y_e^{\mbox{\footnotesize Re}\{\nu\}}\right)
\end{equation}
where $\nu$ is the order of the Hankel functions expressing the mode functions, which is related to the masses of the states via $\nu=\sqrt{9/4-(2m\ell_1)^2}$.

These statements can be proved via the cosmological optical theorem, and the right-hand-sides crucially depend on the behaviour of the bulk-to-boundary propagators as $y_e\longrightarrow0^{+}$.


\subsection{Causality}\label{subsec:Caus}

Causality in quantum mechanical observables in cosmology is perhaps the least understood of all the basic properties which can characterise and constrain them. If on one side such observables live on a space-like boundary at future infinity for which time evolution is completely integrated out, on the other, precisely because  time evolution is integrated out, causality must reflect somehow in their analytic structure. However, it is not trivial how this has to happen: the usual way in which causality is understood is via conditions on the commutation relations on fields, while we have been considering observables computed at equal time, which are naively blind to such conditions. Hence, there are two possible strategies: the first one is to explore the explicit time history for a given observable and try to guess how the causality conditions reflect on the final equal-time observable; or look for new observables, no longer defined at equal time, where causality can be manifest.

While the second route has been not pursued yet to our knowledge, the first one led to a manifestly causal reformulation of the in-in formalism ~\cite{Weinberg:2005vy, Baumgart:2019clc, Baumgart:2020oby}~. Let $\hat{\mathcal{O}}$ be a certain operator, its expectation value in such a formalism can be written as ~\cite{Weinberg:2005vy, Baumgart:2020oby}
\begin{equation}\label{eq:MCinin}
    \begin{split}
        \langle\hat{\mathcal{O}}(\eta_{\circ})\rangle\:=\:
            \lim_{\varepsilon\longrightarrow0}\sum_{v=0}^{\infty}(-i)^v\,&\int_{-\infty}^{\eta_{\circ}}d\eta_v\,
                 \prod_{j=1}^{v-1}\int_{-\infty}^{\eta_{j+1}}d\eta_j\,\times\\
                &\times\langle0|\left[\left[\ldots\left[\hat{\mathcal{O}},\,H_{\varepsilon}(\eta_v)\right]\ldots,
                    \,H_{\varepsilon}(\eta_2)\right],\,H_{\varepsilon}(\eta_1)\right]|0\rangle.
    \end{split}
\end{equation}
The presence of the $R$-products between $H_{\varepsilon}$ and $\mathcal{O}$ make causality manifest. The $R$-products were introduced precisely to study causality in flat space ~\cite{Lehmann:1957zz}~. In the flat-space case, the $R$-products are defined as ~\cite{Lehmann:1957zz}
\begin{equation}\label{eq:Rprod}
    \begin{split}
        R(x;\,x_1,\ldots,x_n)\:=\:&(-1)^n\sum_{\pi\in S_n} \vartheta(\eta_{\circ}-\eta_{\pi(n)})\ldots\vartheta(\eta_{\pi(2)}-\eta_{\pi(1)})\times\\
        &\times\left[\left[\ldots\left[\phi(x),\,\phi(x_{\pi(n)})\right]\ldots,
                        \,\phi(x_{\pi(2)})\right],\,\phi(x_{\pi(1)})\right]
    \end{split}
\end{equation}
Using the standard notion of causality, {\it i.e.} the commutators have to vanish for space-like separations, on expectation values of the $R$-products, relations among them in different regions of kinematic space have been found, which go under the name of Steinmann relations ~\cite{Steinmann:1960soa, Steinmann:1960sob, Araki:1961hb, Ruelle:1961rd}~. Such relations can be translated into the statement about scattering amplitudes that double discontinuities across partially overlapping channels have to vanish in the physical region ~\cite{Stapp:1971hh, Cahill:1973px, Cahill:1973qp, Lassalle:1974jm}
\begin{equation}\label{eq:SteinmRels}
    \mbox{Disc}_{s_{\mathcal{I}}}\left(\mbox{Disc}_{s_{\mathcal{J}}}\mathcal{A}\right)\:=\:0,\qquad
    \left\{
\begin{array}{l}
            \mathcal{I}\nsubseteq\mathcal{J},\\
            \mathcal{J}\nsubseteq\mathcal{I},\\
            \mathcal{I}\cap\mathcal{J}\,\neq\,\varnothing.
        \end{array}
    \right.
\end{equation}
where $\mathcal{A}$ is the scattering amplitude, $s_{\mathcal{I}}$ and $s_{\mathcal{J}}$ are the Mandelstam invariants involving the momenta of particles in the sets $\mathcal{I}$ and $\mathcal{J}$, with $\mathcal{I},\,\mathcal{J}\,\subset\,\{1,\,2,\,\ldots,n\}$.

\begin{figure}[t]
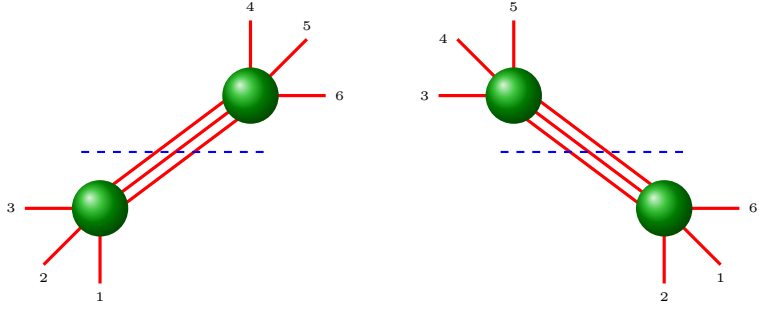

    \centering
    %
    \caption{Steinmann relations for scattering amplitudes. The two graphs represent two partially overlapping channels, and the dashed line the discontinuity along them. The double discontinuities in such pairs of channels ought to vanish in the physical region.}
    \label{fig:SteimnRels}
\end{figure}

As we will see in full glory detail in Section \ref{subsec:FSwf}, Steinmann-like relations of the form \eqref{eq:SteinmRels} hold also for the perturbative wavefunction, at least for a large class of scalar models with a non-trivial flat-space limit and any other process which can be written in terms of these scalar integrands ~\cite{Benincasa:2020aoj}~. It is important to remark that while the Steinmann relations for flat-space scattering amplitudes are a direct consequence of causality, the Steinmann-like relations for the wavefunction coefficients have been proven directly as a statement on the double discontinuities and their eventual connection to causality still needs to be investigated.


\newpage

\addcontentsline{toc}{section}{\underline{Part I: A Boundary Perspective On The Wavefunction}}
\section*{\hfill \underline{Part I: A Boundary Perspective On The Wavefunction}\hfill}

The wavefunction of the universe is a functional of field configurations at the future space-like boundary of an expanding universe and it is assumed to be the result of a causal and unitary evolution. However we do not have a direct access to the processes that happened in the past and lead to what we can compute at the boundary. Rather than playing with assumptions about what could have happened in the past, we can aim to understand how general principles, such as the ones discussed in Section \ref{sec:FunProp} can constrain the wavefunction of the universe, at least in perturbation theory, and consequently the processes which could happen in the early universe. In this sense a useful approach is to think about the wavefunction of the universe directly as a boundary object without making any reference to an explicit time evolution.

In this first part, we will review the recent progress in {\it bootstrapping} the wavefunction coefficients in perturbation theory from basic principles. There are two important disclaimers to be made: first, as the whole review, we will focus on scalar theories only; secondly, the analysis of the wavefunction coefficients relies on its representation in terms of Feynman graphs and will deal with the individual graphs, {\it i.e.} we will study $\psi_{\mathcal{G}}$ as defined in \eqref{eq:Feyn1} but bypassing the time integration. Focusing on individual graphs is sensible for scalar theories as they can be considered as physical. This does not hold in the case of different states for which generally an individual Feynman graph is not gauge invariant. In such cases, the general aim is to develop an approach which is gauge invariant in all its steps, but this is currently not available. The general strategy is to
{
\renewcommand{\theenumi}{\roman{enumi}}
\begin{enumerate}
    \item use symmetries to organise the physical degrees of freedom, conveniently parametrise the kinematic space and constrain (and when it is possible fix) the simplest processes;
    \item use factorisation theorems, unitarity and causality to fix the wavefunction coefficients for arbitrary processes via further constraints or a precise prescription for {\it gluing} lower-point/lower-level wavefunctions into higher-point/higher-level ones.
\end{enumerate}
}
%


\section{Bootstrapping the wavefunction from boundary data}\label{sec:Boot}

Let us begin with a first way of understanding and computing the wavefunction of the universe without making time evolution explicit. The general idea is to {\it bootstrapping} it from constraints due to symmetries and fundamental principles.

\subsection{Symmetries and states}\label{subsec:Sym}

Physical states are irreducible representations of the isometry group of the space-time where the processes occur. In the more familiar case of flat-space processes, such an isometry group is the Poincar{\'e} group $\Pi^{\mbox{\tiny $(1,d)$}}\:=\:\mathbb{R}^{1,d}\rtimes SO(1,d)$, {\it i.e.} the semi-direct product between the space-time translations $\mathbb{R}^{1,d}$ and the Lorentz group $SO(1,d)$. The particles are then its unitary irreducible representations, which are classified by the Casimirs of the group, {\it i.e.} the squared mass $\hat{P}^2=m^2\hat{\mathbb{I}}$ and the spin $\hat{W}^2\:=\:-\rho^2\hat{\mathbb{I}}$, and are typically realised as induced representations from the Lorentz group and diagonalising the space-time translations\footnote{As the Poincar{\'e} group has the semi-direct product structure $\Pi^{\mbox{\tiny $(1,d)$}}\:=\:\mathbb{R}^{1,d}\rtimes SO(1,d)$, the representations can be realised by diagonalising either the momentum operator, as it is usually done, or the Casimirs of the Lorentz group -- see for example ~\cite{Macdowell:1972ef}~ for the $d=3$ case and a first attempt to compute scattering amplitudes of massive states in such representation, and ~\cite{Pasterski:2021rjz}~ for a review its use in the modern context of celestial amplitudes.} (see ~\cite{Benincasa:2013faa}~ and ~\cite{Oblak:2016eij}~). It also has a unique {\it finite-dimensional} representation for zero momenta, the trivial representation, which is taken to be vacuum.

In a general FRW cosmology, the isometry group is simply the $d$-dimensional Euclidean group $ISO(d):=\mathbb{R}^d\rtimes SO(d)$ and, similarly, the states are classified via the Casimir operators $\hat{P}^2$ and $\hat{P}\cdot\hat{L}$, $\hat{P}$ and $\hat{L}$ being the generators of the space translations $\mathbb{R}$ and space rotations $SO(d)$ respectively. This group can be also extended by dilatations, $D\times ISO(d)$, or to the full $dS_{1+d}$ group $SO(1,d+1)$ which includes also $d$-dimensional boosts. Notice that $D\times ISO(d)$ then is the subgroup of $SO(1,d+1)$ which leaves a light-like subspace of $dS_{1+d}$ invariant ~\cite{Patera:1975si}~.

As we will be focusing just on scalar states, we will not discuss in detail the unitary irreducible representations for all these groups\footnote{See Appendix A of ~\cite{Pajer:2020wxk}~ for the unitary irreducible representations of $ISO(3)$ and ~\cite{Dobrev:1977har, Sun:2021thf}~ for the $dS_{1+d}$ case.}. Rather, first it was important to mention {\it what} is a state in an expanding universe, and secondly we will just state that scalars can be labelled by a complex quantum number $\nu$ which can be either real, corresponding to massless/light states in FRW cosmologies, or purely imaginary for massive states in $dS$\footnote{As we will see in Section \ref{sec:ToySc}, the quantum number $\nu$ will reflects into the mode functions as the order parameter for the Hankel functions realising them.}. If the scaling symmetry is present, as for $D\times ISO(d)$ and $SO(1,d+1)$, the quantum number $\nu$ is related to the conformal dimension $\Delta$.


\subsection{Three-point interactions}\label{subsec:3pci}

Let us now consider the simplest possible processes, which involve just three-state. They also should reflect the basic symmetries of our space-time. Using a realisation of the states which diagonalises the spatial-translation operators, then spatial translations implies that the wavefunction coefficients have support on a spatial momentum conserving delta function, while rotational invariance reflects into their dependence on rotational invariant combination of the momenta. In a general FRW cosmology, these are the only symmetries we can rely upon. For three external states, the three-point wavefunction coefficient is just a function of the moduli of the external momenta $\{E_j:=|\vec{p}_j|,\,j=1,2,3\}$, that we will refer to as {\it energies} with a bit of abuse of language:
\begin{equation}\label{eq:psi3E}
    \psi_{3}(\vec{p}_1,\vec{p}_2,\vec{p}_3)\:=\:\delta^{\mbox{\tiny $(d)$}}(\vec{p}_1+\vec{p}_2+\vec{p}_3)\psi'_3(E_1,E_2,E_3).
\end{equation}
Without further assumptions, just the $ISO(d)$ group leaves a big deal of freedom. Promoting the symmetry group to be the $dS_{1+d}$ group $SO(1,d+1)$, then the three-point wavefunction can be determined up to an overall constant. Actually, the $dS$ three-point scalar correlation functions $\langle\phi\phi\phi\rangle$ have been found via symmetries ~\cite{Bzowski:2013sza}~ rather than the three-point wavefunction coefficients:
\begin{equation}\label{eq:3ptCF}
    \langle\phi(\vec{p}_1)\phi(\vec{p}_2)\phi(\vec{p}_3)\rangle'\:=\:\lambda_3\,\left(\prod_{j=1}^3 E_j^{\Delta_j-\frac{d}{2}}\right)
        \int_{0}^{+\infty}dz\,z^{\frac{d}{2}-1}\,\prod_{j=1}^3\,K_{\Delta_{j}-\frac{d}{2}}(E_j z)
\end{equation}
where $\Delta_j$ is the conformal dimension of the state $j$ and $K_{\Delta_j-d/2}$ is a Bessel-$K$ function. However, the three-point wavefunction is related to the correlator via
\begin{equation}\label{eq:3ptCorr3ptWF}
    \langle\phi(\vec{p}_1)\phi(\vec{p}_2)\phi(\vec{p}_2)\rangle'\:=\:\frac{\mbox{Re}\{2\psi'_3(E_1,E_2,E_3)\}}{\prod_{j=1}^3\mbox{Re}\{2\psi'_2(E_j)\}}
\end{equation}
where the prime ``$'$'' indicates that the spatial momentum conserving $\delta$-function has been stripped off.

Going back to the more general FRW cosmologies, we can restrict \eqref{eq:psi3E} by imposing the Bunch-Davies condition, {\it i.e.} no particle production or particle decay is possible in the physical region $\{E_j\in\mathbb{R}_+,\,j=1,2,3\}$. This implies that $\psi'_3$ cannot have singularities as combinations such as $E_1+E_2-E_3$ and $E_1-E_2-E_3$ approach zero, but can have at most a singularity in $E_1+E_2+E_3=0$: while the first type of singularity can be reached for positive values of the energies, for the latter, staying in the physical region would automatically require all the energies to vanish. 

In the case of identical external states, the kinematic space can be also conveniently parametrised in terms of symmetric polynomials $X^{\mbox{\tiny $(1)$}}:=\sum_{j=1}^3E_j$, $X^{\mbox{\tiny $(2)$}}:=\sum_{i<j}E_iE_j$, and $X^{\mbox{\tiny $(3)$}}:=E_1E_2E_3$:
\begin{equation}\label{eq:psi3Eb}
    \psi_3\:=\:\delta^{\mbox{\tiny $(d)$}}(\vec{p}_1+\vec{p}_2+\vec{p}_3)
        \psi'_3(X^{\mbox{\tiny $(1)$}},\,X^{\mbox{\tiny $(2)$}},\,X^{\mbox{\tiny $(3)$}})
\end{equation}

Also if the states have a flat-space counter-part, {\it e.g.} in the $dS$, the states of interest are $SO(1,d+1)$-representations with a Inonu-Wigner contraction to Poincar{\'e} representations, then the leading Laurent coefficient in the expansion as the total energy $E_1+E_2+E_3$ goes to zero is the flat-space amplitude, as we discussed in Section \ref{subsubsec:EtotFS}. In the $dS$ case, such an amplitude turns out to be Lorentz invariant, while for FRW cosmologies with isometry group $D\,\times\,ISO(d)$ Lorentz invariance does not have necessarily to be restored in the flat-space limit, and the related scattering amplitudes do not have Lorentz boost invariance. Such amplitudes have been classified and studied in ~\cite{Pajer:2020wnj}~.

For massless states, the three-point wavefunction is a rational function and can be fixed by these constraints. As for the $dS$ case, these arguments have been used to fix three-point correlation functions rather than the wavefunction for the curvature perturbation $\zeta$ in $d=3$. In this case, one can make use of further constraints coming from the soft limits ~\cite{Maldacena:2002vr, Hinterbichler:2012nm, Creminelli:2012ed,  Hinterbichler:2013dpa, Pajer:2017hmb}~ to obtain the three-point correlation function of $\zeta$ with a two-derivative interaction ~\cite{Pajer:2020wxk}
\begin{equation}\label{eq:Bzz}
    \begin{split}
        \langle\zeta(\vec{p}_1)\zeta(\vec{p}_2)\zeta(\vec{p}_3)\rangle'\:&=\:
        \frac{1}{2}\left(\frac{1}{2\sqrt{\epsilon}\ell M_{\mbox{\tiny pl}}}\right)^4\,\times\\
        &\hspace{-2.5cm}\times\frac{8\epsilon(X^{\mbox{\tiny $(2)$}})^2+(3\eta-22\epsilon)X^{\mbox{\tiny $(1)$}}X^{\mbox{\tiny $(3)$}}+
        (4\epsilon-3\eta)(X^{\mbox{\tiny $(1)$}})^2X^{\mbox{\tiny $(2)$}}+(\eta-\epsilon)(X^{\mbox{\tiny $(1)$}})^4}{X^{\mbox{\tiny $(1)$}}(X^{\mbox{\tiny $(3)$}})^3}
    \end{split}
\end{equation}
where $\epsilon$ and $\eta$ are two free parameter which are left unfixed. They turn out to correspond to
\begin{equation}\label{eq:et}
    \epsilon\: =\:\left(\frac{\ell\dot{\phi}}{\sqrt{2}M_{\mbox{\tiny Pl}}}\right)^2,\qquad
    \eta-2\epsilon\: =\:\frac{2\ell\Ddot{\phi}}{\dot{\phi}}
\end{equation}
and hence they are not independent on each other: this procedure seems not to capture a further constraint.

This can be extended to interactions with higher derivative interactions  ~\cite{Pajer:2020wxk}~ appearing in the EFT of inflation ~\cite{Cheung:2007st}~, even if leaving some unfixed parameters.


\subsection{Four-point interactions from symmetries and singularities}\label{subsec:4ptcdS}

In de Sitter space-time, the states can be labelled via an integer or half-integer number, encoding the spin, and a complex number $\Delta$, the conformal dimension. The larger symmetry group makes the physics much more constrained: together with spatial translations and spatial rotations further constraints come from the special conformal transformations ({\it i.e.} the $dS$ boosts).  It is possible to focus on the so-called conformally coupled scalar in $dS_{1+3}$, which has conformal dimension $\Delta\,=\,2$, as external states. Interestingly, the wavefunction for such states can function as a seed to compute wavefunctions with external states with different spin and different conformal dimensions ~\cite{Arkani-Hamed:2018ahb, Benincasa:2019vqr, Baumann:2019oyu, Baumann:2020dch}~.

The $dS_{1+3}$ wavefunction coefficients $\psi_n$ have to be annihilated by the dilatation and special conformal transformation generators $\hat{\mathcal{D}}$ and $\hat{\mathcal{K}}_j$:
\begin{equation}\label{eq:dScnt}
    \begin{split}
        &0\:=\:\hat{\mathcal{D}}\psi_n\::=\:\left[\sum_{r=1}^n\hat{\mathcal{D}}^{\mbox{\tiny $(r)$}}\right]\psi_n\::=\:
            \sum_{r=1}^n\left[p_j^{\mbox{\tiny $(r)$}}\frac{\partial}{\partial p_j^{\mbox{\tiny $(r)$}}}+3-\Delta_r\right]\psi_n,\\
        &0\:=\:\hat{\mathcal{K}}_{j}\psi_n\::=\:\left[\sum_{r=1}^n\hat{\mathcal{K}}_j^{\mbox{\tiny $(r)$}}\right]\::=\:
            \sum_{r=1}^n
            \left[
                p_j\frac{\partial^2}{\partial p_k^{\mbox{\tiny $(r)$}}\partial p^k_{\mbox{\tiny $(r)$}}}
                -2\hat{\mathcal{D}}^{\mbox{\tiny $(r)$}}\frac{\partial}{\partial p_{\mbox{\tiny $(r)$}}^j}
            \right]\psi_n
    \end{split}
\end{equation}
For four point it is convenient to parametrise the kinematic space in terms of the partial energies associated to a given channel. For example, in the $s$-channel let $y_s:=|\vec{p}_1^{\mbox{\tiny $(1)$}}+\vec{p}^{\mbox{\tiny $(2)$}}|$ be the energy of the internal state, then one can introduce dimensionless variables $(u,v)$ associated to the energies at each graph site $X_1:=|\vec{p}^{\mbox{\tiny $(1)$}}|+|\vec{p}^{\mbox{\tiny $(2)$}}|$ and $X_2:=|\vec{p}^{\mbox{\tiny $(3)$}}|+|\vec{p}^{\mbox{\tiny $(4)$}}|$:
\begin{equation}\label{eq:UV}
    u\::=\:\frac{y_s}{X_1},\qquad v\::=\:\frac{y_s}{X_2}.
\end{equation}
Notice that in these variables, the total energy vanishes if $u+v\,\longrightarrow\,0$, while the partial energies are conserved when $u,\,v\longrightarrow\,-1$.

After a suitable manipulation, the equations \eqref{eq:dScnt} lead to the following differential equation in terms of operators in terms of the kinematic variables $(u,\,v)$ as defined in \eqref{eq:UV}
\begin{equation}\label{eq:DiffUV}
    0\:=\:(\hat{\Delta}_u-\hat{\Delta}_v)\tilde{\psi}_4\::=\:(\hat{\Delta}_u-\hat{\Delta}_v)y_s^{-1}\psi_4,\qquad
    \hat{\Delta}_z\::=\:z^2(1-z^2)\partial_z^2-2z^3\partial_z
\end{equation}
The equation \eqref{eq:DiffUV} has to be satisfied by any four-point wavefunction, irrespectively that it is describes a contact interaction or exchanges of states, as they come from a manipulation of the symmetry constraints. It is possible to distinguish among the different processes specifying different requirements for the singularity structure of the solution. For example, in the case of contact interactions, it is possible to require that $\left.\psi_4\right|_{\mbox{\tiny cont}}$ has a single singularity corresponding to the total energy singularity, which can be reached just outside the physical region. The order as a pole depends on the number derivative of the interactions and can be generally written as ~\cite{Arkani-Hamed:2015bza, Arkani-Hamed:2018kmz}
\begin{equation}\label{eq:tpsi4d}
    \left.\tilde{\psi}_4^{\mbox{\tiny $(n)$}}\right|_{\mbox{\tiny cont}}\:=\:
        \hat{\Delta}_u^n\left.\tilde{\psi}_4^{\mbox{\tiny $(0)$}}\right|_{\mbox{\tiny cont}}\:=\:
        \hat{\Delta}_u^n\frac{uv}{u+v},
\end{equation}
$\left.\tilde{\psi}_4^{\mbox{\tiny $(n)$}}\right|_{\mbox{\tiny cont}}$ and 
$\left.\tilde{\psi}_4^{\mbox{\tiny $(0)$}}\right|_{\mbox{\tiny cont}}$ are the wavefunctions for interactions with and without derivatives.

In the case of the tree-level four-point wavefunction, the equation \eqref{eq:DiffUV} can be written as two ordinary differential equations
\begin{equation}\label{eq:4pttl}
    \left(\hat{\Delta}_u+M^2\right)\tilde{\psi}_4\:=\:\tilde{C}_4(u,v)\:=\:\left(\hat{\Delta}_v+M^2\right)\tilde{\psi}_4,
\end{equation}
with $\tilde{C}_4(u,v)$ satisfying the equation \eqref{eq:DiffUV}. Notice that as $M^2$ becomes large, the interaction reduces to a contact interaction and hence $\tilde{C}_4(u,v)$ is just the contact interaction solution \eqref{eq:tpsi4d} for fixed $n$. The differential equations \eqref{eq:4pttl} can be solved requiring that the solution factorises into a product of a three-particle amplitude and a three-point wavefunction as the partial energy singularities are approached, {\it i.e.} $u,\,v\,\longrightarrow\,-1$, while no folded singularity arises {\it i.e.} it is regular as $u,\,v\,\longrightarrow\,1$. The solution for $M^2=0$ corresponds to a propagating conformally coupled scalar and can be expressed in terms of logarithms and dilogarithms, while for $M^2\neq0$, {\it i.e.} with a massive propagative state, the solution has been found in form of a power series ~\cite{Arkani-Hamed:2018kmz}~.

The solution $\tilde{\psi}_4$ of \eqref{eq:4pttl}, which is for external conformally coupled scalars ($\Delta=2$), can be mapped to a wavefunction for external massless scalars ($\Delta=3$) via differential operators ~\cite{Arkani-Hamed:2018kmz, Baumann:2019oyu}~, the so-called weight-shifting operators ~\cite{Costa:2011dw, Karateev:2017jgd}
\begin{equation}\label{eq:4ptmsl}
    \left.\tilde{\psi}_4\right|_{\Delta=3}\:=\:y_s^3
        \left[
        \left(
            1-\frac{E_1E_2}{\sigma_{\mbox{\tiny L}}}\partial_{\sigma_{\mbox{\tiny L}}}
        \right)\frac{1-u^2}{u^2}\partial_u u
        \right]
        \left[
        \left(
            1-\frac{E_3E_4}{\sigma_{\mbox{\tiny R}}}\partial_{\sigma_{\mbox{\tiny R}}}
        \right)\frac{1-v^2}{v^2}\partial_v v
        \right]
        \tilde{\psi}_4
\end{equation}
$\sigma_{\mbox{\tiny L}}:=X_1+y_s$ and $\sigma_{\mbox{\tiny R}}:=y_s+X_2$ are the partial energies.

The computation of wavefunctions and correlators in de Sitter space-time is important not just for a merely theoretical standpoint, but also for a phenomenological perspective: starting from a wavefunction or correlator in de Sitter is possible to softly break the de Sitter symmetries and take the soft limit to obtain the inflationary observable ~\cite{Kundu:2014gxa, Kundu:2015xta, Arkani-Hamed:2018kmz, Baumann:2019oyu}~. In particular, the three-point inflationary functions with external massless states can be computed considering the de Sitter four-point one with three-massless states and the fourth state with conformal dimension $\Delta=3-\epsilon$. Then, the three-point inflationary correlator with external massless states can be obtained by taking the soft limit for the momentum associated to the state of dimension $\Delta=3-\epsilon$ and performing an expansion in $\epsilon$, which plays the role of the slow-roll parameter. This procedure allowed to recover the inflationary three-point correlator of ~\cite{Maldacena:2002vr}~.


\subsection{A recursive approach}\label{subsec:Rec}

So far we have been discussed a boundary approach which aimed to directly reconstruct the three- and four-point observables from general principles by-passing the time integration. If we were able to obtain a full classification of the three-point wavefunctions with arbitrary external states for a given expanding background, principles such as unitarity and causality might instruct us to glue them together to obtain four- and higher- point wavefunctions. This approach has been successfully pursued for flat-space scattering amplitudes and lead to the BCFW recursion relations ~\cite{Britto:2005fq}~ which are valid at tree level for any consistent theory ~\cite{Britto:2005fq, Benincasa:2007qj, Cheung:2008dn, Cohen:2010mi, Benincasa:2011kn} and at all loops for supersymmetric Yang-Mills integrands ~\cite{Arkani-Hamed:2010kv, Benincasa:2015zna}~. They build upon the physical interpretation of the behaviour of any scattering amplitude (or the integrands in the case of loop-level processes) as its poles are approached and a direct relation between them and the amplitude itself obtained via a deformation of momentum space in an arbitrarily chosen complex direction ~\cite{Britto:2005fq, Benincasa:2013faa}~.

In line of principle, the idea of introducing a one-parameter deformation of the kinematic space and exploiting the analytic structure of our observable of interest as a function of such a parameter, can be used also for the wavefunction, but with some complication. Unlike the scattering amplitude case, the wavefunction of the universe in FRW cosmologies shows a singularity structure more involved that just simple poles. As we will see in the next Section \ref{sec:ToySc}, it is possible to introduce a {\it wavefunction universal integrand} ~\cite{Arkani-Hamed:2017fdk}~ from which it is possible to extract the wavefunction coefficients for conformally-coupled scalars in FRW cosmologies via integral or differential operators ~\cite{Arkani-Hamed:2017fdk, Benincasa:2019vqr}~, and extend to more general scalars ~\cite{Arkani-Hamed:2018kmz, Benincasa:2019vqr, Baumann:2019oyu}~ or propagating spinning states ~\cite{Arkani-Hamed:2018kmz, Baumann:2019oyu}~ via differential operators.

Such wavefunction universal integrands turn out to be the wavefunction coefficients for scalars in flat-space. In this case, all the singularities are simple poles and are associated to the vanishing of the total energies of any subprocess in the wavefunction. One strategy is therefore to introduce a one-parameter deformation of the kinematic space for the wavefunction universal integrands, obtaining a recursive formula relating higher- and lower- wavefunction coefficients and then applying the relevant operators to map the result to the FRW cosmology and the states of interest. 

Let $\psi_n$ be an $n$-point tree-level wavefunction for scalars in flat space. It can be in principle computed by summing all the relevant tree-level graphs with $n$ external states. Let $\{E_j:=|\vec{p}^{\mbox{\tiny $(j)$}}|\,|\,j=1,\ldots,n\}$ and $\{y_{\mathcal{I}} := |\sum_{j\in\mathcal{I}}\vec{p}^{\mbox{\tiny $(j)$}}| \, \Big| \,\mathcal{I}\cup\overline{\mathcal{I}}=\{1,\ldots,n\}\}$ be the set of energies of external and internal states respectively. The most general, linear, one-parameter energy space deformation is given by ~\cite{Arkani-Hamed:2017fdk}
\begin{equation}\label{eq:Edef}
    E_j\:\longrightarrow\:E_j+\alpha_j\zeta,\qquad y_{\mathcal{I}}\:\longrightarrow\:y_{\mathcal{I}}+\beta_{\mathcal{I}}\zeta
\end{equation}
$\forall\,j=1,\ldots,n$ and $\forall\,\mathcal{I}\,|\,\mathcal{I}\cup\overline{\mathcal{I}}=\{1,\ldots,n\}\,\&\,\mbox{dim}\{\mathcal{I}\}\in[2,n-2]$. In \eqref{eq:Edef} $\zeta$ is the deformation parameter and $\{\alpha_j\,|\,j=1,\ldots,n\}$ and $\{\beta_{\mathcal{I}}\,|\,\mathcal{I}\cup\overline{\mathcal{I}}=\{1,\ldots,n\}\}$ are numerical coefficients. 

Importantly, there is a plethora of choices for the coefficients $\alpha$'s and $\beta$'s: different choices can make different poles to be (in)dependent on $\zeta$. However, a comment is in order. If we want to consider the full tree-level $n$-point wavefunction rather than individual graphs, the spatial momentum conservation imposes constraints in energy space and hence the coefficients $\alpha$'s and $\beta$'s in \eqref{eq:Edef} need to be chosen in such a way that such constraints are still satisfied. In order to fix ideas, let us consider a tree-level four-point wavefunction. In energy space, the spatial momentum conservation can be written as
\begin{equation}\label{eq:MomConsE}
    \sum_{j=1}^4E_j^2\:=\:y_s^2+y_t^2+y_u^2
\end{equation}
where $y_s:=|\vec{p}^{\mbox{\tiny $(1)$}}+\vec{p}^{\mbox{\tiny $(2)$}}|$, $y_t:=|\vec{p}^{\mbox{\tiny $(2)$}}+\vec{p}^{\mbox{\tiny $(3)$}}|$ and $y_u:=|\vec{p}^{\mbox{\tiny $(3)$}}+\vec{p}^{\mbox{\tiny $(1)$}}|$. Thus, when performing the energy space deformation \eqref{eq:Edef}, requiring that the deformed version of \eqref{eq:MomConsE} holds implies the following constraints on the coefficients
\begin{equation}\label{eq:abconst}
    \sum_{j=1}^4\alpha_j E_j\:=\:\beta_s y_s+\beta_t y_t +\beta_u y_u,\qquad
    \sum_{j=1}^4\alpha_j^2\:=\:\beta_s^2+\beta_t^2+\beta_u^2.
\end{equation}
If rather than considering $\psi_n\:=\:\sum_{\mathcal{G}}\psi_{\mathcal{G}}$ we were to consider the individual wavefunction contribution $\psi_{\mathcal{G}}$ associated to the graph $\mathcal{G}$, then no such constraints can be imposed.

In any case, any one-parameter deformation \eqref{eq:Edef} maps the wavefunction into a one-parameter family of wavefunctions $\psi_n(\zeta)$ which can be analysed as function of $\zeta$. Then, integrating it over the Riemann sphere, the wavefunction $\psi_n(\zeta=0)$ we would like to compute can be related to its own residues
\begin{equation}\label{eq:RRwf}
    0\:=\:\frac{1}{2\pi i}\oint_{\hat{\mathbb{C}}}\frac{d\zeta}{\zeta}\psi_n(\zeta)\:=\:
        \psi_n(\zeta=0)+\sum_{\{\zeta_a\}}\mbox{Res}\left\{\frac{\psi_n(\zeta)}{\zeta},\zeta=\zeta_a\right\}+\mathcal{C}_n
\end{equation}
where $\{\zeta_a\}$ is the set of poles of $\psi_n(\zeta)$ at finite location, and $\mathcal{C}_n$ is the residue from the pole at infinity. 

If $\psi_n(\zeta)$ vanishes as $\zeta$ is taken to infinity, then $\mathcal{C}_n=0$. This is indeed the case for interactions with polynomial interactions. Solving \eqref{eq:RRwf} with respect to $\psi_n(\zeta=0)$ and assuming that $\mathcal{C}_n=0$, we obtain a representation for the wavefunction in terms of the residues of its poles. Also, we do know the physical meaning of these residues: for the total energy pole, the residue is the flat-space scattering amplitudes now with states whose energies are computed at the location of the poles: $\hat{E}_j^{\mbox{\tiny $(a)$}}:=E_j(\zeta=\zeta_a)$ and 
$\hat{y}_{\mathcal{I}}^{\mbox{\tiny $(a)$}}:=y_{\mathcal{I}}(\zeta=\zeta_a)$. Hence
\begin{equation}\label{eq:RRwf2}
    \psi_n\:=\:\frac{\mathcal{A}_n(\zeta_{\mbox{\tiny tot})}}{\sum_{j=1}^n E_j}+
    \sum_{a}\frac{\mathcal{A}_{\mathcal{I}_a}(\zeta_{a})\times\tilde{\psi}_{\overline{\mathcal{I}}_a}(\zeta_a)}{\sigma_a}
\end{equation}
where the sum over all the poles except the total energy pole $\zeta_{\mbox{\tiny tot}}$ and $\sigma_a$ is a partial energy which picks the $\zeta$-dependence under the chosen deformation\footnote{A recursion relation of this type discussed in ~\cite{Baumann:2021fxj}~ for four-point wavefunction. In that case, the deformation was chosen to be simply $E_1\longrightarrow E_1+\zeta$. However, despite there is no energy conservation, spatial momentum conservation imposes the constraint \eqref{eq:MomConsE} in energy space, that translates in \eqref{eq:abconst} under a deformation. The deformation $E_1\longrightarrow E_1+\zeta$ does not satisfy \eqref{eq:abconst}, nevertheless it provides the right result. Performing this type of deformation can be seen as adding (or modifying) the mass for state $1$. However, in flat-space massless and massive scalars have the same number of degrees of freedom ($1$) and the same three-point observables, which provides an explanation for the right result. For the full wavefunction of spinning state, this difference would matter and, in fact, recursion relations of the type \eqref{eq:RRwf2} can be found for individual graphs in the axial gauge, but they are not yet available for the full wavefunction.} -- obviously, if the deformation does not pick a certain pole, the related residue does not appears in \eqref{eq:RRwf2}. 

Notice that if such a vanishing condition for $\psi_n(\zeta)$ is not satisfied, $\mathcal{C}_n$ can be computed from the knowledge of the zeroes of $\psi_n(\zeta)$ ~\cite{Benincasa:2011kn}~ -- despite this idea appeared first in the context of scattering amplitudes, the resulting recursion relation holds for any rational function with simple poles and hence for the flat-space wavefunction. 
\begin{equation}\label{eq:RRwf3}
    \psi_n\:=\:\sum_{\{\zeta_a\}}
        \left[
            \frac{\mbox{Res}\{\psi_n(\zeta),\,\zeta=\zeta_a\}}{-\zeta_a}
            \prod_{m=1}^{\nu+1}\frac{-\zeta_{\circ}^{\mbox{\tiny $(m)$}}}{\zeta_a-\zeta_{\circ}^{\mbox{\tiny $(m)$}}}
        \right],
\end{equation}
where $\nu$ is the leading behaviour of $\psi_n(\zeta)$ as $\zeta$ is taken to infinity: 
$\psi_n(\zeta)\,\overset{\mbox{\tiny $\zeta\longrightarrow\infty$}}{\sim}\,\zeta^{\nu}$. It can be further generalised by considering, rather than the zeroes, a set of generic points $\{\zeta_{\star}^{\mbox{\tiny $(m)$}},\,m=1,\ldots,\nu+1\,|\,\zeta_{\star}^{\mbox{\tiny $(m)$}}\,\neq\,\zeta_a,\;\forall\,m,\,a\}$ different than the poles ~\cite{Kampf:2013vha}~
\begin{equation}\label{eq:RRwf4}
    \psi_n\:=\:\sum_{\{\zeta_a\}}
        \left[
            \frac{\mbox{Res}\{\psi_n(\zeta),\,\zeta=\zeta_a\}}{-\zeta_a}
            \prod_{m=1}^{\nu+1}\frac{-\zeta_{\star}^{\mbox{\tiny $(m)$}}}{\zeta_a-\zeta_{\star}^{\mbox{\tiny $(m)$}}}
        \right]+\sum_{m=1}^{\nu+1}\psi(\zeta_{\star}^{\mbox{\tiny $(m)$}})
        \prod_{r\neq m}\frac{-\zeta_{\star}^{\mbox{\tiny $(l)$}}}{\zeta_{\star}^{\mbox{\tiny $(m)$}}-\zeta_{\star}^{\mbox{\tiny $(l)$}}}
\end{equation}
The recursion relations \eqref{eq:RRwf3} and \eqref{eq:RRwf4} are relevant for scalars with derivative interactions, such as the non-linear sigma model or the shift-symmetric interaction $(\partial\phi)^4$ which satisfy soft theorems ~\cite{Bittermann:2022nfh}~.

It interesting to remark that the recursion relations \eqref{eq:RRwf2}, \eqref{eq:RRwf3} and \eqref{eq:RRwf4} are also valid for individual graphs $\psi_{\mathcal{G}}$ and loop integrands.

In order to map this structure to general FRW cosmologies, one has to integrate \eqref{eq:RRwf2} over the external energies with a suitable measure ~\cite{Arkani-Hamed:2017fdk, Benincasa:2019vqr}~. We will discuss this map in Sections \ref{sec:ToySc} and \ref{sec:BWF}. One can also use the cutting rules \eqref{eq:CCR}, exploiting the fact that $\psi^{\dagger}$ is regular as any of the partial energies is taken to zero ~\cite{Baumann:2021fxj}~. For massless states in $dS$ ~\cite{Benincasa:2019vqr, Hillman:2021bnk} and propagating light states ~\cite{Benincasa:2019vqr}~, the wavefunction can be obtained from the flat-space ones via certain differential operators.

In the massless case, where the wavefunction is still a rational function but shows high order poles, the combination of recursion relation, cutting rules and manifest locality can allow to compute a single four-point exchange diagram ~\cite{Jazayeri:2021fvk}~. Deforming the energy space in such a way that the total energy does not depend on the parameter deformation, the four-point wavefunction can be written in terms of the coefficients of its Laurent expansion around the total energies. Such coefficients are fixed by the cosmological optical theorem, which will force a boundary term in order for their sum to satisfy it. Such a boundary term gets fully fixed via manifest locality.

A similar relation to \eqref{eq:RRwf} can be obtained by introducing a one-parameter deformation directly in momentum space rather than in energy space ~\cite{Bittermann:2022nfh}~:
\begin{equation}\label{eq:momdef}
	\vec{p}^{\mbox{\tiny $(j)$}}(\zeta)\: =\:\vec{p}^{\mbox{\tiny $(j)$}}+\zeta\vec{q}^{\mbox{\tiny $(j)$}},\qquad
	\vec{p}^{\mbox{\tiny $(j)$}}\cdot\vec{q}^{\mbox{\tiny $(j)$}}\:=\:p^{\mbox{\tiny $(j)$}}q^{\mbox{\tiny $(j)$}},\qquad
	\sum_{j}\vec{q}^{\mbox{\tiny $(j)$}}(z)\:=\:0.
\end{equation}
This definition guarantees that the energies get deformed by the modulus of $\vec{q}^{\mbox{\tiny $(j)$}}$. Under such a deformation, the one-parameter deformed wavefunction turns out to have branch cuts in the partial energies. The recursive structure under such deformation then becomes ~\cite{Bittermann:2022nfh}~
\begin{equation}\label{eq:RRwfmom}
	\resizebox{0.9\hsize}{!}{$\displaystyle
	\psi_n\:=\:-\frac{1}{2\pi i}\oint_{\gamma_{\mbox{\tiny tot}}}\frac{d\zeta}{\zeta}\,\frac{\mathcal{A}(\zeta)}{\sum_{j=1}^nE_j(\zeta)}
		   -\sum_a\frac{1}{2\pi i}\oint_{\gamma_a}\frac{d\zeta}{\zeta}\,y_a(\zeta)\,\tilde{\psi}_{\mathcal{I}_a}(\zeta)\times
			\tilde{\psi}_{\overline{\mathcal{I}}_a}(\zeta)-\mathcal{C}_n
	$}
\end{equation}
where $\gamma_{\mbox{\tiny tot}}$ encircles the point $\sum_{j=1}^nE_j(\zeta)\:=\:0$, while $\gamma_a$ is a contour encircling the partial energy branch cut in the channel $\mathcal{I}_a$, and the sum in the second term runs over all the $\zeta$-dependent partial energies.

The recursion formula \eqref{eq:RRwfmom} has been used in the context of exceptional scalar theories, for which the deformation \eqref{eq:momdef} has been chosen in such a way to be able to make use of the information from the soft limits.


\section{The wavefunction universal integrand}\label{sec:ToySc}

A great deal of features for processes involving scalars in cosmology are captured by a flat-space model with time-dependent couplings, including the mass, and polynomial interactions ~\cite{Arkani-Hamed:2017fdk, Benincasa:2019vqr}~
\begin{equation}\label{eq:S}
    S\:=\:-\int_{-\infty}^{0}\int d^dx\,
        \left[
            \frac{1}{2}\left(\partial\phi\right)^2-\frac{1}{2}m^2(\eta)\phi^2-\sum_{k\ge3}\frac{\lambda_k(\eta)}{k!}\phi^k
        \right].
\end{equation}
It contains an interactive massive scalar in FRW cosmologies\footnote{One can start with the standard action for a massive scalar in a conformally-flat space-time with polynomial interaction and perform the following field redefinition ~\cite{Arkani-Hamed:2017fdk, Benincasa:2019vqr}~ $\phi\,\longrightarrow\,[a(\eta)]^{-(d-1)/2}\phi$.}, provided we take the time-dependent couplings $m^2(\eta)$ and $\lambda_k(\eta)$ to have the following functional forms ~\cite{Arkani-Hamed:2017fdk, Benincasa:2019vqr}~:
\begin{equation}\label{eq:m2lk}
    \begin{split}
        & m^2(\eta)\: :=\:m^2 a^2(\eta) + 2d\left(\xi-\frac{d-1}{4d}\right)
            \left[\partial_{\eta}\left(\frac{\dot{a}}{a}\right)+\frac{d-1}{2}\left(\frac{a}{a}\right)^2\right],
        \\
        &\lambda_k(\eta)\: := \:\lambda_k\left[a(\eta)\right]^{2+\frac{(d-1)(2-k)}{2}}
    \end{split}
\end{equation}
were $m$ and $\lambda_k$ are the {\it bare} couplings, $\xi$ is a parameter such that for $\xi\,=\,0$ the scalar is minimally coupled, while for $\xi\,=\,(d-1)/4d$ it is conformally coupled. Finally, ``$\dot{\phantom{a}}$'' indicates the derivative with respect to the conformal time $\eta$, and $a(\eta)$ is the time-dependent warp factor for a conformally flat metric
\begin{equation}\label{eq:ccm}
    ds^2\:=\:a^2(\eta)\left[-d\eta^2+dx_j dx^j\right].
\end{equation}
The equations of motion for the mode functions do not have an explicit and exact solution for general values of the $m^2(\eta)$. Exact solutions can be found for $m^2(\eta)\sim\eta^{-1}$ and $m^2(\eta)\sim\eta^{-2}$. The last case corresponds to a massless minimally-coupled scalar in a FRW cosmology of the type $a(\eta)\:=\:(-\ell_\gamma\eta^{-1})^{\gamma}$, or to a generic scalar propagating in a $dS$ background (which has the same power-law warp factor with $\gamma\,=\,1$). In all such cases, the mode functions with the Bunch-Davies condition at early times, are given in terms of Hankel functions
\begin{equation}\label{eq:MFsoln}
    \phi_{+}\:=\:\sqrt{-E\eta}\,H_{\nu}^{\mbox{\tiny $(2)$}}(-E\eta),
\end{equation}
where the order parameter $\nu$ of the Hankel function is related to the mass $m$ and the space-time through its spatial dimension and acceleration parameter $\gamma$. In the case of the conformal coupling, $\nu\,=\,\sqrt{1/4-(2m\ell_1)^2}$ in $dS$ ($\gamma\,=\,1$) and $\nu\,=\,1/2$ for a massless state in a FRW cosmology with a generic $\gamma\in\mathbb{R}_+$. For a minimally coupled state, $\nu\,=\,\sqrt{d^2/4-(2m\ell_{1})^2}$ in $dS$ and $\nu\,=\,1/2+(d-1)\gamma/2$ for a massless state in FRW. 

It is also possible to treat the mass perturbatively: the free propagation is the same as for a massless particle in flat space (or, which the same, for a conformally-coupled scalar in FRW cosmologies). Despite a first analysis of this type has been carried out ~\cite{Benincasa:2019vqr}~, deeper insights are needed and we don't discuss it here, focusing on the cases for which $m^2(\eta)=m_{\gamma}^2\eta^{-2}$.

With the mode functions \eqref{eq:MFsoln} at hand, which allow us to fix also the bulk-to-bulk propagator \eqref{eq:G}, we can compute the graph contributions to the wavefunction coefficients via the time integrals \eqref{eq:Feyn1}:
\begin{equation}\label{eq:Feyn2}
    \psi_{\mathcal{G}}\:=\:\delta^{\mbox{\tiny $(d)$}}\left(\sum_{j=1}^k\vec{p}_j\right)
        \int_{-\infty}^0\prod_{v\in\mathcal{V}}\left[d\eta_v\phi_{+}^{\mbox{\tiny $(v)$}}V_v\right]\prod_{e\in\mathcal{E}}G(y_e,\,\eta_{v_e},\eta_{v'_e})
\end{equation}
where $\phi_{+}^{\mbox{\tiny $(v)$}}$ is the product of the mode functions at the vertex $v$ and vertex function $V_v$ is just the time dependent coupling $\lambda_k(\eta_v)$ at the vertex $v$.

Let us notice that the mode function $\phi_{+}^{\mbox{\tiny $(\nu)$}}$ for a state with a generic order parameter $\nu$ is related to the one of conformally coupled scalar $\phi_{+}^{\mbox{\tiny $(1/2)$}}\,=\,e^{iE\eta_v}$ through an operator $\hat{\mathcal{O}}(E)$ acting in energy space ~\cite{Benincasa:2019vqr}
\begin{equation}\label{eq:mfnu12}
    \phi_{+}^{\mbox{\tiny $(\nu)$}}\:=\:\frac{1}{(-E\eta_v)^{\nu-\frac{1}{2}}}\hat{\mathcal{O}}_{\nu}(E)e^{iE\eta_v}
\end{equation}
where
\begin{equation}\label{eq:OnuE}
    \hat{\mathcal{O}}_{\nu}(E)\: :=\:\frac{E^{2\nu}}{\Gamma\left(\frac{1}{2}+\nu\right)\Gamma\left(\frac{1}{2}-\nu\right)}
        \int_{0}^{+\infty}dt\,t^{\nu-\frac{1}{2}}\int_{0}^{+\infty}ds\,s^{-\nu-\frac{1}{2}}\,
        e^{-\left(Et+is\frac{\partial}{\partial E}\right)}
\end{equation}
Interestingly, the operator \eqref{eq:OnuE} reduces to a simple differential operator for $\nu\,=\,l+1/2$ ($l\in\mathbb{Z}_+$)
\begin{equation}\label{eq:OnuEhi}
    \hat{\mathcal{O}}_{l+1/2}(E)\: =\:\prod_{r=1}^{l}\left[E\frac{\partial}{\partial E}-(2r-1)\right].
\end{equation}

For later convenience let us introduce the notation $\psi_{\mathcal{G}}^{\mbox{\tiny $(\{\nu_j\},\{\nu_e\})$}}:=\psi_{\mathcal{G}}$, where the labels $\{\nu_j\}$ and $\{\nu_e\}$ respectively identify the external and internal states.  Using \eqref{eq:mfnu12} we can write the contribution $\psi_{\mathcal{G}}^{\mbox{\tiny $(\{\nu_j\},\{\nu_e\})$}}$ of an arbitrary graph $\mathcal{G}$ to the wavefunction coefficients with arbitrary external states in terms of the the contribution $\psi_{\mathcal{G}}^{\mbox{\tiny $(\{\frac{1}{2}\},\{\nu_e\})$}}$ of the same graph but with external conformally coupled scalars via operators acting in energy space
\begin{equation}\label{eq:PsiGnu}
    \begin{split}
        &\psi_{\mathcal{G}}^{\mbox{\tiny $(\{\nu_j\},\{\nu_e\})$}} \:=\: \prod_{j=1}^n\left[\frac{1}{E_j^{\nu_j-\frac{1}{2}}}\hat{\mathcal{O}}(E_j)\right]
            \prod_{e\in\mathcal{E}}\left[\frac{1}{y_e^{2(\nu_e-\frac{1}{2})}}\right]
            \psi_{\mathcal{G}}^{\mbox{\tiny $(\{\frac{1}{2}\},\{\nu_e\})$}},\\
        &\phantom{\ldots}\\
        &\psi_{\mathcal{G}}^{\mbox{\tiny $(\{\frac{1}{2}\},\{\nu_e\})$}}\: :=\:
        \int_{-\infty}^0\prod_{v\in\mathcal{V}}\left[d\eta_{\nu}\,\frac{\lambda_{k_v}(\eta_v)}{(-\eta_v)^{\rho_v}}\,e^{iX_v\eta_v}\right]
        \prod_{e\in\mathcal{E}}G'(y_e;\,\eta_{v_e},\eta_{v'_e})
    \end{split}
\end{equation}
where $X_v$ is the sum of the energies of the external states at the vertex $v$, $\rho_v$ depends on the order parameter of both internal and external states, and $G'$ is a redefined bulk-to-bulk propagator
\begin{equation}\label{eq:XvRvGp}
    \begin{split}
        &X_v\: :=\:\sum_{j\subset v} E_j,\quad
         \rho_{\nu}\: :=\:\sum_{j=1}^{k_v}\nu_j+\sum_{e\in\mathcal{E}_v}\nu_e-\frac{k+\mbox{dim}\{\mathcal{E}_v\}}{2},\\
        &G'(y_e;\eta_{v_e},\eta_{v'_e})\: :=\:(-y_e\eta_{v_e})^{\nu_e-\frac{1}{2}}(-y_e\eta_{v'_e})^{\nu_e-\frac{1}{2}}
         G(y_e;\eta_{v_e},\eta_{v'_e})
    \end{split}
\end{equation}
$\mathcal{E}_v\subset\mathcal{E}$ being the subset of the bulk-to-bulk propagators incident in the vertex $v$. Furthermore, we can write the time-dependent functions $\lambda_{k_v}(\eta)\eta_v^{-\rho_v}$ in an integral representation
\begin{equation}\label{eq:lambint}
    \begin{split}
        \frac{\lambda_{k_v}(\eta_v)}{(-\eta_v)^{\rho_v}}\:&=\:
            \int_{-\infty}^{+\infty}d\varepsilon_v\,e^{i\varepsilon_v\eta_v}f(\varepsilon_v)\:=\\
            &=\:i^{\beta_{k_v,\nu}}(i\lambda_{k_v})\ell_{\gamma}^{\gamma[2-\frac{(k-2)(d-1)}{2}]}\int_{-\infty}^{+\infty}d\varepsilon_v\,
            e^{i\varepsilon_{v}\eta_v}\varepsilon_v^{\beta_{k_v,\nu}-1}\vartheta(\varepsilon_v)
    \end{split}
\end{equation}
where the first line is valid for any time-dependent coupling, as its specificity are encoded into the function $f(\varepsilon_v)$, while the second line specifies to FRW cosmologies with warp factor $a(\eta)\,=\,(-\ell_{\gamma}/\eta)^{\gamma}$, for which we explicitly know the mode functions, with $\beta_{k_v,\nu}\,:=\,\rho_{\nu}+\gamma[2-(k-2)(d-1)/2]$. More precisely, the second line in \eqref{eq:lambint} holds for $\mbox{Re}\{\beta_{k_v,\nu}\}>0$. If $\mbox{Re}\{\beta_{k_v,\nu}\}<0$, the integral in the second line is substituted by a derivative operator of order $\mbox{Re}\{\beta_{k_v,\nu}\}$. Hence, $\psi'_{\mathcal{G}}$ can be written as\footnote{Here we did not restrict to any specific class of conformally-flat cosmology. However, as discussed, one has to recall that there are cases in which the integral is substituted by a derivative operator.}
\begin{equation}\label{eq:psiG3}
    \begin{split}
        &\psi^{\mbox{\tiny $(\{\frac{1}{2}\},\{\nu_e\})$}}_{\mathcal{G}}\:=\:\int_{-\infty}^{+\infty}\prod_{v\in\mathcal{V}}\left[dx_v\,f(x_v-X_v)\right]
            \tilde{\psi}_{\mathcal{G}}(x_v,y_e),\\
        &\tilde{\psi}^{\mbox{\tiny $(\{\frac{1}{2}\},\{\nu_e\})$}}_{\mathcal{G}}(x_v,y_e)\,:=\,\int_{-\infty}^0\prod_{v\in\mathcal{V}}\left[d\eta_v\,e^{ix_v\eta_v}\right]
            \prod_{e\in\mathcal{E}}G'(y_e;\eta_{v_e},\eta_{v'_e})
    \end{split}
\end{equation}
where $x_v:=X_v+\varepsilon_v$. Notice that all the details of the specific cosmology are encoded into the functions $f(x_v-X_v)$, which act as measure in the space $\{x_v\,|\,v\in\mathcal{V}\}$, while $\tilde{\psi}_{\mathcal{G}}$ encodes all the features that do not depend on the cosmology or, said differently, are common to all cosmologies in conformally-flat space-times. For this reason, $\tilde{\psi}_{\mathcal{G}}$ is referred to as {\it universal integrand}. Therefore, one can focus on the analysis and computation of $\tilde{\psi}_{\mathcal{G}}$, which is now function the {\it energies} $\{x_v,\,y_e\,|\,v\in\mathcal{V},\,e\in\mathcal{E}\}$, and just after we can extrapolate the dynamics in a precise cosmology integrating $\tilde{\psi}_{\mathcal{G}}$ over $\{x_v\,|\,v\in\mathcal{V}\}$ with a specific integration measure $f(x_v-X_v)$ as well as extend to more general external scalars via the operators $\hat{\mathcal{O}}$ and \eqref{eq:PsiGnu}.

Interestingly enough, wavefunctions $\tilde{\psi}^{\mbox{\tiny $(\{\frac{1}{2}\},\{\nu_e\})$}}_{\mathcal{G}}$ with different propagating states ({\it i.e.} different $\nu_e$'s) can be related via a recursive relation ~\cite{Benincasa:2019vqr}
\begin{equation}\label{eq:RRpsi}
    \begin{split}
        &\tilde{\psi}^{\mbox{\tiny $(\{\frac{1}{2}\},\{\nu_e\})$}}_{\mathcal{G}}\:=\:
            \sum_{e\in\mathcal{E}}\hat{\mathcal{O}}_e\tilde{\psi}^{\mbox{\tiny $(\{\frac{1}{2}\},\nu_e-1,\{\nu_{\bar{e}}\})$}}_{\mathcal{G}},\\
        &\hat{\mathcal{O}}_e\,:=\,\frac{2\left(\nu_e-\frac{3}{2}\right)}{\displaystyle\sum_{v\in\mathcal{V}}x_v}
        \left(\frac{\partial}{\partial x_{v_e}}+\frac{\partial}{\partial x_{v'_e}}\right)-
        \frac{\partial^2}{\partial x_{v_e}\partial x_{v'_e}}
    \end{split}
\end{equation}
This relation can be proven considering the integrand in the definition of $\tilde{\psi}^{\mbox{\tiny $(\{\frac{1}{2}\},\{\nu_e\})$}}_{\mathcal{G}}$, applying the total-time translation operator $-i\sum_{v\in\mathcal{V}}\partial_{\eta_v}$ and integrating over the conformal times. As this integral is zero due to the Bunch-Davies condition as well to the vanishing of the bulk-to-bulk propagators at the boundary, acting with the total-time translation in the integrand provides an equation which, after suitable manipulation, returns \eqref{eq:RRpsi}. Importantly, the action of the total-time translation operator on the bulk-to-bulk propagators relates them to the ones for propagating states with order parameter $\nu_e-1$.

For states with $\nu_e=l+1/2$ ($l\in\mathbb{Z}_+$), the recursion relation can be iterated, connecting the wavefunction contribution having propagating states with general $l\in\mathbb{Z}_+$, to the one with only conformally coupled scalars, {\it i.e.} $\{\nu_e=1/2,\,\forall\,e\in\mathcal{E}\}$.

Some comments are now in order. This discussion showed four important facts:
\begin{enumerate}
    \item any graph contribution to the wavefunction coefficients with external arbitrary scalars can be obtained from the same graph contribution but with external conformally-coupled scalars via differential operators acting in energy space;
    \item the graph contributions different propagating states characterised by order parameters which differ by an integer are related to each other via \eqref{eq:RRpsi}. For the case of heavy mass states for which the order parameter is purely imaginary, \eqref{eq:RRpsi} relates them to states which do no appear to be in the Hilbert space. Hence, for heavy mass states it stays as a purely formal expression, while it becomes meaningful for light states, which are characterised by having $\nu_e$ real.
    \item for propagating states with half-integer order parameters, the recursion relation \eqref{eq:RRpsi} has the fully conformally-coupled case as end-point. Hence, the wavefunction with conformally-coupled scalars encodes much of the structure which characterise the wavefunction for other type of scalars and can be extracted via integral-differential operators acting on either the internal or external states;
    \item there exists a {\it universal integrand} common to any conformally-flat cosmology. While the specificity of a given cosmology can be extracted via an integration on the external energies with suitable measure (or via a differential operator), such a universal integrand encodes much of the analytic structure and physics of the wavefunction for this class of cosmologies. Interestingly, such a universal integrand is equivalent to the wavefunction for a massless scalar in flat-space with a space-like boundary.
\end{enumerate}
Since now on, we will focus on such an universal integrand, to which we will reserve the notation $\psi_{\mathcal{G}}$. We will make reference to the wavefunction integrand for different states or the integrated wavefunction when needed.

In the next subsections we will analyse different ways of organising the perturbation theory and, consequently, of computing the wavefunction. Also we will analyse the analytic structure of the wavefunction and its behaviour as its approached. Importantly, this analysis will be carried out at the level of the {\it universal integrand} just defined but it reflects in the full wavefunction contributions of the different scalars in FRW cosmologies via the integro-differential operators defined earlier and the integration over the external energies.


%
\begin{figure}[t]
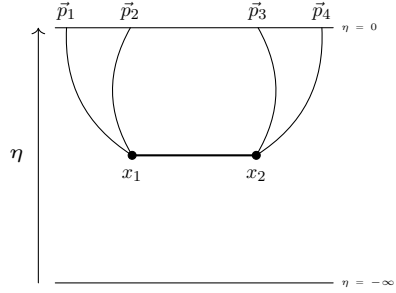

 \centering

 \caption{Feynman graphs and reduced graphs. A Feynman graphs (on the left) that contributes to the wavefunction of the universe, can be mapped into a reduced graph (on the right) by suppressing its external edges.}
 \label{fig:G}
\end{figure}

\subsection{Feynman representation and time diagrammatics}\label{subsec:FRGR}

Let us consider the contribution of a general graph $\mathcal{G}$ to the perturbative wavefunction of the universe and the universal integrand $\psi_{\mathcal{G}}$ associated to it
\begin{equation}\label{eq:UniInt}
    \psi_{\mathcal{G}}(x_v,y_e)\:=\:\int_{-\infty}^0\prod_{v\in\mathcal{V}}
        \left[d\eta_v\,e^{ix_v\eta_v}\right]\prod_{e\in\mathcal{E}}G(y_e;\eta_{v_e},\eta_{v'_e})
\end{equation}
where $\mathcal{V}$ and $\mathcal{E}$ are the set of sites and internal edges of $\mathcal{G}$, while $G(y_e;\eta_{v_e},\eta_{v'_e})$ is the bulk-to-bulk propagator
\begin{equation}\label{eq:UniG}
    \begin{split}
        G(y_e;\eta_{v_e},\eta_{v'_e})\:=\:\frac{1}{2y_e}
            &\left[
                e^{-iy_e(\eta_{v_e}-\eta_{v'_e})}\vartheta(\eta_{v_e}-\eta_{v'_e})+
                 e^{+iy_e(\eta_{v_e}-\eta_{v'_e})}\vartheta(\eta_{v'_e}-\eta_{v_e})
            \right.\\
            &\left.
                -e^{+iy_e(\eta_{v_e}+\eta_{v'_e})}
            \right].
    \end{split}
\end{equation}
Importantly, $\psi_{\mathcal{G}}$ depends on the external states just via the sum of the energies $x_v$ at the vertices $v\in\mathcal{V}$. Hence, we can consider {\it reduced graphs} which are obtained by the original Feynman graphs by suppressing the bulk-to-boundary lines, and assigning the weights $x_v$ and $y_e$ at the vertices and edges respectively.

Because a bulk-to-bulk propagator has three terms, if $n_e$ is the total number of edges $\mathcal{E}$ of the graph $\mathcal{G}$, then the resulting representation from the explicit time integration will have $3^{n_e}$ terms. Let us associate a decoration to the graph that keeps track of the time-ordered contribution from a given bulk-to-bulk propagator and its boundary term. For each term with $\vartheta(\eta_{v_e}-\eta_{v'_e})$, we associate an arrow to the edge $e$ directed from the site $v'_e$ with smaller time $\eta_{v'e}$ to the site with larger time $\eta_{v_e}$ and a dashed edge to indicate the term without time ordering.

\begin{figure}
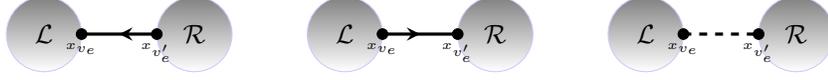

    \centering

    \caption{Time diagrammatics. A given edge can have three decoration assigned: an orientation that provides with a time ordering between the two sites, with the lower time associate to the source and the higher to the sink, while the dashing identifies the term of the bulk-to-bulk propagator without time ordering.}
    \label{fig:TimeDiag}
\end{figure}

The time integrations for the $3^{n_e}$ can be translated into operations on the graph. Each dashed edge is erased and substituted by the inverse of (twice) the corresponding energy $y_e$ and the energies associated to the sites at its endpoints is shifted by $+y_e$:
\begin{equation}\label{eq:TD1}

\end{equation}
For each oriented edge, its endpoints are collapsed onto each other following the direction of the decoration, generating a graph with one edge and one site less, associating the sum of the energies of the two sites to the sink of the erased edge, and with an overall factor of the inverse of the energy external to the source of the erased edge times the inverse of (twice) the energy of the erased edge:
\begin{equation}\label{eq:TD2}
    %
\end{equation}

An ordering for these operations is implied by the directions of the decoration, starting with the most external sources, while erasing the dashed lines commute with the collapsing operation on the directed edges. However, in case the direction on the edges do not imply a unique way of performing these graphical operations and the latter do not commute, one has to sum over all the inequivalent way of collapsing the edges, {\it e.g.}

\begin{equation}\label{eq:TD3}
    %
\end{equation}

Such operations on a graph implement the time integration, and hence explicitly reflect the time evolution of the process ~\cite{Arkani-Hamed:2017fdk}~. Notice that, via the recursion relation  \eqref{eq:RRpsi}, such a structure and operations on graphs persist for more general internal states as well as external states via the operators \eqref{eq:OnuE}.


\subsection{Old-fashioned perturbation theory as a boundary representation}\label{subsec:OFTP}

One price to pay to keep time-evolution manifest is to have a redundant representation con spurious singularities. In fact, because of the Bunch-Davies condition, the physical singularities of the wavefunction occur when sums of energies vanish, while the sheets in kinematic space such as $y_e\,=\,0$ do not correspond to any physical process. A general question then is whether there exist a representation that does not make any reference to an explicit time evolution and simultaneously being characterised by physical singularities only.

Let us consider the general definition \eqref{eq:UniInt} of a graph contribution to the Bunch-Davies wavefunction. Let $\hat{\Delta}$ be the total time-translation operator and let it act on the integrand of \eqref{eq:UniInt}. Because of the Bunch-Davies condition at early times as well as of the requirement that the bulk-to-bulk propagator vanishes at the boundary, the integral of $\hat{\Delta}$ acting on the integrand of \eqref{eq:UniInt} vanish:
\begin{equation}\label{eq:OFPTeq}
    0\:=\hspace{-.25cm}\int\limits_{-\infty(1-i\varepsilon)}^0\left[\prod_{v\in\mathcal{V}}d\eta_v\right]\hat{\Delta}
    \left[
        \prod_{v\in\mathcal{V}}e^{ix_v\eta_v}\prod_{e\in\mathcal{E}}G(y_e;\eta_{v_e},\eta_{v'_e})
    \right],
    \qquad
    \hat{\Delta}:=-i\sum_{v\in\mathcal{V}}\partial_{\eta_v}
\end{equation}
Importantly, when the operator $\hat{\Delta}$ acts on the external states, the terms is mapped into the product between the total energy and $\psi_{\mathcal{G}}$. When instead it acts on a bulk-to-bulk propagator $G$, just the boundary term survives and we obtain the sum of factorised integrals: each of these terms correspond to erasing one edge of the graph $\mathcal{G}$ and shift the energies associated to the sites at its endpoints by $+y_e$:
\begin{equation}\label{eq:OFPT1}
    \resizebox{0.9\hsize}{!}{$\displaystyle
    \left(\sum_{v\in\mathcal{V}}x_v\right)\psi_{\mathcal{G}}\:=\:\sum_{\centernot{e}\in\mathcal{E}}
        \int\limits_{-\infty(1-i\varepsilon)}^0
        \left[\prod_{v\in\tilde{\mathcal{V}}}d\eta_v\,e^{ix_v\eta_v}\right]
        e^{i(x_{v_{\centernot{e}}}+y_{\centernot{e}})\eta_{v_{\centernot{e}}}}
        e^{i(x_{v'_{\centernot{e}}}+y_{\centernot{e}})\eta_{v'_{\centernot{e}}}}
        \prod_{e\in\mathcal{E}\setminus\{\centernot{e}\}}G(y_{e};\eta_{v_e},\eta_{v'_e})
    $}
\end{equation}
Graphically, it can be represented as
\begin{equation}\label{eq:OFPT2}

\end{equation}
with the dashed red edges indicating the edge that get erased and the blobs represent the wavefunction associated to the subgraphs which the original graph reduces to upon the edge deletion.

An important remark is that this recursion relation generates a representation for the wavefunction with physical poles only, each of which is associated to a subgraph of $\mathcal{G}$. Furthermore, given a graph $\mathcal{G}$ with $n_e$ edges, it produced a perturbative expansion with $n_e!$ terms. A comparison with the time-diagrammatics discussed earlier, the latter shows spurious poles in correspondence of (twice) the energy of the internal states and it has $3^n_e$ terms. Interestingly, as long as $n_e\le6$, OFPT provides a lower number of terms, arriving to have $123$ terms less then the Feynman representation for $n_e=5$. However, as $n_e\ge7$, the number of terms of OFPT become rapidly bigger (already for $n_e=7$ it shows $5040$ terms more than the Feynman expansion),


\subsection{Frequency representation for tree level graphs}\label{subsec:FR}

Suitable manipulations on the bulk-to-bulk propagator leads us to two different representations, the Feynman and the OFPT, the former obtained maintaining its time-ordered + boundary term structure ({\it i.e.} with no manipulation), and the other one which is a purely boundary representation. As we will see in Section \ref{subsec:STrReps}, any graph contribution $\mathcal{G}$ to the wavefunction can be expressed as a plethora of representations, which can be easily obtained once the combinatorial structure underlying them is understood. For the time being, a further manipulation of the bulk-to-bulk propagator can allow us to obtain one more representation, which holds just for tree-graphs or, in any case, graphs with some tree-structure.

Let us focus on the most external edges of a given graph $\mathcal{G}$ and write it in terms of time integrals:
\begin{equation}\label{eq:GRRa}
    \begin{tikzpicture}[ball/.style = {circle, draw, align=center, anchor=north, inner sep=0}]
        \coordinate[label=below:{\tiny $x_1$}] (x1) at (0,0);
        \coordinate[label=below:{\tiny $x_2$}] (x2) at ($(x1)+(1,0)$);
        \coordinate (cc) at ($(x2)+(.5,0)$);
        \draw[thick] (x1) --node[midway, above] {\tiny $y_{12}$} (x2);
        \draw[fill,shade, color=blue!20] (cc) circle (.5cm);
        \draw[fill] (x1) circle (2pt);
        \draw[fill] (x2) circle (2pt);
        \node[anchor=center] at (cc) {$\mathcal{G}'$};
        \node[right=.5cm of cc] (tr) {$\displaystyle:=\:\int\limits_{-\infty}^0\prod_{s\in\mathcal{V}\setminus\{s_2\}}
            \hspace{-.25cm}
            \left[d\eta_s\,e^{ix_s\eta_s}\right]\int\limits_{-\infty}^0 d\eta_2\,e^{ix_2\eta_2}
            \hspace{-.25cm}
            \prod_{e\in\mathcal{E}\setminus\{e_{12}\}}
            \hspace{-.25cm}
            G(y_e;\eta_{s_e},\eta_{s'_e})I(y_{12},\eta_2)$};
    \end{tikzpicture}
\end{equation}
where $\mathcal{G}':=\mathcal{G}\setminus\{s_1,\,e_{12}\}\,\subset\,\mathcal{G}$ is the subgraph obtained without with the most external site $s_1$ and edge $e_{12}$, $\mathcal{V}':=\mathcal{V}\setminus\{s_1\}$ is the set of vertices of $\mathcal{G}'$ and
\begin{equation}\label{eq:I}
    I(y_{12},\eta_2):=\int\limits_{-\infty}^0d\eta_1\,e^{ix_1\eta_1}G(y_{12};\eta_1,\eta_2).
\end{equation}
The bulk-to-bulk propagator $G(y_{12};\eta_1,\eta_2)$ can be written as an integral over frequency
\begin{equation}\label{eq:Gw}
    G(y_{12};\,\eta_1,\eta_2)\:=\:\int\limits_{-\infty}^{+\infty}\frac{d\omega}{2\pi i}\,
        \left[
            \frac{e^{i\omega(\eta_1-\eta_2)}}{\omega^2-y_{12}^2+i\varepsilon}-
            \frac{e^{i\omega(\eta_1+\eta_2)}}{\omega^2-y_{12}^2+i\varepsilon}
        \right]
\end{equation}
Performing the integration over $\eta_1$, $I(y_{12},\eta_1)$ becomes a contour integral in the frequency plane $\omega$:
\begin{equation}
    \begin{split}
        I(y_{12},\eta_2)&\:=\:\int\limits_{-\infty}^{+\infty}\frac{d\omega}{2\pi i}\,\frac{1}{\omega-(-x_1+i\varepsilon)}
            \left[
                \frac{e^{-i\omega\eta_2}}{\omega^2-y_{12}^2+i\varepsilon}-\frac{e^{+i\omega\eta_2}}{\omega^2-y_{12}^2+i\varepsilon}
            \right]\:=\\
            &=\:\frac{1}{y_{12}^2-x_1^2}\left[e^{ix_1\eta_2}-e^{iy_{12}\eta_2}\right].
    \end{split}
\end{equation}
with the second line has been obtained by closing the contour of integration in the upper-half $\omega$-plane for the first term and in the lower-half-plane for the second one. Consequently, a tree level graph acquires the form 

\begin{equation}\label{eq:RRw}
    \begin{tikzpicture}[ball/.style = {circle, draw, align=center, anchor=north, inner sep=0}]
        \begin{scope}
            \coordinate[label=below:{\tiny $x_1$}] (x1) at (0,0);
            \coordinate[label=below:{\tiny $x_2$}] (x2) at ($(x1)+(1,0)$);
            \coordinate (cc) at ($(x2)+(.5,0)$);
            \draw[thick] (x1) --node[midway, above] {\tiny $y_{12}$} (x2);
            \draw[fill,shade, color=blue!20] (cc) circle (.5cm);
            \draw[fill] (x1) circle (2pt);
            \draw[fill] (x2) circle (2pt);
            \node[anchor=center] at (cc) {$\mathcal{G}'$};
            \node[right=.5cm of cc] (tr) {$\displaystyle\:=\:\frac{1}{y_{12}^2-x_1^2}\Bigg[$};
            \coordinate (ccA) at ($(tr)+(1.875,0)$);
            \coordinate (ccB) at ($(ccA)+(2,0)$);
            \draw[fill,shade, color=blue!20] (ccA) circle (.5cm);
            \node at (ccA) {$\displaystyle\mathcal{G}'$};
            \draw[fill,shade, color=blue!20] (ccB) circle (.5cm);
            \node at (ccB) {$\displaystyle\mathcal{G}'$};
            \node (mn) at ($(ccA)!.5!(ccB)$) {$\displaystyle -$};
            \node[right=.5cm of ccB] (fin) {$\displaystyle\Bigg]$};
            \coordinate[label=below:{\tiny $x_1+x_2$}] (s2l) at ($(ccA)-(.5,0)$);
            \coordinate[label=below:{\tiny $y_{12}+x_2$}] (s2r) at ($(ccB)-(.5,0)$);
            \draw[fill] (s2l) circle (2pt);
            \draw[fill] (s2r) circle (2pt);
        \end{scope}
    \end{tikzpicture}
\end{equation}
The equation \eqref{eq:RRw} provides a recursion relation that relates a graph $\mathcal{G}$ to a graph $\mathcal{G}'$ with one edge and one site less. Iteratively, it provides a perturbative representation with $2^{n_e}$ terms, showing more term than OFPT just for graphs with up to three edges, while shows constantly less terms than the Feynman representation.


\subsection{Analytic properties of the wavefunction}\label{subsec:AnPropWF}

The general arguments presented in \ref{subsec:SingFact} reveal the location of the singularities and the behaviour of the wavefunction as such singularities are approached. The universal wavefunction integrand turns out to be a rational function with simple poles only that vanishes as any of its energies are taken to infinity in any direction. This implies that its residues are linked to each other. First, let us recall that the singularities of the wavefunction are identified by the sum of the energies of its subprocesses. As such singularities are approached, energy conservation for the particular subprocess get restored and the wavefunction factorises into a flat-space scattering amplitude associated to the subprocess and the wavefunction associated to the complementary subprocess summed on the positive and negative solutions for the energies flowing in the edges which link the two factorised terms. As discussed in \ref{subsec:SingFact}, such a factorisation can be understood from the time integral picture, in which for each bulk-to-bulk propagator connecting the two subprocesses, there are just two of its three components which contribute to the wavefunction in the factorisation.

Let us get a more purely boundary perspective on the factorisation properties of the wavefunction, via its universal integrand. Let us $\mathcal{G}$ a generic graph and $\psi_{\mathcal{G}}$ the wavefunction universal integrand associated to it. It is a function of the energies of the external and internal states, encoded into the weights $\{x_s\,|\,s\in\mathcal{V}\}$ and $\{y_e\,|\,e\in\mathcal{E}\}$ respectively associated to the sites and edges of $\mathcal{G}$. Analytically continuing these variables to become complex, we can view the $\psi_{\mathcal{G}}$ as a function of one of these variables and integrating over the Riemann sphere $\hat{\mathbb{C}}$. Indeed, such an integral vanishes. On the other hand, it can be expressed in terms of sums of residues of all the poles in this variable
\begin{equation}\label{eq:CTPsi}
    0\:=\:\frac{1}{2\pi i}\oint_{\hat{\mathbb{C}}}dx_j\psi_{\mathcal{G}}(x_j)\:=\:
          \sum_{\bar{x}_j\in\mathfrak{P}_{x_j}}\mbox{Res}\{\psi_{\mathcal{G}}(x_j),\,x_j=\bar{x}_j\}
\end{equation}
where $\mathfrak{P}_{x_j}\,=\,\{\bar{x}_j\,|\,\psi_{\mathcal{G}}^{-1}(x_j=\bar{x}_j)=0,\,j\,\mbox{fixed}\}$ is the set of poles of $\psi_{\mathcal{G}}$ in the variable $x_j$. This is the first suggestion that the residues are not independent of each other and, consequently, neither should their physical interpretation.

Let us now consider a subgraph $\mathcal{G}_{\mathcal{I}}\subset\mathcal{G}$ and take the residue of $\psi_{\mathcal{G}}$ associated to it. Without making any assumption about the physical interpretation of the residue, the graph factorises and such factorisation can be depicted as
\begin{equation}\label{eq:GfactRes}

\end{equation}
where $\mathcal{G}_{\mathcal{I}},\,\mathcal{G}_{\centernot{J}}\,\ldots$ are the subgraphs in which $\mathcal{G}$ factorises, the dashed red circles indicate that the total energy associated to the encircled subgraph is taken to vanish, $x_{\mathcal{I}}=\sum_{s\in\mathcal{V}_{\mathcal{J}}}x_s$ ($\mathcal{V}_j$ being the set of sites in $\mathcal{G}_{\mathcal{I}}$). Hence, \eqref{eq:GfactRes} states that when the residue of $\psi_{\mathcal{G}}$ for the total energy pole for $\mathcal{G}_{\mathcal{I}}$ is taken, then $\psi_{\mathcal{G}}$ factorises into the flat-space scattering amplitude associated to the subprocess $\mathcal{G}_{\mathcal{I}}$, and some subprocess associated to the complementary graph $\bar{\mathcal{G}}_{\mathcal{I}}$ -- in the pictorial case shown in \eqref{eq:GfactRes}, $\bar{\mathcal{G}}_{\mathcal{I}}\,=\,\mathcal{G}_{\mathcal{J}}\cup\mathcal{G}_{\mathcal{K}}$ -- which can be represented as $\bar{\mathcal{G}}_{\mathcal{I}}\cup\centernot{\mathcal{E}}\cup\{s_{\centernot{e}}\}$, where $\centernot{\mathcal{E}}$ is the set of edges connecting $\mathcal{G}_{\mathcal{I}}$ and $\bar{\mathcal{G}}_{\mathcal{I}}$ and the sites $\{s_{\centernot{e}}\,|\,\centernot{e}\in\centernot{\mathcal{E}}\}$ have weights $\{x_{\mathcal{I}}+y_{\centernot{e}}\,|\,\centernot{e}\in\centernot{\mathcal{E}}\}$ and are all encircled, {\it i.e.} the residue of the wavefunction associated to $\bar{\mathcal{G}}_{\mathcal{I}}\cup\centernot{\mathcal{E}}\cup\{s_{\centernot{e}}\}$ in the pole $x_{\mathcal{I}}+\sum_{\centernot{e}\in\centernot{\mathcal{E}}}y_{\centernot{e}}=0$ is taken. Therefore, we need to establish the physical interpretation of such a residue and its relation to the others.

We can proceed inductively by considering the smallest graphs at tree and loop level, {\it i.e.} the two-site line graph and the tadpole. In the former case, let us take its integral over the Riemann sphere in the weight associated to any of the two sites. The poles that get picked are the total energy pole and one of the partial energy ones
\begin{equation}\label{eq:Res2sl}

\end{equation}
Consequently, the residue of the partial energy pole is the same, up to a sign, of the residue of the total energy pole, {\it i.e.} the scattering amplitude associated to the full process. The general argument outlined in Section \ref{subsec:SingFact} using the time integral formulation, suggested that the second residue on the right-hand-side of \eqref{eq:Res2sl} should be given by the wavefunction associated to a single site summed over the possible sign for the internal energy $y_{12}$. In fact
\begin{equation}\label{eq:Resx1}
    %
\end{equation}
where $\mathcal{A}_{s_1}=1$ is the flat-space amplitude associated to the individual site $s_1$, while $\psi_{s_2}$ is the (universal integrand of the) wavefunction associated to the site $s_2$. Not surprisingly, a reasoning based on the time integral returns a representation for the residue of the pole in question which contain a spurious pole associated to the energy of the internal state, while a purely boundary argument such as the one based on the contour integral \eqref{eq:Res2sl} returns a representation which is free from spurious poles as well as a direct connection with another residue. This has the important implication that there is just one independent information encoded in the two-site line graph and it is the flat-space amplitude.

A similar analysis can be carried out for the tadpole graph. It also has just two poles in the external energy $x$: the total energy pole and the one associated to the conservation of energy of the subprocess associated to the only site. Hence, the contour integral on the Riemann sphere implies that the two residues are equal, up to a sign:
\begin{equation}\label{eq:ResX}
    \begin{tikzpicture}[ball/.style = {circle, draw, align=center, anchor=north, inner sep=0}]
        \coordinate[label=left:{\tiny $x$}] (x) at (0,0);
        \coordinate (cc) at ($(x)+(.5,0)$);
        \draw[thick] (cc) circle (.5cm);
        \draw[fill] (x) circle (2pt);
        \draw[dashed, red, thick] (cc) circle (.625cm);
        \node[left=.25cm of x] (eq) {$\displaystyle 0\:=\:\frac{1}{2\pi i}\oint_{\hat{\mathbb{C}}}dx\,\psi_{\mathcal{G}}(x)\:=\:$};
        \node[right=.75cm of cc] (pl) {$\displaystyle +$};
        \coordinate[label=left:{\tiny $x$}] (xb) at ($(pl)+(.5,0)$);
        \coordinate (ccb) at ($(xb)+(.5,0)$);
        \draw[thick] (ccb) circle (.5cm);
        \draw[fill] (xb) circle (2pt);
        \draw[dashed, red, thick] (xb) circle (4pt);
    \end{tikzpicture}
\end{equation}
With these relations at hand, we can interpret in terms of scattering amplitudes all the residues of the universal integrand: when we consider a pole corresponding to a certain subgraph $\mathcal{G}_{\mathcal{I}}$, the relation given by the contour integral \eqref{eq:CTPsi} allow to express the residue of the pole associated to a subgraph containing a single site as a linear combination of lower point scattering amplitudes. To fix the ideas we can consider the three-site line and two-site one-loop graphs. In the first case, all the residues can be expressed as products of lower point-scattering amplitudes, except for the ones of the poles associated to single-site subgraphs which instead appear as a linear combination of all the other residues
\begin{equation}\label{eq:Res3sl}

\end{equation}
and similarly for the residues with respect to the poles in $y_{12}+x_2+x_3=0$ and $y_{23}+x_3=0$. The residue in the first two lines turn out to be the product of two lower-point amplitudes, while the one with respect to the poles associated to the subgraphs defined as one of the most external site only, can be expressed as linear combination of the amplitude associated to the full process and the product of two lower-point amplitudes. This procedure that makes use of the Cauchy theorem can be inductively iterated for the arbitrary graphs with three structures. For loop graphs, the situation is not at all dissimilar:
\begin{equation}\label{eq:3slpg}

\end{equation}
The first terms in the right-hand-side of both lines are directly connected with flat-space scattering process: the Cauchy theorem relates the first one to the total energy residue of the tadpole, while the second one to the sum of the residues of the poles associated to the subgraphs of the two-site one-loop graph containing both sites -- there are three of them: one providing the flat-space scattering amplitude for the full graph, and the other two which can be written as the product of the flat-space amplitudes associated to the two-site line graph and to the tadpole.
 

\subsection{Loops and trees}\label{subsec:LTs}

The analysis of the residues of the previous section revealed that for loop graphs they are related to three-graphs. For flat-space processes such relations have been exploited with the general aim of determining the loop integrands from their poles. At one-loop the relation between loops and tree graphs are encoded in the Feynman tree theorem ~\cite{Feynman:1963ax}~, which could be extended via causality to more general loop graphs ~\cite{CaronHuot:2010zt}~. The question is therefore if similar relations also hold for the wavefunction.

Let $\mathcal{G}$ be a graph and $\psi_{\mathcal{G}}$ be the associated wavefunction universal integrand. Any graph with $n_s$ sites and $n_e$ edges can be obtained from a graph with the same number of edges and $n_s+1$ site by suitably identifying two of its sites. Let us label such sites as $s_i$ and $s_j$ with weights $x_i$ and $x_j$ respectively. We can introduce a $1$-parameter family of deformations on $x_i$ and $x_j$ related to the two sites, while keeping all the other weights fixed and the total energy invariant ~\cite{Benincasa:2018ssx}~:
\begin{equation}\label{eq:Def}
    x_i\:\longrightarrow\:x_i+\varepsilon,\qquad
    x_j\:\longrightarrow\:x_j-\varepsilon,\qquad
    x_k\:\longrightarrow\:x_k,\;\forall\,k\,\neq\,i,\,j.
\end{equation}
Such an energy space deformation maps the wavefunction into a one-parameter family of wavefunctions, which can be examined as a function of a deformation parameter $\varepsilon$.

First, notice that the only poles in $\varepsilon$ correspond to those which do not contain simultaneously $x_i$ and $x_j$. In second instance, as the energies $x$'s and $y$'s are all positive, then all the poles whose location depends on $x_i$ are on the negative real axis, while the ones which depend on $x_j$ are located in the positive real axis. We will refer to these poles as {\it left} and {\it right} poles, respectively. Complexifying $\varepsilon$ and taking the contour along the imaginary axis closed equivalently either in the left- or right-half plane, the integral is then the sum of the residues in the relevant half-plane. Such residues turns out to be either functions of $x_i+x_j$ or as sums and differences of $x$'s and $y$'s and they are independent of $x_i$ and $x_j$. The latter turns out to be a set of spurious poles and they disappear upon summation over the residues of the left/right poles. The result is the wavefunction associated to the graph $\mathcal{G}'$ obtained by identifying the sites with weights $x_i$ and $x_j$ and associating to it the weight $x_i+x_j$ -- $\mathcal{G}'$ is a graph with the same number of edges, one site less and one loop more than $\mathcal{G}$:
\begin{equation}\label{eq:CosmFTT}
    \psi_{\mathcal{G}'}(x_s,y_e;\,n_s-1,n_e,L+1)\:=\:\int_{-i\infty}^{+i\infty}\frac{d\varepsilon}{2\pi i}\,\psi_{\mathcal{G}}(x_i+\varepsilon,x_j-\varepsilon,x_{s'},y_e;\,n_s,n_e,L)
\end{equation}
where in the notation for the wavefunction we added the number of sites, edges and loops, while $\{y_e\,|\,e\in\mathcal{E}\}$, $\{x_s\,|\,s\in\mathcal{V}\}$ and $\{x_{s'}\,|\,s'\in\mathcal{V}\setminus\{s_i,s_j\}\}$. This can be thought of as cosmological version of the (generalisation at all loops of the) Feynman-tree theorem. In the latter, the loop-momentum time component is integrated in the amplitude and the terms arising as residues of the poles re-sum into a forward tree-amplitude. In our case, we proceed in a graph-by-graph fashion by introducing a one-parameter deformation and integration along the imaginary axis in this parameter: the residues of the relevant poles re-sums to provide a graph with one-loop more, one site less and the same number of edges, linking a $L+1$-loop graph to the integral along the energies in some complex direction of a $L$-loop graphs. Importantly, the $L$-loop graph can be related to $(L+1)$-loop graphs with different topologies upon taking the integral along different complex energy directions, {\it i.e.} performing different deformations of the energy space.

\begin{figure}[t]
    \centering

    \caption{Examples of relations between graphs $\mathcal{G}'$ with $(L+1)$-loops, $n_s$ sites and $n_e$ edge and a graph $\mathcal{G}$ with $L$-loops, $n_e-1$ sites and $n_e$-edges. In this particular case, the very same three-site line graph $\mathcal{G}$ is related to two $1$-loop graphs with different topologies.}
    \label{fig:CFTT}
\end{figure}


\section{Back to the integrated wavefunction}\label{sec:BWF}

In the previous sections we saw how the analytic structure of the wavefunction universal integrand $\psi_{\mathcal{G}}$ is related to flat-space processes. If on one side it allows us to extract a big deal of physical information, $\psi_{\mathcal{G}}$ is nonetheless still an {\it integrand} and not the object we are ultimately interested in, except in the conformal case: when the theory is conformal, the universal integrand is the actual wavefunction.

In the more general case, the final wavefunction is given by the integral of $\psi_{\mathcal{G}}$ in the space of the weights $\{x_s\,|\,s\in\mathcal{V}\}$ associated to the sites $\mathcal{V}$ of $\mathcal{G}$, with measure $\{\lambda(x_s)\,|\,\forall\,s\in\,\mathcal{V}\}$, and domain of integration $\{x_s\in\,[X_s,\,+\infty[\,|\,\forall\,s\in\mathcal{V}\}$. The specific choice of the measures distinguishes among the different cosmologies. In any case, the integrated wavefunction inherits its singularity structure from the one of the integrand and how it gets intersected by the integration domain.

Interestingly enough, the integration with measure $\{\lambda(x_s)\,|\,\forall\,s\in\,\mathcal{V}\}$ maps $\psi_{\mathcal{G}}$, which is a rational function, into a transcendental wavefunction. This occurs even at tree-level. Let us compare this situation with the conformal one. In the latter, the time evolution is trivial: conformality makes time evolution the same irrespectively of whether the background space-time is expanding, flat or contracting. This is precisely the case in which the universal integrand {\it is} the actual wavefunction and it is a rational function. When instead the time evolution is non-trivial, the wavefunction is a transcendental function starting at tree-level. Transcendentality is the foot-print for having a non-trivial time evolution.

In this section we are going to discuss the structure of the integrated wavefunction $\tilde{\psi}_{\mathcal{G}}$ and how we can extract information about it from the universal integrand $\psi_{\mathcal{G}}$. It is important to emphasise that we are referring to the integration over the external energies which implements the details of the particular cosmology. While $\tilde{\psi}_{\mathcal{G}}$ is the full-fledge wavefunction for any tree-level graph $\mathcal{G}$, for loop graphs it still provides an integrand which needs to be integrated against the loop momenta with measure $\prod_{l=1}^L d^d\ell_l$. Despite its importance for understanding the early time physics, in what follows we will not discuss the loop integration. 

We will focus on the specific case of cubic couplings in $dS_{1+3}$ for which $\lambda_3(x_s)\:=\:\lambda\,\vartheta(x_s-X_s)$. The aim is to extract information about the contribution of an arbitrary graph $\mathcal{G}$ to the integrated wavefunction $\tilde{\psi}_{\mathcal{G}}$, which can be done in terms of the so-called {\it symbols} ~\cite{Chen:1977oja, Goncharov:2005sla, Goncharov:2009lql, Brown:2009qja, Goncharov:2010jf, Duhr:2011zq, Duhr:2014woa}~.


\subsection{Symbology}\label{subsec:Symb}

Let us begin with a very short and self-contained introduction of the so-called {\it symbols} associated to transcendental functions. First, a function $f_k$ is said to be a transcendental function of transcendental degree $k$ if it can be written as a linear combination of $k$-fold iterated integrals of the form
\begin{equation}\label{eq:fk}
    f_k\:=\:\int_{a}^b\,d\log{R}_1\circ\cdots\circ d\log{R}_k
\end{equation}
with $a$ and $b$ being rational numbers, $\{R_j(t)\,|\,j=1,\ldots,k\}$ being a set of $k$ rational functions with rational coefficients and
\begin{equation}\label{eq:ItInt}
    \int_{a}^b\,d\log{R}_1\circ\cdots\circ d\log{R}_k\: :=\:\int_{a}^b\left[\int_a^t d\log{R}_1\circ\cdots\circ d\log{R}_{k-1}\right]
    d\log{R_k}(t).
\end{equation}
Then, the symbol $\mathcal{S}(f_k)$ associated to the transcendental function $f_k$ is given by
\begin{equation}\label{eq:Sfk}
    \mathcal{S}(f_k)\: :=\:R_1\otimes\cdots\otimes R_k
\end{equation}
and it is defined as an element of the k-fold tensor product of the multiplicative group of rational functions (modulo constants) ~\cite{Goncharov:2009lql}~. The set $\{R_j\,|\,j=1,\ldots,k\}$ of entries of the symbol \eqref{eq:Sfk} is typically referred to as {\it alphabet} of $f_k$, and the elements $R_j$ of such a set as {\it letters}. The symbol $\mathcal{S}(f_k)$ satisfies the following properties
\begin{itemize}
    \item[\ding{111}] linearity;
    \item[\ding{111}] additivity:
          \begin{equation}\label{eq:Sfkp2}
                \begin{split}
                    &R_1\otimes\cdots\otimes(R_a R_b)\otimes\cdots\otimes R_k\:=\:
                     R_1\otimes\cdots\otimes R_a\otimes\cdots\otimes R_k +\\ 
                    &\hspace{2cm}R_1\otimes\cdots\otimes R_b\otimes\cdots\otimes R_k,\\
                    &R_1\otimes\cdots\otimes(R_a^n)\otimes\cdots\otimes R_k\:=\:
                    n\left[R_1\otimes\cdots\otimes R_a\otimes\cdots\otimes R_k\right],\\
                    &R_1\otimes\cdots\otimes\rho\otimes\cdots\otimes R_k\:=\:0,\qquad\rho^n=1
                \end{split}
          \end{equation}
          (this property is a consequence of the additivity of the logarithm);
    \item[\ding{111}] invariance under the multiplication of a letter of the alphabet by a constant
          \begin{equation}\label{eq:Sfkp3}
                R_1\otimes\cdots\otimes(c\,R_a)\otimes\cdots\otimes R_k\:=\:
                    R_1\otimes\cdots\otimes R_a\otimes\cdots\otimes R_k;
          \end{equation}
    \item[\ding{111}] the symbol of the product of two functions is the shuffle product $\dsqcup$ of the two functions
          \begin{equation}\label{eq:Sfkp4}
                \mathcal{S}(f_k^{\mbox{\tiny $(1)$}}f_k^{\mbox{\tiny $(2)$}})\:=\:
                    \mathcal{S}(f_k^{\mbox{\tiny $(1)$}})\dsqcup\,\mathcal{S}(f_k^{\mbox{\tiny $(2)$}})
          \end{equation}
          where $\dsqcup$ is defined as
          \begin{equation}\label{eq:shp}
                \begin{split}
              (R_1\otimes R_2)\dsqcup(R_3\otimes R_4)\: :=\:
                &R_1\otimes R_2\otimes R_3\otimes R_4 + \\
                &\hspace{-4cm}R_1\otimes R_3\otimes R_2\otimes R_4 + R_3\otimes R_1\otimes R_2\otimes R_4 +
                R_1\otimes R_3\otimes R_4\otimes R_2 +\\ 
                &\hspace{-4cm}R_3\otimes R_1\otimes R_4\otimes R_2 + R_3\otimes R_4\otimes R_1\otimes R_2
                \end{split}
          \end{equation}
\end{itemize}

Importantly, the knowledge of the symbols determines a transcendental function up to lower degree functions multiplied by a number of appropriate degree of transcendentality.

Finally, one can turn the table around and consider an alphabet and an arbitrary tensor of its letters
\begin{equation}\label{eq:Sanz}
    S\:=\:\sum_{i_1,\ldots,i_k}c_{i_1\ldots i_k}\,R_{i_1}\otimes\cdots\otimes R_{i_k}
\end{equation}
and ask whether it is always possible to find a function $f_k$ whose symbol $\mathcal{S}(f_k)$ is precisely $S$ as given in \eqref{eq:Sanz}. It turns out that there exists a function $f_k$ such that $\mathcal{S}(f_k)\,=\,S$ if and only if the following integrability condition holds ~\cite{Chen:1977oja}
\begin{equation}\label{eq:SfkIntCond}
    \sum_{i_1,\ldots,i_k}c_{i_1\ldots i_k}\,d\log{R_{i_j}}\wedge d\log{R_{i_{j+1}}}\,R_{i_1}\otimes\cdots\otimes R_{i_{j-1}}
        \otimes R_{i_{j+2}}\otimes R_{i_k}\:=\:0
\end{equation}
$\forall\:j\in[1,\,k-1]$.

As an example of a transcendental function and its symbols, let us consider the classic polylogarithms
\begin{equation}\label{eq:ClPlogs}
    \mbox{Li}_k(z)\:=\:\int_0^z\,\mbox{Li}_{k-1}(t)\,d\log{t},\qquad
    \mbox{Li}_1(z)\:=\:-\log{(1-z)}
\end{equation}
then their symbol is given by
\begin{equation}\label{eq:SClPlogs}
    \mathcal{S}(\mbox{Li}_k(z))\:=\:-(1-z)\otimes\underbrace{z\otimes\cdots\otimes z}_{k-1\mbox{ times}}.
\end{equation}
%


\subsection{Universal integrands, symbols and the integrated wavefunction}\label{subsec:ExtrSymb}

The universal integrand $\psi_{\mathcal{G}}$ is a function of the weights associated to the sites $\{x_s\,|\,\forall\, s\in\mathcal{V}\}$ and the edges $\{y_e\,|\,\forall\,e\in\mathcal{E}\}$. However, the map leading from $\psi_{\mathcal{G}}$ to the integrated wavefunction $\tilde{\psi}_{\mathcal{G}}$ involves an integration over $\{x_s\,|\,\forall\, s\in\mathcal{V}\}$ only. Furthermore, as the multiple residue of degree $n_e$ along compatible channels is $\prod_{e\in\mathcal{E}}(2y_e)^{-1}$, we can consider the function
\begin{equation}\label{eq:fgdef}
    f_{\mathcal{G}}\: :=\:\left[\prod_{e\in\mathcal{E}}(2y_e)\right]\psi_{\mathcal{G}}
\end{equation}
Any singularity corresponds to a subgraph $\mathfrak{g}\subseteq\mathcal{G}$ and when such a singularity is approached, the wavefunction universal integrand factorises into a flat-space amplitude $\mathcal{A}_{\mathfrak{g}}$ times a contribution $\hat{\psi}_{\overline{\mathfrak{g}}}$ which can be written as a linear combination of the wavefunction associated to the complementary subgraph $\overline{\mathfrak{g}}$ of $\mathfrak{g}$, with the sum running over all the possible combination of signs associated to the energies of the cut edges, and with coefficients given by $\mp\prod_{\centernot{e}\in\centernot{\mathcal{E}}}(2y_{\centernot{e}})^{-1}$ -- $\centernot{\mathcal{E}}$ is the set of cut edges, {\it i.e.} the edges connecting $\mathfrak{g}$ and $\overline{\mathfrak{g}}$:
\begin{equation}
    \Res_{E_{\mathfrak{g}}}\{\psi_{\mathcal{G}}\}\:=\:\mathcal{A}_{\mathfrak{g}}\times\hat{\psi}_{\overline{\mathfrak{g}}}.
\end{equation}
As noticed in Section \ref{subsec:AnPropWF}, the residues associated to the subgraphs containing a single site only can be rewritten in terms of a linear combination of other residues, providing an alternative way of writing $\hat{\psi}_{\mathcal{G}}$.

Then, the contribution to the symbol of the discontinuity associated to the subgraph $\mathfrak{g}\subseteq\mathcal{G}$ is given by
\begin{equation}\label{eq:SymbCont1}
    \left.\mathcal{S}(f_{\mathcal{G}})\right|_{\mathfrak{g}}\:=\:E_{\mathfrak{g}}\otimes
        \mathcal{S}(\mathcal{A}_{\mathfrak{g}}\times\hat{f}_{\overline{\mathfrak{g}}})\:=\:
        E_{\mathfrak{g}}\otimes\left[\mathcal{S}(\mathcal{A}_{\mathfrak{g}})\dsqcup\,\mathcal{S}(\hat{f}_{\overline{\mathfrak{g}}})\right]
\end{equation}
where $\hat{f}_{\overline{\mathfrak{g}}}:=\prod_{e\in\mathcal{E}_{\overline{\mathfrak{g}}}}(2y_e)\hat{\psi}_{\overline{\mathfrak{g}}}$

The symbol for the full function $f_{\mathcal{G}}$ is then given by summing over all the discontinuities, {\it i.e.} the subgraphs $\mathfrak{g}\subseteq\mathcal{G}$:
\begin{equation}\label{eq:SymbCont2}
    \mathcal{S}(f_{\mathcal{G}})\:=\:\sum_{\mathfrak{g}\subseteq\mathcal{G}}E_{\mathfrak{g}}\otimes\left[\mathcal{S}(\mathcal{A}_{\mathfrak{g}})\dsqcup\,\mathcal{S}(\hat{f}_{\overline{\mathfrak{g}}})\right]
\end{equation}
Notice that the contribution to the symbol coming from the discontinuity along the total energy singularity can be expressed as a sum of series of letters each of which appear in the symbol contribution from the other discontinuities. Hence, \eqref{eq:SymbCont2} can be reorganised to acquire the following form ~\cite{Hillman:2019wgh}~
\begin{equation}\label{eq:SymbContFin}
    \mathcal{S}(f_{\mathcal{G}})\:=\:\sum_{\mathfrak{g}\subset\mathcal{G}}\frac{E_{\mathfrak{g}}}{E_{\mathcal{G}}}\otimes
        \left[\mathcal{S}(\mathcal{A}_{\mathfrak{g}})\dsqcup\,\mathcal{S}(\hat{f}_{\overline{\mathfrak{g}}})\right]
\end{equation}
where now the sum runs over the set of all subgraphs but $\mathcal{G}$. The formula \eqref{eq:SymbContFin} provides a recursion relation to extract the symbols.

Let us end this section by commenting on the extraction of the actual wavefunction from the symbols. As we stated earlier, the symbols do not determine a given function uniquely, but up to transcendental functions of lower degree multiplied by a number of appropriate degree of transcendentality. There is no general procedure to solve this ambiguity. However, it can be done in some explicit case by requiring ~\cite{Hillman:2019wgh}~
\begin{equation}\label{eq:SymbFunct}
    \lim_{y_e\longrightarrow0}f_{\mathcal{G}}\:=\:0,\qquad\forall\;e\in\mathcal{E}.
\end{equation}
For the $2$-site line and $2$-site one-graphs, the symbols define $f_{\mathcal{G}}$ to be a linear combination of dilogarithms and products of logarithms, up to a constant. The condition \eqref{eq:SymbFunct} fixes the such a constant to be $+\pi^2/6$ and $-\pi^2/6$ respectively. Indicating the $f_{\mbox{\tiny $(n_s,L)$}}:=f_{\mathcal{G}}$, where the pair $(n_s,L)$ in the subscript specifies the numbers of sites and loops respectively, then the integrated functions can be written as ~\cite{Hillman:2019wgh}~
\begin{equation}\label{eq:Funct2s}
    \begin{split}
        &\resizebox{0.9\hsize}{!}{$\displaystyle
         f_{\mbox{\tiny $(2,0)$}}\:=\:
            -\mbox{Li}_2\left\{\frac{X_1-Y_{12}}{X_1+Y_{12}}\right\}
            -\mbox{Li}_2\left\{\frac{X_2-Y_{12}}{X_2+Y_{12}}\right\}
            -\mbox{Li}_1\left\{\frac{X_1-Y_{12}}{X_1+Y_{12}}\right\}\mbox{Li}\left\{\frac{X_2-Y_{12}}{X_2+Y_{12}}\right\}
            +\frac{\pi^2}{6}$}\\
        &\resizebox{0.9\hsize}{!}{$\displaystyle
         \phantom{f_{\mbox{\tiny $(2,0)$}}}\:=\:
            -\mbox{Li}_2\left\{\frac{X_1-Y_{12}}{X_1+Y_{12}}\right\}
            -\mbox{Li}_2\left\{\frac{X_2-Y_{12}}{X_2+Y_{12}}\right\}
            +\mbox{Li}_2\left\{\frac{(X_1-Y_{12})(X_2-Y_{12})}{(X_1+Y_{12})(X_2+Y_{12})}\right\}+\frac{\pi^2}{6}$},\\
        &\resizebox{0.9\hsize}{!}{$\displaystyle
         f_{\mbox{\tiny $(2,1)$}}\:=\:
            -\mbox{Li}_2\left\{\frac{X_1-Y_a-Y_b}{X_1+X_2}\right\}-\mbox{Li}_2\left\{\frac{X_2-Y_a-Y_b}{X_1+X_2}\right\}
            -\mbox{Li}_1\left\{\frac{X_1-Y_a-Y_b}{X_1+X_2}\right\}\mbox{Li}_1\left\{\frac{X_2-Y_a-Y_b}{X_1+X_2}\right\}$}\\
        &\resizebox{0.9\hsize}{!}{$\displaystyle
         \phantom{f_{\mbox{\tiny $(2,1)$}}\:}
            +\mbox{Li}_2\left\{\frac{X_1-Y_a+Y_b}{X_1+X_2+2Y_b}\right\}+\mbox{Li}_2\left\{\frac{X_2-Y_a+Y_b}{X_1+X_2+2Y_b}\right\}
            +\mbox{Li}_1\left\{\frac{X_1-Y_a+Y_b}{X_1+X_2+2Y_b}\right\}\mbox{Li}_1\left\{\frac{X_2-Y_a+Y_b}{X_1+X_2+2Y_b}\right\}
         $}\\
        &\resizebox{0.9\hsize}{!}{$\displaystyle
         \phantom{f_{\mbox{\tiny $(2,1)$}}\:}
            +\mbox{Li}_2\left\{\frac{X_1+Y_a-Y_b}{X_1+X_2+2Y_a}\right\}+\mbox{Li}_2\left\{\frac{X_2+Y_a-Y_b}{X_1+X_2+2Y_b}\right\}
            +\mbox{Li}_1\left\{\frac{X_1+Y_a-Y_b}{X_1+X_2+2Y_a}\right\}\mbox{Li}_1\left\{\frac{X_2+Y_a-Y_b}{X_1+X_2+2Y_a}\right\}
            -\frac{\pi^2}{6}
         $}
    \end{split}
\end{equation}
with $f_{\mbox{\tiny $(2,0)$}}$ being in agreement with previous computations performed via standard ~\cite{Arkani-Hamed:2015bza}~ and bootstrap ~\cite{Arkani-Hamed:2018kmz}~ methods. 

Besides these two example, it is a priori possible to reconstruct the final function out of the symbol for transcendental degree $k\,<\,4$. Such a function can be written in terms of classical polylogarithms $\mbox{Li}_k$ only ~\cite{Goncharov:1994}~. This provides a set of functions in terms of which the functions we are interested in can be expressed. Such functions can be organised according to their symmetry properties under the exchanges of letters in the symbols ~\cite{Goncharov:2010jf}~. This procedure allows to fix $f_{\mbox{\tiny $(3,0)$}}$ for the three-site line graph ~\cite{Hillman:2019wgh}~.


\newpage

\addcontentsline{toc}{section}{\underline{Part II: A Combinatorial Origin For The Wavefunction}}
\section*{\hfill \underline{Part II: A Combinatorial Origin For The Wavefunction}\hfill}\label{sec:CombWF}

In the second part of this review, we are going to change gear a provide a first principle mathematical construction which turns out to have all the characteristics we ascribed to the wavefunction of the universe. Starting with such a definition, we will exploit its structure and we will link it to physical properties of the wavefunction.


\section{A crash course on projective polytopes}\label{sec:Polyt}

Before getting into to core of the geometrical-combinatorial description of the wavefunction of the universe, it is worth to summarise some basic ideas about the mathematical objects which will accompany us for the rest of this review: the projective polytopes. The reader who is already familiar with the subject can skip this section and jump directly to the following one. However, for a more in depth discussion of the more general subject of positive geometries we refer to ~\cite{Arkani-Hamed:2017tmz}~. In this section we will review the basic notions of projective polytopes and their duals as well as their characterisation via canonical forms.


\subsection{Defining projective polytopes}\label{subsec:PP}

Let us consider a set of vectors $Z_k^I\,\in\,\mathbb{R}^{N+1}$, with $k\,=\,1,\ldots,\nu$ and $I\,=\,1,\ldots,N+1$ being the $\mathbb{R}^{N+1}$ index. A projective polytope is then defined as the convex hull $\mathcal{P}\,\subset\,\mathbb{P}^{N}(\mathbb{R})$ of the vertices $Z_k^I$:
\begin{equation}\label{eq:PYZ}
    \mathcal{P}(\mathcal{Y},Z)\: :=\:
    \left\{
        \mathcal{Y}^I\,=\,\sum_{k=1}^{\nu}c_k\,Z_k^I\:\in\:\mathbb{P}^N(\mathbb{R})\:\big|\:c_k\,\ge\,0,\:\forall\:k\,=\,1,\ldots,\nu
    \right\},
\end{equation}
with $\mathcal{Y}$ vanishing if and only if $c_k=0,\;\forall\:k\,=\,1,\ldots,\nu$. Any projective polytope is invariant under the $GL(1)$-transformations $\mathcal{Y}\,\longrightarrow\,\lambda\,\mathcal{Y}$ and $Z_k\,\longrightarrow\,\lambda\,Z_k$ ($\lambda\,\in\,\mathbb{R}_+$). If the number $\nu$ of vertices of $\mathcal{P}$ is equal to $N+1$ and all the vertices $\{Z_k\}_{k=1}^{N+1}$ are linearly independent -- thus, they form a basis for $\mathbb{R}^{N+1}$ --, then $\mathcal{P}$ is a {\it simplex} in $\mathbb{P}^N$.

Rather than defining a polytope $\mathcal{P}$ as the convex hull of a set of vertices, it can be also defined via a set of homogeneous (linear) polynomial inequalities $q_j(\mathcal{Y})\,\equiv\,\mathcal{Y}^I\mathcal{W}^{\mbox{\tiny $(j)$}}_I\,\ge\,0$ ($j\,=\,1,\ldots,\tilde{\nu}$), with $\mathcal{W}^{\mbox{\tiny $(j)$}}_I$ being co-vectors in $\mathbb{R}^{N+1}$. Each co-vector $\mathcal{W}_I^{\mbox{\tiny $(j)$}}$ defines a hyperplane in $\mathbb{R}^{N+1}$ and its intersection with the convex hull $\mathcal{P}(\mathcal{Y},Z)$ identifies a codimension-$1$ boundary -- a {\it facet} -- of the polytope $\mathcal{P}$. A vertex $Z^I_k$ is on a given facet $\mathcal{W}^{\mbox{\tiny $(j)$}}$ if and only if $Z^I_k\mathcal{W}^{\mbox{\tiny $(j)$}}_I\,=\,0$. If $Z_{a_{j+1}},\ldots,Z_{a_{j+N}}$ is a subset of vertices of $\mathcal{P}$ on a certain facet $\mathcal{W}^{\mbox{\tiny $(j)$}}_I$ and form a basis of $\mathbb{R}^{N}$, then the co-vector identifying the facet can be written in terms of such subset of vertices as
\begin{equation}\label{eq:WZ}
    \mathcal{W}^{\mbox{\tiny $(j)$}}_I\:=\:(-1)^{(j-1)(N-1)}\varepsilon_{\mbox{\tiny $I\,K_1\ldots K_N$}}Z^{\mbox{\tiny $K_1$}}_{a_{j+1}}\ldots Z^{\mbox{\tiny $K_N$}}_{a_{j+N}},
\end{equation}
where $\varepsilon_{\mbox{\tiny $I\,K_1\ldots K_N$}}$ is the totally anti-symmetric $(N+1)$-dimensional Levi-Civita symbol.


\subsection{Projective polytopes and canonical forms}\label{subsec:PPCF}

Given the definition of a projective polytope $\mathcal{P}$ in $\mathbb{P}^N$ via the inequalities $q_{j}(\mathcal{Y})\,\ge\,0$ ($j\,=\,1,\ldots\tilde{\nu}$), it is possible to associate a differential form $\omega(\mathcal{Y},\,\mathcal{P})$ to it with (logarithmic) singularities only along the boundaries of $\mathcal{P}$, {\it i.e.} where one or more homogeneous linear polynomials $q_j(\mathcal{Y})$ vanish:
\begin{equation}\label{eq:w}
    \omega(\mathcal{Y},\,\mathcal{P})\:=\:\frac{\mathfrak{n}(\mathcal{Y})\langle\mathcal{Y}d^N\mathcal{Y}\rangle}{\prod_{j=1}^{\tilde{\nu}}q_j(\mathcal{Y})},
\end{equation}
where $\langle\mathcal{Y}d^N\mathcal{Y}\rangle$ is the standard measure in $\mathbb{P}^N$:
\begin{equation}\label{eq:PNmeas}
    \langle\mathcal{Y}d^N\mathcal{Y}\rangle\: :=\:\varepsilon_{\mbox{\tiny $I_1 I_2\ldots I_{N+1}$}}\mathcal{Y}^{\mbox{\tiny $I_1$}}d\mathcal{Y}^{\mbox{\tiny $I_2$}}\ldots d\mathcal{Y}^{\mbox{\tiny $I_{N+1}$}}.
\end{equation}
The differential form \eqref{eq:w} is said to be the {\it canonical form} attached to the polytope $\mathcal{P}$ if it satisfies the following requirements:
\begin{enumerate}
    \item given a constant $\lambda\,\in\,\mathbb{R}_+$, $\omega(\mathcal{Y},\,\mathcal{P})$ has to be invariant under the $GL(1)$ transformations     
          $\mathcal{Y}\,\longrightarrow\,\lambda\,\mathcal{Y}$ and $Z_k\,\longrightarrow\,\lambda\,Z_k$ ($\forall\:k\,=\,1,\ldots \nu$). This requirement fixes the numerator $\mathfrak{n}(\mathcal{Y})$ to be a polynomial in $\mathcal{Y}$ of degree $\tilde{\nu}-N-1$;
    \item the residue of $\omega(\mathcal{Y},\,\mathcal{P})$ along any of the poles $q_j(\mathcal{Y})\,=\,0$ is the canonical form of the related facet of     
          $\mathcal{P}$, which is still a polytope $\partial\mathcal{P}^{\mbox{\tiny $(j)$}}$ living in $\mathbb{P}^{N-1}$:
          \begin{equation}\label{eq:Res}
              \Res_{\partial\mathcal{P}^{\mbox{\tiny $(j)$}}}\left\{\omega(\mathcal{Y},\,\mathcal{P})\right\}\:=\:
              \omega(\mathcal{Y}_{\partial},\,\partial\mathcal{P}^{\mbox{\tiny $(j)$}}).
          \end{equation}
          where $\mathcal{Y}_{\partial}\,\in\,\mathbb{P}^{N-1}$, and $\Res$ is the {\it residue operator} applied to $\omega(\mathcal{Y},\,\mathcal{P})$ along the facet $\partial\mathcal{P}^{\mbox{\tiny $(j)$}}$. If we parametrise $\mathbb{P}^N$ with a set of local homogeneous coordinates $(y_j,\,h_j)$ such that the locus $h_j\,=\,0$ identifies the facet $\partial\mathcal{P}^{\mbox{\tiny (j)}}$ and $y_j$ collectively indicates the remaining local coordinates, then the canonical form $\omega(\mathcal{Y},\,\mathcal{P})$ shows a simple pole in $h_j\,=\,0$:
          \begin{equation}\label{eq:Res2}
              \omega(\mathcal{Y},\,\mathcal{P})\:=\:\omega(y_j)\wedge\,\frac{dh_j}{h_j}+\tilde{\omega},
          \end{equation}
          with $\tilde{\omega}$ being analytic in $h_j$ (and, thus, not contributing to the residue), while $\omega(y_j)$ is a codimension-$1$ differential form depending on the collective coordinates $y_j$ and constituting the canonical form of the facet $\partial\mathcal{P}^{\mbox{\tiny $(j)$}}$:
          \begin{equation}\label{eq:Res3}
              \Res_{\partial\mathcal{P}^{\mbox{\tiny $(j)$}}}\left\{\omega(\mathcal{Y},\,\mathcal{P})\right\}\:=\:
              \Res_{h_j\,=\,0}\left\{\omega(\mathcal{Y},\,\mathcal{P})\right\}\:=\:
              \omega(y_j)\:=\:\omega(\mathcal{Y}_{\partial},\,\partial\mathcal{P}^{\mbox{\tiny $(j)$}}),
          \end{equation}
          where all the equalities are valid locally;
    \item for $\mathbb{P}^0$, the canonical form is a constant. Such a constant can be set to $1$, up to a sign which        depends on the orientation. This implies that, taking the residue operator $\Res$ on the canonical form 
          $\omega(\mathcal{Y},\,\mathcal{P})$ ($\mathcal{P}\,\subset\,\mathbb{P}^N$) iteratively $N$-times, it should yield $\pm\,1$ depending on the orientation, or $0$. These highest codimension singularities, which are related to the vertices of the polytope $\mathcal{P}\,\subset\,\mathbb{P}^N$ are called {\it leading singularities}.
\end{enumerate}
The canonical form $\omega(\mathcal{Y},\,\mathcal{P})$ provides a characterisation for the polytope $\mathcal{P}$, associating the facets $\partial\mathcal{P}^{\mbox{\tiny $(j)$}}$ of the latter to its singularities. It can be written both in terms of the co-vectors $\mathcal{W}_I^{\mbox{\tiny $(j)$}}$ identifying the facets of $\mathcal{P}$ as well as in terms of its vertices $Z_k^I$:
\begin{equation}\label{eq:wWZ}
    \omega(\mathcal{Y},\,\mathcal{P})\:=\:\frac{\mathfrak{n}(\mathcal{Y})\langle\mathcal{Y}d^N\mathcal{Y}\rangle}{
                                                \prod_{j=1}^{\tilde{\nu}}(\mathcal{Y}\cdot\mathcal{W}^{\mbox{\tiny $(j)$}})}
                                     \:=\:\frac{\mathfrak{n}(\mathcal{Y})\langle\mathcal{Y}d^N\mathcal{Y}\rangle}{
                                                \prod_{j=1}^{\tilde{\nu}}\langle\mathcal{Y}Z_{a_{j+1}}\ldots Z_{a_{j+N}}\rangle}
\end{equation}
where $\mathcal{Y}\cdot\mathcal{W}^{\mbox{\tiny $(j)$}}\, :=\,\mathcal{Y}^I\mathcal{W}_I^{\mbox{\tiny $(j)$}}$ -- the last equality in \eqref{eq:wWZ} has been obtained via the relation \eqref{eq:WZ} between vertices of $\mathcal{P}$ and hyperplanes containing its facets and, consequently, the notation $\langle\ldots\rangle$ indicates the contraction of its arguments via an $(N+1)$-dimensional Levi-Civita symbol. 

Furthermore, the non-vanishing multiple residues of $\omega(\mathcal{Y},\mathcal{P})$ describe higher co-dimension faces of $\mathcal{P}$ and are determined by the intersection of multiple facets of $\mathcal{P}$ {\it inside} $\mathcal{P}$ itself. The hypersurface identified by the intersections of the facets of $\mathcal{P}$ {\it outside} $\mathcal{P}$ determines the numerator $\mathfrak{n}(\mathcal{Y})$ of $\omega(\mathcal{Y},\,\mathcal{P})$ ~\cite{Arkani-Hamed:2014dca}~. 

The canonical form can be written as a coefficient $\Omega(\mathcal{Y},\mathcal{P})$, named {\it canonical coefficient}, times the standard measure $\langle\mathcal{Y}d^N\mathcal{Y}\rangle$ of $\mathbb{P}^N$
\begin{equation}\label{eq:cfun}
    \omega(\mathcal{Y},\,\mathcal{P})\:=\:\Omega(\mathcal{Y},\,\mathcal{P})\langle\mathcal{Y}d^N\mathcal{Y}\rangle,
\end{equation}
with the canonical function $\Omega(\mathcal{Y},\,\mathcal{P})$ having the following contour integral representation ~\cite{Arkani-Hamed:2014dca}~:
\begin{equation}\label{eq:cfunint}
    \Omega(\mathcal{Y},\,\mathcal{P})\:=\:\frac{1}{N!(2\pi i)^{\nu-N-1}}\int_{\mathbb{R}^{\nu}}
                \prod_{k=1}^{\nu}\frac{dc_k}{c_k-i\varepsilon_k}\,\delta^{\mbox{\tiny $(N+1)$}}\left(\mathcal{Y}-\sum_{k=1}^{\nu}c_k Z_k\right)
\end{equation}
as $\varepsilon_k\,\longrightarrow\,0$, $\forall\; k\,=\,1,\ldots,\nu$. All the possible contours along which the integral \eqref{eq:cfunint} can be performed provide a different triangulation of the polytope $\mathcal{P}$.


\subsection{Dual polytopes}\label{subsec:DP}

Let us consider a polytope $\mathcal{P}\,\subset\,\mathbb{P}^N$ with its vertices given by the vectors $Z_k^I$ ($k\,=\,1,\ldots,\nu$) and its facets by the co-vectors $\mathcal{W}^{\mbox{\tiny $(j)$}}_I$ ($j\,=\,1,\ldots,\tilde{\nu}$) defined via \eqref{eq:WZ}. Then the dual polytope $\tilde{\mathcal{P}}\,\subset\,\mathbb{P}^N$ is defined as the convex hull identified by the vertices $\mathcal{W}^{\mbox{\tiny $(j)$}}_I$ ($j\,=\,1,\ldots,\tilde{\nu}$) in the dual space $\mathbb{P}^N$ of $\mathbb{P}^N$:
\begin{equation}\label{eq:Pdual}
    \tilde{\mathcal{P}}(\mathcal{Y},\,\mathcal{W})\: :=\:
    \left\{
        \mathcal{Y}_I\:=\:\sum_{j=1}^{\tilde{\nu}}c_j\,\mathcal{W}_I^{\mbox{\tiny $(j)$}}\:\in\:\mathbb{P}^N(\mathbb{R})\,\big|\,c_j\,\ge\,0,\:
        \forall\:j\,=\,1,\ldots,\tilde{\nu}
    \right\}
\end{equation}
Because of the relation \eqref{eq:WZ} between $Z_k$'s and $\mathcal{W}^{\mbox{\tiny $(j)$}}$, in the dual space the $Z_k$'s identify the facets of $\tilde{\mathcal{P}}$: $\tilde{q}_k(\mathcal{Y})\,=\,\mathcal{Y}_I\,Z_k^I\,\ge\,0$ ($k\,=\,1,\ldots,\nu$). Hence, the canonical function $\Omega(\mathcal{Y},\,\mathcal{P})$ of the polytope $\mathcal{P}$ can be interpreted as the volume of the dual polytope $\tilde{\mathcal{P}}$ if we consider the co-vectors $\mathcal{W}^{\mbox{\tiny $(j)$}}$ as vertices in the dual space:
\begin{equation}\label{eq:VolPd}
    \Omega(\mathcal{Y},\,\mathcal{P})\:=\:\frac{\mathfrak{n}(\mathcal{Y})}{
                                                \prod_{j=1}^{\tilde{\nu}}\langle\mathcal{Y}Z_{a_{j+1}}\ldots Z_{a_{j+N}}\rangle}
                                     \:=\:\frac{\mathfrak{n}(\mathcal{Y})}{
                                                \prod_{j=1}^{\tilde{\nu}}(\mathcal{Y}\cdot\mathcal{W}^{\mbox{\tiny $(j)$}})}
                                     \:=\:\mbox{Vol}\{\tilde{\mathcal{P}}(\mathcal{Y},\mathcal{W})\}.
\end{equation}
Vice versa, the canonical function $\Omega(\mathcal{Y},\,\tilde{\mathcal{P}})$ associated to $\tilde{\mathcal{P}}$ can be interpreted as the volume of $\mathcal{P}$ if we consider the vectors $Z_k$ as vertices:
\begin{equation}\label{eq:VolPd2}
    \Omega(\mathcal{Y},\,\tilde{\mathcal{P}})\:=\:\frac{\tilde{\mathfrak{n}}(\mathcal{Y})}{
                                    \prod_{k=1}^{\nu}\langle\mathcal{Y}\mathcal{W}^{\mbox{\tiny $(b_{k+1})$}}\ldots\mathcal{W}^{\mbox{\tiny $(b_{k+N})$}}\rangle}
                               \:=\:\frac{\tilde{\mathfrak{n}}(\mathcal{Y})}{
                                    \prod_{k=1}^{\nu}(\mathcal{Y}\cdot Z_{k})}
                               \:=\:\mbox{Vol}\{\mathcal{P}(\mathcal{Y},\,Z)\}.
\end{equation}
%


\subsection{Signed triangulations}\label{eq:ST}

Let us consider a collection of projective polytopes $\mathcal{P}^{\mbox{\tiny $(j)$}}\,\subset\,\mathbb{P}^N$ ($j\,=\,1,\ldots,n$) and a polytope $\mathcal{P}\,\subset\,\mathbb{P}^N$. Then the former constitutes a {\it triangulation} of the latter if the following conditions hold:
{
\renewcommand{\theenumi}{\roman{enumi}}
\begin{enumerate}
    \item $\mathcal{P}^{\mbox{\tiny $(j)$}}\,\subset\,\mathcal{P}, \;\forall\: j\,=\,1,\ldots\,n$ with compatible orientations;
    \item their interiors are disjoint;
    \item $\mathcal{P}$ is the union of the $\mathcal{P}^{\mbox{\tiny $(j)$}}$'s: $\displaystyle\mathcal{P}\,=\,\bigcup_{j=1}^n\mathcal{P}^{\mbox{\tiny $(j)$}}$
\end{enumerate}
}
From the perspective of the canonical forms associated to $\mathcal{P}$ and to the $\mathcal{P}^{\mbox{\tiny $(j)$}}$'s, these requirements translates into the property of the canonical form of $\mathcal{P}$ that can be expressed as the sum of the canonical forms of the $\mathcal{P}^{\mbox{\tiny $(j)$}}$ in such a way that the singularities related to the common facets of the $\mathcal{P}^{\mbox{\tiny $(j)$}}$ become spurious, {\it i.e.} they cancel:
\begin{equation}\label{eq:wsum}
    \omega(\mathcal{Y},\,\mathcal{P})\:=\:\sum_{j=1}^n\omega(\mathcal{Y},\,\mathcal{P}^{\mbox{\tiny $(j)$}}).
\end{equation}
The notion of triangulation can be generalised \cite{Arkani-Hamed:2017tmz}. Let us consider a collection of projective polytopes $\mathcal{P}^{\mbox{\tiny $(j)$}}\,\subset\,\mathbb{P}^N$ ($j\,=\,1,\ldots\,n+1$) and let $\mathcal{Y}\,\in\,\mathbb{P}^N$ be a generic point in $\mathbb{P}^N$. Then our collection $\{\mathcal{P}^{\mbox{\tiny $(j)$}}\}_{j=1}^{n+1}$ {\it interior triangulates} the empty set if for any point $\mathcal{Y}\,\in\,\mathbb{P}^N$ belonging the union of all the $\mathcal{P}^{\mbox{\tiny $(j)$}}$'s, the number of $\mathcal{P}^{\mbox{\tiny $(j)$}}$'s with positive and negative orientation at $\mathcal{Y}$ is the same. Hence, any element of the collection $\{\mathcal{P}^{\mbox{\tiny $(j)$}}\}_{j=1}^{n+1}$ taken with opposite orientation is interior triangulated by the other elements. The notion of interior triangulation reduces to the standard triangulation introduced earlier if any point $\mathcal{Y}\,\in\,\mathbb{P}^N$ is contained exactly in one of the elements of the collection of polytopes $\{\mathcal{P}^{\mbox{\tiny $(j)$}}\}_{j=1}^{n+1}$.

A collection of polytopes instead provides a {\it canonical form triangulation} of the empty set if the sum of their canonical form vanishes:
\begin{equation}\label{eq:CfTr}
    \sum_{j=1}^{n+1}\omega(\mathcal{Y},\,\mathcal{P}^{\mbox{\tiny $(j)$}})\:=\:0
\end{equation}
As for the interior triangulation, any element of such a collection with its orientation reversed is canonical-form triangulated by the other elements, {\it e.g.}
\begin{equation}\label{eq:CfTr2}
    \omega(\mathcal{Y},\,\mathcal{P}_{-}^{\mbox{\tiny $(n+1)$}})\:=\:\sum_{j=1}^n\omega(\mathcal{Y},\,\mathcal{P}^{\mbox{\tiny $(j)$}}),
\end{equation}
where $\mathcal{P}_{-}^{\mbox{\tiny $(n+1)$}}$ denotes the polytope $\mathcal{P}^{\mbox{\tiny $(n+1)$}}$ with reversed orientation.


\subsection{Differential forms with non-logarithmic singularities}\label{eq:NonLog}

\begin{figure}[t]
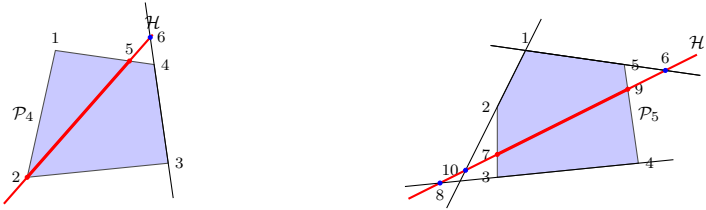

              \centering

              \caption{Examples of hyperplanes intersecting polytopes in $\mathbb{P}^2$ both on the convex hulls and outside. The covariant restriction of the 
                       canonical forms of the polytopes onto such hyperplanes will be in covariant pairing with $\mathcal{P}_{\mathcal{H}}$ as they are associated to the specific signed triangulation which goes through these points outside $\mathcal{P}_{\mathcal{H}}$.}
              \label{fig:PH2}
          \end{figure}

Let us now consider a projective polytope $\mathcal{P}\,\subset\,\mathbb{P}^N$ and its associated canonical form $\omega(\mathcal{Y},\,\mathcal{P})$. Let $\mathcal{F}_{\mathcal{P}}\, :=\,\{\mathcal{W}_I^{\mbox{\tiny $(j)$}}\,\in\,\mathbb{P}^N,\:j\,=\,1,\ldots,\tilde{\nu}\}$ be the set of the facets of $\mathcal{P}$, and let $\mathcal{H}\,\subset\,\mathbb{P}^N$ be an hyperplane of co-dimension $N-M$, with a non-vanishing intersection with $\mathcal{P}$ but such that it does not belong to $\mathcal{F}_{\mathcal{P}}$, {\it i.e.} it contains none of the facets of $\mathcal{P}$:
\begin{equation}\label{eq:H}
    \begin{split}
    \mathcal{H}\: :=\: 
        &\left\{
            \mathcal{Y}\,\in\,\mathbb{P}^N(\mathcal{R})\:\big|\: h_l(\mathcal{Y})\, :=\,\mathcal{Y}^I\mathfrak{h}_I^{\mbox{\tiny $(l)$}}\,=\,0,\;
                \mathcal{P}\cap\mathfrak{h}_I^{\mbox{\tiny $(l)$}}\,\neq\,\varnothing,\:
                \mathfrak{h}_I^{\mbox{\tiny $(l)$}}\,\notin\,\mathcal{F}_{\mathcal{P}},
         \right.\\
        &\hspace{3cm}\forall\,l=1,\ldots,N-M\Big\}.
    \end{split}
\end{equation}
Let $\mathcal{P}_{\mathcal{H}}\, :=\,\mathcal{P}\cap\mathcal{H}$ be the restriction of $\mathcal{P}$ onto $\mathcal{H}$. It is then possible to define the {\it covariant restriction} of the canonical form $\omega(\mathcal{Y},\,\mathcal{P})$ associated to $\mathcal{P}$ onto $\mathcal{H}$ as the differential form ~\cite{Benincasa:2020aoj}
\begin{equation}\label{eq:wH}
    \omega(\mathcal{Y}_{\mathcal{H}})\: :=\:\mathcal{L}^{\mbox{\tiny $(0)$}}_{\mathcal{H}}\left\{\omega(\mathcal{Y},\mathcal{P})\right\}\:=\:
        \frac{1}{(2\pi i)^{N-M}}\oint_{\mathcal{H}}\,\frac{\omega(\mathcal{Y},\,\mathcal{P})}{\prod_{l=1}^{N-M}h_l(\mathcal{Y})}
\end{equation}
where $\mathcal{L}^{\mbox{\tiny $(0)$}}$ is the Laurent operator acting along a co-dimension $N-M$ hyperplane and extracting the zero-th order coefficient. If we parametrise $\mathbb{P}^N$ via the local homogeneous coordinates $(y,\,h)$, with $h\,:=\,\{h_1,\ldots,h_{N-M}\}$ being the collective coordinate such that $h\,=\,0$ locally identifies $\mathcal{H}$ and $y$ collectively indicating the remaining local coordinates, then the canonical form $\omega(\mathcal{Y},\,\mathcal{P})$ can be written as
\begin{equation}
    \omega(\mathcal{Y},\,\mathcal{P})\:=\:\omega^{\mbox{\tiny $(N-M)$}}(y)\wedge dh+\tilde{\omega}
\end{equation}
where $\tilde{\omega}$ depends polynomially on $h$ with degree higher or equal to $1$. Consequently, the form $\omega^{\mbox{\tiny $(N-M)$}}(y)$ provides locally the covariant restriction \eqref{eq:wH}
\begin{equation}\label{eq:wHloc}
    \mathcal{L}^{\mbox{\tiny $(0)$}}_{\mathcal{H}}\left\{\omega(\mathcal{Y},\mathcal{P})\right\}\:=\:
    \mathcal{L}^{\mbox{\tiny $(0)$}}_{h=0}\left\{\omega(\mathcal{Y},\mathcal{P})\right\}\:=\:
    \omega^{\mbox{\tiny $(N-M)$}}(y)\:=\:
    \omega^{\mbox{\tiny $(N-M)$}}(\mathcal{Y}_{\mathcal{H}}).
\end{equation}
Notice that the covariant restriction \eqref{eq:wH} of a canonical form along any hyperplane $\mathcal{H}$ has some remarkable properties:
{
\renewcommand{\theenumi}{\roman{enumi}}
\begin{enumerate}
    \item it is no longer $GL(1)$-invariant, rather it is $GL(1)$-covariant -- under the $GL(1)$-transformation, it scales as $\lambda^{\mbox{\tiny $-(N-M)$}}$:
          \begin{equation}\label{eq:GL1cov}
              \omega^{\mbox{\tiny $(N-M)$}}(\mathcal{Y}_{\mathcal{H}})\:
                    \xrightarrow{\mathcal{Y}_{\mathcal{H}}\,\longrightarrow\,\lambda\mathcal{Y}_{\mathcal{H}}}\:
                    \lambda^{\mbox{\tiny $-(N-M)$}}\omega^{\mbox{\tiny $(N-M)$}}(\mathcal{Y}_{\mathcal{H}});
          \end{equation}
    \item if $\mathcal{W}^{\mbox{\tiny $(j_1\ldots j_{m_j})$}}\,:=\,\bigcap_{r=1}^{m_j}\mathcal{W}^{\mbox{\tiny $(j_r)$}}$ is the intersection of $m_j$ facets 
          and $\mathcal{H}\cap\mathcal{W}^{\mbox{\tiny $(j_1\ldots j_{m_j})$}}\,\neq\,\varnothing$, then the linear homogeneous polynomials $q_{j_r}(\mathcal{Y})\,=\,\mathcal{Y}\cdot\mathcal{W}^{\mbox{\tiny $(j_r)$}}$ ($r\,=\,1,\ldots,m_{j}$) which identify a subset of the poles in the canonical form $\omega(\mathcal{Y},\mathcal{P})$, collapse into each other onto the covariant restriction on $\mathcal{H}$, producing a higher multiplicity pole in $\omega^{\mbox{\tiny $(N-M)$}}(\mathcal{Y}_{\mathcal{H}})$;
    \item if $\mathcal{H}$ intersects the boundaries of the polytope $\mathcal{P}$ just on the polytope itself, then the covariant restriction of the canonical 
          form $\omega(\mathcal{Y},\mathcal{P})$ onto $\mathcal{H}$ has poles only along the boundaries of the restriction  $\mathcal{P}_{\mathcal{H}}$ of $\mathcal{P}$ onto $\mathcal{H}$ and their multiplicity is given by the co-dimension of the boundary of $\mathcal{P}$ that $\mathcal{H}$ intersects, which is lowered if such an intersection lies on the locus of the zeroes of the canonical form $\omega(\mathcal{Y},\mathcal{P})$. 
          \begin{figure}[t]
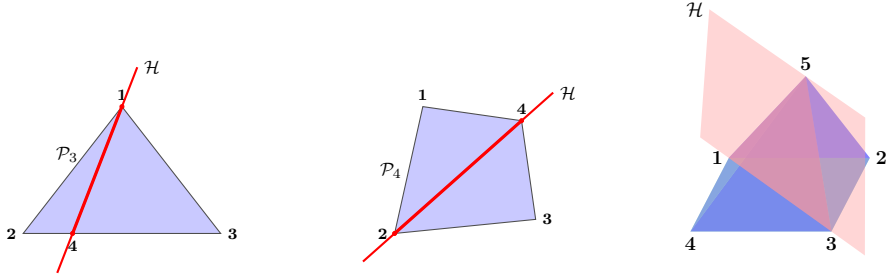

              \centering

              \caption{Examples of hyperplanes $\mathcal{H}$ intersecting polytopes in $\mathbb{P}^2$ and $\mathbb{P}^3$ only on the convex hulls themselves. In 
                       all three cases, the covariant restrictions of the canonical forms onto such hyperplanes are covariant forms associated to the restrictions of the original polytopes. Such restrictions for the triangle, quadrilateral and square pyramid are, respectively, the segments $(\mathbf{14})$ and $\mathbf{(24)}$, and the triangle $(\mathbf{135})$, with the covariant forms having: one double pole and one simple pole for $(\mathbf{14})$; two double poles for $(\mathbf{14})$; and two double poles and one simple poles for $(\mathbf{135})$.}
              \label{fig:PH}
          \end{figure}
          In this case, $\omega^{\mbox{\tiny $(N-M)$}}(\mathcal{Y}_{\mathcal{H}})$ can be naturally associated to the polytope $\mathcal{P}_{\mathcal{H}}\,=\,\mathcal{P}\cap\mathcal{H}$ and it is said to be a {\it covariant form} ~\cite{Benincasa:2020aoj} (see Figure \ref{fig:PH});
    \item if $\mathcal{H}$ intersects the boundaries of the polytope $\mathcal{P}$ both inside and outside, then the covariant 
          restriction 
          $\omega^{\mbox{\tiny $(N-M)$}}(\mathcal{Y}_{\mathcal{H}})$ of the canonical form $\omega(\mathcal{Y},\mathcal{P})$ onto $\mathcal{H}$, show poles both along the boundaries of $\mathcal{P}_{\mathcal{H}}$ and outside. It can be naturally understood as coming from the restriction of a signed triangulation of $\mathcal{P}$ through such points, which translates in a signed triangulation of $\mathcal{P}_{\mathcal{H}}$. Such covariant restrictions are said to be in {\it covariant pairing} with $\mathcal{P}_{\mathcal{H}}$ ~\cite{Benincasa:2020aoj} (see Figure \ref{fig:PH2}).
\end{enumerate}
}
Now we are ready to focus on a special class of polytopes, called {\it cosmological polytopes} ~\cite{Arkani-Hamed:2017fdk, Benincasa:2019vqr}~, which provide a combinatorial first-principle definition for the perturbative wavefunction of the universe for a large class of toy models with propagating conformally-coupled ~\cite{Arkani-Hamed:2017fdk} and massless ~\cite{Benincasa:2019vqr} in FRW cosmologies, and more generally for scalars with polynomial interactions and time-dependent couplings.
          %
          

\section{Cosmological polytopes}\label{sec:}

Let us begin with considering a general connected graph $\mathcal{G}$ with $n_s$ sites\footnote{In order to avoid language clash, we will refer to the vertices of a graph as {\it sites}, while reserving the term {\it vertices} for the polytopes.} and $n_e$ edges. Let us assign the weights $x_j$ and $y_{kl}$ to the $j$-th site and to the edge between the $k$-th and $l$-th, for the sites and edges. Notice that any connected graph $\mathcal{G}$ with $n_s$ sites and $n_e$ edges can be seen as a collection of $n_e$ graphs with two sites and one edge and $2n_e-n_s$ sites identified:
\begin{equation*}
  
\end{equation*}
Let us focus on one of such two-site graphs with one edge. It is possible to identify a projective space $\mathbb{P}^2$ with local coordinates given by the weights of the two-site graph, so that a generic point $\mathcal{Y}\,\in\,\mathbb{P}^2$ is parametrised as $\mathcal{Y}\,=\,(x_1,\,y_{12},\,x_2)$, as well as a triangle $\triangle$ embedded in it with the midpoints of its sides given by the vectors $\mathbf{x}_1\,=\,(1,0,0)$, $\mathbf{y}_{12}\,=\,(0,1,0)$ and $\mathbf{x}_2\,=\,(0,0,1)$: the triangle is defined as convex hull of the vectors $\{\mathbf{x}_1-\mathbf{y}_{12}+\mathbf{x}_2,\:\mathbf{x}_1+\mathbf{y}_{12}-\mathbf{x}_2,-\mathbf{x}_1+\mathbf{y}_{12}+\mathbf{x}_2\}$ or equivalently by the inequalities $x_1+x_2\,\ge\,0$, $x_1+y_{12}\,\ge\,0$ and $y_{12}+x_2\,\ge\,0$:
\begin{equation*}\label{eq:2sTr}
    \begin{tikzpicture}[ball/.style = {circle, draw, align=center, anchor=north, inner sep=0}, cross/.style={cross out, draw, minimum size=2*(#1-\pgflinewidth),
                        inner sep=0pt, outer sep=0pt}]
        \coordinate[label=below:{\tiny $x_1$}] (x1) at (0,0);
        \coordinate[label=below:{\tiny $x_2$}] (x2) at ($(x1)+(1.5,0)$);
        \draw[-,thick] (x1) --node[above] {\tiny $y_{12}$} (x2);
        \draw[fill] (x1) circle (2pt);
        \draw[fill] (x2) circle (2pt);
        \begin{scope}[scale={.8}, transform shape]
        \coordinate[label=above:{\footnotesize $\mathbf{x}_1-\mathbf{y}_{12}+\mathbf{x}_2$}] (Z1) at ($(x2)+(5,1.125)$);
        \coordinate[label=below:{\footnotesize $\mathbf{x}_1+\mathbf{y}_{12}-\mathbf{x}_2$}] (Z2) at ($(Z1)+(-1.5,-2.25)$);
        \coordinate[label=below:{\footnotesize $-\mathbf{x}_1+\mathbf{y}_{12}+\mathbf{x}_2$}] (Z3) at ($(Z1)+(+1.5,-2.25)$);
        \draw[-,thick,color=blue] (Z1) -- (Z2);
        \draw[-,thick,color=blue] (Z1) -- (Z3);
        \draw[-,thick,color=red] (Z2) -- (Z3);
        \coordinate[label=left:{\footnotesize $\mathbf{x}_1$}] (X1) at ($(Z1)!0.5!(Z2)$);
        \coordinate[label=right:{\footnotesize $\mathbf{x}_2$}] (X2) at ($(Z1)!0.5!(Z3)$);
        \coordinate[label=below:{\footnotesize $\mathbf{y}_{12}$}] (X3) at ($(Z2)!0.5!(Z3)$);
        \draw[color=blue,fill] (X1) circle (1pt);
        \draw[color=blue,fill] (X2) circle (1pt);
        \draw[color=red,fill] (X3) circle (1pt);
        \end{scope}
        \coordinate (a1) at ($(x2)+(1.5,0)$);
        \coordinate (a2) at ($(X1)-(1.5,0)$);
        \draw[solid,color=red!90!blue,line width=.1cm,shade,preaction={-triangle 90,thin,draw,color=red!90!blue,shorten >=-1mm}] (a1) -- (a2);
        \draw[solid,color=red!90!blue,line width=.1cm,shade,preaction={-triangle 90,thin,draw,color=red!90!blue,shorten >=-1mm}] (a2) -- (a1);
    \end{tikzpicture}
\end{equation*}
and a generic point inside such a triangle can be expressed as
\begin{equation}\label{eq:Ytr}
    \mathcal{Y}\:=\:x_1\mathbf{x}_1+y_{12}\mathbf{y}_{12}+x_2\mathbf{x}_2,
\end{equation}
with the coefficients $x_1$, $y_{12}$ and $x_2$ precisely satisfying the inequalities defining the facets of the triangle.

A collection of $n_e$ two-sites line graph is then described by a collection of $n_e$ triangles in $\mathbb{P}^{3n_e-1}$. Identifying $2n_e-n_s$ sites of the collection of graphs in order to obtain a connected graph $\mathcal{G}$ with $n_s$ sites and $n_e$ edges, from the projective polytope perspective corresponds to intersecting the collection of triangles in the midpoints of at most two out of its three sides: the convex hull $\mathcal{P}_{\mathcal{G}}$ in $\mathbb{P}^{n_s+n_e-1}$ of the $3n_e$ vertices of the intersected triangles is the cosmological polytope associated to $\mathcal{G}$
\begin{equation}\label{eq:CPdef}
    \mathcal{P}_{\mathcal{G}}\:=\:
    \scalebox{.75}{
    $\displaystyle
        \left\{
            \mathcal{Y}=\sum_{j=1}^{n_e}\sum_{k=1}^3c_k^{\mbox{\tiny $(j)$}}Z_k^{\mbox{\tiny $(j)$}}\,\in\,\mathbb{P}^{n_s+n_e-1}\,
            \left|
            \begin{array}{l}
                c_k^{\mbox{\tiny $(j)$}}\,\ge\,0,\\
                \{Z_{k-1}^{\mbox{\tiny $(j)$}} + Z_k^{\mbox{\tiny $(j)$}}\,\sim\,Z_{k-1}^{\mbox{\tiny $(j')$}} + Z_k^{\mbox{\tiny $(j')$}}\}_r,
            \end{array}
            \begin{array}{l}
                 \forall\, k=1,2,3,\:\forall\, j\in[1,n_e]  \\
                 k=1,2,\:j\neq j'\in[1,n_e] 
            \end{array}
            \right.
        \right\},
    $
        }
\end{equation}
where $Z_k^{\mbox{\tiny $(j)$}}$ ($k\,=\,1,2,3$) indicates the vertices of the $j$-th triangle, while $\{Z_{k-1}^{\mbox{\tiny $(j)$}} + Z_k^{\mbox{\tiny $(j)$}}\,\sim\,Z_{k-1}^{\mbox{\tiny $(j')$}} + Z_k^{\mbox{\tiny $(j')$}}\}_r$ are the conditions on the vertices imposed by intersecting the triangles in the midpoints of their sides, with $r\,=\,2n_e-n_s\,\in\,[n_e-1,\,2(n_e-1)]$ being the number of relation between pairs of vertices of different triangles: imposing that two segments intersect each other in their midpoint is equivalent to say that their four vertices belong to the same $2$-plane.

\begin{figure*}[t]
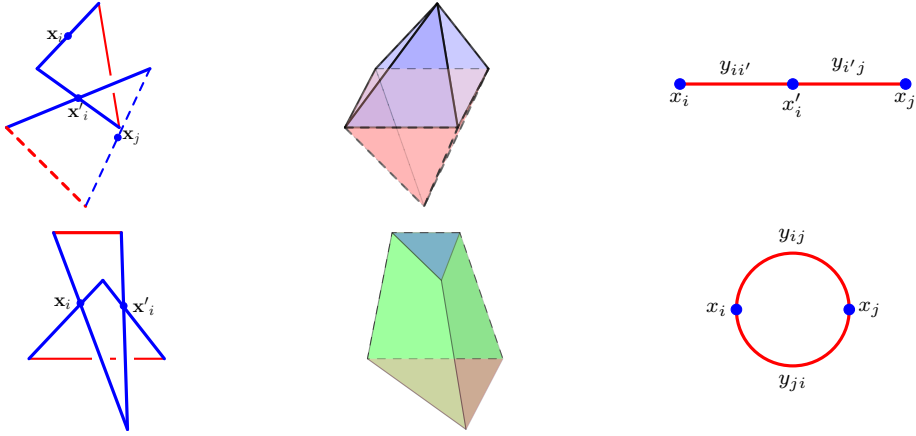


\caption{Cosmological polytopes obtained by intersecting two triangles. The two rows show the cosmological polytopes generated via the intersection on one and two midpoints, respectively, with their convex hulls and the associated graphs.}
 \label{fig:cp}
\end{figure*}

Given a cosmological polytope $\mathcal{P}_{\mathcal{G}}\,\subset\,\mathbb{P}^{n_s+n_e-1}$, the associated canonical form
\begin{equation}\label{eq:cfCP}
    \omega(\mathcal{Y},\mathcal{P}_{\mathcal{G}})\:=\:\Omega(\mathcal{Y},\mathcal{P}_{\mathcal{G}})\langle\mathcal{Y}d^{n_s+n_e-1}\mathcal{Y}\rangle
\end{equation}
has logarithmic singularities on, and only on, all of its boundaries ~\cite{Arkani-Hamed:2017fdk}~. Its canonical function $\Omega(\mathcal{Y},\mathcal{P}_{\mathcal{G}})$ is precisely the universal integrand associated to the related graph $\mathcal{G}$
\begin{equation}\label{eq:CPWF}
    \Omega(\mathcal{Y},\mathcal{P}_{\mathcal{G}})\:=\:\psi_{\mathcal{G}}(x_s,\, y_e).
\end{equation}
Importantly, as the canonical form $\omega(\mathcal{Y},\mathcal{P}_{\mathcal{G}})$ is singular only along the boundaries of the cosmological polytope $\mathcal{P}_{\mathcal{G}}$, such boundaries precisely encode the residues of the wavefunction $\psi_{\mathcal{G}}$.


\subsection{Facet structure of the cosmological polytopes}\label{subsec:FSCP}

The definition of the cosmological polytopes as intersection of a collection of triangles, allows for an immediate characterisation of their face structure. 

Let $\mathcal{P}_{\mathcal{G}}\,\subset\,\mathbb{P}^{n_s+n_e-1}$ be the cosmological polytope associated to a graph $\mathcal{G}$. Let the vectors $\{Z_k^I\,\in\,\mathbb{P}^{n_s+n_e-1}\}_{k=1}^{3n_e}$ be the set of vertices of $\mathcal{P}_{\mathcal{G}}$. As for any other polytope, the facets of the cosmological polytopes are identified by those hyperplanes given by the co-vectors $\mathcal{W}_I^{\mbox{\tiny $(j)$}}\,\in\,\mathbb{P}^{n_s+n_e-1}$ such that $\mathcal{P}_{\mathcal{G}}\,\cap\,\mathcal{W}^{\mbox{\tiny $(j)$}}_I\,\neq\,\varnothing$ and $Z_k^I\mathcal{W}_I^{\mbox{\tiny $(j)$}}\,\ge\,0$. Characterising the facets means establishing their vertex structure. In order to identify which vertices belong to a given facet $\mathcal{W}^{\mbox{\tiny $(j)$}}$, we need to establish the set $\mathcal{V}_{\mathcal{W}^{\mbox{\tiny $(j)$}}}$ of all the vertices $Z_k$'s such that $Z_k^I\mathcal{W}_I^{\mbox{\tiny $(j)$}}\,=\,0$ , while keeping $Z_{k'}^I\mathcal{W}_I^{\mbox{\tiny $(j)$}}\,\ge\,0$ for all $Z_{k'}\,\not\in\,\mathcal{V}_{\mathcal{W}^{\mbox{\tiny $(j)$}}}$.

Given a graph $\mathcal{G}$, every edge $e_{ij}$ connecting two sites $s_i$ and $s_j$ has three vertices of the cosmological polytope $\mathcal{P}_{\mathcal{G}}$ associated to it:
\begin{equation}\label{eq:Ve}
    \{\mathbf{x}_i-\mathbf{y}_{ij}+\mathbf{x}_j,\; \mathbf{x}_i+\mathbf{y}_{ij}-\mathbf{x}_j,\; -\mathbf{x}_i+\mathbf{y}_{ij}+\mathbf{x}_j\}
\end{equation}
Let us introduce a basis $\{\mathbf{\tilde{x}}_I^{\mbox{\tiny $(i)$}},\,\mathbf{\tilde{y}}_I^{\mbox{\tiny $(ij)$}},\,\mathbf{\tilde{x}}_I^{\mbox{\tiny $(j)$}}\}$ for the co-vectors, such that 
\begin{equation}\label{eq:bscv}
\mathbf{x}^I_i\mathbf{\tilde{x}}_I^{\mbox{\tiny $(j)$}}\,=\,\delta_i^{\phantom{i}j},\qquad
\mathbf{y}^I_{ij}\mathbf{\tilde{y}}_I^{\mbox{\tiny $(kl)$}}\,=\,\delta_{ij}^{\phantom{ij}kl},\qquad 
\mathbf{x}^I_i\mathbf{\tilde{y}}_I^{\mbox{\tiny $(jk)$}}\,=\,0, \qquad
\mathbf{y}^I_{ij}\mathbf{\tilde{x}}_I^{\mbox{\tiny $(k)$}}\,=\,0.
\end{equation}
A generic hyperplane $\mathcal{W}_I$ can be therefore expanded in such a basis as
\begin{equation}\label{eq:hypexp}
    \mathcal{W}_I\:=\:\sum_{v\in\mathcal{V}}\tilde{x}_v\mathbf{\tilde{x}}_v+\sum_{e\in\mathcal{E}}\tilde{y}_{e}\mathbf{\tilde{y}}_e
\end{equation}
where $\mathcal{V}$ and $\mathcal{E}$ are respectively the sets of vertices and edges of the graph $\mathcal{G}$. Requiring the inequalities $Z_k^I\mathcal{W}_I\,\ge\,0$ for each edge $e_{ij}$ can be translated into conditions on the coefficients in \eqref{eq:hypexp}:
\begin{equation}\label{eq:ZWc}
    {\scalebox{.8}{$\displaystyle
    \alpha_{\mbox{\tiny $(ij,ij)$}}\, :=\,\tilde{x}_i-\tilde{y}_{ij}+\tilde{x}_j\,\ge\,0,\qquad
    \alpha_{\mbox{\tiny $(ij,i)$}}\, :=\,\tilde{x}_i+\tilde{y}_{ij}-\tilde{x}_j\,\ge\,0,\qquad
    \alpha_{\mbox{\tiny $(ij,j)$}}\, :=\,-\tilde{x}_i+\tilde{y}_{ij}+\tilde{x}_j\,\ge\,0
    $
    }}
\end{equation}
Notice that, given two edges $e_{ij}$ and $e_{jk}$ sharing one of the sites, namely $s_{j}$, the related $\alpha$'s defined in \eqref{eq:ZWc} satisfy the very same linear relation which defines a cosmological polytope via the intersection of triangles in the midpoints of at most two of their sides:
\begin{equation}\label{eq:ZWcc}
    \alpha_{\mbox{\tiny $(ij,ij)$}}\,+\,\alpha_{\mbox{\tiny $(ij,j)$}}\:=\:
    \alpha_{\mbox{\tiny $(jk,jk)$}}\,+\,\alpha_{\mbox{\tiny $(jk,j)$}}
\end{equation}
In terms of the $\alpha$'s, a vertex of the polytope $\mathcal{P}_{\mathcal{G}}$ associated to a given edge of the graph $\mathcal{G}$ is on the hyperplane $\mathcal{W}_I$ if the related $\alpha$ is zero. So the problem of finding the full set of vertices of $\mathcal{P}_{\mathcal{G}}$ which are on the hyperplane $\mathcal{W}_I$ translates into the question of finding the maximal set of $\alpha$'s which can be set to zero, compatibly with the relations \eqref{eq:ZWcc} and without setting all of them to zero.

In order to be able to identify the vertices of a cosmological polytope $\mathcal{P}_{\mathcal{G}}$ which are on a given hyperplane, it is useful to introduce markings on the associated graph $\mathcal{G}$ that can keep track of both the relations \eqref{eq:ZWc} and the $\alpha$'s which are set to zero. For any edge $e_{ij}$ between the sites $s_i$ and $s_j$, we associate a cross 
$

$
for any $\alpha\,>\,0$ while no marking is associated to $\alpha\,=\,0$:
\begin{equation*}
 %
\end{equation*}
In other words, the vertices of $\mathcal{P}_{\mathcal{G}}$ which are associated to the marked $\alpha$'s \underline{\it are not} on the hyperplane $\mathcal{W}$. Explicitly, the three markings above identify $\alpha_{\mbox{\tiny $(ij,ij)$}}\,>\,0$, $\alpha_{\mbox{\tiny $(ij,j)$}}>\,0$ and $\alpha_{\mbox{\tiny $(ij,i)$}}\,>\,0$ respectively. Given that 
$\alpha_{\mbox{\tiny $(ij,ij)$}}\,=\,\mathcal{W}\cdot(\mathbf{x}_i-\mathbf{y}_{ij}+\mathbf{x}_j)$,
$\alpha_{\mbox{\tiny $(ij,j)$}}\,=\,\mathcal{W}\cdot(-\mathbf{x}_i+\mathbf{y}_{ij}+\mathbf{x}_j)$, and
$\alpha_{\mbox{\tiny $(ij,i)$}}\,=\,\mathcal{W}\cdot(\mathbf{x}_i+\mathbf{y}_{ij}-\mathbf{x}_j)$, 
then the three markings indicate that the vertices
$\mathbf{x}_i-\mathbf{y}_{ij}+\mathbf{x}_j$, $-\mathbf{x}_i+\mathbf{y}_{ij}+\mathbf{x}_j$ and $\mathbf{x}_i+\mathbf{y}_{ij}-\mathbf{x}_j$ respectively \underline{\it are not} on the hyperplane.

Notice now that the relation \eqref{eq:ZWcc} shows a sum of positive terms in both the sides of the equation. This implies that, if in any of the sides both $\alpha$'s are zero, the ones on the other side need to vanish as well. Graphically, this observation translates in a series of markings which \underline{\it are not allowed} on edges sharing one site:
\begin{equation*}
    \begin{tikzpicture}[ball/.style = {circle, draw, align=center, anchor=north, inner sep=0}, 
                        cross/.style={cross out, draw, minimum size=2*(#1-\pgflinewidth), inner sep=0pt, outer sep=0pt}]
        \begin{scope}
            \coordinate (s) at (0,0);
            \coordinate (sl) at ($(s)-(1.25,0)$);
            \coordinate (sr) at ($(s)+(1.25,0)$);
            \draw[-,thick] (sl) -- (sr);
            \draw[fill] (s) circle (2pt);
            \node[very thick, cross=4pt, rotate=0, color=blue] at ($(s)!.25!(sr)$) {};
        \end{scope}
        \begin{scope}[shift={(4,0)}, transform shape]
            \coordinate (s) at (0,0);
            \coordinate (sl) at ($(s)-(1.25,0)$);
            \coordinate (sr) at ($(s)+(1.25,0)$);
            \draw[-,thick] (sl) -- (sr);
            \draw[fill] (s) circle (2pt);
            \node[very thick, cross=4pt, rotate=0, color=blue] at ($(s)!.5!(sr)$) {};
        \end{scope}
        \begin{scope}[shift={(8,0)}, transform shape]
            \coordinate (s) at (0,0);
            \coordinate (sl) at ($(s)-(1.25,0)$);
            \coordinate (sr) at ($(s)+(1.25,0)$);
            \draw[-,thick] (sl) -- (sr);
            \draw[fill] (s) circle (2pt);
            \node[very thick, cross=4pt, rotate=0, color=blue] at ($(s)!.25!(sr)$) {};
            \node[very thick, cross=4pt, rotate=0, color=blue] at ($(s)!.5!(sr)$) {};
        \end{scope}
    \end{tikzpicture}
\end{equation*}
Thus, the problem of finding the maximum number of vanishing $\alpha$'s on a given hyperplane $\mathcal{W}_I$, {\it i.e.} the vertices of $\mathcal{P}_{\mathcal{G}}$ on the facet $\mathcal{P}_{\mathcal{G}}\cap\mathcal{W}$ is translated at graph level in looking for the allowed configurations of marking such that removing further markings would either return one of the configurations which are not allowed, like the ones just illustrated, or it would remove all of them.

These considerations acquires the form of a very simple rule on the graph $\mathcal{G}$: any facet $\mathcal{W}$ of $\mathcal{P}_{\mathcal{G}}$ is identified by a subgraph $\mathfrak{g}\,\subseteq\,\mathcal{G}$, with its vertex configuration given by marking in the middle all the edges which are fully contained in $\mathfrak{g}$ as well as marking those edges departing from $\mathfrak{g}$ close to their sites contained in $\mathfrak{g}$:
\begin{equation*}

\end{equation*}
So, given a subgraph $\mathfrak{g}\,\subseteq\,\mathcal{G}$, the related facet is identified by the co-vector $\mathcal{W}_I^{\mbox{\tiny $(\mathfrak{g})$}}$ 
\begin{equation}\label{eq:WIg}
    \mathcal{W}^{\mbox{\tiny $(\mathfrak{g})$}}\:=\:\sum_{v\in\mathcal{V}_{\mathfrak{g}}}\mathbf{\tilde{x}}_v + 
                                                    \sum_{e\in\mathcal{E}^{\mbox{\tiny ext}}_{\mathfrak{g}}}\mathbf{\tilde{y}}_e
\end{equation}
where $\mathcal{V}_{\mathfrak{g}}\,\subseteq\,\mathcal{V}$ is the set of sites of the subgraph $\mathfrak{g}\,\subseteq\,\mathcal{G}$, while $\mathcal{E}^{\mbox{\tiny ext}}_{\mathfrak{g}}$ is the subset of edges departing from $\mathfrak{g}$. Hence, any facet of $\mathcal{P}_{\mathcal{G}}$ is identified by the equation
\begin{equation}\label{eq:WYeq}
    0\:=\:\mathcal{Y}^I\mathcal{W}_I^{\mbox{\tiny $(\mathfrak{g})$}}\:=\:
          \sum_{s\in\mathcal{V}_{\mathfrak{g}}}x_s+\sum_{e\in\mathcal{E}^{\mbox{\tiny ext}}_{\mathfrak{g}}}y_e
\end{equation}
which precisely coincide with the physical singularities of the (universal integrand of the) wavefunction of the universe $\psi_{\mathcal{G}}(x_s,y_e)$. This implies also that the facets of the cosmological polytopes encode the residues of the poles of $\psi_{\mathcal{G}}(x_s,y_e)$ and, hence, their physical information.

The marking described above allows us to determine via pretty straightforward graphical rules the vertices of the cosmological polytope which {\it are not} on the facet $\mathcal{P}_{\mathcal{G}}\cap\mathcal{W}^{\mbox{\tiny $(\mathfrak{g})$}}$ associated to a certain subgraph $\mathfrak{g}$. It is useful to introduce a marking 
$

$
for those vertices which {\it are} on the facet $\mathcal{W}^{\mbox{\tiny $(\mathfrak{g})$}}$:
\begin{equation*}
    %
\end{equation*}
As we will explicitly see in the next sections, such markings will turn out to be extremely useful to understand the structure of the facets and extract its physical interpretation.

It is appropriate to end this section spending a few words on a facet that, from a physics perspective, has high relevance. We have just established a $1-1$-correspondence between subgraphs $\mathfrak{g}\subseteq\mathcal{G}$ and facets $\mathcal{P}_{\mathcal{G}}\cap\mathcal{W}^{\mbox{\tiny $(\mathfrak{g})$}}$. Let us consider $\mathfrak{g}\,=\,\mathcal{G}$, {\it i.e.} the full graph. Notice that such subgraph contains all sites and vertices and, hence, there are not edges departing from it. This implies that:
\begin{enumerate}
    \item the hyperplane containing the facet is given by the co-vector 
          \begin{equation}\label{eq:Wetot}
            \mathcal{W}^{\mbox{\tiny $(\mathcal{G})$}}\,=\,\sum_{s\in\mathcal{V}}\mathbf{\tilde{x}}_s
          \end{equation}
          and, consequently, it is identified by the equation 
          \begin{equation}\label{eq:WYetot}
            0\,=\,\mathcal{Y}^I\mathcal{W}_I^{\mbox{\tiny $(\mathcal{G})$}}\,=\,\sum_{s\in\mathcal{V}}x_s,
          \end{equation}
          which is nothing but the conservation of the total energy. In other words, this facet encodes the flat-space scattering amplitude. We will refer to it as {\it scattering facet};
    \item the scattering facet turns out to have a very specific vertex structure. Since the related subgraph contains all the edges and, thus, there are no 
          edges departing from it, the vertices which {\it do not} belong to it are the ones marked by 
          $
          \begin{tikzpicture}[cross/.style={cross out, draw, minimum size=2*(#1-\pgflinewidth), inner sep=0pt, outer sep=0pt}]
            \node[very thick, cross=4pt, rotate=0, color=blue] at (0,0) {};
          \end{tikzpicture}
          $ 
          in the middle of each edge, or, equivalently the vertices which {\it do} belong to it are marked 
          $
          \begin{tikzpicture}[ball/.style = {circle, draw, align=center, anchor=north, inner sep=0}]
  	        \node[ball,text width=.18cm,thick,color=blue, anchor=center, scale=.75] at (0,0) {};
          \end{tikzpicture}
          $
          close to the two sites connected by each edge. The scattering facet
          $\mathcal{S}_{\mathcal{G}}\,=\,\mathcal{P}_{\mathcal{G}}\cap\mathcal{W}^{\mbox{\tiny $(\mathcal{G})$}}\,\subset\,\mathbb{P}^{n_s+n_e-2}$ 
          is therefore the convex hull defined by all the vertices of the cosmological polytope $\mathcal{P}_{\mathcal{G}}$ of the form
          \begin{equation}\label{eq:SFv}
              \{\mathbf{x}_i+\mathbf{y}_{ij}-\mathbf{x}_j,\; -\mathbf{x}_i+\mathbf{y}_{ij}+\mathbf{x}_j\}.
          \end{equation}
\end{enumerate}
Such specific vertex structure of the scattering facet $\mathcal{S}_{\mathcal{G}}$ will play a crucial role in the combinatorial proof of the cutting rules ~\cite{Arkani-Hamed:2018ahb} and Steinmann relations ~\cite{Benincasa:2020aoj} for scattering amplitudes, which we will discuss in Section \ref{subsec:Un} and Section \ref{subsec:Caus} respectively.


\subsection{Higher-codimension faces}\label{subsec:FShc}

The markings introduced above to characterise the vertex structure of the facets also allows to characterise the vertex structure of the faces of any co-dimension as well. A codimension-$k$ face is defined as the intersection of $k$ hyperplanes 
$\mathcal{W}^{\mbox{\tiny $\mbox{\tiny $(\mathfrak{g}_1\ldots\mathfrak{g}_k)$}$}}\,:=\,\bigcap_{j=1}^k\mathcal{W}^{\mbox{\tiny $(\mathfrak{g}_j)$}}$ 
containing a facet each on the convex hull $\mathcal{P}_{\mathcal{G}}$.

Having to determine the faces of codimension-$k$ of a cosmological polytope $\mathcal{P}_{\mathcal{G}}$, we need to understand which sets of $k$ facets intersect on a codimension-$k$ subspace. Given a set of $k$ facets identified by the hyperplanes $\mathcal{W}^{\mbox{\tiny $(\mathfrak{g}_j)$}}$ ($j\,=\,1,\ldots k$), the intersection of its elements $\mathcal{W}^{\mbox{\tiny $\mbox{\tiny $(\mathfrak{g}_1\ldots\mathfrak{g}_k)$}$}}\,\neq\,\varnothing$ exists in a codimension-$k$ subspace if and only if the vectors of the vertices shared by all the facets span it. In other words, we can identify if a set of facets intersect into a codimension-$k$ face, if there are at least $n_s+n_e-k$ {\it linearly independent} vectors among the ones representing the vertices on such an intersection.

For the sake of concreteness and illustration, let us begin with the case $k\,=\,2$, {\it i.e.} codimension-$2$ faces, and consider the one for general $k$ later on. For $k\,=\,2$, we need to understand whether two facets identified by the hyperplanes $\mathcal{W}^{\mbox{\tiny $(\mathfrak{g}_1)$}}$ and $\mathcal{W}^{\mbox{\tiny $(\mathfrak{g}_2)$}}$ intersect into a $n_s+n_e-3$ subspace. 

\paragraph{Codimension-$2$ faces and outer intersections.} The vertices of the cosmological polytope $\mathcal{P}_{\mathcal{G}}$ on the intersection are given by the union of the two collections of markings
$

$
identifying the two facets $\mathcal{P}_{\mathcal{G}}\,\cap\,\mathcal{W}^{\mbox{\tiny $(\mathfrak{g}_1)$}}$ and $\mathcal{P}_{\mathcal{G}}\,\cap\,\mathcal{W}^{\mbox{\tiny $(\mathfrak{g}_2)$}}$. For example, if we take $\mathfrak{g}_1\,=\,\mathcal{G}$ and $\mathfrak{g}_2\,=\,\mathfrak{g}$:
\begin{equation*}
    %
\end{equation*}
where the first two markings separately identify the vertices which {\it are not} on the facets, while the last one represents the union of the two markings. Hence, the number of vertices which are on $\mathcal{P}_{\mathcal{G}}\cap\mathcal{W}^{\mbox{\tiny $(\mathfrak{g}_1)$}}\cap\mathcal{W}^{\mbox{\tiny $(\mathfrak{g}_2)$}}$ is given by $3n_e-\mathfrak{d}_{\mbox{\tiny $\mathfrak{g}_1\cap\mathfrak{g}_2$}}$, {\it i.e.} by subtracting the dimension $\mathfrak{d}_{\mbox{\tiny $\mathfrak{g}_1\cap\mathfrak{g}_2$}}$ of the collection of markings 
$
\begin{tikzpicture}[cross/.style={cross out, draw, minimum size=2*(#1-\pgflinewidth), inner sep=0pt, outer sep=0pt}]
    \node[very thick, cross=4pt, rotate=0, color=blue, scale=.75] at (0,0) {};
\end{tikzpicture}
$
from the total number of vertices of the cosmological polytope $\mathcal{P}_{\mathcal{G}}$, which is $3n_e$. Indeed, if 
\begin{equation}\label{eq:cnns}
    3n_e-\mathfrak{d}_{\mbox{\tiny $\mathfrak{g}_1\cap\mathfrak{g}_2$}}\:<\:n_s+n_e-3,
\end{equation}
{\it i.e.} the number of vertices on $\mathcal{P}_{\mathcal{G}}\cap\mathcal{W}^{\mbox{\tiny $(\mathfrak{g}_1)$}}\cap\mathcal{W}^{\mbox{\tiny $(\mathfrak{g}_2)$}}$ is less than the dimension of the space the vertices live in, then they cannot span it and, consequently, they cannot identify a codimension-$2$ face. One example in which the condition \eqref{eq:cnns} is fulfilled on the graph $\mathcal{G}$ depicted above is\footnote{As we will see in Section \ref{subsec:Un}, the vertices configuration depicted in this example, identifies a codimension-$3$ face $\mathcal{P}_{\mathcal{G}}\cap\mathcal{W}^{\mbox{\tiny $(\mathcal{G})$}}\cap\mathcal{W}^{\mbox{\tiny $(\mathfrak{g}_1)$}}\cap\mathcal{W}^{\mbox{\tiny $(\mathfrak{g}_2)$}}$ which is related to sequential cuts on the flat space scattering amplitudes, on which the two-loop amplitude factorises into a product of three tree-level amplitudes.}
\begin{equation*}

\end{equation*}
The opposite condition $3n_e-\mathfrak{d}_{\mbox{\tiny $\mathfrak{g}_1\cap\mathfrak{g}_2$}}\:\ge\:n_s+n_e-3$, however, constitutes a {\it necessary but not sufficient condition} for $\mathcal{P}_{\mathcal{G}}\cap\mathcal{W}^{\mbox{\tiny $(\mathfrak{g}_1)$}}\cap\mathcal{W}^{\mbox{\tiny $(\mathfrak{g}_2)$}}\,\neq\,\varnothing$: even if the number of vertices is enough for them to be able to span the subspace of the right codimension, not necessarily they are linearly independent. In this sense, it is not enough to count how many vertices would be on the intersection of interest, but we would need to count how many of them are {\it linearly independent}.  However there is a case in which the condition \eqref{eq:cnns} becomes also sufficient, {\it i.e.} when one of the two subgraphs is also subgraph of the other ~\cite{Benincasa:2020aoj, Benincasa:2021qcb}~. As a matter of example, let us consider $\mathfrak{g}_{2}\subset\mathfrak{g}_1$ and let $L_{\mathfrak{g}_1}$ and $n_{\mathfrak{g}_2}$ be the number of loops of $\mathfrak{g}_1$ and the number of edges departing from $\mathfrak{g}_2$ respectively. Using the relation $n_s\,=\,n_e-L+1$ between the number of sites, edges and loops of any graph, the condition \eqref{eq:cnns} can be rewritten as
\begin{equation}\label{eq:cnns2}
    n_{\mathfrak{g}_2}\:>\:L_{\mathfrak{g}_1}+1,
\end{equation}
which is more transparent. This is the case for the example depicted above.

More generally, when we consider the face $\mathcal{P}_{\mathcal{G}}\cap\mathcal{W}^{\mbox{\tiny $(\mathfrak{g}_1)$}}\cap\mathcal{W}^{\mbox{\tiny $(\mathfrak{g}_2)$}}$, the graph $\mathcal{G}$ factorises as follows ~\cite{Benincasa:2020aoj}~:
\begin{equation}\label{eq:Gdec}
    \left.\mathcal{G}\right|_{\mathcal{P}_{\mathcal{G}}\cap\mathcal{W}^{\mbox{\tiny $(\mathfrak{g}_1)$}}\cap\mathcal{W}^{\mbox{\tiny $(\mathfrak{g}_2)$}}}
            \:=\:(\mathfrak{g}_1\cap\mathfrak{g}_2)\,\cup\,(\mathfrak{g}_1\cap\bar{\mathfrak{g}}_2)\,\cup\,(\mathfrak{g}_2\cap\bar{\mathfrak{g}}_1)\,\cup\,
                  \mathfrak{g}_c\,\cup\,\centernot{\mathcal{E}},
\end{equation}
where $\bar{\mathfrak{g}}_j$ indicates the complementary subgraph of $\mathfrak{g}_j$, $\mathfrak{g}_c\,:=\,(\bar{\mathfrak{g}}_1\cap\bar{\mathfrak{g}}_2)\cup\bar{\centernot{\mathcal{E}}}$ is the union between the intersection of the two complementary graphs and the collection $\bar{\centernot{\mathcal{E}}}$ departing from it\footnote{Notice that strictly speaking $\mathfrak{g}_c\,:=\,(\bar{\mathfrak{g}}_1\cap\bar{\mathfrak{g}}_2)\cup\bar{\centernot{\mathcal{E}}}$ is not a graph in the ordinary sense as it includes the edges in $\bar{\centernot{\mathcal{E}}}$ but not the sites outside of $\bar{\mathfrak{g}}_1\cap\bar{\mathfrak{g}}_2$ such edges end on. Similarly, the decomposition \eqref{eq:Gdec} is not a graph theory statement in the ordinary sense either, as contains both $\mathfrak{g}_c$ and $\centernot{\mathcal{E}}$.}, while $\centernot{\mathcal{E}}$ is given by the set of edges connecting $\mathfrak{g}_1\cap\mathfrak{g}_2$, $\mathfrak{g}_1\cap\bar{\mathfrak{g}}_2$ and $\mathfrak{g}_2\cap\bar{\mathfrak{g}}_1$ to each other and not contained in any of them. Importantly, each of the four subgraphs as well as the collection $\centernot{\mathcal{E}}$ identifies five lower-dimensional polytopes living in five different subspaces ~\cite{Benincasa:2020aoj}~:
\begin{equation*}

\end{equation*}
In particular, each of the three subgraphs  $\mathfrak{g}_1\cap\mathfrak{g}_2$, $\mathfrak{g}_1\cap\bar{\mathfrak{g}}_2$ and $\mathfrak{g}_2\cap\bar{\mathfrak{g}}_1$  identifies a polytope whose vertices correspond to the two markings
$
 %
$
close to the sites connected by each edge of the subgraph itself. Consequently, the polytopes associated to such subgraphs are lower-dimensional scattering facets $\mathcal{S}_{\tilde{\mathfrak{g}}}\,\subset\,\mathbb{P}^{n_s^{\mbox{\tiny $\tilde{\mathfrak{g}}$}}+n_e^{\mbox{\tiny $\tilde{\mathfrak{g}}$}}-2}$, where
$\tilde{\mathfrak{g}}\,=\,\{\mathfrak{g}_1\cap\mathfrak{g}_2,\,\mathfrak{g}_1\cap\bar{\mathfrak{g}}_2,\,\mathfrak{g}_2\cap\bar{\mathfrak{g}}_1\}$, while $n_s^{\mbox{\tiny $\tilde{\mathfrak{g}}$}}$ and $n_e^{\mbox{\tiny $\tilde{\mathfrak{g}}$}}$ are respectively the number of sites and edges in $\tilde{\mathfrak{g}}$.

Furthermore, $\mathfrak{g}_c$ identifies a polytope living in $\mathbb{P}^{n_s^{\mbox{\tiny $\mathfrak{g}_c$}}+n_e^{\mbox{\tiny $\mathfrak{g}_c$}}+n_{\bar{\centernot{\mathcal{E}}}}-1}$, where $n_s^{\mbox{\tiny $\mathfrak{g}_c$}}$, $n_e^{\mbox{\tiny $\mathfrak{g}_c$}}$ are respectively the number of sites and edges of the subgraph $\bar{\mathfrak{g}}_1\cap\bar{\mathfrak{g}}_2$, while $n_{\bar{\centernot{\mathcal{E}}}}$ is the number of edges departing from $\bar{\mathfrak{g}}_1\cap\bar{\mathfrak{g}}_2$. Finally, the vertices on $\centernot{E}$ identify a simplex in $\mathbb{P}^{n_{\centernot{\mathcal{E}}}-1}$.

Having identified such a factorised structure, then the requirement that $\mathcal{P}_{\mathcal{G}}\cap\mathcal{W}^{\mbox{\tiny $(\mathfrak{g}_1)$}}\cap\mathcal{W}^{\mbox{\tiny $(\mathfrak{g}_2)$}}$ identifies a codimension-$2$ faces can be rephrased into the requirement that the dimensions of these five subspace add up to the dimension of a codimension-$2$ space, {\it i.e.} $n_s+n_e-3$. Explicitly, the dimension of $\mathcal{P}_{\mathcal{G}}\mathcal{W}^{\mbox{\tiny $(\mathfrak{g}_1)$}}\cap\mathcal{W}^{\mbox{\tiny $(\mathfrak{g}_2)$}}$ is given by ~\cite{Benincasa:2020aoj}
\begin{equation}\label{eq:dimC2}
    \begin{split}
        \mbox{dim}\{\mathcal{P}_{\mathcal{G}}\cap\mathcal{W}^{\mbox{\tiny $(\mathfrak{g}_1)$}}\cap\mathcal{W}^{\mbox{\tiny $(\mathfrak{g}_2)$}}\}\:&=\:
            \sum_{\mathcal{S}_{\tilde{\mathfrak{g}}}}(n_s^{\mbox{\tiny $\tilde{\mathfrak{g}}$}}+n_e^{\mbox{\tiny $\tilde{\mathfrak{g}}$}}-1)+n_{\centernot{\mathcal{E}}}+
            (n_s^{\mbox{\tiny $\mathfrak{g}_c$}}+n_s^{\mbox{\tiny $\mathfrak{g}_c$}}-1)-1\:=\\
        &=\:n_s+n_e-1-\sum_{\mathcal{S}_{\mbox{\tiny $\tilde{\mathfrak{g}}$}}}1
    \end{split}
\end{equation}
where the sum is over the scattering facets related to the subgraphs $\mathfrak{g}_1\cap\mathfrak{g}_2$, $\mathfrak{g}_1\cap\bar{\mathfrak{g}}_2$ and
$\mathfrak{g}_2\cap\bar{\mathfrak{g}}_1$. Notice that it coincides with the dimension of a codimension-$2$ space if and only if the sum in the last line in \eqref{eq:dimC2} contains just two elements, {\it i.e.} one among $\mathfrak{g}_1\cap\mathfrak{g}_2$, $\mathfrak{g}_1\cap\bar{\mathfrak{g}}_2$ and
$\mathfrak{g}_2\cap\bar{\mathfrak{g}}_1$ is the empty set. This occurs in three cases. If $\mathfrak{g}_1\,\subset\,\mathfrak{g}_2$, then $\mathfrak{g}_1\cap\mathfrak{g}_2\,=\,\mathfrak{g}_1$, $\mathfrak{g}_1\cap\bar{\mathfrak{g}}_2\,=\,\varnothing$, and $\bar{\mathfrak{g}}_{1}\,\cap\,\bar{\mathfrak{g}}_2\,=\,\bar{\mathfrak{g}}_2$ and hence the graph factorises into
\begin{equation*}
    \mathcal{G}\:=\:\mathfrak{g}_1\,\cup\,(\mathfrak{g}_2\cap\bar{\mathfrak{g}}_1)\,\cup\,(\bar{\mathfrak{g}}_2\cup\bar{\centernot{\mathcal{E}}})\,\cup\,\centernot{\mathcal{E}}
\end{equation*}
The second case is given by $\mathfrak{g}_1\cap\mathfrak{g}_2\,=\,\varnothing$, which implies that $\mathfrak{g}_i\subset\bar{\mathfrak{g}}_j$ and hence
\begin{equation*}
    \mathcal{G}\:=\:\mathfrak{g}_1\,\cup\,\mathfrak{g}_2\,\cup\, (\mathfrak{\mathfrak{g}}_1\cap\bar{\mathfrak{g}}_2\cup\bar{\centernot{\mathcal{E}}})\,\cup\,\centernot{\mathcal{E}}
\end{equation*}
The last one is obtained for $\bar{\mathfrak{g}}_1\cap\bar{\mathfrak{g}}_2\,=\,\varnothing$, {\it i.e.} $\bar{\mathfrak{g}}_i\subset\mathfrak{g}_j$.

Importantly, both the conditions \eqref{eq:cnns} and \eqref{eq:dimC2} which prevents two $k$-dimensional hyperplanes in projective space to have a $(k-1)$-dimensional intersection has a beautiful geometrical description and reflects in the structure of the canonical form of the cosmological polytope: the intersection of the facets satisfying the conditions \eqref{eq:cnns} and \eqref{eq:dimC2} turns out to occur {\it outside} of the cosmological polytope $\mathcal{P}_{\mathcal{G}}$ ~\cite{Benincasa:2020aoj} and the locus defined by such intersections determines the zeros of the canonical function $\Omega(\mathcal{Y},\mathcal{P}_{\mathcal{G}})$. Hence, the intersection among $k$ facets occur on a codimension-$k$ surface outside $\mathcal{P}_{\mathcal{G}}$ determining a subspace of such a locus and the canonical form develops a zero in it so that there is no multiple residue ~\cite{Benincasa:2020aoj}~. Said differently and more generally, the locus of the intersections $\mathcal{C}(\mathcal{P}_\mathcal{G})$ of the facets of $\mathcal{P}_{\mathcal{G}}$ {\it outside} $\mathcal{P}_{\mathcal{G}}$ identifies the locus of the zeroes, and hence the numerator, of the canonical form.

This extensive discussion about codimension-$2$ faces allows to systematically explore more generally the codimension-$k$ faces $\mathcal{P}\cap\mathcal{W}^{\mbox{\tiny $(\mathfrak{g}_1\ldots\mathfrak{g}_k)$}}$: given a cosmological polytope $\mathcal{P}_{\mathcal{G}}$, we can immediately investigate the structure of the codimension-$3$ faces via the same procedure just outlined on the intersection between any codimension-$2$ face $\mathcal{P}_{\mathcal{G}}\cap\mathcal{W}^{\mbox{\tiny $(\mathfrak{g}_i\mathfrak{g}_j)$}}$ and any other hyperplane $\mathcal{W}^{\mbox{\tiny $(\mathfrak{g}_l)$}}$, for $l\,\neq\,i,\,j$. Thus, knowing the structure of codimension-$(k-1)$ faces, the procedure we just discussed allows us to determine the structure of the codimension-$k$ ones. Let us discuss some aspects of it and the conditions that allow to have a codimension-$k$ intersection among facets on/outside a cosmological polytope.

\paragraph{Codimension-$k$ faces and outer intersections.} Let us now consider the hyperplane $\mathcal{W}^{\mbox{\tiny $(\mathfrak{g}_1\ldots\mathfrak{g}_k)$}}$ defined above as the intersections of $k\,\le\,n_s+n_e$ hyperplanes $\{\mathcal{W}^{\mbox{\tiny $(\mathfrak{g}_j)$}},\,j=1,\ldots,k\}$ identified by the subgraphs $\{\mathfrak{g}_j,\,j=1,\ldots,k\}$. Then, the graph $\mathcal{G}$ gets decomposed into $2^k$ subgraphs identified by the intersections among $\{\mathfrak{g}_j,\,j=1,\ldots,k\}$ and/or their complementaries $\{\bar{\mathfrak{g}}_j,\,j=1,\ldots,k\}$:
\begin{equation}\label{eq:Gdec2}
    \left.\mathcal{G}\right|_{\mathcal{P}_{\mathcal{G}}\cap\mathcal{W}^{\mbox{\tiny $(\mathfrak{g}_1\ldots\mathfrak{g}_k)$}}}
        \:=\:\bigcup_{j=0}^k\,\bigcup_{\sigma(j)\in \tilde{S}_k}
    \mathfrak{g}_{\sigma(1)}\cap\ldots\cap\mathfrak{g}_{\sigma(j)}\cap
    \bar{\mathfrak{g}}_{\sigma(j+1)}\cap\ldots\bar{\mathfrak{g}}_{\sigma(k)}
\end{equation}
where 
\begin{equation*}
    \tilde{S}_k\,=\,\{1,\ldots,k\,|\,\sigma(r)<\sigma(r+1),\,r=1,\ldots,j-1\:\&\:\sigma(s)<\sigma(s+1),\,s=j,\ldots,k\}. 
\end{equation*}
There are two important remarks that deserve to be made:
\begin{enumerate}
   \item not all the terms in \eqref{eq:Gdec2} necessarily need to be non-empty;
   \item the terms in \eqref{eq:Gdec2} for $j=1,\ldots\,k$ identify $2^k-1$ lower dimensional scattering facets.
\end{enumerate}
Also, let $\centernot{\bar{\mathcal{E}}}$ the set of $n_{\centernot{\bar{\mathcal{E}}}}$ cut edges edges departing from $\bigcap_{j=1}^k\bar{\mathfrak{g}}_j$, then the vertices $\mathfrak{g}_c :=(\mathfrak{g}_1\cap\ldots\cap\mathfrak{g}_k)\cup\centernot{\bar{\mathcal{E}}}$ identify a polytope $\mathcal{P}_{\mathfrak{g}_c}\,\subset\,\mathbb{P}^{\bar{N}-1}$, with $\bar{N}:=n_s^{\mbox{\tiny $\bar{g}$}}+n_e^{\mbox{\tiny $\bar{g}$}}+n_{\centernot{\bar{\mathcal{E}}}}$, where $n_s^{\mbox{\tiny $\bar{g}$}}$ and $n_e^{\mbox{\tiny $\bar{g}$}}$ are the number of sites and edges in $\mathfrak{g}_c$ respectively. Finally, the $n_{\centernot{\mathcal{E}}}$ vertices associated to the set of cut edges $\centernot{\mathcal{E}}$ among the lower-dimensional scattering facets, identify a simplex $\Sigma_{\centernot{\mathcal{E}}}\subset\mathbb{P}^{n_{\centernot{\mathcal{E}}}-1}$. 

\begin{figure}
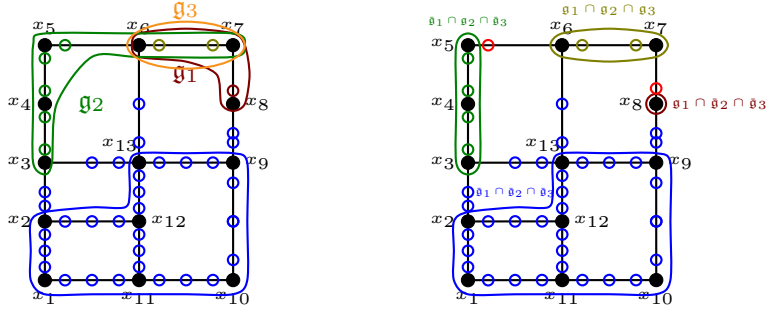

 \centering
 \vspace{-.25cm}

\caption{Example of intersections among several facets. We depict a codimension-$3$ face $\mathcal{P}_{\mathcal{G}}\cap\mathcal{W}^{\mbox{\tiny $(\mathfrak{g}_1\mathfrak{g}_2\mathfrak{g}_3)$}}$. As $\mathfrak{g}_1\cap\mathfrak{g}_2\cap\bar{\mathfrak{g}}_3$, $\mathfrak{g}_1\cap\bar{\mathfrak{g}}_2\cap\mathfrak{g}_3$, $\bar{\mathfrak{g}}_1\cap\mathfrak{g}_2\cap\mathfrak{g}_3$ and $\bar{\mathfrak{g}}_1\cap\bar{\mathfrak{g}}_2\cap\mathfrak{g}_3$ are all empty, it factorises into three lower-dimensional scattering facets, $\mathcal{S}_{\mathfrak{g}_1 \cap \mathfrak{g}_2 \cap \mathfrak{g}_3}$, $\mathcal{S}_{\mathfrak{g}_1 \cap \bar{\mathfrak{g}}_2 \cap \bar{\mathfrak{g}}_3}$, and $\mathcal{S}_{\bar{\mathfrak{g}}_1 \cap \mathfrak{g}_2 \cap\bar{\mathfrak{g}}_3}$. The markings
\circle{red!50!black},
\circle{red!50!green},
\circle{green!50!black},
\redcircle{}
and
\circle{blue}
respectively tdenote the vertices of such three lower-dimensional scattering facets, the simplex $\Sigma_{\centernot{\mathcal{E}}}$ and $\mathcal{P}_{\mathfrak{g}_c}$.
}
 \label{fig:ldint}
\end{figure}

Hence, the dimension of the intersection $\mathcal{P}_{\mathcal{G}}\cap\mathcal{W}^{\mbox{\tiny $(\mathfrak{g}_1\ldots\mathfrak{g}_k)$}}$ is given by ~\cite{Benincasa:2021qcb}~:
\begin{equation}\label{eq:PgWgk}
    \begin{split}
        \mbox{dim}\{\mathcal{P}_{\mathcal{G}}\cap\mathcal{W}^{\mbox{\tiny $(\mathfrak{g}_1\ldots\mathfrak{g}_k)$}}\}\:&=\:
        \sum_{\mathcal{S}_{\mathfrak{g}}}(n_s^{(\mathfrak{g})}+n_e^{(\mathfrak{g})}-1)+n_{\centernot{\mathcal{E}}}+
            n_s^{(\bar{\mathfrak{g}})}+n_e^{(\bar{\mathfrak{g}})}+n_{\centernot{\bar{\mathcal{E}}}}-1\\
        &=\:n_s+n_e-1-\sum_{\mathcal{S}_{\mathfrak{g}}}1,
    \end{split}
\end{equation}
with the sum running over all the non-empty intersections that identify the lower dimensional scattering facets. Now, in order for $\mathcal{P}_{\mathcal{G}}\cap\mathcal{W}^{\mbox{\tiny $(\mathfrak{g}_1\ldots\mathfrak{g}_k)$}}$ to have codimension-$k$, as one would expect, then one should have
\begin{equation}\label{eq:PgWgknon0}
    \sum_{\mathcal{S}_{\mathfrak{g}}}1\:=\:k
\end{equation}
{\it i.e.}, just $k$ out of the $2^k-1$ intersections which can give a scattering facet has to be non-empty. Indeed, if all the graphs are mutually partial overlapping, then all the terms in \eqref{eq:Gdec2} would be non-empty and hence the sum on the left-hand-side in \eqref{eq:PgWgknon0} would be equal to $2^k$ and hence \eqref{eq:PgWgknon0} would become $2^k=k+1$, which does not have solution for $k\in\mathbb{Z}_{+}$. However, notice in \eqref{eq:PgWgk} that the sum of the number of sites over all the subsets always returns the total number $n_s$ of sites of $\mathcal{G}$, while this is not necessarily true for the number of edges, as on $\mathcal{P}_{\mathcal{G}}\cap\mathcal{W}^{\mbox{\tiny $(\mathfrak{g}_1\ldots\mathfrak{g}_k)$}}$ there might be no vertex associated to some of the cut edges. Let $\centernot{n}$ the number of edges of $\mathcal{G}$ with no vertex on  $\mathcal{P}_{\mathcal{G}}\cap\mathcal{W}^{\mbox{\tiny $(\mathfrak{g}_1\ldots\mathfrak{g}_k)$}}$, then $\mathcal{P}_{\mathcal{G}}\cap\mathcal{W}^{\mbox{\tiny $(\mathfrak{g}_1\ldots\mathfrak{g}_k)$}}\,\neq\varnothing$ if ~\cite{Benincasa:2021qcb}
\begin{equation}\label{eq:PgWgknon0gen}
    \sum_{\mathcal{S}_{\mathfrak{g}}}1+\centernot{n}\:\le\:k.
\end{equation}

A comment is now in order. Let us consider two graphs $\mathfrak{g}_!$ and $\mathfrak{g}_2$ such that the hyperplanes they identify intersect each other outside $\mathcal{P}_{\mathcal{G}}$, {\it i.e.} $\mathcal{P}_{\mathcal{G}}\cap\mathcal{W}^{\mbox{\tiny $(\mathfrak{g}_1)$}}\cap\mathcal{W}^{\mbox{\tiny $(\mathfrak{g}_2)$}}\,=\,\varnothing$. Recall that in this case, the graph $\mathcal{G}$ is divided into three lower dimensional scattering facets and the remaining set of vertices, associated to the $\mathfrak{g}_c :=(\bar{\mathfrak{g}}_1\cap\bar{\mathfrak{g}_2})\cup\centernot{\bar{\mathcal{E}}}$, define a polytope $\mathcal{P}_{\mathfrak{g}_c}$. The dimensional counting reveal that this intersection does not live in $n_s+n_e-3$, but rather it would live in $n_s+n_e-4$. Then, if it possible to choose a third subgraph $\mathfrak{g}_3\subseteq\mathcal{G}$ such that no further marking is added, then $\mathcal{P}_{\mathcal{G}}\cap\mathcal{W}^{\mbox{\tiny $(\mathfrak{g}_1\mathfrak{g}_2\mathfrak{g}_3)$}}\,\neq\,\varnothing$ as now this intersection lives in $n_s+n_e-4$ dimensions as required. In general, given a set of $k$ subgraphs $\{\mathfrak{g}_j\}_{j=1}^k$ such that $\mathcal{P}_{\mathcal{G}}\cap\mathcal{W}^{\mbox{\tiny $(\mathfrak{g}_1\ldots\mathfrak{g}_k)$}}\,=\,\varnothing$, {\it i.e.} the sum of the lower-dimensional scattering facets that this intersection would generate is greater than $k$. Let $n_{\mathcal{S}}$ be such a sum. If it is possible to consider further $n_{\mathcal{S}}-k$ subgraphs such that no further scattering facet is added, then the new intersection is non-empty and hence represent a codimension-$k$ face of $\mathcal{P}_{\mathcal{G}}$.


\subsection{Factorisations and Steinmann-like relations}\label{subsec:FSwf}

The cosmological polytopes $\mathcal{P}_{\mathcal{G}}$ provide a first principle definition for the (integrand of the) Bunch-Davies wavefunction of the universe for a large class of scalar models, whose functional form is returned by the canonical function $\Omega(\mathcal{Y},\,\mathcal{P}_{\mathcal{G}})\,=\,\psi_{\mathcal{G}}(x,y)$. Because of such relation, each facet encodes the residue of a given pole of the wavefunction, while the structure of higher codimension faces encodes the physical information contained in the multiple residues of the wavefunction. In the previous section we focus on how identify both the facets and the higher codimension faces of the cosmological polytopes and relating them to graphical rules:
\begin{enumerate}
	\item the {\it facets} are in $1-1$ correspondence with the subgraphs $\mathfrak{g}\,\subseteq\,\mathcal{G}$, and they are the intersection 
		$\mathcal{P}_{\mathcal{G}}\cap\mathcal{W}^{\mbox{\tiny $(\mathfrak{g})$}}$ of the cosmological polytope $\mathcal{P}_{\mathcal{G}}$ and
		hyperplanes $\mathcal{W}^{\mbox{\tiny $(\mathfrak{g})$}}_I$ of the form
		\begin{equation}\label{eq:Wdef2}
			\mathcal{W}^{\mbox{\tiny $(\mathfrak{g})$}}_I\:=\:\sum_{s\in\mathcal{V}_{\mathfrak{g}}}\mathbf{\tilde{x}}_s+
									  \sum_{e\in\mathcal{E}_{\mathfrak{g}}^{\mbox{\tiny ext}}}\mathbf{\tilde{y}}_e
		\end{equation}
		and such that
		\begin{equation}\label{eq:WY2}
			0\:=\:\mathcal{Y}^I\mathcal{W}_I^{\mbox{\tiny $(\mathfrak{g})$}}\:=\:\sum_{s\in\mathcal{V}_{\mathfrak{g}}}x_s+
											     \sum_{e\in\mathcal{E}_{\mathfrak{g}}^{\mbox{\tiny ext}}}y_e.
		\end{equation}
	\item the codimension-$k$ faces are intersections $\mathcal{P}_{\mathcal{G}}\cap\mathcal{W}^{\mbox{\tiny $(\mathfrak{g}_1\ldots\mathfrak{g}_k)$}}$ of
		the cosmological polytope $\mathcal{P}_{\mathcal{G}}$ and multiples hyperplanes $\mathcal{W}^{\mbox{\tiny $(\mathfrak{g}_1\ldots\mathfrak{g}_k)$}}$
		which can be determined considering the associated subgraphs and the requirement that the graph factorisation structure define lower-dimensional polytopes embedded into $\mathbb{P}^{n_s+n_e-k-1}$.
\end{enumerate}
In this section we pause on the physical information encoded into the faces of the cosmological polytopes and how their structures reflect on the wavefunction.
\\

\paragraph{Facets, amplitudes and individual cuts.} As already shown, each facet $\mathcal{P}_{\mathcal{G}}\cap\mathcal{W}^{\mbox{\tiny $(\mathfrak{g})$}}$ of the cosmological polytopes, corresponding to a subgraph $\mathfrak{g}\subseteq\mathcal{G}$ describes a residue of the wavefunction universal integrand $\psi_{\mathcal{G}}(x,y)$. In a nutshell, the facets are identified by hyperplanes \eqref{eq:Wdef2}, whose defining equation \eqref{eq:WY2} is one of the poles of $\psi_{\mathcal{G}}(x,y)$, and the associated polytope $\mathcal{P}_{\mathcal{G}}\cap\mathcal{W}^{\mbox{\tiny $(\mathfrak{g})$}}$ is described by a canonical function which is precisely the residue of the wavefunction integrand at that pole.

A special facet is identified by $\mathfrak{g}\,=\,\mathcal{G}$ and encodes the flat-space scattering amplitude, as it is identified by the vanishing of the total energy (see equation \eqref{eq:Wetot}), with its canonical form returning the flat-space scattering amplitude itself. As already emphasised, such scattering facet has the distinctive feature of being the convex hull of all the vertices of $\mathcal{P}_{\mathcal{G}}$ of the form 
$\{\mathbf{x}_i+\mathbf{y}_{ij}-\mathbf{x}_j,\,-\mathbf{x}_i+\mathbf{y}_{ij}+\mathbf{x}_j\}$. which are marked by
$

$
close to sites $s_i$ and $s_j$, on the edge $e_{ij}$ connecting them (for all $i,\,j$). 

Notice that the scattering facet is the only facet whose related subgraph does not have any edge departing from it. Let us focus on the other facets and let $\mathfrak{g}$ be the connected subgraph identifying them. They are the convex hull of the remaining vertices of $\mathcal{P}_{\mathcal{G}}$ after deleting the vertices marked in the middle of the edges contained in $\mathfrak{g}$ as well as the ones closest to $\mathfrak{g}$ in the edges departing from $\mathfrak{g}$:
\begin{equation*}
    %
\end{equation*}
On this class of facets, the graph $\mathcal{G}$ factorises into $\mathcal{G}\,=\,\mathfrak{g}\cup\bar{\mathfrak{g}}_c$, where $\bar{\mathfrak{g}}_c\,:=\,\bar{\mathfrak{g}}\cup\centernot{\mathcal{E}}$, and consequently its canonical function can be written as:
\begin{equation}\label{eq:CFWg}
    \Omega(\mathcal{P}_{\mathcal{G}}\cap\mathcal{W}^{\mbox{\tiny $(\mathfrak{g})$}})\:=\:
    \Omega(\mathcal{S}_{\mathfrak{g}})\times\Omega(\mathcal{P}_{\mathfrak{g}_c})
\end{equation}
Notice that the factor related to the subgraph $\mathfrak{g}$ has the typical vertex configuration of a scattering facet: it defines a lower-dimensional polytope $\mathcal{S}_{\mathfrak{g}}\,\subset\,\mathbb{P}^{n_s^{\mbox{\tiny $\mathfrak{g}$}}+n_e^{\mbox{\tiny $\mathfrak{g}$}}-2}$, which is the convex hull of the vertices $\{\mathbf{x}_i+\mathbf{y}_{ij}-\mathbf{x}_j,\,-\mathbf{x}_i+\mathbf{y}_{ij}+\mathbf{x}_j\:|\:\forall\,x_i,\,x_j\in\mathfrak{g}\}$. 

The second factor related to $\mathfrak{g}_c$ shows the very same vertex configuration of the cosmological polytope $\mathcal{P}_{\bar{\mathfrak{g}}}$ with two extra vertices for each edge in $\centernot{\mathcal{E}}$, defining a polytope living in $\mathbb{P}^{n_s^{\mathfrak{g}_c}+n_e^{\mathfrak{g}_c}+n_{\centernot{\mathcal{E}}}-1}$ with $3n_e^{\mbox{\tiny $\mathfrak{g}_c$}}+2n_{\centernot{\mathcal{E}}}$ vertices. Importantly, such a polytope can be triangulated\footnote{Notice that this is a triangulation in a general sense, as it does not involve necessarily simplices} via all the polytopes defined as convex hull of all the $3n_e$ vertices attached to the subgraph $\bar{\mathfrak{g}}$ and one of the two vertices attached to each edge in $\centernot{\mathcal{E}}$. The canonical function $\Omega(\mathcal{P}_{\mathfrak{g}_c})$ can then be written as the sum of the canonical function of such polytopes:
\begin{equation}\label{eq:WFfact}
    \Omega(\mathcal{P}_{\mathfrak{g}_c})\:=\:\sum_{\{v_{ij}\,=\,\mathcal{Z}_{ij}\}}\Omega(\mathcal{P}_{\bar{\mathfrak{g}},\{v_{ij}\}})
\end{equation}
where $\mathcal{Z}_{ij}\,:=\,\{\mathbf{x}_{i}-\mathbf{y}_{ij}+\mathbf{x}_{j},\,\mathbf{x}_{i}+\mathbf{y}_{ij}-\mathbf{x}_{j}\}$ is the set of two vertices of the facet $\mathcal{P}_{\mathcal{G}}\cap\mathcal{W}_{\mathfrak{g}}$ associated to the edge $e_{ij}\,\in\,\centernot{\mathcal{E}}$, while $\mathcal{P}_{\bar{\mathfrak{g}},\{v_{ij}\}}$ is the convex hull of all the vertices of the facet associated to the edges in $\bar{\mathfrak{g}}$ and one of the two vertices in $\mathcal{Z}_{ij}$. Each term in the sum \eqref{eq:WFfact} is nothing but the product of the inverse of the energies of all the edges in $\centernot{\mathcal{E}}$ times the wavefunction related to the subgraph $\bar{\mathfrak{g}}$ with energies associated to each sites connected to the cut edges $\centernot{\mathcal{E}}$ shifted by ($\pm$) the energy of the relevant cut edge. Hence, the canonical function $\Omega(\mathcal{P}_{\mathfrak{g}_c})$ can be written as
\begin{equation}\label{eq:WFfact2}
    \Omega(\mathcal{P}_{\mathfrak{g}_c})\:=\:\sum_{\{\sigma_{ij}=\pm1\}}\frac{\psi_{\mathfrak{\bar{g}}}(x_s(\sigma_{ij}),y_{ij})}{\displaystyle\prod_{e_{ij}\in\centernot{\mathcal{E}}}2y_{ij}}
\end{equation}
where 
\begin{equation}\label{eq:WFfact2se}
    x_s(\sigma_{ij})\:=\:x_s+\sum_{e_{ij}\in\centernot{\mathcal{E}}\cap\mathcal{E}_v}\sigma_{ij}y_{ij}
\end{equation}
Consequently, the facet $\mathcal{P}_{\mathcal{G}}\cap\mathcal{W}_{\mathfrak{g}}$ has the structure of a product of a lower dimensional scattering facet $\mathcal{S}_{\mathfrak{g}}$ times another polytope encoding the wavefunction of the universe related to the subprocess $\bar{\mathfrak{g}}$ computed at both positive and negative energies for each cut edge. This is completely general phenomenon valid for each facet $\mathcal{P}_{\mathcal{G}}\cap\mathcal{W}_{\mathfrak{g}}$ with $\mathfrak{g}\subset\mathcal{G}$ and which provides a combinatorial proof of the factorisation of the wavefunction of the universe into a lower-point scattering amplitudes and a sum of lower-point wavefunctions computed at both negative and positive energies for the cut edge. This is the basic statement that as the total energy of any subprocess of a given way function is taken to zero via analytic continuation outside of the physical region, the conditions for a (high energy) flat-space scattering get restored together with Lorentz invariance for the subprocess in question, and each of the cut propagators gets two contributions out of the three in which it can be decomposed -- one from the advanced/retarded part and one from the boundary term.

\paragraph{Codimension-$2$ faces, sequential cuts and Steinmann-like relations.} In the previous paragraph we saw how the structure of a facet $\mathcal{P}_{\mathcal{G}}\cap\mathcal{W}^{\mbox{\tiny $(\mathfrak{g})$}}$ associated to any subgraph $\mathfrak{g}\subsetneq\mathcal{G}$ factorises into two lower-dimensional polytopes, implying a factorisation of the wavefunction as the condition $\mathcal{Y}\cdot\mathcal{W}_{\mathfrak{g}}\,=\,0$ associated to the subgraph $\mathfrak{g}$ is imposed. In the present paragraph, we are going to investigate the structure of the codimension-$2$ faces.

In the general discussion about the higher-codimension faces, we have already analysed the conditions for which the intersection $\mathcal{P}_{\mathcal{G}}\cap\mathcal{W}^{\mbox{\tiny $(\mathfrak{g}_1)$}}\cap\mathcal{W}^{\mbox{\tiny $(\mathfrak{g}_2)$}}$ defining a codimension-$2$ face exists. Let us focus on a pair of subgraphs $\mathfrak{g}_1$ and $\mathfrak{g}_2$ such that
\begin{equation}\label{eq:gpoc}
    \mathfrak{g}_1\cap\mathfrak{g}_2\,\neq\,\varnothing,\qquad
    \mathfrak{g}_1\cap\bar{\mathfrak{g}}_2\,\neq\,\varnothing,\qquad
    \bar{\mathfrak{g}}_1\cap\mathfrak{g}_2\,\neq\,\varnothing,\qquad
    \bar{\mathfrak{g}}_1\cap\bar{\mathfrak{g}}_2\,\neq\,\varnothing
\end{equation}
where, as usual, $\bar{\mathfrak{g}}_j$ denotes the complement graph of $\mathfrak{g}_j$. All the pairs of graphs that satisfy such conditions correspond to partially-overlapping momentum channels. As we saw earlier, the canonical function of this type of face should factorise into three lower-dimensional scattering facets associated to the subgraphs $\mathfrak{g}_1\cap\mathfrak{g}_2$, $\mathfrak{g}_1\cap\bar{\mathfrak{g}}_2$ and $\bar{\mathfrak{g}}_1\cap\mathfrak{g}_2$, a lower-dimensional polytope associated to $\mathfrak{g}_c\,=\,(\mathfrak{g}_1\cap\mathfrak{g}_2)\cup\centernot{\bar{\mathcal{E}}}$ (where $\centernot{\bar{\mathcal{E}}}$ are the edges departing from $\bar{\mathfrak{g}}_1\cap\bar{\mathfrak{g}}_2$) as well as a simplex associated to the edges $\centernot{\mathcal{E}}$ connecting the three subgraphs $\mathfrak{g}_1\cap\mathfrak{g}_2$, $\mathfrak{g}_1\cap\bar{\mathfrak{g}}_2$ and $\bar{\mathfrak{g}}_1\cap\mathfrak{g}_2$ among each other. This intersection should live in $\mathbb{P}^{n_v+n_e-3}$, but the dimension of the polytopes the canonical function factorises into is given by equation \eqref{eq:dimC2}, which we rewrite here 
\begin{equation}\label{eq:dimC2b}
    \mbox{dim}\{\mathcal{P}_{\mathcal{G}}\cap\mathcal{W}^{\mbox{\tiny $(\mathfrak{g}_1)$}}\cap\mathcal{W}^{\mbox{\tiny $(\mathfrak{g}_2)$}}\}\:=\:
        n_v+n_e-1-\sum_{\mathcal{S}_{\mathfrak{g}}}1\:=\:n_v+n_e-4,
\end{equation}
{\it i.e.} the intersection has lower dimension and the corresponding sequential cuts vanish. Consequently, the (universal integrand of the) wavefunction of the universe satisfies Steinmann-like relations. In particular, we can write:
\begin{equation}\label{eq:SteinRel}
    \Res_{E_{\mathfrak{g}_1}}\Res_{E_{\mathfrak{g}_2}}\psi_{\mathcal{G}}\:=\:0,\qquad
    \mbox{ for }
    \left\{
        \begin{array}{l}
            \mathfrak{g}_1\:\centernot{\subseteq}\:\mathfrak{g}_2,\\
            \mathfrak{g}_2\:\centernot{\subseteq}\:\mathfrak{g}_1,\\
            \mathfrak{g}_1\cap\mathfrak{g}_2\,\neq\,\varnothing
        \end{array}
    \right.
\end{equation}
where the energies $E_{\mathfrak{g}_j}$ are defined as
\begin{equation}\label{eq:Egj}
    E_{\mathfrak{g}_j}\: :=\:\sum_{s\in\mathfrak{g}_j}x_s+\sum_{e\in\mathcal{E}_{\mathfrak{g}_j}^{\mbox{\tiny ext}}}y_e
\end{equation}
Such restriction on the integrand can be promoted to statement about the discontinuity for the tree-level wavefunction $\Psi_\mathcal{G}^{\mbox{\tiny tree}}$ as the integration over the energies is known to give rise to polylogarithms ~\cite{Arkani-Hamed:2017fdk, Hillman:2019wgh}~. At loop level, while there is no enough information about the analytic structure of the loop integrated wavefunction, the integration over the energies related to the sites of the graph generates the actual loop integrand which is still given in terms of polylogarithms as for the tree-level wavefunction ~\cite{Arkani-Hamed:2017fdk, Hillman:2019wgh}~. Hence, for such loop integrands, which are no longer rational functions, \eqref{eq:SteinRel} implies that the double discontinuities in partially-overlapping channels vanish.

Finally, let us consider two subgraphs $\mathfrak{g}_1,\,\mathfrak{g}_2\,\subseteq\,\mathcal{G}$, such that one is a subgraph of the other, {\it e.g.} $\mathfrak{g}_2\,\subset\,\mathfrak{g}_1$. In the previous subsection we saw that if the number of edges $n_{\mathfrak{g}_2}$ departing from $\mathfrak{g}_2$ is greater than $L_{\mathfrak{g}_1}+1$, $L_{\mathfrak{g}_1}$ being the number of loops of $\mathfrak{g}_1$, then also this intersection occurs. in codimension-$2$, outside $\mathcal{P}_{\mathcal{G}}$. This can be also translates into the statement that the double residue of $\psi_{\mathcal{G}}$ with respect to the energies associated to the subprocesses identified by $\mathfrak{g}_1$ and $\mathfrak{g}_2$, vanish:
\begin{equation}\label{eq:dref}
    \Res_{E_{\mathfrak{g}_1}}\Res_{E_{\mathfrak{g}_2}}\psi_{\mathcal{G}}\,=\,0,\qquad
    \mbox{for }
    \left\{
        \begin{array}{l}
            \mathfrak{g}_j\subset\mathfrak{g}_i,\\
            n_{\mathfrak{g}_j}\,>\,L_{\mathfrak{g}_i}+1
        \end{array}
    \right.
\end{equation}
Importantly, while the subgraph $\mathfrak{g}_1$ identifies a lower-dimensional scattering amplitudes, $\mathfrak{g}_2\,\subset\,\mathfrak{g}_1$ such that $n_{\mathfrak{g}_2}\,>\,L_{\mathfrak{g}_1}+1$ perform cuts on such a scattering amplitude with {\it no defined energy flow} on the cut edges. Hence, the vanishing double residues \eqref{eq:dref} -- and, thus, the incompatibility between such pairs of channels -- imply that the wavefunction $\psi_{\mathcal{G}}$ correctly encodes such a defined energy flow for the flat-space amplitudes. 

\paragraph{Higher codimension constraints and sequential cuts.} We can repeat a similar discussion for higher codimension faces. In the previous discussion we found that the generic condition on $k$ hyperplanes, each of which contains one facet of $\mathcal{P}_{\mathcal{G}}$, for them to intersect outside $\mathcal{P}_{\mathcal{G}}$ in codimension $k$, is that the number of lower-dimensional scattering facets on such an intersection would be greater than $k-1$. Or, in a more refined way, that such a number is greater than $k-\centernot{n}_{\centernot{\mathcal{E}}}$, see equation \eqref{eq:PgWgknon0gen}.

This combinatorial statement on the higher-codimension face structure of the cosmological polytope, can be translated as constraints on the analytic properties of the wavefunction. Given a set of $k$ channels identified by the subgraphs $\{\mathfrak{g}_j\,|\,j=1,\ldots,k\}$, they are incompatible if the wavefunction would factorise in more than $k$ lower-dimensional scattering amplitudes:
\begin{equation}\label{eq:wfgmc}
    \Res_{E_{\mathfrak{g}_1}}\Res_{E_{\mathfrak{g}_2}}\cdots\Res_{E_{\mathfrak{g}_k}}\psi_{\mathcal{G}}\,=\,0.
\end{equation}
Turning the table around, as the residues along multiple channels are taken, the wavefunction has to factorise in {\it precisely} $k$ lower-dimensional scattering amplitudes, and a linear combination of a lower dimensional wavefunction associated to the intersection of all the complementary graphs, which is the sum over all the possible combination of the energy signs along the cut edges.


\subsection{Signed-triangulations and perturbative expansions.}\label{subsec:STrReps}

The understanding of the face structure of the cosmological polytopes for arbitrary codimensions outlined above, provide us with further insights on the properties that the canonical form, and consequently the wavefunction coefficients, ought to satisfy. In particular, given a graph $\mathcal{G}$  and the associated cosmological polytope $\mathcal{P}_{\mathcal{G}}$, the condition \eqref{eq:PgWgknon0gen} allows to characterise the locus $\mathcal{C}(\mathcal{P}_{\mathcal{G}})$ of the intersections of the hyperplanes containing the facets of $\mathcal{P}_{\mathcal{G}}$, outside of $\mathcal{P}_{\mathcal{G}}$. As already remarked, such a locus $\mathcal{C}(\mathcal{P}_{\mathcal{G}})$ {\it is} the locus of the zeros of the canonical form $\omega(\mathcal{Y},\,\mathcal{P}_{\mathcal{G}})$, and the condition \eqref{eq:PgWgknon0gen} may allow to fix the numerator of the canonical form in an invariant way, at least in principle.

The general structure of the canonical form $\omega(\mathcal{Y},\mathcal{P}_{\mathcal{G}})$, associated to $\mathcal{P}_{\mathcal{G}}\,\subset\,\mathbb{P}^{n_s+n_e-1}$, can be written as:
\begin{equation}\label{eq:wgen}
    \omega(\mathcal{Y},\mathcal{P}_{\mathcal{G}})\,=\,
        \frac{\mathfrak{n}_{\delta}(\mathcal{Y})}{\displaystyle\prod_{\mathfrak{g}\subseteq\mathcal{G}}
            q_{\mathfrak{g}}(\mathcal{Y})}\,\langle\mathcal{Y}d^{n_s+n_e-1}\mathcal{Y}\rangle,
\end{equation}
where $q_{\mathfrak{g}}\,:=\,\mathcal{Y}^I\mathcal{W}^{\mbox{\tiny $(\mathfrak{g})$}}_I$ is the linear polynomial which identifies the facet $\mathcal{P}_{\mathcal{G}}\cap\mathcal{W}^{\mbox{\tiny $(\mathfrak{g})$}}\,\neq\,\varnothing$ associated to the subgraph $\mathfrak{g}$, and $\mathfrak{n}_{\delta}(\mathcal{Y})$ is an homogeneous polynomial in $\mathcal{Y}$ of degree $\delta$. The $GL(1)$ invariance under rescaling of $\mathcal{Y}$ fixes the degree $\delta$ of the polynomial $\mathfrak{n}_{\delta}(\mathcal{Y})$ to be $\delta\,=\,\tilde{\nu}-n_s-n_e$, $\tilde{\nu}$ being the number of facets of $\mathcal{P}_{\mathcal{G}}$. Importantly, one can think of $\mathfrak{n}_{\delta}(\mathcal{Y})$ to be fixed by a symmetric $\delta$-tensor $\mathcal{C}_{\mbox{\tiny $I_1\ldots I_{\delta}$}}$:
\begin{equation}\label{eq:ndc}
    \mathfrak{n}_{\delta}(\mathcal{Y}):=\mathcal{C}_{I_{1}\ldots I_{\delta}}\mathcal{Y}^{I_1}\ldots\mathcal{Y}^{I_{\delta}},
\end{equation}
with $\mathcal{C}_{I_1\ldots I_{\delta}}$ having
\begin{equation}\label{eq:cdofs}
    \Delta\: :=\:
    \begin{pmatrix}
        n_s+n_e+\delta-1 \\
        \delta
    \end{pmatrix}
    -1
\end{equation}
degrees of freedom. 

Despite the analysis of the face structure outlined in the previous subsections provides a complete characterisation of the locus $\mathcal{C}(\mathcal{P}_{\mathcal{G}})$, using this information to fix the $\Delta$ degrees of freedom of the tensor $\mathcal{C}_{I_1\ldots I_{\delta}}$ is not an easy task for arbitrary $\delta$ -- see the appendix of ~\cite{Benincasa:2021qcb}~ for a simple example where it can actually be done.

If this completely invariant way of fixing the canonical form $\omega(\mathcal{Y},\mathcal{P}_{\mathcal{G}})$ is in practice not easily feasible, at least currently, the knowledge of $\mathcal{C}(\mathcal{P}_{\mathcal{G}})$ can still allow us to fix $\omega(\mathcal{Y},\mathcal{P}_{\mathcal{G}})$. Firstly, recall that $\mathcal{C}(\mathcal{P}_{\mathcal{G}})$ is fixed by the intersections of the hyperplanes containing the facets of $\mathcal{P}_{\mathcal{G}}$ outside $\mathcal{P}_{\mathcal{G}}$, each of which determines a lower-dimensional hyperplane in $\mathcal{C}(\mathcal{P}_{\mathcal{G}})$. Secondly, for some of such intersections, we can think of triangulating $\mathcal{P}_{\mathcal{G}}$ through one of them:  as they are fixed by facets of $\mathcal{P}_{\mathcal{G}}$, such triangulations do not add spurious facets and, consequently, the canonical form of $\mathcal{P}_{\mathcal{G}}$ can be written as
\begin{equation}\label{eq:phystrgen}
    \omega(\mathcal{Y},\mathcal{P}_{\mathcal{G}})\:=\:\sum_{j=1}^n\omega(\mathcal{Y},\mathcal{P}^{\mbox{\tiny $(j)$}}_{\mathcal{G}}),
\end{equation}
where $\{\mathcal{P}^{\mbox{\tiny $(j)$}}_{\mathcal{G}},\,j=1,\ldots n\}$ is the collection of simplices triangulating $\mathcal{P}_{\mathcal{G}}$ and $\omega(\mathcal{Y},\mathcal{P}^{\mbox{\tiny $(j)$}}_{\mathcal{G}})$ is the canonical form of the $j$-th elements of such a collection. As no spurious facets are added, this means that {\it all} the hyperplanes containing the facets of $\mathcal{P}^{\mbox{\tiny $(j)$}}_{\mathcal{G}}$, for all $j=1,\ldots,n$, are also hyperplanes containing the facets of $\mathcal{P}_{\mathcal{G}}$. Consequently, the wavefunction $\psi_{\mathcal{G}}$ gets represented as a sum of terms with physical poles only ~\cite{Benincasa:2021qcb}~. Therefore, all subspaces of $\mathcal{C}(\mathcal{P}_{\mathcal{G}})$ that allow for such signed triangulations, offer the possibility of having representations of $\psi_{\mathcal{G}}$ with no spurious singularity added. Said differently, this class of signed triangulations of $\mathcal{P}_{\mathcal{G}}$ provides a systematic classification and algorithm for writing down the organisation of the perturbative expansion, all of them characterised by having physical poles only ~\cite{Benincasa:2021qcb}~. 

Let $\mathfrak{G}_{\circ}:=\{\mathfrak{g}_j,\,j=1,\ldots,k\}$ be the set of subgraphs that identifies the hyperplane $\mathcal{W}^{\mbox{\tiny $(\mathfrak{g}_1\ldots\mathfrak{g}_k)$}}$ such that $\mathcal{P}_{\mathcal{G}}\cap\mathcal{W}^{\mbox{\tiny $(\mathfrak{g}_1\ldots\mathfrak{g}_k)$}}\,=\,\varnothing$. It is a subspace of $\mathcal{C}(\mathcal{P}_{\mathcal{G}})$. Let us assume we can triangulate $\mathcal{P}_{\mathcal{G}}$ through it -- later on we will come back on how to identify the subspaces of $\mathcal{C}(\mathcal{P}_{\mathcal{G}})$ that allow for a signed triangulation through just one of them. Such a signed triangulation is characterised by a collection of simplices, each of which is identified by the $k$ inequalities $\{q_{\mathfrak{g}_j}(\mathcal{Y})\,>\,0,\,|\,j=1,\ldots,k\}$ associated to $\mathfrak{G}_{\circ}$ as well as further $n_s+n_e-k$ inequalities associated to subgraphs $\mathfrak{g}\,\notin\,\mathfrak{G}_0$ such that they identify a codimension-$(n_s+n_e-k)$ face of the full cosmological polytope $\mathcal{P}_{\mathcal{G}}$, {\it i.e.} 
\begin{equation}\label{eq:STsimpl}
    \mbox{Res}_{\mathcal{W}^{\mbox{\tiny $(\mathfrak{g}_{\sigma(1)})$}}}
    \mbox{Res}_{\mathcal{W}^{\mbox{\tiny $(\mathfrak{g}_{\sigma(2)})$}}}
    \ldots
    \mbox{Res}_{\mathcal{W}^{\mbox{\tiny $(\mathfrak{g}_{\sigma(n_s+n_e-k)})$}}} \omega(\mathcal{Y},\mathcal{P}_{\mathcal{G}})\:\neq\:0
\end{equation}
Therefore, given $\mathfrak{G}_{\circ}$, each possible set $\mathfrak{G}_c$ with $(n_s+n_e-k)$ elements $\mathfrak{g}\notin\mathfrak{G}_{\circ}$ satisfying the condition \eqref{eq:STsimpl} define the collection of simplices signed-triangulating $\mathcal{P}_{\mathcal{G}}$ through $\mathcal{W}^{\mbox{\tiny $(\mathfrak{g}_1\ldots\mathfrak{g}_k)$}}$, and the canonical form of $\omega(\mathcal{Y},\mathcal{P}_{\mathcal{G}})$ can then be written as
\begin{equation}\label{eq:wst}
    \omega(\mathcal{Y},\mathcal{P}_{\mathcal{G}})\:=\:
        \sum_{\{\mathfrak{G}_c\}}\prod_{\mathfrak{g}'\in\mathfrak{G}_c}\frac{1}{q_{\mathfrak{g'}}(\mathcal{Y})}
        \frac{\langle\mathcal{Y}d^{n_s+n_e-1}\mathcal{Y}\rangle}{\displaystyle
            \prod_{\mathfrak{g}\in\mathfrak{G}_{\circ}}q_{\mathfrak{g}}(\mathcal{Y})}
\end{equation}
Notice that the formula \eqref{eq:wst} provides a general expression for any signed triangulation through a specific class of subspaces of $\mathcal{C}(\mathcal{P}_{\mathcal{G}})$ identified by all the possible ways in which $\mathfrak{G}_{\circ}$ can be chosen. However, how can $\mathfrak{G}_{\circ}$ be chosen, and consequently how can the subspaces of $\mathcal{C}(\mathcal{P}_{\mathcal{G}})$ through which we can signed-triangulate $\mathcal{P}_{\mathcal{G}}$ as in \eqref{eq:wst}, be determined?

\begin{figure}
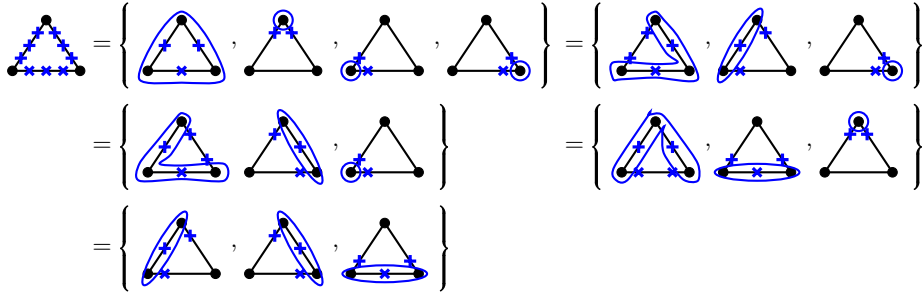

    \centering
    \vspace{-.25cm}

    \caption{Higher codimension intersections of the facets outside $\mathcal{P}_{\mathcal{G}}$ for the triangle graph. The set of subgraphs in each curly brackets represent a certain $\mathfrak{G}_{\circ}$ and covers completely the graph via complementary markings.}
    \label{fig:Zeros}
\end{figure}

As reviewed in Section \ref{subsec:FSCP}, the vertex structure for a certain facet $\mathcal{P}_{\mathcal{G}}\cap\mathcal{W}^{\mbox{\tiny $(\mathfrak{g})$}}\,\neq\,\varnothing$ can be graphically obtained considering the associated connected subgraph $\mathfrak{g}\subseteq\mathcal{G}$ and marking with \bluecross{}  all the edges contained in $\mathfrak{g}$ as well as the edges departing from $\mathfrak{g}$ close to the sites in $\mathfrak{g}$. Then, subspaces of $\mathcal{C}(\mathcal{P}_{\mathcal{G}})$ are given by all those markings associated to $k\le n_s+n_e$ subgraphs $\mathfrak{g}\subseteq\mathcal{G}$ that cover $\mathcal{G}$ completely: this means that no vertex of the cosmological polytope is associated to $\mathcal{P}_{\mathcal{G}}\cap\mathcal{W}^{\mbox{\tiny $(\mathfrak{g}_1\ldots\mathfrak{g}_k)$}}$, which is consequently empty. A special class of such set of subgraphs is identified by those with {\it complementary markings}, {\it i.e.} the elements in a given set introduce markings which do no overlap with each other (see Figure \ref{fig:Zeros}).

Given a subspace of $\mathcal{C}(\mathcal{P}_{\mathcal{G}})$ identified by the set of subgraphs $\mathfrak{G}_{\circ}$ and the $(n_s+n_e-k)$ compatible channels related to the set $\mathfrak{G}_c$ of subgraphs not contained in $\mathfrak{G}_{\circ}$, $\mathfrak{G}_c$ identifies a codimension-$(n_s+n_e-k)$ face of $\mathcal{P}_{\mathcal{G}}$. Defining $\mathcal{W}^{\mbox{\tiny $(\mathfrak{G}_{\circ})$}}:=\bigcap_{\mathfrak{g}\in\mathfrak{G}_{\circ}}\mathcal{W}^{\mbox{\tiny $(\mathfrak{g})$}}$ and  $\mathcal{W}^{\mbox{\tiny $(\mathfrak{G}_c)$}}:=\bigcap_{\mathfrak{g}\in\mathfrak{G}_c}\mathcal{W}^{\mbox{\tiny $(\mathfrak{g})$}}$, such a face is given by $\mathcal{P}_{\mathcal{G}}\cap\mathcal{W}^{\mbox{\tiny $(\mathfrak{G}_c)$}}$. Interestingly, as such a face is taken, the subspace identified by $\mathfrak{G}_{\circ}$ gets projected onto it and $\mathcal{P}_{\mathcal{G}}\cap\mathcal{W}^{\mbox{\tiny $(\mathfrak{G}_c)$}}\cap\mathcal{W}^{\mbox{\tiny $(\mathfrak{G}_{\circ})$}}\,\subseteq\,\mathbb{P}^0$ is not vanishing, with its canonical form being $\pm\,1$. Consequently, the subgraphs in the set $\mathfrak{G}_{\circ}\cup\mathfrak{G}_{c}$ identifies a simplex in $\mathbb{P}^{n_s+n_e-1}$. Then, the sum over all the possible choices of $\mathfrak{G}_c$ covers the full cosmological polytope $\mathcal{P}_{\mathcal{G}}$, with $\mathcal{P}_{\mathcal{G}}\cap\mathcal{W}^{\mbox{\tiny $(\mathfrak{G}_{\circ})$}}\,=\,\varnothing$.

It is important to stress once more the physical relevance of the class of signed-triangulations expressed by the formula \eqref{eq:wst} and by their combinatorial origin. They correspond different perturbative representations of the wavefunctions, all of them showing physical singularities only. Importantly, while they look in principle very different, they are treated on the same footing, not only from the combinatorial perspective (they are all part of a special class of signed-triangulations), but also (and especially) from a physical point of view, as they make manifest different subset of compatibility conditions on multiple channels. In fact, the analysis of the face structure of the cosmological polytope as well as of the locus of the intersections of its facets outside it, leads to novel constraints, given by equations \eqref{eq:wfgmc} and \eqref{eq:STsimpl}, which generalise the Steinmann-like relations ~\cite{Benincasa:2020aoj}~. While these constraints have been proven for the universal integrand $\psi_{\mathcal{G}}$, they extend as statement on the multiple discontinuities for the tree-level wavefunction as well as for the loop integrand in $dS$ space-time, as they are expected to be polylogarithms ~\cite{Arkani-Hamed:2017fdk, Hillman:2019wgh}~. More generally, they are statement on graph contributions to the wavefunction, in generic theories involving states with flat-space counterparts, despite strictly speaking they are related to conformally coupled scalar with non-conformal polynomial interactions: firstly, both more general scalars and spinning states via differential operators ~\cite{Benincasa:2019vqr, Baumann:2019oyu, Baumann:2020dch}~; secondly, tensor integrals can be analysed through the scalar ones using relations at both integrand and integral level\footnote{Such relations have been extensively exploited in the context of scattering amplitudes, while it is still an unexplored territory for the wavefunction.}

\subsubsection{An explicit example}\label{subsec:EEtr}

As a matter of illustration, let us consider the explicit example of a one-loop three-site graph, as in Figure \ref{fig:Zeros}. The associated cosmological polytope is the convex hull of $9$ vertices in $\mathbb{P}^5$.

Let us label the three sites of the graphs as $1$, $2$ and $3$, while the edges connecting sites $(1,2)$, $(2,3)$ and $(3,1)$ with the latin letters $a$, $b$ and $c$ respectively. The hyperplanes containing the facets are identified by all the possible connected subgraphs\footnote{For a generic polygon graph, which is characterised by $n_e$ edges and $n_s\,=\,n_e$ sites, the associated cosmological polytope lives in $\mathbb{P}^{2n_e-1}$, and it has $1+n_e^2$ facets and the locus $\mathcal{C}(\mathcal{P}_{\mathcal{G}})$ of the intersections of the facets outside $\mathcal{P}_{\mathcal{G}}$ is determined by a symmetric tensor with $(n_e-1)^2$ indices and 
$$
\Delta = 
\begin{pmatrix}
    n_e^2 \\
    (n_e-1)^2
\end{pmatrix}
-1
$$
degrees of freedom.
} 
$\{\mathcal{G},\,\{\mathfrak{g}_e\,|\,e=a,b,c\},\,\{\mathfrak{g}_{ij}\,|\,i,j=1,2,3,\,i\neq j\},\,\{\mathfrak{g}_j\,|\,j=1,2,3\}\}$, where $\mathfrak{g}_{e}$ contains all the sites as well as all the edges but $e$, $\mathfrak{g}_{ij}$ is the two-site line graph containing the sites $i$ and $j$ as well as the edge $e_{ij}$ connecting them, and finally $\mathfrak{g}_j$ contains the site $j$ only:
\begin{equation}\label{eq:qgs}
    \begin{split}
        &q_{\mathcal{G}}=\sum_{j=1}^3 x_j,\quad q_{\mathfrak{g}_e}=\sum_{j=1}^3 x_j+2y_e,\quad
         q_{\mathfrak{g}_{ij}}=\sum_{\substack{k=i,j \\ l\neq i,j}}(x_k+y_{e_{kl}}),\quad
         q_{\mathfrak{g}_{j}}=x_j+\sum_{k\neq j}y_{e_{jk}}
    \end{split}
\end{equation}
All the possible combinations of subgraphs which completely mark the graph $\mathcal{G}$ via complementary markings and fix all the possible $\mathfrak{G}_{\circ}$ are illustrated in Figure \ref{fig:Zeros}. 

The signed-triangulations for all the possible choices of the set $\mathfrak{G}_{\circ}$, are expressed by the formula \eqref{eq:wst} for the canonical form of $\mathcal{P}_{\mathcal{G}}\,\subset\,\mathbb{P}^5$:
\begin{equation}\label{eq:cftr}
    \omega(\mathcal{Y},\mathcal{P}_{\mathcal{G}})\:=\:\sum_{\{\mathfrak{G}_c\}}\prod_{\mathfrak{g}'\in\mathfrak{G}_c}\frac{1}{q_{\mathfrak{g}'}(\mathcal{Y})}
        \frac{\langle\mathcal{Y}d^5\mathcal{Y}\rangle}{\displaystyle\prod_{g\in\mathfrak{G}_{\circ}}q_{\mathfrak{g}}(\mathcal{Y})}.
\end{equation}
We can straightforwardly write them explicitly
\begin{itemize}
    \item $\mathfrak{G}_{\circ}=\{\mathcal{G},\,\{\mathfrak{g}_j\}_{j=1}^3\}$. As $\mathcal{P}_{\mathcal{G}}$ lives in $\mathbb{P}^5$, the sets $\mathfrak{G}_c$ contains two elements only. Then the compatibility conditions on pair of channels identified by the subgraphs $\mathfrak{g},\,\mathfrak{g}'\,\notin\,\mathfrak{G}_{\circ}$, {\it i.e.} that they cannot be partially overlapping, allow to identify the sets $\{\mathfrak{G}_c\}$ as
    \begin{equation}\label{eq:Gc1}
        \{\mathfrak{G}_c\}\,=\,\{\{\mathfrak{g}_a,\,\mathfrak{g}_{23}\},\,\{\mathfrak{g}_{a},\mathfrak{g}_{31}\},\,
                                   \{\mathfrak{g}_b,\,\mathfrak{g}_{31}\},\,\{\mathfrak{g}_{b},\mathfrak{g}_{12}\},\,
                                   \{\mathfrak{g}_c,\,\mathfrak{g}_{12}\},\,\{\mathfrak{g}_{c},\mathfrak{g}_{23}\}\}
    \end{equation}
    and hence the canonical form can be readily written as
    \begin{equation}\label{eq:trOFPT}
        \begin{split}
            \omega(\mathcal{Y},\mathcal{P}_{\mathcal{G}})\:=&\:
                \left[
                    \frac{1}{q_{\mathfrak{g}_a}}
                    \left(
                        \frac{1}{q_{\mathfrak{g}_{23}}}+\frac{1}{q_{\mathfrak{g}_{31}}}
                    \right)+
                    \frac{1}{q_{\mathfrak{g}_b}}
                    \left(
                        \frac{1}{q_{\mathfrak{g}_{31}}}+\frac{1}{q_{\mathfrak{g}_{12}}}
                    \right)+
                \right.\\
                &\left.\hspace{1cm}+
                    \frac{1}{q_{\mathfrak{g}_c}}
                    \left(
                        \frac{1}{q_{\mathfrak{g}_{12}}}+\frac{1}{q_{\mathfrak{g}_{23}}}
                    \right)
                \right]
                \frac{\langle\mathcal{Y}d^{5}\mathcal{Y}\rangle}{q_{\mathcal{G}}
                    q_{\mathfrak{g}_1}q_{\mathfrak{g}_2}q_{\mathfrak{g}_3}}.
        \end{split}
    \end{equation}
    This is nothing but that OFPT representation for the wavefunction.
    \item $\mathfrak{G}_{\circ}=\{\mathfrak{g}_a,\,\mathfrak{g}_{12},\,\mathfrak{g}_{3}\}$. As in the previous case, the invariance under the $GL(1)$
          transformation $\mathcal{Y}\,\longrightarrow\,\lambda\,\mathcal{Y}$, implies that each set $\mathfrak{G}_c$ identifies a codimension-$3$ boundary of the cosmological polytope, involving facets other than the ones associated to the subgraphs in $\mathfrak{G}_{\circ}$. The compatibility conditions on the codimension-$3$ faces for subgraphs $\mathfrak{g}\notin\mathfrak{G}_{\circ}$ return the following list of $\mathfrak{G}_c$'s:
         \begin{equation}
            \begin{split}
                \{\mathfrak{G}_c\}\:=\:&
                \{\{\mathcal{G},\,\mathfrak{g}_b,\,\mathfrak{g}_{31}\},\,\{\mathcal{G},\,\mathfrak{g}_b,\,\mathfrak{g}_2\},\,
                \{\mathcal{G},\,\mathfrak{g}_c,\,\mathfrak{g}_{23}\},\,\{\mathcal{G},\,\mathfrak{g}_c,\,\mathfrak{g}_1\},\\
                &\{\mathcal{G},\,\mathfrak{g}_{23},\,\mathfrak{g}_2\},\{\mathcal{G},\,\mathfrak{g}_{31},\,\mathfrak{g}_1\},\,
                \{\mathfrak{g}_{b},\,\mathfrak{g}_{31},\,\mathfrak{g}_2\},\,\{\mathfrak{g}_c.\,\mathfrak{g}_{23},\,\mathfrak{g}_1\}\}
            \end{split}
        \end{equation}
        and the canonical form can be written as
        \begin{equation}\label{eq:cfa123}
            \begin{split}
                \omega(\mathcal{Y},\mathcal{P}_{\mathcal{G}})\:=\:&
                \left\{
                    \frac{1}{q_{\mathcal{G}}}
                    \left[
                        \frac{1}{q_{\mathfrak{g}_b}}\left(\frac{1}{q_{\mathfrak{g}_{31}}}+\frac{1}{\mathfrak{g}_2}\right)+
                        \frac{1}{q_{\mathfrak{g}_c}}\left(\frac{1}{q_{\mathfrak{g}_{23}}}+\frac{1}{\mathfrak{g}_1}\right)+
                    \right.
                \right.\\
                &\left.\left.+
                        \frac{1}{q_{\mathfrak{g}_{23}}q_{\mathfrak{g}_2}}+\frac{1}{q_{\mathfrak{g}_{31}}q_{\mathfrak{g}_1}}
                    \right]+
                    \frac{1}{q_{\mathfrak{g}_b}q_{\mathfrak{g}_{31}}q_{\mathfrak{g}_2}}+
                    \frac{1}{q_{\mathfrak{g}_c}q_{\mathfrak{g}_{23}}q_{\mathfrak{g}_1}}
                \right\}
                \frac{\langle\mathcal{Y}d^5\mathcal{Y}\rangle}{q_{\mathfrak{g}_a}q_{\mathfrak{g}_{12}}q_{\mathfrak{g}_3}}
            \end{split}
        \end{equation}
          Interestingly, this signed triangulation returns a new way of organising the perturbative expansion.
    \item $\mathfrak{G}_{\circ}=\{\mathfrak{g}_b,\,\mathfrak{g}_{23},\,\mathfrak{g}_1\}$ and             
          $\mathfrak{G}_{\circ}=\{\mathfrak{g}_c,\,\mathfrak{g}_{31},\,\mathfrak{g}_2\}$. The signed triangulations corresponding to thee two choices of the subspace of $\mathcal{C}(\mathcal{P}_{\mathcal{G}})$ have the very same structure of the one just discussed, and they can be obtained from it via the following label-shift:
          \begin{equation}\label{eq:cflsh}
              a\:\longrightarrow\:b\:\longrightarrow\:c,\qquad
              i\:\longrightarrow\:i+1\:\longrightarrow\:i+2.
          \end{equation}
    \item $\mathfrak{G}_{\circ}\:=\:\{\mathfrak{g}_{12},\,\mathfrak{g}_{23},\,\mathfrak{g}_{31}\}$. Again, each $\mathfrak{G}_c$ identifies a codimension-$3$ 
          face of $\mathcal{P}_{\mathcal{G}}$. Again, the compatibility conditions on the hyperplanes related to subgraphs which are not in $\mathfrak{G}_{\circ}$, determines all the sets $\mathfrak{G}_c$:
          \begin{equation}\label{eq:cfijGc}
                \begin{split}
                    \{\mathfrak{G}_c\}\:=\:&\{\{\mathcal{G},\,\mathfrak{g}_a,\mathfrak{g}_1\},\,\{\mathcal{G},\,\mathfrak{g}_a,\mathfrak{g}_2\},
                                             \{\mathcal{G},\,\mathfrak{g}_b,\mathfrak{g}_2\},\,\{\mathcal{G},\,\mathfrak{g}_b,\mathfrak{g}_3\},
                                             \{\mathcal{G},\,\mathfrak{g}_c,\mathfrak{g}_3\},\\ 
                                            &\{\mathcal{G},\,\mathfrak{g}_c,\mathfrak{g}_1\},\{\mathfrak{g}_a,\,\mathfrak{g}_1,\,\mathfrak{g}_2\},
                                             \{\mathfrak{g}_b,\,\mathfrak{g}_2,\,\mathfrak{g}_3\},\{\mathfrak{g}_c,\,\mathfrak{g}_3,\,\mathfrak{g}_1\},
                                             \{\mathfrak{g}_1,\,\mathfrak{g}_2,\,\mathfrak{g}_3\}
                                             \}
                \end{split}
          \end{equation}
          and the signed-triangulation for canonical form then becomes
          \begin{equation}\label{eq:cftr123}
                \begin{split}
                    \omega(\mathcal{Y},\mathcal{P}_{\mathcal{G}})\:=\:&
                    \left\{
                        \frac{1}{q_{\mathcal{G}}}
                        \left[
                            \frac{1}{q_{\mathfrak{g}_a}}\left(\frac{1}{q_{\mathfrak{g}_1}}+\frac{1}{q_{\mathfrak{g}_2}}\right)+
                            \frac{1}{q_{\mathfrak{g}_b}}\left(\frac{1}{q_{\mathfrak{g}_2}}+\frac{1}{q_{\mathfrak{g}_3}}\right)+
                        \right.
                    \right.\\
                    &\left.
                        +\frac{1}{q_{\mathfrak{g}_c}}\left(\frac{1}{q_{\mathfrak{g}_3}}+\frac{1}{q_{\mathfrak{g}_1}}\right)
                        \right]
                        +\frac{1}{q_{\mathfrak{g}_a}q_{\mathfrak{g}_1}q_{\mathfrak{g}_2}}+\frac{1}{q_{\mathfrak{g}_b}q_{\mathfrak{g}_2}q_{\mathfrak{g}_3}}+\\
                    &\left.
                        +\frac{1}{q_{\mathfrak{g}_c}q_{\mathfrak{g}_3}q_{\mathfrak{g}_1}}+\frac{1}{q_{\mathfrak{g}_1}q_{\mathfrak{g}_2}q_{\mathfrak{g}_3}}
                    \right\}\frac{\langle\mathcal{Y}d^5\mathcal{Y}\rangle}{q_{\mathfrak{g}_{12}}q_{\mathfrak{g}_{23}}q_{\mathfrak{g}_{31}}}.
                \end{split}
          \end{equation}
          Such a representation is novel as well ~\cite{Benincasa:2021qcb}~.
\end{itemize}
This example is meant to work just as an illustration, as the procedure holds for any graph $\mathcal{G}$. 


\section{Generalised cosmological polytopes}\label{sec:GenCP}

Cosmological polytopes have their own first principle definition in terms of the intersection of $n_e$ triangles, originally defined in $\mathbb{P}^{3n_e-1}$, in the midpoints of at most two out of their three edges. This idea can be generalised if we extend our building blocks to include triangles and segments ~\cite{Benincasa:2019vqr}~. 

\begin{figure}
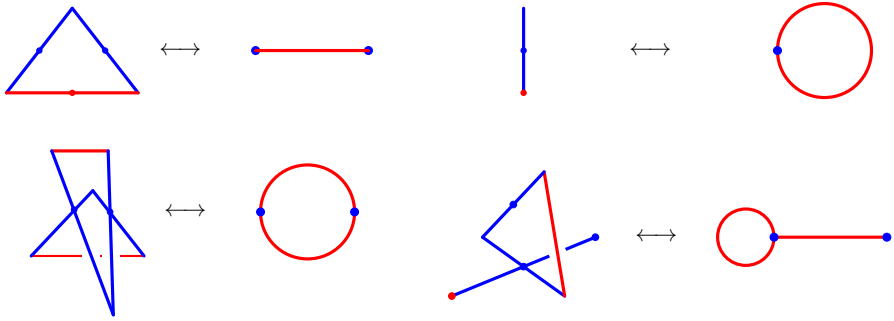


 \caption{Generalised cosmological polytopes. Considering a collection of triangles and segments, they are defined by intersecting the elements of such a collection in the midpoints of their sides, and taking the convex hull of their vertices. As each triangle and each segment are in $1-1$ correspondence with a two-site line graph and a tadpole respectively, each generalised cosmological polytope is in $1-1$ correspondence with a graph built by merging in their sites such building block graphs.}
\end{figure}

Let us consider a collection of $n_e$ triangles and $n_h$ segments in $\mathbb{P}^{3n_e+2n_h-1}$ with local coordinates 
$\mathcal{Y}\,=\,(\{x_i\},\,\{y_i\},\,\{x'_i\},\,\{x''_j\},\,\{h_j\})$
, each of which is identified by the vertices 
$\{\mathbf{x}_i-\mathbf{y}_{i}+\mathbf{x'}_i,\,\mathbf{x}_i+\mathbf{y}_{i}-\mathbf{x'}_i,\, -\mathbf{x}_i+\mathbf{y}_{i}+\mathbf{x'}_i\}_{i=1}^{n_e}$
and
$\{2\mathbf{x''}_j-\mathbf{h}_j,\,\mathbf{h}_j\}_{j=1}^{n_h}$
respectively. We can construct new polytopes by intersecting both triangles and segments in their midpoints and taking the convex hull of their vertices. The resulting polytope then lives in $\mathbb{P}^{3n_e+2n_h-r-1}$, $r$ being the number of intersections among the building blocks, and it is characterised by having $3n_e+2n_h$ vertices. 

Interestingly, also this new class of polytopes is in $1-1$ correspondence with weighted graphs. As we discussed for the original construction of the cosmological polytopes, each triangle is in $1-1$ correspondence with $2$-sites line graphs and, because of the relation between the intersectable sides of the triangles and the sites of these graph, the cosmological polytope constructed as an intersection of triangles turns out to be in $1-1$ correspondence with the graph obtained by merging accordingly the related $2$-sites graphs in their sites. What about a segment $\{2\mathbf{x}-\mathbf{h},\,\mathbf{h}\}$? Does it have a graph interpretation?

It turns out that the answer to this last question is affirmative. Let us consider the projective line $\mathbb{P}^1$ with local coordinates $\mathcal{Y}\,=\,(x,h)$, and the segment $\{2\mathbf{x}-\mathbf{h},\,\mathbf{h}\}$. The canonical form associated to such a segment is
\begin{equation}\label{eq:CFseg}
    \omega(\mathcal{Y},\,\mathcal{P})\:=\:\frac{1}{x(x+2h)}\,\frac{dx\wedge dh}{\mbox{Vol}\{GL(1)\}}.
\end{equation}
To such a segment we can associate a tadpole, with weights $x$ and $h$ on its site and hedge respectively. The recursive rules of OFPT \eqref{eq:OFPT2} for computing the wavefunction contribution of such a graph, precisely returns the canonical function in \eqref{eq:CFseg}. Also in this case, the facets of the segments are identified by the subgraphs. Recall that given a hyperplane $\mathcal{W}^{\mbox{\tiny $(\mathfrak{g})$}}$, it contains a facet $\mathcal{P}_{\mathcal{G}}\cap\mathcal{W}^{\mbox{\tiny $(\mathfrak{g})$}}\,\neq\,\varnothing$ of the polytope $\mathcal{P}_{\mathcal{G}}$, and $\mathcal{Z}^I_j\mathcal{W}^{\mbox{\tiny $(\mathfrak{g})$}}_I\,=\,0$ if the vertex $\mathcal{Z}_j$  of the polytope is on $\mathcal{P}_{\mathcal{G}}\cap\mathcal{W}^{\mbox{\tiny $(\mathfrak{g})$}}$ and $\mathcal{Z}^I\mathcal{W}^{\mbox{\tiny $(\mathfrak{g})$}}_I\,>\,0$ if it is not:
\begin{equation*}
 \begin{tikzpicture}[ball/.style = {circle, draw, align=center, anchor=north, inner sep=0}, cross/.style={cross out, draw, minimum size=2*(#1-\pgflinewidth), inner sep=0pt, outer sep=0pt}, scale={1.125}, transform shape]
  \begin{scope}[shift={(2.25,-2.25)}, scale={.5}]
   \coordinate (LC) at (,0); 
   \coordinate [label=left:{$\displaystyle x$}] (x) at ($(LC)+(-1.25cm,0)$);
   \coordinate [label=right:{$\displaystyle h$}] (h) at ($(LC)+(1.25cm,0)$);    
   \tikzset{point/.style={insert path={ node[scale=2.5*sqrt(\pgflinewidth)]{.} }}} 

   \draw[very thick] (LC) circle (1.25cm); 
   \draw[fill] (x) circle (3pt);
   \node[very thick, cross=6pt, rotate=0, color=blue] (X2) at (h) {};   
  
   \coordinate [label={$\displaystyle\mathcal{W}\cdot(2\mathbf{x}-\mathbf{h})>0$}] (hyp1) at ($(LC)-(0,2)$);
  \end{scope}
  \begin{scope}[shift={(7,-2.25)}, scale={.5}]
   \coordinate (LC) at (,0); 
   \coordinate [label=left:{$\displaystyle x$}] (x) at ($(LC)+(-1.25cm,0)$);
   \coordinate [label=right:{$\displaystyle h$}] (h) at ($(LC)+(1.25cm,0)$);    
   \tikzset{point/.style={insert path={ node[scale=2.5*sqrt(\pgflinewidth)]{.} }}} 

   \draw[very thick] (LC) circle (1.25cm); 
   \draw[fill] (x) circle (3pt);
   \coordinate (va) at ($(x)+(.02,.375)$);
   \coordinate (vb) at ($(x)+(.02,-.375)$);
   \node[very thick, cross=6pt, rotate=0, color=blue] (X1a) at (va) {};      
   \node[very thick, cross=6pt, rotate=0, color=blue] (X1b) at (vb) {};  
  
   \coordinate [label={$\displaystyle\mathcal{W}\cdot\mathbf{h}>0$}] (hyp1) at ($(LC)-(0,2)$);
  \end{scope}
 \end{tikzpicture}
\end{equation*}
In the last graph the double marking indicates the absence of the same vertex: the tadpole can be seen obtained from a two-site graph by merging the two sites. From the polytope perspective, this is equivalent in identifying the two intersectable sides (squeezing the non-intersectable one) ~\cite{Benincasa:2019vqr}~ and, consequently, the two vertices of the triangle. Hence, the characterisation of the face structure of the generalised cosmological polytopes  proceeds precisely as described in \ref{subsec:FSCP}.

This construction is pretty general. However, if consider a graph $\mathcal{G}$ with its site having precisely as many tadpoles as open edges incident in it, the related polytope encodes the physics of the wavefunction with propagating massless states in FRW cosmologies ~\cite{Benincasa:2019vqr}~. More precisely, the wavefunction contribution from this graph is the {\it covariant form} obtained from the canonical form of the graph via the covariant restriction along the hyperplane defined by $\mathcal{H}\,:=\,\{h_j\,=\,0,\:\forall\,j=1,\ldots,2n_e\}$.

\begin{figure}
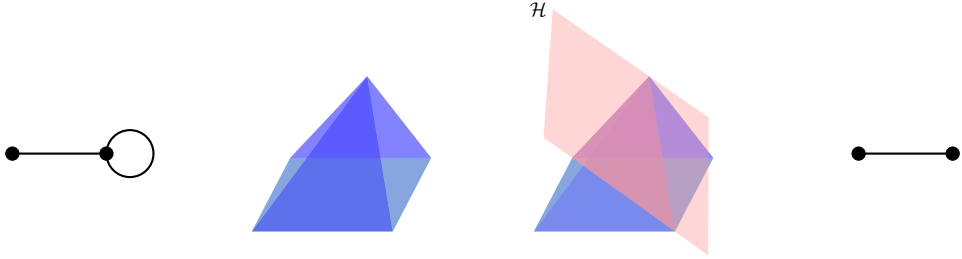

    \centering

    \caption{A visualisable example of a restriction of generalised cosmological polytope $\mathcal{P}_{\mathcal{G}_t}$ to a cosmological polytope $\mathcal{P}_{\mathcal{G}}$. The map $\mathcal{G}_{t}\,\longrightarrow\,\mathcal{G}$ by shrinking the tadpole in $\mathcal{G}_t$ corresponds to restrict $\mathcal{P}_{\mathcal{G}_t}$ onto a specific hyperplane $\mathcal{H}$. The covariant restriction of the canonical form of $\mathcal{P}_{\mathcal{G}_t}$, as defined in \eqref{eq:wH}, associates a covariant form of covariant degree-$1$ to the cosmological polytope $\mathcal{P}_{\mathcal{G}}$.}
    \label{fig:GCP}
\end{figure}

\subsection{(Generalised) cosmological polytopes, covariant forms and the wavefunction for massless states}

Let us consider a generalised cosmological polytope $\mathcal{P}_{\mathcal{G}^{\mbox{\tiny $(n_h)$}}}\subset\mathbb{P}^{n_s+n_e+n_h-1}$ related to a graph $\mathcal{G}^{\mbox{\tiny $(n_h)$}}$ with $n_s$ sites, $n_e$ edges connecting different sites, and $n_h$ tadpole edges. Let $k\,\in\,[1,\,n_h]$ be an integer. Then, the restriction $\left.\mathcal{P}_{\mathcal{G}}^{\mbox{\tiny $(n_h)$}}\right|_{\mathcal{H}^{\mbox{\tiny $(k)$}}}:=\mathcal{P}_{\mathcal{G}}^{\mbox{\tiny $(n_h)$}}\cap\mathcal{H}^{\mbox{\tiny $(k)$}}\,\subset\,\mathbb{P}^{n_s+n_e+n_h-k-1}$ of the generalised cosmological polytope $\mathcal{P}_{\mathcal{G}}^{\mbox{\tiny $(n_h)$}}$ onto the hyperplane $\mathcal{H}^{\mbox{\tiny $(k)$}}$ defined as
\begin{equation}\label{eq:Hk}
    \resizebox{0.9\hsize}{!}{$\displaystyle
    \mathcal{H}^{\mbox{\tiny $(k)$}}=
    \left\{
        \mathcal{Y}\in\mathbb{P}^{n_s+n_e+n_h-1}\,\Big|\,\mathcal{Y}\cdot\tilde{\mathbf{h}}_l=0,\;
        \left\{
            \begin{array}{l}
                 \mathbf{h}_h\cdot\tilde{\mathbf{h}}_l=\delta_{kl},\qquad\forall\,l\in[1,k],\,h\in[1,n_h]\\
                  (\mathbf{x}_s,\,\mathbf{y}_e)\cdot\tilde{h}_l=0,\quad\forall\,s\in[1,n_s],\,e\in[1,n_e]
            \end{array}
        \right.
    \right\}$},
\end{equation}
is still a generalised cosmological polytope related to a graph $\mathcal{G}^{\mbox{\tiny $(n_h-k)$}}$ which is obtained from $\mathcal{G}^{\mbox{\tiny $(n_h)$}}$ by suppressing $k$ tadpoles, and the covariant restriction of the canonical form of $\mathcal{P}_{\mathcal{G}}^{\mbox{\tiny $(n_h)$}}$ is a covariant form of degree-$k$ associated to $\left.\mathcal{P}_{\mathcal{G}}^{\mbox{\tiny $(n_h)$}}\right|_{\mathcal{H}^{\mbox{\tiny $(k)$}}}$~\cite{Benincasa:2020uph}~. 

If $k\,=\,n_h\,=2n_e$, with a tadpoles at a site for each edge attached to it, the covariant restriction $\left.\mathcal{P}_{\mathcal{G}}^{\mbox{\tiny $(2n_e)$}}\right|_{\mathcal{H}^{\mbox{\tiny $(2n_e)$}}}$ is nothing but the cosmological polytope $\mathcal{P}_{\mathcal{G}}$ obtained from $\mathcal{P}_{\mathcal{G}}^{\mbox{\tiny $(2n_e)$}}$ by suppressing all the tadpoles, and the covariant function -- {\it i.e.} the rational function obtained from the covariant form by stripping out the standard measure of the projective space where the associated polytope lives in -- provides the universal integrand for propagating massless scalars with polynomial interactions in FRW cosmologies ~\cite{Benincasa:2019vqr}~:
\begin{equation}\label{eq:WFmsls}
    \omega^{\mbox{\tiny $(2n_e)$}}(\mathcal{Y}_{\mathcal{H}^{\mbox{\tiny $(2n_e)$}}})\:=\:(-1)^{n_e}\psi_{\mathcal{G}}
        \langle\mathcal{Y}_{\mathcal{H}^{\mbox{\tiny $(2n_e)$}}}d^{n_s+n_e-1}\mathcal{Y}_{\mathcal{H}^{\mbox{\tiny $(2n_e)$}}}\rangle
\end{equation}
%


\section{Flat-space physics from the cosmological wavefunction}\label{sec:Flat}

Flat-space scattering amplitudes are encoded into the wavefunction of the universe in the sheet of kinematic space where the conservation of total energy holds \cite{Maldacena:2011nz, Raju:2012zr}. Such a sheet lies outside of the physical region and can be reached upon analytic continuation in such a way that some of the states has positive energy and the others have negative energy. As such analytic continuation is performed, a general question is how basic principles and symmetries of flat-space amplitudes, absents in cosmology, get restored: the wavefunction, as the other observables that can be constructed from it, is defined on a space-like surface at fixed time, and therefore Lorentz invariance is manifestly broken, while time evolution is integrated out, and hence it is not obvious as the crucial principles ruling flat-space physics, such as Lorentz invariance, unitarity and causality, should be emerge.

For all theories with a description in terms of cosmological polytopes, as well as those which can be expressed or obtained from the graphs related to the cosmological polytopes, this question can be addressed in a sharp way.

Flat-space physics is encoded in the scattering facet of the cosmological polytope, whose combinatorial structure is a direct consequence of the first principle definition of the cosmological polytope. It is identified by the subgraph $\mathfrak{g}\,=\,\mathcal{G}$ and it is the convex hull of the set of $2n_e$ vertices %
$$\mathcal{V}_{\mathcal{G}}\: :=\{\mathbf{x}_i+\mathbf{y}_i-\mathbf{x'}_i,\, -\mathbf{x}_i+\mathbf{y}_i+\mathbf{x'}_i\,|\,j=1,\ldots,n_e\}$$.
Therefore, all the questions about flat-space physics and how it emerges from the more fundamental context of an expanding universe can be formulated as questions on the scattering facet. In the next subsections we will show how Lorentz invariance as well as flat-space unitarity and causality emerge from its combinatorial structure.


\subsection{Emergence of Lorentz invariance}\label{subsec:LI}

In the usual formulation of scattering amplitudes, Lorentz invariance is made manifest through the parametrisation of the kinematic space via Mandelstam invariants. The kinematic space for the wavefunction is instead parametrised in terms of variables which are rotationally invariant, the moduli of the spatial momenta. Hence, a first way of rephrasing the question of the emergence of Lorentz invariance is asking how the rotational invariant poles of the wavefunction can pair up to form the usual Lorentz invariant propagators as we go on the total energy conservation sheet. Importantly, the wavefunction depends only on spatial momenta, so it is in principle not obvious for loop contributions how the time-component $\ell_{\circ}$ in the loop integration, which is needed for the integrands to be manifestly Lorentz invariant.

Given any projective polytope $\mathcal{P}\subset\mathbb{P}^{N-1}$, its canonical function has a contour integral representation given by equation \eqref{eq:cfunint}, as discussed in Section \ref{subsec:PPCF}. Let us rewrite it here for later convenience:
\begin{equation}\label{eq:cfunint2}
    \Omega(\mathcal{Y},\,\mathcal{P})\:=\:\frac{1}{(N-1)!(2\pi i)^{\nu-N}}\int_{\mathbb{R}^{\nu}}
                \prod_{k=1}^{\nu}\frac{dc_k}{c_k-i\varepsilon_k}\,\delta^{\mbox{\tiny $(N)$}}\left(\mathcal{Y}-\sum_{k=1}^{\nu}c_k Z_k\right),
\end{equation}
where $\{Z_k^I\,\in\,\mathbb{P}^{N-1}\,|\,k=1,\ldots,\nu\}$ identifies the set of $\nu$ vertices of $\mathcal{P}$, while $\mathcal{Y}$ parametrises a generic point in $\mathbb{P}^{N-1}$.

Taking $\mathcal{P}$ to be the scattering facet $\mathcal{S}_{\mathcal{G}}\,=\,\mathcal{P}_{\mathcal{G}}\cap\mathcal{W}^{\mbox{\tiny $(\mathcal{G})$}}$ of the cosmological polytope $\mathcal{P}_{\mathcal{G}}\subset\mathbb{P}^{n_s+n_e-1}$ associated to a graph $\mathcal{G}$ and therefore $N=n_s+n_e-1$, then it has $\nu\,=\,2n_e\,=\,n_s+n_e-1+L$ vertices, where $L$ is the number of loops of $\mathcal{G}$ and we have used the relation $n_s\,=\,n_e+1-L$ relating sites, edges and loops of $\mathcal{G}$, and is embedded in $\mathbb{P}^{n_s+n_e-2}$. Notice that at tree level, {\it i.e.} for $L\,=\,0$, the scattering facet $\mathcal{S}_{\mathcal{G}}\subset\mathbb{P}^{n_s+n_e-2}$ has $n_s+n_e-1$ vertices and consequently it is always a simplex in $\mathbb{P}^{n_s+n_e-2}$. For $L\,>\,0$ the scattering facet is never simplicial.

Let us then consider the contour representation \eqref{eq:cfunint2} for the canonical function of the scattering facet. The $\delta$-function fixes $n_s+n_e-1$ integration variables, leaving precisely $L$ of the $c_k$'s unfixed. Indeed, if $L\,=\,0$, then the integrations are completely localised and return a simplex. Going back to the general case, there is some freedom in the implementation of the constraints introduced by the $\delta$-function and, consequently, in the choice of which $c_k$'s to fix. Interestingly, these choices have a beautiful geometrical interpretation: each of them corresponds to the selection of all those hyperplanes that do not contain the vertices of $\mathcal{S}_{\mathcal{G}}$ associated to the unfixed $c_k$'s \cite{Arkani-Hamed:2018ahb}. 

\begin{figure}[t]
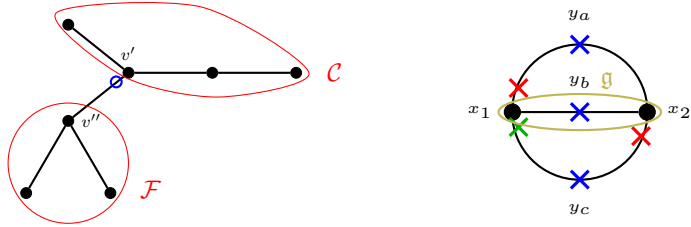

    \centering

    \caption{Lorentz invariance on the scattering facet.}
    \label{fig:LorInv}
\end{figure}
It is possible to keep track of such a choice by marking the graph $\mathcal{G}$ with a red cross \redcross~ for each vertex associated to a $c_k$ chosen not to be localised by the $\delta$-function -- see Figure \ref{fig:LorInv}. The codimension-$1$ hyperplane which does not contain such vertices is then identified by further marking the other vertices which are not on by a green cross \greencross, and corresponds to the subgraph $\mathfrak{g}$ enclosed by the red/green marking
\begin{equation}\label{eq:HypCs}
    \mathcal{W}^{\mbox{\tiny $(\mathfrak{g})$}}\:=\:
        \sum_{s\in\mathcal{V}_{\mathfrak{g}}}\mathbf{\tilde{x}}_s+
        \sum_{e\in\mathcal{E}_{\mathfrak{g}}^c}\mathbf{\tilde{y}}_e-
        \sum_{e\in\mathcal{E}_{\mathfrak{g}}^u}\mathbf{\tilde{y}}_e
\end{equation}
where $\mathcal{E}_{\mathfrak{g}}^c$ and $\mathcal{E}_{\mathfrak{g}}^u$ are the sets of edges departing from $\mathfrak{g}$ with and without marking -- see Figure \ref{fig:LorInv}. The solution for the $c_k$ related to the vertex marked by the green cross \greencross~ is then given by
\begin{equation}\label{eq:cksol}
    c_k(\greencross)\:=\:\tilde{\mathcal{Y}}^I\mathcal{W}^{\mbox{\tiny $(\mathfrak{g})$}}_I,\qquad
    \mbox{with }
    \tilde{\mathcal{Y}}\: :=\:\mathcal{Y}-\sum_{j\in\mathcal{V}_r}c_jZ_j
\end{equation}
where $\mathcal{V}_r$ is the set of vertices associated to the unfixed $c_k$'s, which are marked by a red cross \redcross. Each of this type contribution provides a factor of the form
\begin{equation}\label{eq:ckintf}
    \prod_{j=1}^L\frac{1}{c_j-y_{e_j}+i\varepsilon_{k_j}},
\end{equation}
where $\{c_j\,|\,j=1,\ldots,L\}$ is the set of unfixed integration variables. Interestingly, the green cross either of the two ends
of the other edges: the contributions associated to a different location of the green cross on the same edge have the same dependence on the local variables $x$'s and $y$'s but differs for the sign of the $y_e$ associated to the edge marked by the green cross. Consequently, the canonical function of the scattering facet acquires the following form:
\begin{equation}\label{eq:LorInvCF}
    \resizebox{0.9\hsize}{!}{$\displaystyle
        \Omega(\mathcal{S}_{\mathcal{G}})\:=\:
        \int\prod_{j=1}^L
        \left[
            \frac{dc_j}{\left(c_j-\frac{y_{e_j}}{2}\right)^2-\left(\frac{y_{e_j}}{2}-i\varepsilon_j\right)^2}
        \right]
        \prod_{s=1}^{n_e-L}
        \frac{1}{\left(\sum_{r_s}\sigma_{r_s}c_r-\frac{\mathfrak{y}_s}{2}\right)^2-
        \left(\frac{y_s}{2}-i\varepsilon_s\right)^2},
    $}
\end{equation}
with $\sigma_{r_s}$ and $\mathfrak{y}_s$ being respectively a suitable sign and a combination of the local coordinates $x$'s and $y$'s. 

Interpreting the integration variables $c_j$'s as the time component $\ell_{\circ}$ of the loop momenta, each quadratic factor above is a Lorentz invariant propagator. Hence, the canonical function \eqref{eq:LorInvCF}, together with the $d$-dimensional loop measure $d^d\ell^{\mbox{\tiny $(j)$}}$ constitutes the Lorentz invariant integrand for the graph $\mathcal{G}$.

It is important to stress that the contour integral representation of the scattering facet makes Lorentz invariance manifest, with the $i\varepsilon$ of the Lorentz invariant propagator inherited from the $i\varepsilon$ prescription in the definition of the contour integration.

Indeed, once we obtain a representation for the canonical form as in \eqref{eq:LorInvCF}, we can still perform the integration over the leftover $c_j$'s. As each of the $c_j$'s runs on $\mathbb{R}$, we can compute this integration by closing the contour of integration either in lower-half or upper-half $c_j$-plane. Given that we have $L$ of such variables, there are several ways in which we can perform the integration over all of the $c_j$'s. All such different ways of performing the $c_j$'s integration translates in several ways of triangulating the scattering facet -- this is a general statement for any polytope: if the polytope is not a simplex, the contour integral representation provides an analytic way of computing its triangulations ~\cite{Arkani-Hamed:2017tmz}~. Also, as the $c_j$'s are interpreted as the time component $\ell_{\circ}^{\mbox{\tiny $(j)$}}$ of the loop momenta, such integrations corresponds to integrating the $\ell_{\circ}^{\mbox{\tiny $(j)$}}$'s via contour integration using the $i\varepsilon_j$'s to pick up the relevant poles, as for the Feynman-tree theorem ~\cite{Feynman:1963ax, CaronHuot:2010zt}~. 


It is useful to illustrate both the emergence of Lorentz invariance as well as the correspondence between triangulations of the scattering facet and the $\ell_{\circ}^{\mbox{\tiny $(j)$}}$'s with a simple example.


\subsubsection{An illustrative and visualisable example: the $2$-site $2$-loop graph}\label{subsubsec:2s2lG}

\begin{figure}[t]
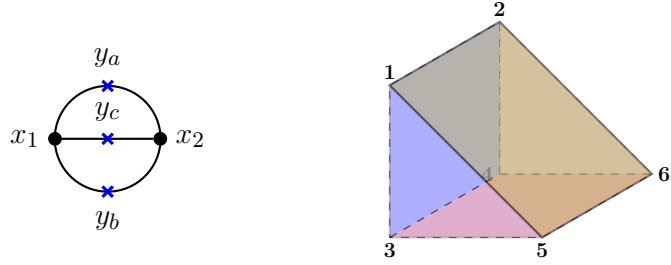

    \centering

    \caption{Scattering facet associated to the $2$-site $2$-loop graph.}
    \label{fig:SF2s2l}
\end{figure}

Let us consider the scattering facet associated to the $2$-site $2$-loop graph, which turns out to be a prism in $\mathbb{P}^3$ identified as the convex hull of the following vertices -- see Figure \ref{fig:SF2s2l}:
\begin{equation}
    \begin{split}
        \{
            &{\bf x}_1+{\bf y}_{a}-{\bf x}_2,\;-{\bf x}_1+{\bf y}_{a}+{\bf x}_2,\\
            &{\bf x}_1+{\bf y}_{c}-{\bf x}_2,\;-{\bf x}_1+{\bf y}_{c}+{\bf x}_2,\\
            &{\bf x}_1+{\bf y}_{b}-{\bf x}_2,\;-{\bf x}_1+{\bf y}_{b}+{\bf x}_2
        \}    
    \end{split}
\end{equation}

Let us assign the labels $\mathbf{1}$, $\mathbf{2}$, $\mathbf{3}$, $\mathbf{4}$, $\mathbf{5}$, $\mathbf{6}$ to the vertices of the scattering facet, following the ordering in the list above. Then, the contour integral representation for the canonical function is given by
\begin{equation}\label{eq:2s2lcf}
    \Omega(\mathcal{S}_{\mathcal{G}})\:\sim\:\int_{\mathbb{R}^6}\prod_{j=1}^6\frac{dc_j}{c_j-i\varepsilon_j}\,
    \delta^{\mbox{\tiny $(4)$}}\left(\mathcal{Y}-\sum_{j=1}^6 c_jZ_j\right),
\end{equation}
where the $Z_j$ are the vectors of $\mathbb{R}^4$ representing the $j$-th vertex, according to the label assignment above, while $\sim$ indicates that we will be sloppy with the overall factor. Hence, localising the variables $\{c_k\,|\,k=2,\ldots,5\}$ we obtain:
\begin{equation}\label{eq:2s2lcf2}
    \Omega(\mathcal{S}_{\mathcal{G}})\:\sim\:\int_{\mathbb{R}^2}\frac{dc_6}{c_6-i\varepsilon_6}\frac{dc_1}{c_1-i\varepsilon_1}\,
    \frac{1}{\langle3452\rangle}\prod_{k=2}^6\frac{1}{c_k^{\star}(c_6,c_1)-i\varepsilon_k}
\end{equation}
where the $\{c_k^{\star}(c_6,c_1)\,|\,k=2,\ldots,5\}$ are the solutions of the $\delta$-functions and they depend on the integration variables $(c_6,\,c_1)$, while $\langle3452\rangle$ is the Jacobian coming from solving the $\delta$-functions.

The $\delta$-function fixes $4$ of the $6$ integration variables, leaving two of them to be unfixed as predicted by our counting in the general proof presented above. We need to decide which variables we want to localise via the $\delta$-function and which of them leaving free. Let us choose the latter to be the $c_1$ and $c_6$. It is convenient to assign the marking \redcross~ to the two vertices $Z_1$ and $Z_6$ associated to them -- see Figure \ref{fig:LorInv}. Then, the hyperplane corresponding to the solution $c_k^{\star}(c_6,c_1)$ is spanned by the vertices $\{Z_m\,|\,m=2,\ldots,5,\:m\,\neq\,k\}$:
\begin{equation}\label{eq:cksolexp}
    c_k^{\star}(c_6,c_1)\:=\:\tilde{\mathcal{Y}}^I\mathcal{W}^{\mbox{\tiny $(\mathfrak{g}_k)$}}_I,\qquad
    \left\{
        \begin{array}l
             \mathcal{W}^{\mbox{\tiny $(\mathfrak{g}_k)$}}_I\,:=\,\epsilon_{\mbox{\tiny $IJKL$}}Z_{k+1}^J Z_{k+2}^K Z_{k-1}^L.\\
             \phantom{\ldots}\\
             \tilde{\mathcal{Y}}\,:=\,\mathcal{Y}-c_6 Z_6-c_1 Z_1
        \end{array}
    \right.
\end{equation}
which precisely can be obtained graphically by associating a \greencross~ to the vertex $Z_k$ whose associated $c_k$ we need to solution of, and considering the graph $\mathfrak{g}_k$ enclosed by the red/green crosses -- see Figure \ref{fig:LorInv}, where the solution for $k=5$ is shown. In a completely explicit form:
\begin{equation}\label{eq:cksolexp2}
    \left\{
        \begin{array}{l}
            \displaystyle
            c_2\:=\:\frac{\langle\mathcal{Y}345\rangle}{\langle2345\rangle}-
                    c_1\frac{\langle1345\rangle}{\langle2345\rangle}-
                    c_6\frac{\langle6345\rangle}{\langle2345\rangle},
            \quad
            c_3\:=\:\frac{\langle\mathcal{Y}452\rangle}{\langle3452\rangle}-
                    c_1\frac{\langle1452\rangle}{\langle3452\rangle}-
                    c_6\frac{\langle6452\rangle}{\langle3452\rangle},\\
            \phantom{\ldots}\\
            \displaystyle
            c_4\:=\:\frac{\langle\mathcal{Y}523\rangle}{\langle4523\rangle}-
                    c_1\frac{\langle1523\rangle}{\langle4523\rangle}-
                    c_6\frac{\langle6523\rangle}{\langle4523\rangle},
            \quad
            c_5\:=\:\frac{\langle\mathcal{Y}234\rangle}{\langle5234\rangle}-
                    c_1\frac{\langle1234\rangle}{\langle5234\rangle}-
                    c_6\frac{\langle6234\rangle}{\langle5234\rangle}.
        \end{array}
    \right.
    \\
\end{equation}
Notice the following relations among the vertices
\begin{equation}\label{eq:2s2lZrels}
    Z_1\,+\,Z_4\:\sim\:Z_2\,+\,Z_4,\qquad
    Z_3\,+\,Z_6\:\sim\:Z_4\,+\,Z_5    
\end{equation}
which imply that $c_2$ and $c_5$ only depend on $c_1$ and $c_6$ respectively. Hence, the canonical form can be written explicitly in terms of the integration variables $c_1$ and $c_6$ as well as the kinematics, as
\begin{equation}\label{eq:2s2lcf3}
    \begin{split}
        \Omega(\mathcal{S}_{\mathcal{G}})\:&\sim\:
        \int_{\mathbb{R}^2}\frac{dc_6}{c_6-i\varepsilon_6}\,\frac{dc_1}{c_1-i\varepsilon_1}\,
        \frac{1}{(c_1-y_a+i\varepsilon_2)(c_6-c_1+\frac{y_c+y_a-y_b+x_1}{2}-i\varepsilon_3)}\times\\
        &\hspace{.5cm}\times\,\frac{1}{(c_1-c_6+\frac{y_c-y_a+y_b-x_1}{2}-i\varepsilon_4)(c_6-y_b+i\varepsilon_5)}\:=\\
        &=\:\int_{\mathbb{R}^2}\frac{dc_6}{(c_6-\frac{y_b}{2})^2-(\frac{y_b}{2}-i\varepsilon_6)^2}\,
        \frac{dc_1}{(c_1-\frac{y_a}{2})^2-(\frac{y_a}{2}-i\varepsilon_1)^2}\,\times\\
        &\hspace{.5cm}\times\,\frac{1}{(c_6-c_1+\frac{y_a-y_b+x_1}{2})^2-(\frac{y_c}{2}-i\varepsilon_{34})^2},
    \end{split}
\end{equation}
where in the second equality the poles related to $(c_5,\,c_6)$, $(c_1,\,c_2)$ and $(c_3,\,c_3)$, upon the identifications $\varepsilon_5\,=\,\varepsilon_6$, $\varepsilon_1\,=\,\varepsilon_2$, $\varepsilon_3\,=\,\varepsilon_4\,\equiv\,\varepsilon_{34}$ and interpreting $c_6$ and $c_1$ as the time component of the loop momenta, pair up to form the Lorentz invariant propagators.

\begin{figure}[t]
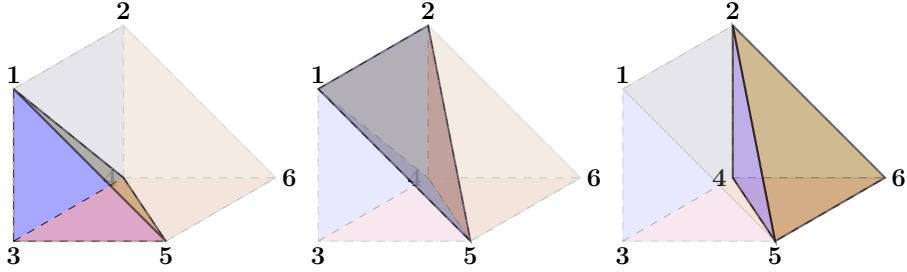

    \centering

    \caption{One possible triangulation of the scattering facet associated to the $2$-site $2$-loop graph. It corresponds in integrating both the $\ell_{\circ}$'s in the upper-half plane.}
    \label{fig:SFtr2s2l}
\end{figure}

We can think to integrate over the leftover variables $c_6$ and $c_1$. Each of these integrations can be performed by closing the contour of integration either in the lower-half or upper-half $c_j$-plane. Let us perform first the integration over $c_6$. In such a variable, the integrand has $4$ poles and the $i\varepsilon_j$-prescription splits them by placing two of them in the lower-half and the other two in the upper-half plane. Notice that, when this integration is performed, the new integrand in the $c_1$ variable can show differences between the $\varepsilon$'s. Then, in principle, there is the issue of which sign for them need to be chosen. It can be shown that both choices return the same result ~\cite{Arkani-Hamed:2017tmz}~. Performing both the integrations in the upper-half plane, we obtain the following representation for $\Omega(\mathcal{S}_{\mathcal{G}})$ and, hence, for the amplitude integrand:
\begin{equation}\label{eq:CF45}
   \begin{split}
    \Omega(\mathcal{S}_{\mathcal{G}})\:&=\:
                \frac{1}{2y_a}\frac{1}{2y_{b}}\frac{1}{y_c^2-(y_a+y_b-x_1)^2}\:+\:
                \frac{1}{2y_b}\frac{1}{2y_c}\frac{1}{y_a^2-(y_b-y_c-x_1)^2}\:+\\
            &+\:\frac{1}{2y_c}\frac{1}{2y_a}\frac{1}{y_b^2-(y_c+y_a+x_1)^2},
   \end{split}
\end{equation}
which corresponds to the triangulation of the scattering facet shown in Figure \ref{fig:SFtr2s2l}.


\subsection{Emergence of flat-space unitarity}\label{subsec:Un}

The avatar of flat-space unitarity in scattering amplitudes is given by the optical theorem which relates the discontinuities along the amplitude branch cuts to its factorisation in lower-loop scattering amplitudes. As the scattering facet $\mathcal{S}_{\mathcal{G}}\:=\mathcal{P}_{\mathcal{G}}\cap\mathcal{W}^{\mbox{\tiny $(\mathcal{G})$}}$ is given by the intersection of the cosmological polytope $\mathcal{P}_{\mathcal{G}}$ with the hyperplane $\mathcal{W}^{\mbox{\tiny $(\mathcal{G})$}}$ identified by the subgraph $\mathfrak{g}=\mathcal{G}$, looking at the discontinuity along branch cuts then means looking at the codimension-$1$ face structure of the scattering facets, {\it i.e.} at the non-vanishing intersections $\mathcal{P}_{\mathcal{G}}\cap\mathcal{W}^{\mbox{\tiny $(\mathcal{G})$}}\cap\mathcal{W}^{\mbox{\tiny $(\mathfrak{g})$}}\,\neq\,\varnothing$, with $\mathcal{W}^{\mbox{\tiny $(\mathfrak{g})$}}$ being the hyperplane identified by any subgraph $\mathfrak{g}\subset\mathcal{G}$ such that the number $n_{\mathfrak{g}}$ of edges departing from $\mathfrak{g}$ is less or equal $L+1$, where $L$ is the number of loops of $\mathcal{G}$.

These faces are polytopes defined as the convex hull of the vertices $\{\mathbf{x}_j+\mathbf{y}_{ij}-\mathbf{x}_i\,|\,x_i\,\notin\,\mathfrak{g}\:\vee\:x_j\,\in\,\mathfrak{g}\}$, which graphically correspond to adding the marking \bluecross\ on $\mathcal{S}_{\mathcal{G}}$ closest to $\mathfrak{g}$ on the edges departing from $\mathfrak{g}$, {\it i.e.} its vertices are obtained from the ones of $\mathcal{S}_{\mathcal{G}}$ by eliminating the ones corresponding to the additional marking.

\begin{figure}[t]
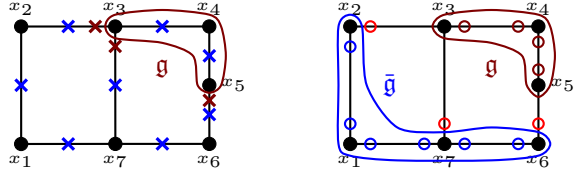

\centering

\caption{Codimension-one faces of the scattering facet and individual cuts of the scattering amplitude. 
On the left, the marked vertices \bluecross\ do not appear on the scattering facet $\mathcal{S}_{\mathcal{G}}$, while the ones marked by \purplecross\ get eliminated on $\mathcal{S}_{\mathcal{G}}\cap\mathcal{W}^{\mbox{\tiny $(\mathfrak{g})$}}$. On the right, the markings depict the vertices which {\it do} contribute to $\mathcal{S}_{\mathcal{G}} \cap \mathcal{W}_{\mathfrak{g}}$: \purplecircle\ and \bluecircle\ are respectively associated with the scattering facets $\mathcal{S}_{\mathfrak{g}}$ and $\mathcal{S}_{\bar{\mathfrak{g}}}$. The remaining vertices, marked by \redcircle, are associated with the simplex $\Sigma_{\centernot{\mathcal{E}}}$.}
\label{fig:individual_cut}
\end{figure}

On this face, the canonical function manifestly factorises into a pair of lower-dimensional scattering facets and a simplex, which encodes the Lorentz-invariant phase-space measure of the cut propagators:
\begin{equation}\label{eq:CutR}
    \Omega(\mathcal{Y},\,\mathcal{S}_{\mathcal{G}}\cap\mathcal{W}^{\mbox{\tiny $(g)$}})\:=\:
    \Omega(\mathcal{Y}_{\mathfrak{g}},\,\mathcal{S}_{\mathfrak{g}})\times
    \Omega(\mathcal{Y}_{\mathfrak{\bar{g}}},\,\mathcal{S}_{\mathfrak{\bar{g}}})\times
    \Omega(\mathcal{Y}_{\centernot{\mathcal{E}}},\,\Sigma_{\centernot{\mathcal{E}}})
\end{equation}
where $\mathcal{S}_{\mathfrak{g}}$ and $\mathcal{S}_{\mathfrak{\bar{g}}}$ are the lower-dimensional scattering facets associated to $\mathfrak{g}$ and $\bar{\mathfrak{g}}$ respectively, where $\bar{\mathfrak{g}}$ is the complement of $\mathfrak{g}$, and $\Sigma_{\centernot{\mathcal{E}}}$ is the simplex identified by the vertices associated to the cut edges $\centernot{\mathcal{E}}$ ~\cite{Arkani-Hamed:2018ahb}~. Interestingly, the vertex configuration of $\Sigma_{\centernot{\mathcal{E}}}$ encodes the energy flow along the cut edges $\centernot{\mathcal{E}}$, whose direction depends on whether the marking \redcircle\ is close to $\mathfrak{g}$ or $\bar{\mathfrak{g}}$.

The factorisation \eqref{eq:CutR} returns nothing but the cutting rules, and can be more explicitly written as
\begin{equation}\label{eq:CutRexp}
    \Omega(\mathcal{Y},\,\mathcal{S}_{\mathcal{G}}\cap\mathcal{W}^{\mbox{\tiny $(g)$}})\:=\:
    \left(\prod_{e\in\centernot{\mathcal{E}}}\frac{1}{2y_e}\right)
    \mathcal{A}[\mathfrak{g}]\times\mathcal{A}[\bar{\mathfrak{g}}].
\end{equation}
These cutting rules are a consequence of the optical theorem, which is the imprint of unitarity on scattering amplitudes
\begin{equation}\label{eq:FSOT}
    \begin{split}
        -i\langle\{p\}_{\mbox{\tiny out}}|\hat{T}-\hat{T}^{\dagger}|\{p\}_{\mbox{\tiny in}}\rangle\:&=\:
        \langle\{p\}_{\mbox{\tiny out}}|\hat{T}^{\dagger}\hat{T}|\{p\}_{\mbox{\tiny in}}\rangle\:=\\
        &=\:\sum_{I}\hspace{-.5cm}\int\:\langle\{p\}_{\mbox{\tiny out}}|\hat{T}^{\dagger}|I\rangle
        \langle I|\hat{T}|\{p\}_{\mbox{\tiny in}}\rangle,
    \end{split}
\end{equation}
Considering the energy propagators with the suitable Feynman $i\varepsilon$-prescription, that its imaginary part yields an energy delta-functions
$$
\frac{1}{\pi}\mbox{Im}\left\{\frac{1}{E-i\varepsilon}\right\}\:=\:\delta(E)
$$
and consequently each residue of the scattering amplitude integrand returns a contribution to the imaginary part of the amplitude itself. As a parenthetical remark, in general we are more accustomed to thinking about the cutting rules in terms of Lorentz-invariant propagators. However, notice that they are quadratic in the energies -- one can write $(l-P)^2$ as $(l-P)^2\,=\,(l_{\circ}-P_{\circ})^2-(\vec{l}-\vec{P})^2$ -- and the residue of such propagator has two solutions, the positive and the negative energy one, despite the fact that cutting one propagator is understood to select the positive energy solution. In any case, in the physical region
\begin{equation}
    \frac{1}{\pi}\mbox{Im}\left\{\frac{1}{(l-P)^2-i\varepsilon}\right\}\:=\:
        \delta^{\mbox{\tiny $(+)$}}((l-P)^2)\:=\:\frac{1}{2y}\delta(l_{\circ}-P_{\circ}-y),
\end{equation}
where the superscript ${}^{(+)}$ in the delta-function for the cut-propagator indicates that the positive energy solution is taken\footnote{Otherwise, strictly speaking one should sum over both positive and energy solutions:
$$
    \delta((l-P)^2)\:=\:\frac{1}{2y}\left[\delta(l_{\circ}-P_{\circ}-y)+\delta(l_{\circ}-P_{\circ}+y)\right],
$$
where, as in the text, $y\,=\,|\vec{l}-\vec{P}|$.}, $y\,:=\,|\vec{l}-\vec{P}|$ and the factor $(2y)^{-1}$ precisely contributes to the Lorentz-invariant phase-space measure.


\subsection{Emergence of flat-space causality}\label{subsec:FsCaus}

As discussed in Section \ref{subsec:Caus}, one of the avatars of causality in the scattering amplitudes is provided by the Steinmann relations, which state that its double discontinuities across partially overlapping channels ought to vanish in the physical region -- see Figure \ref{fig:SteimnRels} for a pictorial representation:
\begin{equation}\label{eq:SRfs}
    \mbox{Disc}_{s_{\mathcal{I}}}\left(\mbox{Disc}_{s_{\mathcal{J}}}\mathcal{A}\right)\:=\:0,\qquad
    \left\{
        \begin{array}{l}
            \mathcal{I}\nsubseteq\mathcal{J},\\
            \mathcal{J}\nsubseteq\mathcal{I},\\
            \mathcal{I}\cap\mathcal{J}\,\neq\,\varnothing.
        \end{array}
    \right.
\end{equation}
where $\mathcal{A}$ is the flat-space amplitudes, $\mathcal{I},\,\mathcal{J}\,\subset\,\{1,\ldots,n\}$ label a subset of the external particles, and $s_{\mathcal{I}},\,s_{\mathcal{J}}$ are the Mandelstam invariants constructed out of the momenta in the sets $\mathcal{I},\,\mathcal{J}$.

Combinatorially, looking at double discontinuities across partially overlapping channels translates into the analysis of codimension-$2$ faces of the scattering facet related to pair of graphs $\mathfrak{g}_1$ and $\mathfrak{g}_2$ such that 
$\mathfrak{g}_1\cap\mathfrak{g}_2\,\neq\,\varnothing$, $\mathfrak{g}_1\cap\bar{\mathfrak{g}}_2\,\neq\,\varnothing$, $\bar{\mathfrak{g}}_1\cap\mathfrak{g}_2\,\neq\,\varnothing$, and
$\bar{\mathfrak{g}}_1\cap\bar{\mathfrak{g}}_2\,\neq\,\varnothing$. Codimension-$2$ faces can be thought of as the intersections of the hyperplanes $\mathcal{W}^{\mbox{\tiny $(\mathfrak{g}_1)$}}$ and $\mathcal{W}^{\mbox{\tiny $(\mathfrak{g}_2)$}}$, identified by the graphs $\mathfrak{g}_1$ and $\mathfrak{g}_2$, with the scattering facet $\mathcal{S}_{\mathcal{G}}$. Hence, the vanishing of double discontinuities across partially overlapping channels translates into the intersection $\mathcal{S}_{\mathcal{G}}\cap\mathcal{W}^{\mbox{\tiny $(\mathfrak{g}_1)$}}\cap\mathcal{W}^{\mbox{\tiny $(\mathfrak{g}_2)$}}$ being empty or, which is the same, that the two hyperplanes intersect each other in codimension-$2$ outside the scattering facet. 

Notice that the analysis of the codimension-$2$ faces of the scattering facet is completely analogous to the one performed in Section \ref{subsec:FShc} for the higher codimension faces of the cosmological polytope. Hence, one should check whether the dimension of $\mathcal{S}_{\mathcal{G}}\cap\mathfrak{g}_1\cap\mathfrak{g}_2$, which can be computed from the dimension of the lower-dimensional scattering facets in which it should factorise, is $n_s+n_e-4$, which is the expected dimension for a non-empty codimension-$2$ face of $S_{\mathcal{G}}$.

\begin{figure}
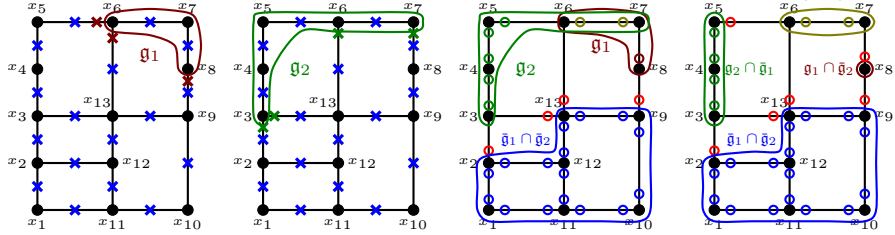

 \centering
 \vspace{-.5cm}

\caption{Pair of subgraphs corresponding to partially overlapping momentum channels. The first two pictures represent their separate realisation as codimension-$1$ faces on the scattering facet, while the last two depict their putative simultaneous intersection with the scattering facet. The dimensional counting shows that such an intersection is actually empty and in codimension-$2$ occurs outside of the scattering facet.}
\label{fig:SRfs}
\end{figure}

In particular, were it to be non-empty, $\mathcal{S}_{\mathcal{G}}\cap\mathcal{W}^{\mbox{\tiny $(\mathfrak{g}_1)$}}\cap\mathcal{W}^{\mbox{\tiny $(\mathfrak{g}_2)$}}$ should factorise into four lower-dimensional scattering facets, which are identified by the intersection among the subgraphs $\mathfrak{g}_j$ and their complements, as well as a simplex $\Sigma_{\centernot{\mathcal{E}}}$ defined as the convex hull of the vertices related to the cut edges $\centernot{\mathcal{E}}$. As the dimension of the scattering facet related to a graph $\mathfrak{g}$ is $n_s^{\mbox{\tiny $(\mathfrak{g})$}}+n_e^{\mbox{\tiny $(\mathfrak{g})$}}-2$ and the dimension of $\Sigma_{\centernot{\mathcal{E}}}$ is $n_{\centernot{\mathcal{E}}}-1$ being a simplex, then
\begin{equation}\label{eq:SRsfdim}
    \begin{split}
        \mbox{dim}
        \left(
            \mathcal{S}_{\mathcal{G}}\cap\mathcal{W}^{\mbox{\tiny $(\mathfrak{g}_1)$}}\cap\mathcal{W}^{\mbox{\tiny $(\mathfrak{g}_2)$}}
        \right)\:&=\:
        \sum_{\mathcal{S}_{\mathfrak{g}}}(n_s^{\mbox{\tiny $(\mathfrak{g})$}}+n_e^{\mbox{\tiny $(\mathfrak{g})$}}-1)+
        n_{\centernot{\mathcal{E}}}-1\:=\\
        &=\:n_s+n_e-1-\sum_{\mathcal{S}_{\mathfrak{g}}}1
    \end{split}
\end{equation}
where the sum runs over the lower-dimensional scattering facets. The last line in \eqref{eq:SRsfdim} is obtained by just considering that the sum over the lower-dimensional scattering facets at graph level covers all the sites of the graph $\mathcal{G}$, while such a sum for the edges together with the cut edges cover all the edges of $\mathcal{G}$. Given that there are four lower-dimensional scattering facet,  \eqref{eq:SRsfdim} yields $n_s+n_e-5$ rather than $n_s+n_e-4$ for partially overlapping channels. Hence $\mathcal{S}_{\mathcal{G}}\cap\mathcal{W}^{\mbox{\tiny $(\mathfrak{g}_1)$}}\cap\mathcal{W}^{\mbox{\tiny $(\mathfrak{g}_2)$}}$ is empty, or, which is the same, the intersection $\mathcal{W}^{\mbox{\tiny $(\mathfrak{g}_1)$}}\cap\mathcal{W}^{\mbox{\tiny $(\mathfrak{g}_2)$}}$ in codimension-$2$ occurs outside the scattering facet $\mathcal{S}_{\mathcal{G}}$.

It is important to emphasise two essential points:
{
\renewcommand{\theenumi}{\roman{enumi}}
\begin{enumerate}
    \item The argument above provides a combinatorial proof for the Steinmann relations in flat-space and holds for any graph and any number of external states. It is well known that there exist non-vanishing double discontinuities for partially overlapping channels, as for example in the box graph ~\cite{Stapp:1971hh}~, {\it outside the physical region of the scattering amplitudes} and it is not visible in Lorentz signature. Being the cosmological polytope intrinsically Lorentzian, it does not allow to access the region where the contribution appears. So, there is no contradiction with the existence of such a contribution neither for Steinmann relations as originally proven for amplitudes ~\cite{Stapp:1971hh}~ nor for the proof highlighted above ~\cite{Benincasa:2020uph}~.
    \item The flat-space Steinmann relations are enforced by the combinatorial structure of the entire cosmological polytope and are produced by the same mechanism which induces the Steinmann-like relations for the wavefunction. This is the sense in which flat-space causality emerges from the cosmological setting.
\end{enumerate}
}
%


\subsection{Physical representations for scattering amplitudes}\label{subsec:Nreps}

As the discontinuities across partially overlapping channels provide constraints on the analytic structure of scattering amplitudes, we can ask whether we can use them to bootstrap the amplitudes themselves or if it is possible to formulate novel constraints which can allow us to do it.

In our combinatorial picture, the statement of the Steinmann relation rephrases into the constrain that the intersection of the hyperplanes identified by partially overlapping subgraphs occur outside the scattering facet. Also we also saw that all the intersections of the hyperplanes identified by the subgraphs occurring outside a polytope -- in our current case, the scattering facet -- form a locus which determines the numerator of the canonical form. The analysis of the higher-codimension face structure discussed for the cosmological polytopes in Section \ref{subsec:FShc}, can be carried out also for the scattering facet and obtain the conditions which fixes the locus $\mathcal{C}(\mathcal{S}_{\mathcal{G}})$ of the intersections of the faces outside the scattering faces, {\it i.e.} the vanishing conditions on the multiple residues which determine the zeroes of the canonical form $\omega(\mathcal{Y},\,\mathcal{S}_{\mathcal{G}})$ and hence its numerator.

\begin{figure}
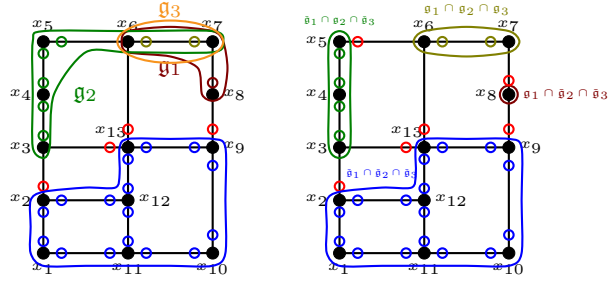

 \centering
 \vspace{-.25cm}

\caption{Intersection of the three facets on $\mathcal{S}_{\mathcal{G}}$. This face factorises in four lower dimensional scattering facets, whose vertices are depicted by the markings
\circle{red!50!black},
\circle{red!50!green},
\circle{green!50!black} and
\circle{blue},
and a simplex $\Sigma_{\centernot{\mathcal{E}}}$ identified by 
\redcircle{}.
}
 \label{fig:ldintsg2}
\end{figure}

Importantly, in the case of the scattering facet $\mathcal{S}_{\mathcal{G}}$, given a graph $\mathcal{G}$ with $L$ loops and the codimension-$k$ hyperplane $\mathcal{W}^{\mbox{\tiny $(\mathfrak{g}_1\ldots\mathfrak{g}_k)$}}$ given by the intersection of the hyperplanes $\{\mathcal{W}^{\mbox{\tiny $(\mathfrak{g}_j)$}}\,|\,j=1,\ldots,k\}$ identified by the subgraphs $\{\mathfrak{g}_j\,|\,j=1,\ldots,k\}$, the analysis of whether $\mathcal{S}_{\mathcal{G}}\cap\mathcal{W}^{\mbox{\tiny $(\mathfrak{g}_1\ldots\mathfrak{g}_k)$}}$ is empty in codimension-$k$ involves only subgraphs $\mathfrak{g}_j\subset\mathcal{G}$ such that the number $n_{\mathfrak{g}_j}$ of cut edges satisfies the condition $n_{\mathfrak{g}_j}\,>\,L+1$: for all the other subgraphs $\mathfrak{g}_m$, $\mathcal{S}_{\mathcal{G}}\cap\mathcal{W}^{\mbox{\tiny $(\mathfrak{g}_m)$}}\,=\,\varnothing$ and hence they do not identify a boundary of the scattering facet, {\it i.e.} the scattering amplitude does not have a singularity associated to this class of subgraphs.

Following the analysis discussed in Section \ref{subsec:FShc}, this questions reduces to a dimension counting of the lower dimensional polytopes which $\mathcal{S}_{\mathcal{G}}\cap\mathcal{W}^{\mbox{\tiny $(\mathfrak{g}_1\ldots\mathfrak{g}_k)$}}$ should factorise into:
\begin{equation}\label{eq:HCBSF}
    \mbox{dim}\left(\mathcal{S}_{\mathcal{G}}\cap\mathcal{W}^{\mbox{\tiny $(\mathfrak{g}_1\ldots\mathfrak{g}_k)$}}\right)\:=\:
        \sum_{\mathcal{S}_{\mathfrak{g}}}\left(n_s^{\mbox{\tiny $(\mathfrak{g})$}}+n_e^{\mbox{\tiny $(\mathfrak{g})$}}-1\right)+n_{\centernot{\mathcal{E}}}-1
\end{equation}
where the sum runs over the lower-dimensional scattering facets as in the above discussion for the flat-space Steinmann relations. Then, $\mathcal{S}_{\mathcal{G}}\cap\mathcal{W}^{\mbox{\tiny $(\mathfrak{g}_1\ldots\mathfrak{g}_k)$}}\,=\,\varnothing$ if $\mbox{dim}\left(\mathcal{S}_{\mathcal{G}}\cap\mathcal{W}^{\mbox{\tiny $(\mathfrak{g}_1\ldots\mathfrak{g}_k)$}}\right)\,<\,n_s+n_e-2-k$. Introducing $\centernot{n}_{\centernot{\mathcal{E}}}$ as the number of edges of $\mathcal{G}$ completely marked on this intersection, {\it i.e.} no vertex associated to these edges is on the intersection, then this condition can be rewritten in a simpler form as
\begin{equation}\label{eq:HCBSF2}
    \sum_{\mathcal{S}_{\mathfrak{g}}}1+\centernot{n}_{\centernot{\mathcal{E}}}\:>\:k+1
\end{equation}
and allow us to identify all the intersection of the hyperplanes containing the facets of $\mathcal{S}_{\mathcal{G}}$ and defining the locus $\mathcal{C}(\mathcal{S}_{\mathcal{G}})$ of the zeroes of the canonical form $\omega(\mathcal{Y},\,\mathcal{S}_{\mathcal{G}})$.

Now, we can also identify those subspaces of $\mathcal{C}(\mathcal{S}_{\mathcal{G}})$ which allow us to signed-triangulate the scattering facet through just one of them without having to introduce new boundaries. As each boundary corresponds to a singularity in the scattering amplitude, this implies that no spurious singularity is introduced and all these signed triangulations return representations of the scattering amplitudes without spurious singularities:
\begin{equation}\label{eq:PRSG}
    \omega(\mathcal{Y},\,\mathcal{S}_{\mathcal{G}})\:=\:
        \sum_{\{\mathfrak{G}_c\}}\prod_{\mathfrak{g}'\in\mathfrak{G}_c}\frac{1}{q_{\mathfrak{g}'}(\mathcal{Y})}\,
        \frac{\langle\mathcal{Y}d^{n_s+n_e-2}\mathcal{Y}\rangle}{\displaystyle\prod_{\mathfrak{g}\in\mathfrak{G}_{\circ}}q_{\mathfrak{g}}(\mathcal{Y})},
\end{equation}
where $\mathfrak{G}_{\circ}$ identifies a subset of graphs such that together they satisfy the condition \eqref{eq:HCBSF2} and define a subspace of $\mathcal{C}(\mathcal{S}_{\mathcal{G}})$ through which we can sign-triangulate $\mathcal{S}_{\mathcal{G}}$, $\mathfrak{G}_{c}$ indicates a set of $n_s+n_e-k-1$ subgraphs which do not belong to $\mathfrak{G}_{\circ}$ and identify a face of $\mathcal{S}_{\mathcal{G}}$, and $q_{\mathfrak{g}}(\mathcal{Y})\,:=\,\mathcal{Y}^I\mathcal{W}^{\mbox{\tiny $(\mathfrak{g})$}}_I$. Importantly, $\mathcal{W}^{\mbox{\tiny $(\mathfrak{G}_{\circ})$}}\, :=\,\bigcap_{\mathfrak{g}\in\mathfrak{G}_{\circ}}\mathcal{W}^{\mbox{\tiny $(\mathfrak{g})$}}$ and $\mathcal{W}^{\mbox{\tiny $(\mathfrak{G}_{c})$}}\, :=\,\bigcap_{\mathfrak{g}\in\mathfrak{G}_{c}}\mathcal{W}^{\mbox{\tiny $(\mathfrak{g})$}}$ are such that $\mathcal{S}_{\mathcal{G}}\cap\mathcal{W}^{\mbox{\tiny $(\mathfrak{G}_{\circ})$}}\cap\mathcal{W}^{\mbox{\tiny $(\mathfrak{G}_{c})$}}\,\subset\,\mathbb{P}^0$ and it is non-empty.

\begin{figure}
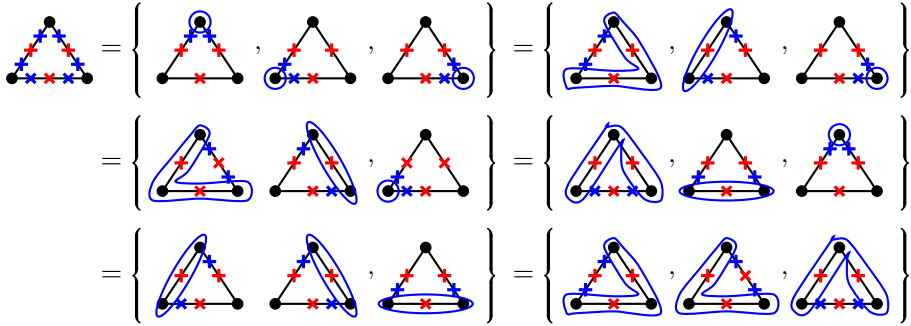

    \centering
    \vspace{-.25cm}

    \caption{Higher codimension intersections of the facets outside $\mathcal{S}_{\mathcal{G}}$. The markings $\redcross$ and $\bluecross$ respectively identifies $\mathcal{S}_{\mathcal{G}}$ and the intersections of the hyperplane identified by a subgraph.}
    \label{fig:ZerosSg}
\end{figure}

Notice that each of the sets $\mathfrak{G}_c$ singles out a subset of compatible channels in codimension-$(n_s+n_e-k-1)$, {\it i.e.} channels such that the multiple residue of the scattering amplitude along them is non-zero, and the sum over $\{\mathfrak{G}_c\}$ sums over all the compatible singularities in codimension-$(n_s+n_e-k-1)$ which can be identified by the subgraphs which are not contained in the particular zero, given by the chosen $\mathfrak{G}_{\circ}$. So these representations make manifest a subset of the (in)compatibility conditions for the channels.

This picture provides the set of all those representations with physical singularities only, most of which are not known. Irrespectively of their quite different appearance, all of them fall under the same hat of the (in)compatibility conditions on the singularities of the scattering amplitudes and the signed-triangulations of the scattering facet we discussed makes this feature manifest. This picture also allows to readily recognise the old-fashioned perturbation theory and the causal representations, providing a proof of the latter ~\cite{Benincasa:2021qcb}~. Old-fashioned perturbation theory is obtained by choosing $\mathfrak{G}_{\circ}$ to be the set of all the subgraphs $\mathfrak{g}_s$ given by just a site $s$ of $\mathcal{G}$ -- see Figure \ref{fig:ZerosSg}:
\begin{equation}\label{eq:OFPTSg}
    \omega(\mathcal{Y},\,\mathcal{S}_{\mathcal{G}})\:=\:
        \sum_{\{\mathfrak{G}_c\}}\prod_{\mathfrak{g}'\in\mathfrak{G}_c}\frac{1}{q_{\mathfrak{g}'}(\mathcal{Y})}\,
        \frac{\langle\mathcal{Y}d^{n_s+n_e-2}\mathcal{Y}\rangle}{\displaystyle\prod_{s\in\mathcal{V}}q_{\mathfrak{g}_s}(\mathcal{Y})}
\end{equation}
where $\mathcal{V}$ is the set of all the sites of $\mathcal{G}$. Indeed \eqref{eq:OFPTSg} has the very same form of \eqref{eq:PRSG}, with $\mathfrak{G}_{\circ}\,=\,\{\mathfrak{g}_s\,|\,s\in\mathcal{V}\}$. Notice that the decomposition depends only on the edges. So, given all the graphs with the same number of edges but which can be differ for the number of sites, the structure \eqref{eq:OFPTSg} is unchanged.

The causal representation is instead picked if we choose $\mathfrak{G}_{\circ}$ to be the set of those subgraphs $\mathfrak{g}_e$ containing all the sites as well as all the edges of $\mathcal{G}$ but one, which we label by $e$ -- see Figure \ref{fig:ZerosSg}:
\begin{equation}\label{eq:CRSg}
   \omega(\mathcal{Y},\,\mathcal{S}_{\mathcal{G}})\:=\:
        \sum_{\{\mathfrak{G}_c\}}\prod_{\mathfrak{g}'\in\mathfrak{G}_c}\frac{1}{q_{\mathfrak{g}'}(\mathcal{Y})}\,
        \frac{\langle\mathcal{Y}d^{n_s+n_e-2}\mathcal{Y}\rangle}{\displaystyle\prod_{e\in\mathcal{E}}q_{\mathfrak{g}_e}(\mathcal{Y})},
\end{equation}
which again has the very same form of \eqref{eq:PRSG}, with $\mathfrak{G}_{\circ}\,=\,\{\mathfrak{g}_e\,|\,e\in\mathcal{E}\}$. The factor $(\prod_{e\in\mathcal{E}}q_{\mathfrak{g}_e}(\mathcal{Y}))^{-1}\,=\,(2y_e)^{-1}$ represents the measure of the Lorentz-invariant phase space, so \eqref{eq:CRSg} makes manifest the Steinmann-relations and their generalisation to higher codimensions. 

\begin{figure}
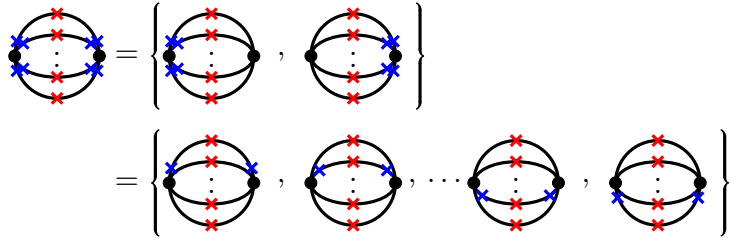

    \centering
    \vspace{-.25cm}

    \caption{Zeroes which identifies the old-fashioned perturbation theory and the causal representations for the $2$-site $L$-loop graph.}
    \label{fig:ZerosLl2sA}
\end{figure}

An interesting feature of this representation is that its structure does not depend on the number of edges, whose dependence is all encoded in the prefactor: given all the connected graphs with the same number of sites but which differ for the number of edges, then the causal representation of the related contribution to the amplitude is constituted by the very same number of terms with the same set of poles.


\section{Physics from the partial energy singularities}\label{subsec:PES}

The in-depth analysis of the scattering facet allow to both understand how flat-space physics emerge from the cosmological context and provide new insights in the structure of the scattering amplitudes as well as novel way of computing them.

The cosmological polytopes encode the coefficients of all the other singularities of the wavefunction in the other facet. We have already shown the existence of multiple-residue conditions on the wavefunction and this indeed constrains such coefficients. It is then worth to dig a bit more into their structure and understand whether we can extract interesting features. It is useful to split the discussion according whether the graphs are trees or loop graphs.
\\

\paragraph{Partial energy singularities and trees.} Let us begin with considering a generic tree graph $\mathcal{G}$. The residues corresponding to the vanishing partial energies are returned by the facets $\mathcal{P}_{\mathcal{G}}\cap\mathcal{W}^{\mbox{\tiny $(\mathfrak{g})$}}$ identified by any subgraph $\mathfrak{g}\subset\mathcal{G}$. Notice that any subgraph which include all the edges but one is a simplex as it is the convex hull of $2n_e$ vertices, two for each edge -- they are the vertices marked by the open circle close to the end points of each edge contained in $\mathfrak{g}$ as well as those marked in the middle and at endpoint furthest from $\mathfrak{g}$ for the edges which get cuts by $\mathfrak{g}$. Said differently, given a subgraph $\mathfrak{g}\subset\mathcal{G}$, if the sum of the number of its edges and of the cut ones is $n_e$, then the associated facet is a simplex, otherwise it is not simplicial:

\begin{equation*}
  
\end{equation*}

\begin{figure}
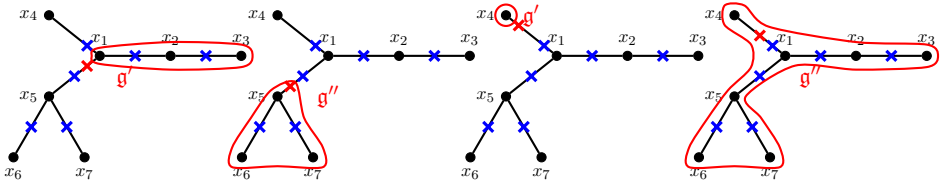

    \centering

    \caption{Subgraphs corresponding to the solutions for the contour integral representation of a facet, such that they produce two propagators which, combined together, acquire a Lorentz-invariant form.}
    \label{fig:Fsolns}
\end{figure}

Let us consider the contour representation for the canonical form of one of the simplicial facets, which we rewrite here for convenience:
\begin{equation}\label{eq:CIR}
    \Omega(\mathcal{Y},\,\mathcal{P}_{\mathcal{G}}\cap\mathcal{W}^{\mbox{\tiny $\mathfrak{g}$}})\:\sim\:
        \int\prod_{j=1}^{2n_e}\frac{dc_j}{c_j-i\varepsilon_j}\,\delta^{\mbox{\tiny $(2n_e)$}}
        \left(\mathcal{Y}-\sum_{j=1}^{2n_e}c_j Z_j\right).
\end{equation}
Being a simplex, the integral above is completely localised by the delta functions. The solution of any$c_j$ is given by the hyperplane which does not contain the vertex $Z_j$ associated to $c_j$ itself and which is identified by a subgraph $\mathfrak{g'}$. Let us mark such a vertex with a red cross \redcross. Let us now consider the solution of another of the $c$'s such that it is given by the hyperplane which does not contain the other vertex on the same edge as $Z_j$ and it is identified by the subgraph $\mathfrak{g''}$ -- see Figure \ref{fig:Fsolns}.

\noindent
Then the solutions for these $c$'s can be written as
\begin{equation}
    \mathcal{Y}^I\mathcal{W}^{\mbox{\tiny $(\mathfrak{g}')$}}_I\:\sim\:
    y_e+\sum_{\bar{e}\in\mathcal{E}^{\mbox{\tiny ext}}_{\mathfrak{gg}'}}y_{\bar{e}}+\sum_{s\in\mathcal{V}_{\mathfrak{g}'}}x_s,
    \quad
    \mathcal{Y}^I\mathcal{W}^{\mbox{\tiny $(\mathfrak{g}'')$}}_I\:\sim\:
    y_e+\sum_{\bar{e}\in\mathcal{E}^{\mbox{\tiny ext}}_{\mathfrak{g}''\mathfrak{g}}}y_{\bar{e}}+\sum_{s\in\mathcal{V}_{\mathfrak{g}''}}x_s
\end{equation}
where $e$ labels the edges marked by \redcross, while $\mathcal{E}^{\mbox{\tiny ext}}_{\mathfrak{g}\mathfrak{g}'}$ and$\mathcal{E}^{\mbox{\tiny ext}}_{\mathfrak{g}\mathfrak{g}''}$ are the edges departing simultaneously from $(\mathfrak{g},\,\mathfrak{g}')$ and $(\mathfrak{g},\,\mathfrak{g}'')$ respectively. Notice that $\mathcal{E}^{\mbox{\tiny ext}}_{\mathfrak{g}\mathfrak{g}'}\cup\mathcal{E}^{\mbox{\tiny ext}}_{\mathfrak{g}\mathfrak{g}''}\,=\,\mathcal{E}^{\mbox{\tiny ext}}_{\mathfrak{g}}$, which is the set of all the edges departing from $\mathfrak{g}$. Furthermore, given that we are considered the facet identified by the graph $\mathfrak{g}$, then these two linear contribution can be recast in the Lorentz-invariant form
$[y_e^2-(\sum_{\bar{e}\in\mathcal{E}^{\mbox{\tiny ext}}_{\mathfrak{g}'}}y_{\bar{e}}^2+\sum_{s\in\mathcal{V}_{\mathfrak{g}'}}x_s)^2]^{-1}$. Notice that when $\mathfrak{g}'$ is the subgraph corresponding to just one of the outer sites, $\mathcal{E}^{\mbox{\tiny ext}}_{\mathfrak{g}'}\,=\,\varnothing$ and the two solutions corresponding to $\mathfrak{g}'$ and $\mathfrak{g}''$ together acquire the form $-[y_e^2-x_{s_e}^2]^{-1}$, where $e$ and $s_e$ label a given outer edges of $\mathfrak{g}$ and $s_e\in\mathcal{V}_{\mathfrak{g}'}$. Hence, the canonical function for any of the simplicial faces can be written as
\begin{equation}\label{eq:CFfs}
    \begin{split}
        \Omega(\mathcal{Y},\mathcal{P}_{\mathcal{G}}\cap\mathcal{W}^{\mbox{\tiny $(\mathfrak{g})$}})\,&\sim\,
        \prod_{\hat{e}\in\mathcal{E}_{\mathfrak{g}}^{\mbox{\tiny ext}}}\frac{-1}{y_{\hat{e}}^2-x_{s_{\hat{e}}}^2}
        \prod_{e\in\mathcal{E}_{\mathfrak{g}}}\frac{1}{\displaystyle y_e^2-\left(\sum_{\bar{e}\in\mathcal{E}_{\mathfrak{g}'}^{\mbox{\tiny ext}}}y_{\bar{e}}+\sum_{s\in\mathcal{V}_{\mathfrak{g'}}}x_s\right)^2}\:=\\
        &=\:(-1)^{\mbox{\tiny dim}\{\mathcal{E}^{\mbox{\tiny ext}}\}}
        \prod_{\hat{e}\in\mathcal{E}_{\mathfrak{g}}^{\mbox{\tiny ext}}} \mathcal{A}[\mathfrak{g}_{s_e}]\times\mathcal{A}[\mathfrak{g}]\:\equiv\:
        (-1)^{\mbox{\tiny dim}\{\mathcal{E}^{\mbox{\tiny ext}}\}}\mathcal{A}[\mathcal{G}],
    \end{split}
\end{equation}
where $\mathcal{A}[\mathfrak{g}_{s_e}]$, $\mathcal{A}[\mathfrak{g}]$ and $\mathcal{A}[\mathcal{G}]$ are the flat space amplitudes associated to the subgraphs $\mathfrak{g}_{s_e}$, which contains just the outer site $s_e$, $\mathfrak{g}$ and $\mathcal{G}$ respectively. This implies that {\it all} the simplicial facets of the cosmological polytope return the flat-space scattering amplitude, up to an overall sign which depends on the orientation of the simplices ~\cite{Benincasa:2018ssx}~.

Let us now turn to the facets which are not simplicial. They are identified by any subgraph $\mathfrak{g}$ such that the sum of the number $n_e^{\mbox{\tiny $(\mathfrak{g})$}}$ of its edges and the number $n_{\mathfrak{g}}$ of the edges departing from it is strictly less than the number $n_e$ of edges of $\mathcal{G}$. The related facet $\mathcal{P}_{\mathcal{G}}\cap\mathcal{W}^{\mbox{\tiny $(\mathfrak{g})$}}$ is then the convex hull of $3n_e-n_e^{\mbox{\tiny $(\mathfrak{g})$}}-n_{\mathfrak{g}}\,>\,2n_e$ and it is not a simplex in $\mathbb{P}^{2n_e-1}$. Its contour integral representation is then not fully localised by the $2n_e$ delta-functions, and there are other further $n_e-n_e^{\mbox{\tiny $(\mathfrak{g})$}}-n_{\mathfrak{g}}$ integrations that can be made. All the inequivalent contours that can be chosen for such integration determine different triangulations for this class of facets. Among all the triangulations there is one which is characterised by simplices with two vertices for each edges of the associated graph. As the $c_k$'s related to the vertices $Z_k$'s associated to the external end of the outer edges do not depend on the $c_j$'s chosen as independent, they are fixed once for all by the delta-functions and hence the $Z_k$'s are common to all the simplices. The set of vertices which are common to all the simplices is completed by the ones whose associated $c_j$ depends only on the $c_k$'s which are fixed once for all. We can conveniently mark all such vertices with a red open circle \redcircle.

\begin{figure}
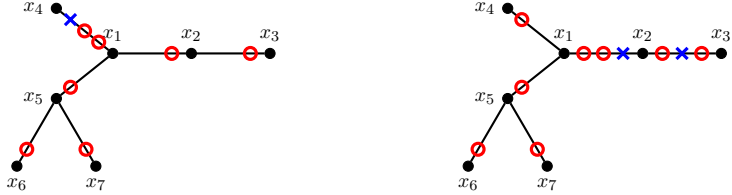

    \centering

    \caption{Example of non-simplicial facet of cosmological polytope associated to a tree graph. The \redcircle~ marks the vertices that are common to all the simplices of the triangulation which expands the wavefunction in terms of flat-space amplitudes.}
    \label{fig:nsimpF}
\end{figure}
Then the simplices which define the triangulation in question are obtained by considering all the possible markings with two vertices per edges in such a way that the ones marked by \redcircle~ are always included. For each of these simplices the very same analysis carried out for the simplicial faces applies: for each of them, the contributions coming from two faces which only differ from each other for a vertex related to the same edge $e$, differ just by the sign of $y_e$ associated to the edge and consequently they group together to form a Lorentz-invariant propagator. The canonical form of these facets can be written as sum of products of Lorentz invariant propagators. Furthermore, a simplex in this triangulation differs respect to one of the simplicial facets by one vertex on the same edge of the related graph. Consequently, they have the same canonical function, up to an overall sign.

Interestingly, we just saw that, given a tree-level graph $\mathcal{G}$, any of its facets associated to a vanishing partial energy is related to the flat-space physics either matching the actual scattering amplitude associated to $\mathcal{G}$ or describing the process in terms of sums of products of lower-point flat-space processes.

\begin{figure}
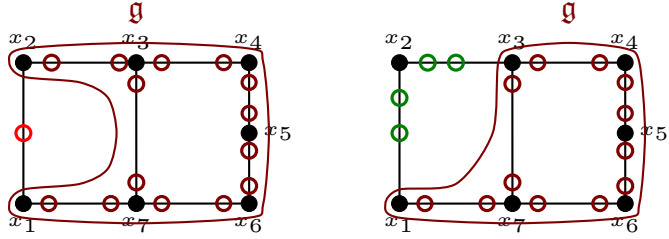

    \centering
  
    \caption{Examples of facets for the loop graphs.}
    \label{fig:LGfac}
\end{figure}

\paragraph{Partial energy singularities and loops.} Let us move to loop graphs. In this case we can distinguish two classes of facets, identified by a subgraph $\mathfrak{g}$ which cuts twice at least one edge, and which cuts the edges just once only -- see Figure \ref{fig:LGfac}. It is useful to begin with considering those facets identified by subgraphs containing all the sites of $\mathcal{G}$ as well as all its edges but one, which is cut twice (see the graph on the left in Figure \ref{fig:LGfac}). 

As we saw in the discussion of the emergence of Lorentz invariance, its contour integral representation has $L$ unfixed integration variables, which we have some freedom to choose. Any other $c_j$'s are determined by the hyperplane which do not contain neither the vertices associated to it nor the other related to the $c$'s chosen as free. Among them, the variable associated to the only vertex of this facet marked on the edge $e$ that is cut twice, turns out to be completely fixed: the related hyperplane does not contain any of the vertices associated to such an edge and it does not depend of any of the possible free integration variables. This implies that this facet factorises into such a vertex, which provides the measure for the Lorentz-invariant phase-space, and the lower-dimensional scattering facet associated to $\mathfrak{g}$:
\begin{equation}\label{eq:pesl1}
    \Omega\left(\mathcal{Y},\,\mathcal{P}_{\mathcal{G}}\cap\mathcal{W}^{\mbox{\tiny $(\mathfrak{g})$}}\right)\:=\:
    \frac{1}{2y_e}\times\mathcal{A}[\mathfrak{g}]
\end{equation}

Let us now focus on a second class of facets, identified by a subgraph $\mathfrak{g}$ such that it includes all sites of $\mathcal{G}$ but one (see the graph on the right in Figure \ref{fig:LGfac}). We can exploit again the contour integral representation for its canonical function. As independent integration variables, it is convenient to choose $L-1$ of them associated to the edges of $\mathfrak{g}$ and one associated to one of the edges departing from $\mathfrak{g}$. Let us indicate the latter with $\hat{c}$. Interestingly, once the delta-functions localise some of the integrations, the denominators either depend on $\hat{c}$ or on the other $L-1$ free integration variables, and the contour integral representation for the canonical form of this facet decouples into a product of a one-dimensional and $(L-1)$-dimensional integrals
\begin{equation}\label{eq:pesl2}
    \resizebox{0.9\hsize}{!}{$\displaystyle
    \begin{split}
        \Omega\left(\mathcal{Y},\,\mathcal{P}_{\mathcal{G}}\cap\mathcal{W}^{\mbox{\tiny $(\mathfrak{g})$}}\right)\:&\sim\: 
        \left(
            \int\frac{d\hat{c}}{
            \left[
                \left(\hat{c}-\frac{\mathfrak{y}_a}{4}\right)^2-\left(\frac{\mathfrak{y}_a}{4}-i\hat{\varepsilon}\right)^2
            \right]
            \left[
                \left(\hat{c}-\frac{\mathfrak{y}_b}{4}\right)^2-\left(\frac{\mathfrak{y}_c}{4}-i\hat{\varepsilon}\right)^2
            \right]}
        \right)\times\\
        &\hspace{-2cm}\times
        \left(
            \int\prod_{j=1}^{L-1}\frac{dc_j}{
            \left(c_j-\frac{y_{e_j}}{2}\right)^2-\left(\frac{y_{e_j}}{2}-i\varepsilon_j\right)^2}
            \prod_{k=1}^{2n_e-1+L}\frac{1}{\left(\sum_r\sigma_rc_r-\mathfrak{y}_r/2\right)^2-\left(\frac{y_k}{2}-i\varepsilon_k\right)^2}
        \right)
    \end{split}
    $}
\end{equation}
where the integral on the first line is related to the part of $\mathcal{G}$ outside $\mathfrak{g}$, the integral in the second line is related to $\mathfrak{g}$, $\mathfrak{y}$ are linear combinations of the local coordinates $x$'s and $y$'s and $\sigma_r$ are suitable signs. Both these integrals have a Lorentz-invariant form for its propagators with the $i\varepsilon$-prescription inherited from the contour integral representation for the canonical function we started with. Also, notice that the second integral returns a lower-dimensional flat-space amplitude $\mathcal{A}[\mathfrak{g}]$, while the first integral is isomorphic to a one-loop scattering facet.

This analysis carries over all the other facets, for which the Lorentz-invariant form for the propagators emerge by looking at a specific triangulation, in a similar fashion as it happens at tree-level.


\section{Symmetries of the wavefunction universal integrands}\label{subsec:SymCP}

So far we have seen how the cosmological polytopes encodes the structure we can ascribe to the wavefunction, the different ways in which we organise its perturbative expansion and how we can extract physics out of it. It is then natural to ask how this combinatorial picture encodes the symmetries of the wavefunction. Strictly speaking, the canonical form of a cosmological polytope returns the contribution of a given graph to what we called the wavefunction universal integrand. It still needs to be integrated over the external energies $x$'s appearing as weights of the relevant graph, with a suitable measure which encodes the features of the cosmology one would like to consider. The only exception is provided by conformal theories, in which case the wavefunction universal integrand is no longer integrand, rather it is the graph contribution to the wavefunction itself -- in the loop case, there are still the loop integration to be performed and hence we still deals with a loop-integrand, which now is cosmology specific. 

While we will come back to the map between universal integrands and integrated functions, it is worth noticing that any transformation which leaves the canonical form invariant, is a symmetry of the wavefunction universal integrand and, consequently, it is independent of the specific cosmology. Secondly, indeed such symmetries do not have to carry over to the integrated wavefunction as the integration might break them -- they will be symmetries of the tree-wavefunction or of the loop-integrand just in the conformal case. However, the type of integrated function we should obtain, depends on both the form and properties of the universal integrand and on the measure of integration that encodes the precise cosmology. Thus, understanding them can be of help to develop a systematic and simple way of extracting the integrated functions without having to do the integrals.

In this section we will focus on the general question: ``which transformations do preserve the canonical form of the cosmological polytopes''. Interestingly, while any of such transformations are indeed symmetries of the wavefunction universal integrand, not all of them are properly symmetries of the relevant cosmological polytope: concretely, the ones whose generators are derivative operators, correspond to transformations which change the hyperplane at infinity leaving the vertices of the cosmological polytope fixed in such a way that the canonical function of the cosmological polytope -- or, which is the same, the volume of its dual -- is preserved. These transformations are of crucial importance because they correspond to continuous symmetries of the wavefunction universal integrand.

A systematic analysis of this question has not been performed yet. However, in the rest of the subsection we provide some interesting examples of transformations which preserve the canonical form.


\subsection{Combinatorial automorphisms and the Bunch-Davies vacuum}\label{subsec:CombAut}

\begin{figure}[t]
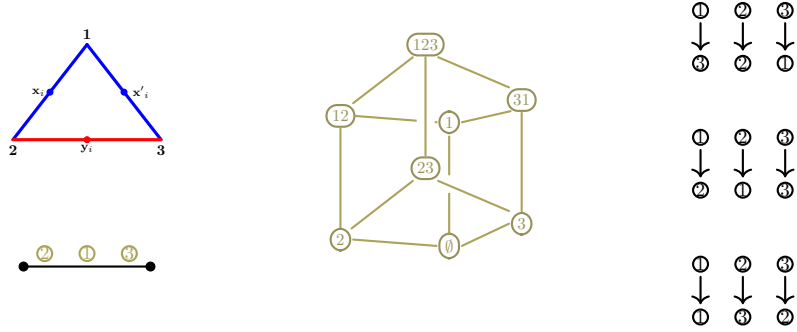

    \centering

    \caption{Combinatorial automorphisms for the cosmological polytope associated to the $2$-site line graph.}
    \label{fig:CombAut}
\end{figure}

One of the symmetry groups of any polytope is the so-called {\it combinatorial automorphisms}. Given a cosmological polytope $\mathcal{P}_{\mathcal{G}}$, the combinatorial automorphisms are defined as the symmetry group which preserves its face lattice. The face lattice has its vertices given by all the faces of $\mathcal{P}_{\mathcal{G}}$, including both $\mathcal{P}_{\mathcal{G}}$ and the empty set, and its edges connecting the vertices are given by the containment relations -- see Figure \ref{fig:CombAut}.

Recall that a cosmological polytope $\mathcal{P}_{\mathcal{G}}$ is in $1-1$ correspondence with a graph $\mathcal{G}$, and any graph $\mathcal{G}$ with $n_e$ edges can be seen as obtained by suitably merging a collection of $2$-site line graphs in some of their sites. As a $2$-site line graph is associated to a triangle, this construction corresponds to the intrinsic definition of the cosmological polytope as the intersection of $n_e$ triangles in the midpoints of at most two out of their three sides. We can thus first understand the combinatorial automorphisms for the triangle and then try to extend to arbitrary cosmological polytopes using their first principle definition.

In the case of the triangle, its face lattice is a cube and its combinatorial automorphism group is given by the transposition of two vertices -- see Figure \ref{fig:CombAut}. Despite there are three possible transpositions, it is two-dimensional as the third one is generated by a composition of the other two. The combinatorial automorphism group for this case is already tamed!

Let us move to more general graphs: when the associated cosmological polytope $\mathcal{P}_{\mathcal{G}}$ is generated as the intersection of triangles, linear relations, $Z_i+Z_j\sim Z_{i'}+Z_{j'}$, are imposed among the vertices of the sides that are intersected, which consequently live on the same hyperplane. Such relations need to be preserved by the combinatorial automorphisms. Precisely these conditions prevent to transpose most of the vertices associated to the internal edges, as these hyperplane would not be preserved. However, they also make clear which vertex transposition do not affect them.

\begin{figure}
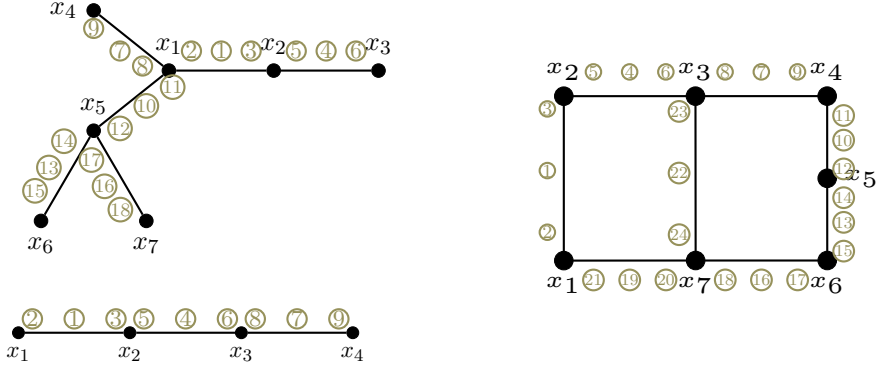

    \centering

    \caption{Combinatorial automorphisms. The vertices of the cosmological polytopes associated to a graph has to satisfy linear relations because of its intrinsic definition. For the graphs depicted, such relations are: $\{Z_1+Z_3\sim Z_4+Z_5,\,Z_1+Z_2\sim Z_7+Z_8\sim Z_{10}+Z_{11},\,Z_{10}+Z_{12}\sim Z_{13}+Z_{14}\sim Z_{16}+Z_{17}\}$ (up left); $\{Z_1+Z_3\sim Z_4+Z_5,\,Z_4+Z_4\sim Z_7+Z_8\})$ (bottom left); $\{Z_1+Z_3\sim Z_4+Z_5,\,Z_4+Z_6\sim Z_7+Z_8\sim Z_{22}+Z_{23},\,Z_7+Z_9\sim Z_{10}+Z_{11},\, Z_{10}+Z_{12}\sim Z_{13}+Z_{14},\, Z_{13}+Z_{15}\sim Z_{16}+Z_{17},\,Z_{16}+Z_{18}\sim Z_{19}+Z_{20}\sim Z_{22}+Z_{24},\, Z_{19}+Z_{21}\sim Z_1+Z_2\}$. The automorphism group is more constraints as more linear relations fix the cosmological polytope. In the upper left graph, it is constituted by the transpositions of all pairs of the vertices $\{Z_6,\,Z_9,\,Z_{15},\,Z_{18}\}$ associated to the outer ends of the external edges, as well as of the pairs $\{(4,5),\,(7,8),\,(13,14),\,(16,17)\}$ each of which is associated to the same edge and appear in the same constraints. For the $4$-site line graph, the automorphism group contains both these classes of transpositions, $\{(2,9)\}$ and$\{(1,3),\,(7,8)\}$, as well as the transformation which swaps the triples $(1,3,5)$ and $(7,8,6)$ element by element keeping $(2,4,9)$ frozen. Finally, the loop graph is very constrained and the combinatorial automorphism group is one dimensional and transposes $(3,5,4,6,23,8,7,9,11,19)$ and $(2,21,19,20,24,18,16,17,15,13)$, keeping $(1,22)$ fixed.}
    \label{fig:CombAutGen}
\end{figure}

Therefore, given a graph the combinatorial automorphism group is given by the transposition between any pair of the unconstrained vertices, by all the transpositions which preserves the individual constrains and the simultaneous exchange of more vertices acting on non-trivially on a larger number of constraints leaving them invariant -- see Figure \ref{fig:CombAutGen}. Notice that the first two classes of transformation can be defined just for tree level graphs or anyhow with graph which show some tree structure.

In our cosmological polytope picture, the vertices are kept fixed while any transformation should act on the hyperplane at infinity. So it can be convenient to translate the combinatorial automorphism group as a transformation acting in $\mathcal{Y}$, {\it i.e.} in the space of local coordinates, which physically are nothing but the energies. Let is begin with considering a graph $\mathcal{G}$ with some external tree structure. Then, a subset of the combinatorial automorphism group is defined by the set transpositions of the two constrained vertices associated to the most outer edges of the graph. Let $(x_i,y_{ij},x_j)$ be the labels associated to one of such edges and its sites and let $j$ be the site in common with the rest of the graph. Hence, the transposition which belongs to the automorphism group maps $\{\mathbf{x_i}-\mathbf{y}_{ij}+\mathbf{x}_j,\,-\mathbf{x}_i+\mathbf{y}_{ij}+\mathbf{x}_j\}$ into each other. In energy space, it corresponds to the exchange $x_i\,\longleftrightarrow\,y_{ij}$. As far as the other combinatorial automorphisms are concerned, they correspond to reflections along a symmetry axis of the graph and in energy space they exchange the $(x_{i},y_{ij},x_{j})$'s in such a way that the vertices along the symmetry axis are kept frozen.

First, notice that these symmetries either leave the poles of the wavefunction invariant or map some of them into each other as they depend just on sums of $x$'s and $y$'s. Secondly, if we go to any of the facets, which are related to the flat-space physics, the very same discrete transformations are no longer a symmetry\footnote{Indeed, any of these facets is also a polytope and, as a such, it has an combinatorial automorphism group. However, it is not the same group as the one acting on the full cosmological polytope: just those transformations which leave the hyperplane that identifies the facet invariant survive.}. Hence these transformations guarantees the Bunch-Davies vacuum condition.


\subsection{Scaling transformations}\label{subsec:Trsf1der}

The combinatorial automorphisms analysed above is a group of discrete transformations. In this section we will start discussing continuous transformations. Concretely, we are going to deal with those transformations which are generated by first-derivative operators which leave the canonical function invariant. Let us consider a tree-level graph $\mathcal{G}$ and its associated cosmological polytope $\mathcal{P}_{\mathcal{G}}\subset\mathbb{P}^{N-1}$, it is straightforward to identify scaling-like symmetries:
\begin{equation}\label{eq:ScaleSym}
    \hat{\mathcal{D}}\: :=\:\frac{\partial}{\partial\mathcal{Y}^I}\mathcal{Y}^I,\qquad
    \hat{\mathcal{O}}^{\mbox{\tiny $(i)$}}\: :=\frac{\partial}{\partial\mathcal{Y}^I}
        \frac{Z_i^I\langle\mathcal{Y} i_2\ldots i_N\rangle}{\langle i i_2\ldots i_N\rangle}
\end{equation}
where $N\,=\,n_s+n_e$, the $Z$'s are the vertices of $\mathcal{P}_{\mathcal{G}}$ and $\{Z_i,Z_{i_2},\ldots Z_{i_N}\}$ constitutes a basis for $\mathbb{R}^N$. Importantly, $\hat{\mathcal{D}}$ is nothing but the dilatation operator, whose form in momentum space is given by:
\begin{equation}\label{eq:DilOp}
    \hat{\mathcal{D}}\: :=\:\sum_{r=1}^n\hat{\mathcal{D}}^{\mbox{\tiny $(r)$}},\qquad
    \hat{\mathcal{D}}^{\mbox{\tiny $(r)$}}\: :=\:-(\Delta_r-d)+p_j^{\mbox{\tiny $(r)$}}
    \frac{\partial}{\partial p_j^{\mbox{\tiny $(r)$}}},
\end{equation}
where $r$ labels the external states, $\hat{\mathcal{D}}^{\mbox{\tiny $(r)$}}$ is the dilatation operator for the state $r$, $\Delta_r$ is the conformal dimension of the state $r$, $d$ is the number of spatial dimensions, and $j$ run over the spatial directions, 

How many symmetries of this type exist depends on the concrete cosmological polytope. However, they exists if the related graph has a tree structure and their number corresponds to the number of external edges in such a tree-structure. So, if the number of such edges is $m$, then there will be $m+1$ of such generators. In this sense, any line graph enjoys three of such scaling-like symmetries, while a star graph four.

For the sake of concreteness, let us ask which operators of this type leave the canonical form of the triangle in $\mathbb{P}^2$ associated to the $2$-line site graph. There are in principle {\it four} operators that satisfy this property
\begin{equation}\label{eq:ScOp2s0l}
    \hat{\mathcal{D}}\: :=\:\frac{\partial}{\partial\mathcal{Y}^I}\mathcal{Y}^I,\quad
    \hat{\mathcal{O}}^{\mbox{\tiny $(i)$}}\: :=\:\frac{\partial}{\partial\mathcal{Y}^I}Z_{i}^I
        \frac{\langle\mathcal{Y}(i+1)(i-1)\rangle}{\langle (i-1)i(i+1)\rangle}
\end{equation}
where $i\,=\,1,2,3$. However not all of them are independent as they turn out to be related by the following linear relation
\begin{equation}\label{eq:ScOp2s0lLinR}
    \hat{\mathcal{D}}\:=\:\sum_{i=1}^3\hat{\mathcal{O}}^{\mbox{\tiny $(i)$}}.
\end{equation}
Notice that the number of operators in \eqref{eq:ScOp2s0l} seem to violate the previous statement that the number of scaling operators $\hat{\mathcal{O}}^{\mbox{\tiny $(i)$}}$ corresponds to the number of external edges. However, for the $2$-site line graph, the existence of these three operators reflects the symmetries of the triangle, while for more generic graph, the existence of constraints on the vertices break such symmetries. 

Let us consider a generic graph $\mathcal{G}$ with $\tilde{n}_e$ external edges, and the associated cosmological polytope $\mathcal{P}_{\mathcal{G}}\,\subset\,\mathbb{P}^{n_s+n_e-1}$. Then the space of scaling symmetries is $\tilde{n}_e+1$ and are given by \eqref{eq:ScaleSym}. However, there are now new operators which can be defined: they preserve the hyperplanes identified by the linear relations among the vertices.

In order to fix ideas, let us consider the $3$-site line graph and the related cosmological polytope, which lives in $\mathbb{P}^4$. It is the convex hull of six vertices, $\{Z_1,Z_2,Z_3,Z_4,Z_5,Z_6\}$, four of which are linearly dependent, namely $Z_1+Z_3\sim Z_4+Z_5$, because of the intrinsic definition of the polytopes. Then one can define the operators \eqref{eq:ScaleSym}, where the operators $\hat{\mathcal{O}}^{\mbox{\tiny $(i)$}}$ are associated to the linearly independent vertices, {\it i.e.} $Z_2$ and $Z_6$. However, one can find other two operators which leave the hyperplane $Z_1+Z_3\sim Z_4+Z_5$ invariant:
\begin{equation}\label{eq:OpsHyp}
    \begin{split}
        &\hat{\mathcal{O}}^{\mbox{\tiny $(15)$}}\:=\:\frac{\partial}{\partial{\mathcal{Y}}^I}
            \left[  
                \frac{Z_1^I}{\langle12346\rangle}-\frac{Z_5^I}{\langle52346\rangle}
            \right]
            \frac{\langle\mathcal{Y}2356\rangle\langle\mathcal{Y}1246\rangle}{\langle\mathcal{Y}1256\rangle},\\
        &\hat{\mathcal{O}}^{\mbox{\tiny $(14)$}}\:=\:\frac{\partial}{\partial{\mathcal{Y}}^I}
            \left[  
                \frac{Z_1^I}{\langle12356\rangle}-\frac{Z_4^I}{\langle42356\rangle}
            \right]
            \frac{\langle\mathcal{Y}2346\rangle\langle\mathcal{Y}1256\rangle}{\langle\mathcal{Y}1246\rangle},
    \end{split}
\end{equation}
where the denominators make these operators $GL(1)$-invariant -- notice that the $\mathcal{Y}$-dependent denominator does not provide any contribution as its derivative vanishes upon contraction with both the $Z_i^I$ in each numerator. Furthermore, the operators \eqref{eq:OpsHyp} are mapped into each other under either of the exchange $1\,\longleftrightarrow\,3$ or $4\,\longleftrightarrow\,5$, and therefore are invariant under the composition of these exchanges. This is just a reflection of the linear relation $Z_1+Z_3\,\sim\,Z_4+Z_5$.


\subsection{Special conformal transformations}\label{subsec:Trsf2der}

The approach outlined in the previous section boils down to look for all those transformations generated by first order derivative operators leave the canonical form of the cosmological polytope invariant, and parametrise their space via a convenient basis. We can extend this idea to transformation with second-derivative generators. This is important to identify the combinatorial origin of the special conformal transformations, which in momentum space are precisely given by a second order operator
\begin{equation}\label{eq:SCTK}
    \hat{\mathcal{K}}_i\: =\: \sum_{r=1}^n \hat{\mathcal{K}}_i^{\mbox{\tiny $(r)$}},\qquad
    \hat{\mathcal{K}}_i^{\mbox{\tiny $(r)$}}\: :=\:
    \left[
        2\hat{\mathcal{D}}^{\mbox{\tiny $(r)$}}\delta_i^{\phantom{i}j}-
        p_i^{\mbox{\tiny $(r)$}}\frac{\partial}{\partial p_j^{\mbox{\tiny $(r)$}}}
    \right]
    \frac{\partial}{\partial p^j_{\mbox{\tiny $(r)$}}}
\end{equation}
where $r$ runs over the states, $i$ is the space index, and $\hat{\mathcal{D}}^{\mbox{\tiny $(r)$}}$ is the dilatation operator associated to the state $r$.

While a systematic classification of symmetries of the canonical form generated by second-derivative operators has not been performed yet, it is possible to identify the spatial conformal generator on a graph-by-graph basis and focusing on $\phi^3$ interactions in $d=5$, which is a conformal theory for which the canonical function represents the actual wavefunction associated to a given graph. We will come back to this point at the very end of this section.


\subsubsection{Two-site line graph}\label{subsubsec:K2s0l}

Let us first ask the question about the space of two-derivative operators that leave invariant the canonical form of the relevant cosmological polytope, and then look for the special conformal generators \eqref{eq:SCTK}. As we already learnt, the cosmological polytope associated to the $2$-site line graph is a triangle in $\mathbb{P}^2$. Interestingly, {\it all} second order operators annihilating its canonical form can be obtained as compositions of the first order scaling operators \eqref{eq:ScOp2s0l} discussed in the previous section. Let us write them here for convenience
\begin{equation}\label{eq:1stOps2s0l}
    \hat{\mathcal{D}}\:=\:\frac{\partial}{\partial\mathcal{Y}^I}\mathcal{Y}^I\:=\:
        \sum_{i=1}^3\hat{\mathcal{O}}^{\mbox{\tiny $(i)$}},\quad
    \hat{\mathcal{O}}^{\mbox{\tiny $(i)$}}\,=\,\frac{\partial}{\partial\mathcal{Y}^I}Z_i^{I}\,
        \frac{\langle\mathcal{Y}(i+1)(i-1)\rangle}{\langle(i-1)i(i+1)\rangle}
\end{equation}
Secondly, there are just three independent special conformal generators: one of them is just a first derivative operator acting on the dilatation operator, while the other two acquire the form\footnote{Here we indicates the special conformal transformation generators as $\hat{\underline{\mathcal{K}}}_j$ as they are proportional to the form one would obtain via a change of variable from the spatial momentum space to the local coordinates $\mathcal{Y}\,=\,(x_1,y,x_2)$.}
\begin{equation}\label{eq:2ndOps2s0l}
    \begin{split}
        &\hat{\underline{\mathcal{K}}}_1\:=\:\frac{\langle\mathcal{Y}Z_{\star}1\rangle\langle Z_{\star}23\rangle}{\langle123\rangle}
            \frac{\partial}{\partial\mathcal{Y}^I}
            \left[
                \frac{\langle123\rangle}{\langle Z_{\star}23\rangle\langle Z_{\star}31\rangle}Z_3^I
                    \hat{\mathcal{O}}^{\mbox{\tiny $(1)$}}-       
                \frac{\langle123\rangle}{\langle Z_{\star}31\rangle\langle Z_{\star}12\rangle}Z_1^I
                    \hat{\mathcal{O}}^{\mbox{\tiny $(3)$}}+
            \right.\\
        &\hspace{1cm}+
            \left.
                \left(
                    \frac{\langle123\rangle}{\langle Z_{\star}31\rangle\langle Z_{\star}12\rangle}Z_1^I+
                    \frac{\langle123\rangle}{\langle Z_{\star}12\rangle\langle Z_{\star}23\rangle}Z_2^I+
                    \frac{\langle123\rangle}{\langle Z_{\star}23\rangle\langle Z_{\star}31\rangle}Z_3^I
                \right)
                \hat{\mathcal{O}}^{\mbox{\tiny $(2)$}}
            \right],\\
        &\hat{\underline{\mathcal{K}}}_2\:=\:\frac{\langle\mathcal{Y}Z_{\star}1\rangle\langle Z_{\star}23\rangle}{\langle123\rangle}
            \frac{\partial}{\partial\mathcal{Y}^I}
            \left[
                \frac{\langle123\rangle}{\langle Z_{\star}23\rangle\langle Z_{\star}12\rangle}Z_2^I
                    \hat{\mathcal{O}}^{\mbox{\tiny $(1)$}}-       
                \frac{\langle123\rangle}{\langle Z_{\star}12\rangle\langle Z_{\star}31\rangle}Z_1^I
                    \hat{\mathcal{O}}^{\mbox{\tiny $(2)$}}+
            \right.\\
        &\hspace{1cm}+
            \left.
                \left(
                    \frac{\langle123\rangle}{\langle Z_{\star}31\rangle\langle Z_{\star}12\rangle}Z_1^I+
                    \frac{\langle123\rangle}{\langle Z_{\star}12\rangle\langle Z_{\star}23\rangle}Z_2^I+
                    \frac{\langle123\rangle}{\langle Z_{\star}23\rangle\langle Z_{\star}31\rangle}Z_3^I
                \right)
                \hat{\mathcal{O}}^{\mbox{\tiny $(3)$}}
            \right]
    \end{split}
\end{equation}
where $Z_{\star}\in\mathbb{P}^2$ is an arbitrary point that we take to be {\it inside} the cosmological polytope in order to have all the $\langle\ldots\rangle$ containing it positive, and the prefactor is chosen in order for the operators to be $GL(1)$-invariant. Notice that, the sum of the operators \eqref{eq:2ndOps2s0l} is just a first order operator acting on the dilatation
\begin{equation}\label{eq:2ndOpsA2s0l}
    \hat{\underline{\mathcal{K}}}_1+\hat{\underline{\mathcal{K}}}_2\:=\:\frac{\langle\mathcal{Y}Z_{\star}1\rangle\langle Z_{\star}23\rangle}{\langle123\rangle}
            \frac{\partial}{\partial\mathcal{Y}^I}
            \left[
                \frac{\langle123\rangle}{\langle Z_{\star}23\rangle\langle Z_{\star}12\rangle}Z_2^I+
                \frac{\langle123\rangle}{\langle Z_{\star}23\rangle\langle Z_{\star}31\rangle}Z_3^I
            \right]
            \hat{\mathcal{D}}.
\end{equation}
Two comments are now in order:
{
\renewcommand{\theenumi}{\roman{enumi}}
\begin{enumerate}
    \item The operators $\hat{\underline{\mathcal{K}}}_j$ -- and, consequently, the special conformal transformations -- annihilate the canonical form of the triangle in a trivial way, via the action of the first order scaling operators $\{\hat{\mathcal{O}}^{\mbox{\tiny $(j)$}}\,|\,j=1,2,3\}$.
    \item The canonical form for the triangle encodes the wavefunction of the universe of the $2$-site line graph {\it for any} polynomial interactions {\it in any} dimensions. These means that the transformations we have been discussing are symmetries {\it for all} these cases. More precisely, the canonical form returns an universal integrand for all these cases {\it in any} FRW cosmologies, so these transformations are a symmetry of this integrand. Indeed the integration that maps it to the actual wavefunction may break these symmetries. Nevertheless, the functional form for integrated wavefunction will crucially depend on the properties of its integrand. Finally, the second order differential operators \eqref{eq:2ndOps2s0l} can be interpreted as special conformal transformations just for specific values of $\Delta$ and $d$, that precisely make the theory conformal. Even so, as already stated, they generates symmetries of the integrand.
\end{enumerate}
}


\subsubsection{Three-site line graph}\label{subsubsec:K3s0l}

Let us now consider the three-site line graph. The associated polytope $\mathcal{P}_{\mathcal{G}}\,\subset\,\mathbb{P}^4$ is the convex hull of the following six vertices
\begin{equation*}
    \begin{split}
        &\{             \mathbf{x}_1-\mathbf{y}_{12}+\mathbf{x}_2,\:\mathbf{x}_1+\mathbf{y}_{12}-\mathbf{x}_2,\:
                        -\mathbf{x}_1+\mathbf{y}_{12}+\mathbf{x}_2\\
        &\phantom{\{}   \mathbf{x}_2-\mathbf{y}_{23}+\mathbf{x}_3,\:\mathbf{x}_2+\mathbf{y}_{23}-\mathbf{x}_3,\:
                        -\mathbf{x}_2+\mathbf{y}_{23}+\mathbf{x}_3\}.
    \end{split}
\end{equation*}
Let us label them as $\{Z_j\,|\,j=1,\ldots,6\}$ in the same order as they appear in the list above. 

As already discussed, the first order operators that annihilate the canonical form of the associated polytope $\mathcal{P}_{\mathcal{G}}$ are two scaling operators $\{\hat{\mathcal{O}}^{\mbox{\tiny $(j)$}}\,|\,j=1,2\}$ and two operators \eqref{eq:OpsHyp} which preserve the hyperplane identified by its definition as intersection of two triangles.

We can classify all the possible second order operators that annihilate the canonical form of $\mathcal{P}_{\mathcal{G}}$. Contrarily to the $2$-site line graph case, not all of them can be written in terms of the first order operators. In particular, there are six of them which turn out to annihilate the canonical forms of the simplices of certain signed triangulations of $\mathcal{P}_{\mathcal{G}}$:
\begin{equation}\label{eq:2ndOps3s0l}
    \begin{split}
        &\hat{\mathcal{O}}^{\mbox{\tiny $(1,3)$}}\:\sim\:
            \frac{\partial^2}{\partial\mathcal{Y}^I\partial\mathcal{Y}^J}Z_1^I Z_3^J
            \langle\mathcal{Y}5264\rangle,\qquad
         \hat{\mathcal{O}}^{\mbox{\tiny $(4,5)$}}\:\sim\:
            \frac{\partial^2}{\partial\mathcal{Y}^I\partial\mathcal{Y}^J}Z_4^I Z_5^J
            \langle\mathcal{Y}3261\rangle,\\
        &\hat{\mathcal{O}}^{\mbox{\tiny $(1,4)$}}\:\sim\:
            \frac{\partial^2}{\partial\mathcal{Y}^I\partial\mathcal{Y}^J}Z_1^I Z_4^J
            \langle\mathcal{Y}2356\rangle,\qquad
         \hat{\mathcal{O}}^{\mbox{\tiny $(3,5)$}}\:\sim\:
            \frac{\partial^2}{\partial\mathcal{Y}^I\partial\mathcal{Y}^J}Z_3^I Z_5^J
            \langle\mathcal{Y}1246\rangle,\\
        &\hat{\mathcal{O}}^{\mbox{\tiny $(1,5)$}}\:\sim\:
            \frac{\partial^2}{\partial\mathcal{Y}^I\partial\mathcal{Y}^J}Z_1^I Z_5^J
            \langle\mathcal{Y}2346\rangle,\qquad
         \hat{\mathcal{O}}^{\mbox{\tiny $(3,4)$}}\:\sim\:
            \frac{\partial^2}{\partial\mathcal{Y}^I\partial\mathcal{Y}^J}Z_3^I Z_4^J
            \langle\mathcal{Y}1256\rangle
    \end{split}
\end{equation}
The operators in the first line, $\hat{\mathcal{O}}^{\mbox{\tiny $(1,3)$}}$ and $\hat{\mathcal{O}}^{\mbox{\tiny $(4,5)$}}$, two annihilate the simplices of the signed triangulations which provides the frequency representations \eqref{eq:RRw}, the ones in the second line annihilate term by term the OFPT expansion, and finally the ones in the third line annihilate the other perturbative expansion with physical poles only.


\section{From flat-space to cosmology}\label{subsec:SMWF}

One remarkable consequences of the combinatorial automorphisms discussed in Section \ref{subsec:CombAut} is the fact that a subset of the singularities of the wavefunction of the universe encode still the flat-space scattering amplitude, while coefficient of the others can be written as sums of products of lower-point wavefunctions. In other words, {\it the whole} physical content of the wavefunction seem to be tied to flat space physics.

A natural question therefore is how much of the structure of the wavefunction of the universe can be fixed from the knowledge of the flat-space scattering amplitudes. As we will shortly show, in the context of the toy models which can be described by the (generalised) cosmological polytopes, given the flat-space scattering amplitudes and the Bunch-Davies vacuum condition, the wavefunction of the universe can be unambiguously reconstructed at tree-level ~\cite{Benincasa:2018ssx}~.

\paragraph{The wavefunction representatives.} A first observation is that in a neighbourhood of the total energy pole, the wavefunction of the universe is nothing but the scattering amplitude divided by the total energy. Turning the table around, taking an $\mathcal{A}_{\mathcal{G}}$ and dividing it by the total energy $\sum_{s\in\mathcal{V}}x_s$ provides a representative $\psi_{\mathcal{A}_{\mathcal{G}}}$ of the wavefunction in a neighbourhood of the vanishing total energy,
\begin{equation}\label{eq:WFrepEtot}
    \psi_{\mathcal{A}_{\mathcal{G}}}\: :=\:\frac{\mathcal{A}_{\mathcal{G}}}{\displaystyle\sum_{s\in\mathcal{V}}x_s}.
\end{equation}
This holds at any order in perturbation theory. Already the singularity structure of \eqref{eq:WFrepEtot} indicates that it cannot be the end of the story: it contains poles of the form $y-x$ which can be reached in the physical region leading to particle production. Hence it cannot be already the final answer for the Bunch-Davies wavefunction.

Let us take the combinatorial point of view. Let us consider the space $\mathbb{P}^2$ with local homogeneous coordinates $\mathcal{Y}\,:=\,(x,\,y,\,x')$ and a triangle $\mathcal{T}$ defined as the convex hull of the vertices
\begin{equation}\label{eq:TrWFrep}
    \{\mathbf{x}+\mathbf{y}-\mathbf{x'},\,-\mathbf{x}+\mathbf{y}+\mathbf{x'},\,\mathbf{x'}\}.
\end{equation}
First, notice that the boundary identified by the two vertices $\{\mathbf{x}+\mathbf{y}-\mathbf{x'},\,-\mathbf{x}+\mathbf{y}+\mathbf{x'}\}$ is precisely the scattering facet of the triangle associated to the $2$-site line graph -- $\mathbf{x'}$ can be any of the two midpoints of its intersectable edges. Its canonical form is 
\begin{equation}\label{eq:TrWFrepCF}
    \omega(\mathcal{Y},\mathcal{T})\:=\:\frac{1}{(x+x')(y^2-x^2)}\frac{dx_1\wedge dy\wedge dx_2}{\mbox{Vol}\{GL(1)\}}
\end{equation}
which is precisely the wavefunction representative $\psi_{\mathcal{A}_{\mathcal{G}}}$ with $\mathcal{G}$ being the $2$-site line graph.

Let us now consider a collection of $n_e$ of such triangles $\{\mathbf{x}_j+\mathbf{y}_j-\mathbf{x'}_j,\,-\mathbf{x}_j+\mathbf{y}_j+\mathbf{x'}_j,\,\mathbf{x'_j}\}$
and let us intersect all of them in their vertices $\{\mathbf{x'}_j\,|\,j=1,\ldots,n_e\}$. The resulting polytope $\underline{\mathcal{P}}$ is the convex hull of $2n_e+1$ vertices in $\mathbb{P}^{2n_e}$: it is thus a simplex with canonical form
\begin{equation}\label{eq:TrWFrepCF2}
    \omega(\mathcal{Y},\underline{\mathcal{P}})\:=\:\frac{1}{\displaystyle x'+\sum_{j=1}^{n_e}x_j}\prod_{j=1}^{n_e}\frac{1}{y^2_j-x_j^2}\,\frac{dx'}{\mbox{Vol}\{GL(1)\}}\bigwedge_{j=1}^{n_e} dx_j\wedge dy_j,
\end{equation}
which is precisely the wavefunction representative for a star graph with $n_e$ edges. Notice that the polytope $\underline{\mathcal{P}}\subset\mathbb{P}^{2n_e+1}$ is just the convex hull of the vertices defining the scattering facet of the star graph with an additional point $\mathbf{x}'$.

Let $\mathcal{G}$ be a generic graph with $n_s$ sites and $n_e$ edges. The associated scattering facet is a polytope with $2n_e$ vertices $\{\mathbf{x}_i+\mathbf{y}_{ij}-\mathbf{x}_j,\,-\mathbf{x}_i+\mathbf{y}_{ij}+\mathbf{x}_j\}$ in $\mathbb{P}^{n_s+n_e-2}$. One can construct a new polytope $\underline{\mathcal{P}}_{\mathcal{G}}$ in one dimension higher by adding an extra vertex $\mathbb{x}_k$, which can be any of the points $\{\mathbf{x}_i,\,\mathbf{x}_j\}$. It is straightforward to show that the canonical function $\Omega(\mathcal{Y},\underline{P}_{\mathcal{G}})$ for $\underline{\mathcal{P}}_{\mathcal{G}}$ is just the canonical function $\Omega(\mathcal{Y},\mathcal{S}_{\mathcal{G}})$ times the total energy pole
\begin{equation}\label{eq:TrWFrepCF3}
    \Omega(\mathcal{Y},\underline{\mathcal{P}}_{\mathcal{G}})\:=\:
        \frac{1}{\displaystyle x_k+\sum_{s\in\mathcal{V}\setminus\{s_k\}}x_s}\times\Omega(\mathcal{Y,\mathcal{S}_{\mathcal{G}}})
\end{equation}
with $\Omega(\mathcal{Y},\mathcal{S}_{\mathcal{G}})$ independent of $x_k$. This construction represents the combinatorial realisation of the wavefunction representatives \eqref{eq:WFrepEtot}, with the different possible choices for $x_k$ providing all the different representatives for the same graph.
\\

\paragraph{From the representatives to the wavefunction.} We just saw how from a graph contribution to a scattering amplitude we could construct a class of wavefunction representatives. In what follows we will see how from the wavefunction representatives we can reconstruct the full graph contribution to the wavefunction. From a physical perspective we need an extra input: we need to know that we are looking for a wavefunction in the Bunch-Davies vacuum, {\it i.e.} the final result should just have singularities outside the physical region given by sums of energies. Interestingly, the Bunch-Davies vacuum condition translates in a discrete symmetry in energy space for tree-level graphs. From the combinatorial perspective, this is equivalent in providing as data the scattering facet as well as the fact that the object in one-dimension higher we would like to construct should be invariant under a specific combinatorial automorphism group $\mathcal{CA}(\mathcal{P}_{\mathcal{G}})$.

Let us focus on tree-level graphs. So, given a tree-level graph $\mathcal{G}$ with $n_e$ edges and $n_s=n_e+1$ sites, we can construct a class of polytopes $\underline{\mathcal{P}}_{\mathcal{A}_{\mathcal{G}}}\,\subset\,\mathbb{P}^{n_s+n_e-1}$ by knowing the existence of a scattering facet $\mathcal{S}_{\mathcal{G}}\,\subset\,\mathbb{P}^{n_s+n_e-2}$ associated to $\mathcal{G}$ by adding an extra vertex $\mathbf{x}_k$ in one dimension higher, with $\mathbf{x}_k\,\in\,\{\mathbf{x}_j\,|\,j=1,\ldots,n_s\}$. Being any of the $\underline{\mathcal{P}}_{\mathcal{A}_{\mathcal{G}}}$ a polytope in its own right, it also has its own combinatorial automorphism group $\mathcal{CA}(\mathcal{P}_{\mathcal{A}_{\mathcal{G}}})$. Therefore, the information we need is just encoded in $\mathcal{CA}(\mathcal{P}_{\mathcal{G}})/
(\mathcal{CA}(\mathcal{P}_{\mathcal{G}})\cap(\mathcal{CA}(\mathcal{P}_{\mathcal{A}_{\mathcal{G}}}))$. Such a group can be realised as a matrix action on $\underline{\mathcal{P}}_{\mathcal{A}_{\mathcal{G}}}$ ~\cite{Benincasa:2018ssx}~
\begin{equation}\label{eq:ExeA}
 T_{e_E}\:=\:
 \resizebox{.35\hsize}{!}{
 \bordermatrix{
  ~         & \bar{v} & 1      & \ldots & v_e    	     & \ldots & n_v    & e_1    & \ldots & e      	     & \ldots & n_e \cr
  {\bar{v}} & 1       & 0      & \ldots & 0      	     & \ldots & 0      & 0      & \ldots & 0      	     & \ldots & 0\cr
  1         & 0       & 1      & \ldots & 0      	     & \ldots & 0      & 0      & \ldots & 0      	     & \ldots & 0\cr
  \ldots    & \ldots  & \ldots & \ldots & \ldots 	     & \ldots & \ldots & \ldots & \ldots & \ldots	     & \ldots & \ldots\cr 
  v_e       & 0       & 0      & \ldots & 0      	     & \ldots & 0      & 0      & \ldots & {\bf\color{red} 1} & \ldots & 0\cr
  \ldots    & \ldots  & \ldots & \ldots & \ldots 	     & \ldots & \ldots & \ldots & \ldots & \ldots 	     & \ldots & \ldots\cr 
  n_v       & 0       & 0      & \ldots & 0      	     & \ldots & 1      & 0      & \ldots & 0      	     & \ldots & 0\cr
  \ldots    & \ldots  & \ldots & \ldots & \ldots 	     & \ldots & \ldots & \ldots & \ldots & \ldots 	     & \ldots & \ldots\cr 
  e_1       & 0       & 0      & \ldots & 0      	     & \ldots & 0      & 1      & \ldots & 0      	     & \ldots & 0\cr
  \ldots    & \ldots  & \ldots & \ldots & \ldots 	     & \ldots & \ldots & \ldots & \ldots & \ldots 	     & \ldots & \ldots\cr 
  e         & 0       & 0      & \ldots & {\bf\color{red} 1} & \ldots & 0      & 0      & \ldots & 0                  & \ldots & 0\cr
  \ldots    & \ldots  & \ldots & \ldots & \ldots 	     & \ldots & \ldots & \ldots & \ldots & \ldots 	     & \ldots & \ldots\cr 
  n_e       & 0       & 0      & \ldots & 0      	     & \ldots & 0      & 0      & \ldots & 0      	     & \ldots & 1\cr  
 }
 },\quad
 T_{e_I}\:=\:
 \resizebox{.35\hsize}{!}{
 \bordermatrix{
  ~         & \bar{v} & 1    	           & \ldots & v_e    & \ldots & n_v    		    & e_1    	         & \ldots & e      & \ldots & n_e \cr
  {\bar{v}} & 1       & 0      	           & \ldots & 0      & \ldots & 0      		    & 0      	         & \ldots & 0      & \ldots & 0\cr
  1         & 0       & {\bf\color{red} 0} & \ldots & 0      & \ldots & {\bf\color{red} -1} & {\bf\color{red} 1} & \ldots & 0      & \ldots & 0\cr
  \ldots    & \ldots  & \ldots 	           & \ldots & \ldots & \ldots & \ldots 		    & \ldots 	         & \ldots & \ldots & \ldots & \ldots\cr 
  v_e       & 0       & 0      	           & \ldots & 0      & \ldots & 0      		    & 0       	         & \ldots & 1      & \ldots & 0\cr
  \ldots    & \ldots  & \ldots 	           & \ldots & \ldots & \ldots & \ldots 		    & \ldots 	         & \ldots & \ldots & \ldots & \ldots\cr 
  n_v       & 0       & 0      	           & \ldots & 0      & \ldots & 1      		    & 0      	         & \ldots & 0      & \ldots & 0\cr
  \ldots    & \ldots  & \ldots 	           & \ldots & \ldots & \ldots & \ldots 		    & \ldots 	         & \ldots & \ldots & \ldots & \ldots\cr 
  e_1       & 0       & {\bf\color{red} 1} & \ldots & 0      & \ldots & {\bf\color{red} 1}  & 0      	         & \ldots & 0      & \ldots & 0\cr
  \ldots    & \ldots  & \ldots             & \ldots & \ldots & \ldots & \ldots 		    & \ldots 	         & \ldots & \ldots & \ldots & \ldots\cr 
  e         & 0       & 0      	           & \ldots & 1      & \ldots & 0      		    & 0      	         & \ldots & 0      & \ldots & 0\cr
  \ldots    & \ldots  & \ldots 	           & \ldots & \ldots & \ldots & \ldots 		    & \ldots 	         & \ldots & \ldots & \ldots & \ldots\cr 
  n_e       & 0       & 0       	   & \ldots & 0      & \ldots & 0      		    & 0      	         & \ldots & 0      & \ldots & 1\cr  
 }
 } 
\end{equation}
with $T_{e_{\mbox{\tiny E}}}$ and $T_{e_{\mbox{\tiny I}}}$ being transformations acting on vertices associated to, respectively, an external and internal edge of $\mathcal{G}$. Specifically, when acting on $\underline{\mathcal{P}}_{\mathcal{A}_{\mathcal{G}}}$, all its vertices are invariant except the one associated to the edge $e_{\mbox{\tiny E/I}}$,
\begin{equation}\label{TeEI}
    -\mathbf{x}_e+\mathbf{y}_e+\mathbf{x'}_e\:\longrightarrow\:\mathbf{x}_e+\mathbf{y}_e-\mathbf{x'}_e.
\end{equation}
Consequently $T_{e}(\underline{\mathcal{P}}_{\mathcal{A}_{\mathcal{G}}})\:=\:\mathcal{P}'_{\mathcal{A}_{\mathcal{G}}}$. We can construct a differential form as the sum of the canonical forms of all the polytopes obtained from $\underline{\mathcal{P}}_{\mathcal{A}_{\mathcal{G}}}$ acting with all the $T_e$'s and their composition. Denoting the identity as $T_{\circ}$, as well as the $n_e$ edges as $\{e_j\,|\,j=1,\ldots,n_e\}$, we can write
\begin{equation}\label{eq:hPG}
    \omega(\mathcal{Y},\hat{\mathcal{P}}_{\mathcal{G}})\:=\:
    \sum\limits_{\substack{\{e_j\in\mathcal{E}\cap\{\circ\}\} \\ e_1<\ldots<e_{n_e} \\ e_j\neq e_k \mbox{ if } e_j,e_j\neq\circ}}
    \omega(\mathcal{Y},\bigcirc_{j=1}^{n_e}T_{e_j}(\underline{\mathcal{P}}_{\mathcal{A}_{\mathcal{G}}}))
\end{equation}
Notice that putting together the polytopes generated by the operators $T_e$ and their inequivalent compositions, we get the same set of vertices as the cosmological polytope $\mathcal{P}_{\mathcal{G}}$ as well as extra vertices $\mathbf{x}_s$ which turn out to live inside the convex hull $\mathcal{P}_{\mathcal{G}}$. It can be straightforwardly shown that all these polytopes provide a triangulation for $\mathcal{P}_{\mathcal{G}}$ and \eqref{eq:hPG} is precisely the canonical form of $\mathcal{P}_{\mathcal{G}}$ represented as the just mentioned triangulation. We have therefore reconstructed a tree cosmological polytope knowing its scattering facet and the combinatorial automorphisms it is supposed to have ~\cite{Benincasa:2018ssx}~.

The loop cosmological polytopes can be reconstructed from the tree-level ones via a projection which intersects the desired points of type $\mathbf{x}_s$ ~\cite{Benincasa:2018ssx}~.


\section{Symbols from the cosmological polytopes}\label{sec:SymbCP}

A cosmological polytope $\mathcal{P}_{\mathcal{G}}$ associated to a graph $\mathcal{G}$ provides a combinatorial description for the universal integrand $\psi_{\mathcal{G}}$ and encodes the information of the singularity structure of the perturbative wavefunction associated to that graph for {\it any} FRW cosmology. It offers also the possibility of extracting information about the integrated wavefunction, concretely the {\it symbols} introduced in Section \ref{subsec:Symb}.

The classical polylogarithms introduced in Section \eqref{subsec:Symb} are a special class of {\it Aomoto polylogarithms} ~\cite{Goncharov:2009lql, Aomoto:1982ahl}~. It is instructive to have a brief but contained discussion about them and their relation to polytopes.


\subsection{Aomoto polylogarithms and polytopes} \label{subsec:AomPl}

The Aomoto polylogarithms can be generally defined as ~\cite{Goncharov:2009lql, Aomoto:1982ahl}
\begin{equation}\label{eq:AomPols}
    \Lambda(\overline{\Delta},\underline{\Delta})\: :=\:\int_{\overline{\Delta}}\omega_{\underline{\Delta}}
\end{equation}
where $\omega_{\underline{\Delta}}$ is the canonical form in $\mathbb{CP}^{n-1}$ associated to a simplex $\underline{\Delta}$, and $\overline{\Delta}$ is another simplex in $\mathbb{CP}^{n-1}$ such that the pair $(\overline{\Delta},\,\underline{\Delta})$ is admissible, {\it i.e.} they do not share any faces of the same dimensions.

Let $\{\underline{\mathcal{Z}}_j\,|\,j=1,\ldots\,n\}$ and $\{\underline{\mathcal{W}}^{\mbox{\tiny $(j)$}}\,|\,j=1,\ldots\,n\}$ be the sets of vertices and facets of $\underline{\Delta}$ respectively. Let also $\{\overline{\mathcal{Z}}_j\,|\,j=1,\ldots\,n\}$ and $\{\overline{\mathcal{W}}^{\mbox{\tiny $(j)$}}\,|\,j=1,\ldots\,n\}$ be the sets of vertices and facets of $\overline{\Delta}$ respectively. Then the canonical form $\omega_{\underline{\Delta}}$ can be written in terms of the facets $\{\underline{\mathcal{W}}^{\mbox{\tiny $(j)$}}\,|\,j=1,\ldots\,n\}$ or the vertices $\{\underline{\mathcal{Z}}_j\,|\,j=1,\ldots\,n\}$ in the usual way as in \eqref{eq:wWZ}, as well as explicitly in a $d\log$ form
\begin{equation}\label{eq:CFL}
    \omega_{\underline{\Delta}}\:=\:\bigwedge_{j=1}^{n-1}
    d\log{\frac{(\mathcal{Y\cdot\mathcal{W}^{\mbox{\tiny $(j)$}}})}{(\mathcal{Y\cdot\mathcal{W}^{\mbox{\tiny $(n)$}}})}}
\end{equation}
$\mathcal{Y}$ being a generic point in $\mathbb{CP}^{n-1}$. What is the symbol $\mathcal{S}(\Lambda)$ for an Aomoto polylogarithm?

Let us consider a vertex $\underline{\mathcal{Z}}_1$ of $\underline{\Delta}$, and let it vary. Then ~\cite{Arkani-Hamed:2017ahv}
\begin{equation}\label{eq:dzO}
    d_{\underline{\mathcal{Z}}_1}\Lambda(\overline{\Delta},\underline{\Delta})\:=\:(n-1)
        \int_{\overline{\Delta}}d_{\mathcal{Y}}\,
        \frac{\langle\delta\underline{\mathcal{Z}}_1\mathcal{Y}d^{n-2}\mathcal{Y}\rangle\langle\underline{\mathcal{Z}}_1\ldots\underline{\mathcal{Z}}_n\rangle}{\langle\underline{\mathcal{Z}}_3\ldots\underline{\mathcal{Z}}_1\rangle\cdots\langle\mathcal{Y}\underline{\mathcal{Z}}_1\cdots\underline{\mathcal{Z}}_{n-1}\rangle}.
\end{equation}
Being a total derivative, the integral localises at the boundaries $\overline{\mathcal{W}}^{\mbox{\tiny $(j)$}}$ of $\overline{\Delta}$ and it reduces to a $d\log{(\underline{\mathcal{Z}}_1\cdot\overline{\mathcal{W}}^{\mbox{\tiny $(j)$}})}$ times a similar integral of one dimension-less
\begin{equation}\label{eq:dz0}
    d_{\underline{\mathcal{Z}}_1}\Lambda(\overline{\Delta},\underline{\Delta})\:=\:(n-1)
    \sum_{j=1}^n
        \left[
            d_{\underline{\mathcal{Z}}_1}\log{(\underline{\mathcal{Z}}_1\cdot\overline{\mathcal{W}}^{\mbox{\tiny $(j)$}})}\,\otimes
            \Lambda(\overline{\Delta}^{\mbox{\tiny $(j)$}},\underline{\Delta}^{\mbox{\tiny $(j)$}})
        \right]
\end{equation}
with $\underline{\Delta}^{\mbox{\tiny $(j)$}}$ being a codimension-$1$ simplex defined as convex hull of vertices obtained from the projection of $\{\underline{\mathcal{Z}}_j\,|\,j=2,\ldots,n\}$ through $\underline{\mathcal{Z}}_1$ onto a chosen hyperplane $\overline{\mathcal{W}}^{\mbox{\tiny $(j)$}}$. Importantly, the variation with respect to $\underline{\mathcal{Z}_1}$ corresponds to a projection through $\underline{\mathcal{Z}}_1$. This procedure can be iterated by varying a vertex in $\underline{\Delta}^{\mbox{\tiny $(j)$}}$ until the highest codimension boundaries are reached, obtaining the following general form for the symbol $\mathcal{S}(\Lambda)$
\begin{equation}\label{eq:SLDD}
    \begin{split}
        \mathcal{S}(\Lambda)\:=\:&\sum_{\rho,\sigma\in S_n}\,\mbox{sgn}(\rho)\mbox{sgn}(\sigma)
            \langle\overline{\mathcal{Z}}_{\rho(2)}\underline{\mathcal{Z}}_{\sigma(2)}\cdots\underline{\mathcal{Z}}_{\sigma(n)}\rangle\otimes\\
        &\otimes\langle\overline{\mathcal{Z}}_{\rho(2)}\overline{\mathcal{Z}}_{\rho(3)}\underline{\mathcal{Z}}_{\sigma(3)}\cdots\underline{\mathcal{Z}}_{\sigma(n)}\rangle\otimes\cdots\otimes
        \langle\overline{\mathcal{Z}}_{\rho(2)}\cdots\overline{\mathcal{Z}}_{\rho(n)}\underline{\mathcal{Z}}_{\sigma(n)}\rangle.
    \end{split}
\end{equation}

Notice that \eqref{eq:dz0}, and consequently the symbol \eqref{eq:SLDD}, captures the singularity structure of $\Lambda(\overline{\Delta},\underline{\Delta})$. In general, it always depends on one side on the structure of the integrand, which is encoded in the simplex $\underline{\Delta}$, and on the other on the integration path, encoded by the simplex $\overline{\Delta}$. A branch cut is then captured by the intersection of the boundaries of $\underline{\Delta}$ with the contour and it is given by the residue of the integrand with respect to the subset boundaries $\underline{\mathcal{W}}_{\overline{\Delta}}\::=\:\{\underline{\mathcal{W}}^{\mbox{\tiny $(j)$}}\,|\,\overline{\Delta}\cap\underline{\mathcal{W}}^{\mbox{\tiny $(j)$}}\,\neq\,\varnothing\}$:
\begin{equation}\label{eq:AomPolsDisc}
    \mbox{Disc}\Lambda(\overline{\Delta},\underline{\Delta})\:=\:\sum_{\underline{\mathcal{W}}^{\mbox{\tiny $(j)$}}\in\underline{\mathcal{W}}_{\overline{\Delta}}}\int_{\overline{\Delta}\cap\underline{\mathcal{W}}^{\mbox{\tiny $(j)$}}}
    \mbox{Res}_{\underline{\mathcal{W}}^{\mbox{\tiny $(j)$}}}\omega_{\underline{\Delta}}
\end{equation}
with $\overline{\Delta}\cap\underline{\mathcal{W}}^{\mbox{\tiny $(j)$}}\,\subset\,\mathbb{CP}^{n-2}$ being a codimension-$1$ polytope. 

It turns out that given an hyperplane $\underline{\mathcal{W}}^{\mbox{\tiny $(j)$}}\subset\underline{\mathcal{W}}_{\overline{\Delta}}$, the sum of the projections through each of the vertices $\{\overline{\mathcal{Z}}_j\,|\,j=1,\ldots,n\}$ of $\overline{\Delta}$ all belonging to the same half-space determined by $\underline{\mathcal{W}}^{\mbox{\tiny $(j)$}}$ provides the discontinuity associated to $\underline{\mathcal{W}}^{\mbox{\tiny $(j)$}}$ ~\cite{Arkani-Hamed:2017ahv}~. A projection through a vertex $\overline{\mathcal{Z}}_j$ is performed if $\overline{\mathcal{Z}}_j$ is deformed to pass the hyperplane $\mathcal{W}^{\mbox{\tiny $(j)$}}$ in such a way that the intersection $\overline{\Delta}\cap\underline{\mathcal{W}}^{\mbox{\tiny $(j)$}}$ is non-empty, identifying a possible branch point at $\overline{\mathcal{Z}}_j\cdot\underline{\mathcal{W}}^{\mbox{\tiny $(j)$}}=0$. Hence, the first entries of the symbol $\mathcal{S}(\Lambda)$ are given by all the possible pairs with such a feature. The procedure is then iterated on the codimension-$1$ resulting polytopes and, later, on all higher codimension ones, arriving again at \eqref{eq:SLDD}.

\paragraph{Generalised Aomoto polylogarithms.} The definition of Aomoto polylogarithm \eqref{eq:AomPols} can be generalised by considering a pair of polytopes $(\overline{\mathcal{P}},\,\underline{\mathcal{P}})$ rather than just simplices:
\begin{equation}\label{eq:AomPols2}
    \Lambda(\overline{\mathcal{P}},\underline{\mathcal{P}})\: :=\:\int_{\overline{\mathcal{P}}}\Omega_{\underline{\mathcal{P}}},
\end{equation}
where now $\Omega_{\underline{\mathcal{P}}}$ is the canonical form of $\underline{\mathcal{P}}$. Then, the symbol of \eqref{eq:AomPols2} can be extracted again as consecutive projections through vertices, with the novel feature that a very same face can be represented via different subsets of vertices which are incident to it ~\cite{Arkani-Hamed:2017ahv}~.


\subsection{Symbols from the cosmological polytopes}\label{subsec:SymbCP1}

Let us go back to the cosmological polytopes. The aim is to have a systematic procedure to extract information about the integrated wavefunction from its geometry:
\begin{equation}\label{eq:PsiGintf}
    \tilde{\psi}_{\mathcal{G}}\:=\:
        \int_{\overline{\mathcal{P}}_{\mathcal{G}}}\omega(\mathcal{Y},\mathcal{P}_{\mathcal{G}})
\end{equation}
where, taking the local coordinates $\mathcal{Y}\,=\,(x_{s_1},\ldots,x_{s_{n_s}},y_{e_1},\ldots,y_{e_{n_e}})$, the region of integration $\overline{\mathcal{P}}_{\mathcal{G}}$ is given by $\{x_s\in[X_s,+\infty]\,|\,\forall\,s\in\mathcal{V}\}$ and $\{y_e=Y_e\,|\,\forall\,e\in\mathcal{E}\}$.

A first observation is that given a cosmological polytope $\mathcal{P}_{\mathcal{G}}$ with canonical form $\omega(\mathcal{Y},\,\mathcal{P}_{\mathcal{G}})$, then the latter has multiple residues equal to 
$\pm\prod_{e\in\mathcal{E}}(2y_e)^{-1}$ and the form $f_{\mathcal{G}}:=\prod_{e\in\mathcal{E}}(2y_e)\omega(\mathcal{Y},\mathcal{P}_{\mathcal{G}})$ integrates to polylogarithms. Importantly, we can consider the $f_{\mathcal{G}}$ rather than the canonical form as the integration we are concerned with is over the weights $\{x_s\,|\,s\in\mathcal{V}\}$ of the sites of graph, {\it i.e.} over the external energies. Furthermore, as the integration range of interest is $x_s\in[X_s,\,+\infty]$ for each site $s$, it is convenient to define hyperplanes $\{\mathcal{W}_e\,|\,\mathcal{Y}\cdot\mathcal{W}_e=0\;\forall\,e\in\mathcal{E}\}$ and $\{\mathcal{W}_s\,|\,\mathcal{Y}\cdot\mathcal{W}_e\ge0,\;s\in\mathcal{V}\}$ such that the former localises the $y_e$ integration on $\{y_e\,=\,Y_e\,|\,\forall\,e\in\mathcal{E}\}$, while the latter constitute the boundaries of the region of integration:
\begin{equation}\label{eq:Wint}
    \left(\mathcal{W}_e\right)_I\: :=\:y_{e_1}\left(\mathbf{\tilde{y}}_e\right)_I-Y_e\left(\mathbf{\tilde{y}}_{e_1}\right)_I,
    \qquad
    \left(\mathcal{W}_s\right)_I\: :=\:y_{e_1}\left(\mathbf{\tilde{x}}_s\right)_I-X_s\left(\mathbf{\tilde{y}}_{e_1}\right)_I
\end{equation}
$\mathcal{W}_{e_1}:=\mathbf{\tilde{y}}_{e_1}$ being the hyperplane associated to the edge $e_1$ -- its choice is completely arbitrary.

In order to extract the symbols, we need to look for subsets of the type $\mathcal{V}':=\{\mathcal{Z}_{j_1},\ldots\,\mathcal{Z}_{j_{n_s+n_e-1}}\}\,\subseteq\,\mathcal{V}$ of vertices of $\mathcal{P}_{\mathcal{G}}$ such that $\mathcal{Z}_{j_r}$ is connected to $\mathcal{Z}_{j_{r+1}}$ via an edge of $\mathcal{P}_{\mathcal{G}}^{\mbox{\tiny $\{j_1,\ldots,j_{r-1}\}$}}$, {\it i.e.} the projection of $\mathcal{P}_{\mathcal{G}}$ through $\{\mathcal{Z}_{j_1},\ldots,\mathcal{Z}_{j_{r-1}}\}$.

Any allowed sequence of vertices $\{\mathcal{Z}_{j_1},\ldots\,\mathcal{Z}_{j_{n_s+n_e-1}}\}$ projects $\mathcal{P}_{\mathcal{G}}$ onto a polytope in $\mathbb{P}^1$, {\it i.e.} a segment. This means that all the vertices $\{\mathcal{Z}_{j_1},\ldots\,\mathcal{Z}_{j_{n_s+n_e-1}}\}$ have been projected down to the same one boundary, while any vertex $\mathcal{Z}_{j_a}$ which does not belong to such a subset are such that all $\langle\mathcal{Z}_{j_a}\mathcal{Z}_{j_1}\ldots\mathcal{Z}_{j_{n_s+n_e-1}}\rangle$ have the same sign. Let us indicate it as $\sigma[\mathcal{Z}_{j_1}\ldots\mathcal{Z}_{j_{n_s+n_e-1}}]$. This specific feature is guaranteed by the choice of the set of vertices $\{\mathcal{Z}_{j_1},\ldots\,\mathcal{Z}_{j_{n_s+n_e-1}}\}$ as connected via an edge of the projected polytope.

Similarly for the hyperplanes, for which one can pick a special collection of $n_s+n_e-1$ hyperplanes $\{\mathcal{W}_{a_1},\ldots,\mathcal{W}_{a_{n_s+n_e-1}}\}$, such that those hyperplanes which do not belong to this set have the same sign $\sigma[\mathcal{W}_{a_1}\ldots\mathcal{W}_{a_{n_s+n_e-1}}]$. Then, the symbol of $f_{\mathcal{G}}$ can be written as
\begin{equation}\label{eq:Simbfg}
    \begin{split}
        \mathcal{S}(f_{\mathcal{G}})=&\hspace{-.5cm}\sum_{\substack{\{a_1,\ldots,a_{n_s-1}\} \\ \{b_1,\ldots,b_{n_s+n_e-1}\}}}
            \hspace{-.5cm}
            \sigma[\mathcal{W}_{e_1}\ldots\mathcal{W}_{e_{n_e}}\mathcal{W}_{s_{a_1}}\ldots\mathcal{W}_{s_{a_{n_s-1}}}]
            \sigma[\mathcal{Z}_{b_1}\ldots\mathcal{Z}_{b_{n_s+n_e-1}}]\times\\
            &\hspace{-1cm}\times[\mathcal{W}_{e_1}\cdots\mathcal{W}_{e_{n_e}};\mathcal{Z}_{b_1}\cdots\mathcal{Z}_{n_{n_e}}]
            \otimes[\mathcal{W}_{e_1}\cdots\mathcal{W}_{e_{n_s}}\mathcal{W}_{s_{a_1}};\mathcal{Z}_{b_1}\cdots\mathcal{Z}_{b_{n_e}}\mathcal{Z}_{b_{n_e+1}}]\otimes\\
            &\hspace{-1cm}\cdots\otimes[\mathcal{W}_{e_1}\cdots\mathcal{W}{e_{n_e}}\mathcal{W}_{s_{a_1}}\cdots\mathcal{W}_{s_{a_{n_s-1}}};
            \mathcal{Z}_{b_1}\cdots\mathcal{Z}_{b_{n_e}}\mathcal{Z}_{b_{n_s+n_e-1}}],
    \end{split}
\end{equation}
where
\begin{equation}\label{eq:sqbkr}
    [\mathcal{W}_1\cdots\mathcal{W}_{k};\mathcal{Z}_{1}\cdots\mathcal{Z}_{k}]\: :=\:\mbox{det}_{a,b}\{\mathcal{W}_a\cdot\mathcal{Z}_b\},\qquad a,\,b\,=\,1,\ldots,k,
\end{equation}
and $\{\mathcal{Z}_{b_1}\ldots\mathcal{Z}_{b_{n_s+n_e-1}}\}$ identify a path of connected vertices under projection.

The symbol entry is therefore given by tensoring together the upper-left adjusted minors of the matrices of the type \eqref{eq:sqbkr} appearing in \eqref{eq:Simbfg}. The geometry of the cosmological polytope completely fixes the symbol associated to its canonical form. As a final remark, notice that the $\mathcal{W}$'s are defined in \eqref{eq:Wint} via a reference edge $e_1$ chosen arbitrarily. As there is nothing special about such a particular choice, the symbol does not depend on it.


\section{Conclusion}\label{sec:Concl}

Recent years saw a great deal of progress in the understanding of cosmological observables and their computation. In particular, a new perspective has been introduced, which imports a two-fold lesson learned in the context of flat-space scattering amplitudes: also cosmological observables on one side can be understood directly as function of the external, physical, data only and, on the other, they can be encoded in novel mathematical objects having their own first principle definition. 
Nevertheless, it is fair to admit that we started just to scratch the surface, and there are still a number of basic questions which need to be addressed. Let us examine them closely.



\paragraph{The full-fledge perturbative wavefunction.} As emphasised along the whole review, all the progress achieved has been made for individual Feynman graphs, {\it i.e.} we have been learning how to understand and compute individual contributions in a concrete way of organising perturbation theory. However, the main lesson coming from the on-shell formulation of scattering amplitudes is precisely the fact that in order to have a full-fledge boundary description of our observable we need to change the way of organising perturbation theory. Recursion relations, which in the case of scattering amplitudes have been the starting point, have been exploited in a very limited amount and, again, in a graph-by-graph fashion. There are two good reasons for that. The first one is that most of the progress has been carried out for scalars: also in the case of scattering amplitudes, the on-shell description of scattering processes of scalars is not that different from the Feynman diagrammatics. Secondly, for spinning states, the helicity-spinor formalism for theories with a $SO(3)$ (sub)group introduced in ~\cite{Maldacena:2011nz}~, and extensively used in ~\cite{Baumann:2020dch, Pajer:2020wnj, Baumann:2021fxj}~, is currently formulated in a precise gauge, the axial gauge. This fact suggests that we still need to find the most suitable variables for dealing with any type of state: most of the progress in scattering amplitudes was precisely due to the fact that parametrising the kinematic space in terms of spinors made manifest the behaviour of the amplitudes under the Lorentz little group. Nevertheless, in the subsequent developments the momentum twistors ~\cite{Hodges:2009hk}~ turned out to play a crucial role in discovering the combinatorial structure underlying amplitudes in $\mathcal{N}=4$ SYM theory and, later on, to be more generically a convenient parametrisation for the kinematic space for general theories. Hence, we should answer the following questions: what is the most suitable set of variables in the cosmological context? Which feature or symmetry transformation do we need to make manifest? The problem of working with an unsuitable parametrisation of the kinematic space is that cancellations and other features might not be easy to read off. Going back to the scalar interactions, also in this case the achieved understanding is not quite satisfactory. Even the combinatorial formulation of the wavefunction in terms of cosmological polytopes is at the individual graph level, and it does not seem straightforward to define a single geometrical object encoding all the contributions from the relevant graphs. In the case of scattering amplitudes, all the positive geometries describing scalar interactions encode all the different channels. Hence, given a wavefunction thought of as a sum of graphs, the scattering facets of the cosmological polytopes associated to these graphs should organise into a single object combinatorially equivalent to the positive geometries for scalar scattering amplitudes. Understanding this issue does not have a simple academic value, rather it can provide the route to finally overcome the individual graph treatment that has been pursued so far, and also will provide unified treatment for both flat-space and cosmological processes.

\paragraph{Consistency conditions on cosmological processes.} One of the most intriguing outcome of the on-shell formulation of scattering amplitudes has been that a great deal of results emerge from consistency with principles such as unitarity and locality ~\cite{Benincasa:2007xk, Benincasa:2011pg, McGady:2013sga}~. In particular, such requirements put strong constraints on the consistency of the interactions. It would be desirable to have such a simple and general argument also for the wavefunction of the universe. 

\paragraph{Function space for the wavefunction.} Most of the results obtained so far deal either with the tree-level wavefunction or with the all-loop wavefunction universal integrand. In the case of conformally-coupled scalars and light states, the tree-level wavefunctions are expected to be given in terms of polylogarithms \cite{Arkani-Hamed:2017fdk, Hillman:2019wgh}, while no general statement can be made for arbitrary states. What about loops? In this case there have been very few insights. Taking the perspective of the wavefunction universal integrand, one has to deal with two integrations: the one over the energy of the external states with a suitable measure, which implements the specificity of the desired cosmology, and the loop integrations. The first type of integration is expected to return a transcendental function which can be expressed in terms of polylogarithms, while no statement can be currently made on the loop integration. So, it is possible to generally ask: what is the function space for the integrated wavefunction? A further lesson that we can take from the scattering amplitude literature is the fact that our observable can be expressed in terms of a basis of scalar integrals whose coefficients are rational functions of the external momenta which can be fixed from the factorisation properties. At one loop, in the scattering amplitude context there is a specific procedure, the Passarino-Veltman reduction ~\cite{Passarino:1978jh, vanNeerven:1983vr, Bern:1993kr}~, that provides such a basis and the coefficients are fixed by unitarity in terms of products of tree-level amplitudes. More generally, one needs to identify a set of master integrals and fix them via unitarity and/or with the aids of further constraints such as the Steinmann relations ~\cite{Bartels:2008ce, Caron-Huot:2016owq, Caron-Huot:2018dsv, Caron-Huot:2019vjl, Caron-Huot:2019bsq, Caron-Huot:2020bkp}~. The question about the functions space for the loop wavefunction can be addressed by extending such an approach: we are endowed with factorisation theorems, cosmological cutting rules as well as Steinmann-like relations (which are proven to be valid as long as the states into consideration have a flat-space counterpart). In this context it would be also interesting to ask whether such Steinmann-like relations extend to wavefunctions involving states with no flat-space counterpart, such as the partially-massless ones.

\paragraph{The IR structure.} Understanding the analytic structure of cosmological observables at loops is of fundamental importance to understand potential issues with the consistency of perturbation theory due to the presence of IR divergences ~\cite{Ford:1984hs, Antoniadis:1985pj, Tsamis:1994ca, Tsamis:1996qm, Tsamis:1997za, Polyakov:2007mm, Polyakov:2009nq, Senatore:2009cf, Giddings:2010nc, Giddings:2010ui, Burgess:2010dd, Marolf:2010nz, Marolf:2010zp, Rajaraman:2010xd, Krotov:2010ma, Giddings:2011zd, Marolf:2011sh, Giddings:2011ze, Senatore:2012nq, Pimentel:2012tw, Senatore:2012ya, Polyakov:2012uc, Beneke:2012kn, Akhmedov:2013vka, Anninos:2014lwa, Akhmedov:2017ooy, Hu:2018nxy, Akhmedov:2019cfd, Gorbenko:2019rza, Baumgart:2019clc, Mirbabayi:2019qtx, Cohen:2020php, Mirbabayi:2020vyt, Baumgart:2020oby, Cohen:2021fzf}~. The flat-space scattering amplitudes for massless states are also plagued with IR divergences. However, their structure is constrained by the fact that they need to cancel when computing the cross-section. Their origin lies in the fact that we typically works with the {\it wrong} asymptotic states, as they are given by plane waves which are non-normalisable modes. In the context of gauge theories, one can define the correct asymptotic states via a dressing with a cloud of soft photons, the so-called Faddeev-Kulish coherent states ~\cite{Kulish:1970ut}~. However, given that the form of the IR divergences is constrained by physical arguments, we are got used to deal with them, even if attempt to formulate a well-defined S-matrix have been recently proposed ~\cite{Hannesdottir:2019rqq, Hannesdottir:2019opa}~. Interestingly, the combinatorial-geometric approach to scattering amplitudes naturally provides a new way of understanding the structure of IR divergences ~\cite{Arkani-Hamed:2022cqe}~ as well as an IR-finite observable ~\cite{Arkani-Hamed:2021iya}~. Going back to the cosmological set-up, there are natural questions to answer: what is the structure of the IR divergences? Is it possible define an observable which is IR finite? 

\paragraph{The imprint of causality.} Steinmann-like relations have been proven for the wavefunction universal integrand ~\cite{Benincasa:2020aoj} as well as their higher-codimension extensions ~\cite{Benincasa:2021qcb}~. While in flat-space they are tied to causality, this is not immediately obvious from their derivation, as they have been proven directly as a statement about partially overlapping channels. It would be interesting to understand whether also the Steinmann-like relation for the wavefunction are a consequence of causality and, more generally, what is its imprint in the analytic structure of the wavefunction. The understanding of the constraints imposed by causality, as well as by unitarity, are of fundamental importance for constraining the coefficients of high-dimension operators in the EFT of inflation. Despite the study of positive bounds dictated by unitarity has been started in the context amplitudes without Lorentz boosts have been initiated ~\cite{Grall:2020tqc, Grall:2021xxm}~, no much is known about such bounds on the wavefunction. It would be interesting to explore whether a structure such as the EFT-hedron and the associated infinitely many constraints on the coefficients of high dimension operators ~\cite{Arkani-Hamed:2020blm}~ emerge in the cosmological context as well. 

\paragraph{What are the correct observables?} One important lesson that the on-shell formulation of flat-space scattering taught is that the language of fields is {\it a} way of talking about the physics at accessible high energies, and that it is in principle possible to prescind on it and have a description directly in terms of observables. An important difference between scattering amplitude and any cosmological observables we have been discussed is that the latter seems to be tied to the notion of fields: the wavefunction of the universe encode the probability distribution of a field configuration, the correlation functions are for operators made out of fields, and the mean square distribution is still in field space. This fact also reflects into the field-redefinition dependence of the cosmological observables, while the scattering amplitudes are field-redefinition invariance. If on one side we can ask whether there exist a way of canonically define these observables, without making any reference to a pre-existent Lagrangian, on the other fields are just a (redundant) way of packaging the degrees of freedom. Also, neither the wavefunction nor the correlators seem to be able to capture all the aspects of the physics in an expanding universe, {\it e.g.} the branched diffusion process in the evolution of massless states, which generates a plethora of late-time configurations. In order to understand this phenomenon and its relation to (the lack of) cluster composition, it was necessary to introduce an intrinsically non-local observable, the mean-square displacement distribution. Is it possible to define a desirable field redefinition invariant and sufficiently non-local observable and, maybe, that it is also IR finite?


\section*{Acknowledgements}

I would like to thank Nima Arkani-Hamed, whom I am in debt with for bringing me into this subject and for extensive and continuous discussions on all the topics related to this review, as well as Dionysios Anninos for introducing me to the connection between expanding universes and spin glasses and for constant interest and support. I would also like to thank everyone who contributed and are contributing to interfacing the two subjects of cosmology and scattering amplitudes, and especially whom I had the chance to benefit from with discussions, correspondence and collaborations. In particular: Soner Albayrak, Dionysios Anninos, Nima Arkani-Hamed, Daniel Baumann, Jacob Bourjaily, Freddy Cachazo, David Daamgard, Alice Di Tucci, Carlos Duaso Pueyo, Livia Ferro, Humberto G{\'o}mez, Tanguy Grall, Aaron Hillman, Lukas K{\"u}hne, Arthur Lipstein, Matteo Maglio, Andrew McLeod, Leonid Monin, Enrico Pajer, Matteo Parisi, Guilherme Pimentel, Gizem Seng{\"o}r, David Stefanyszyn, Jakub Supel, William Torres Bobadilla, Francisco Vaz{\~a}o, Cristian Vergu.

Finally, I would like to thank the developers of {\tt Polymake} ~\cite{polymake:2000, polymake:2017}~, {\tt TOPCOM} ~\cite{Rambau:TOPCOM-ICMS:2002}~, {\tt SageMath} ~\cite{sagemath}~, {\tt Maxima} ~\cite{maxima} and {\tt Tikz} ~\cite{tantau:2013a}~. 

This research received funding from the European Research Council (ERC) under the European Union’s Horizon 2020 research and innovation programme (grant agreement No 725110), {\it Novel structures in scattering amplitudes}.

%

\bibliographystyle{ws-ijmpa}
\bibliography{cprefs}

\end{document}